\documentclass[a4paper,11pt]{article}
\usepackage[utf8]{inputenc}
\usepackage[english]{babel}
\usepackage[toc, page]{appendix} 
\usepackage[margin=1in, includefoot]{geometry}

\usepackage{amsmath, amsfonts, amsthm, amssymb, mathtools}
\usepackage{graphicx}
\usepackage{float}
\usepackage{array}
\newcommand{\PreserveBackslash}[1]{\let\temp=\\#1\let\\=\temp}
\newcolumntype{C}[1]{>{\PreserveBackslash\centering}p{#1}}
\newcolumntype{R}[1]{>{\PreserveBackslash\raggedleft}p{#1}}
\newcolumntype{L}[1]{>{\PreserveBackslash\raggedright}p{#1}}
\usepackage{enumitem}
\usepackage{booktabs}
\usepackage{multirow}
\usepackage{arydshln}
\usepackage{doi}
\usepackage{xcolor}
\usepackage{makecell}
\usepackage{adjustbox}
\usepackage[nameinlink,capitalize,noabbrev]{cleveref}
\usepackage{natbib}
\usepackage{setspace}
\usepackage{algorithm}
\usepackage[noend]{algpseudocode}
\usepackage{tabularx}
\usepackage{tabu}
\usepackage{subcaption}

\AtBeginEnvironment{appendices}{\crefalias{subsection}{appendix}}

\makeatletter
\def\@endtheorem{\endtrivlist}
\makeatother

\makeatletter
\newcommand{\multiline}[1]{%
	\begin{tabularx}{\dimexpr\linewidth-\ALG@thistlm}[!hp]{@{}X@{}}
		#1
	\end{tabularx}
}
\makeatother
\hypersetup{
	colorlinks=true,
	allcolors=blue
}

\newtheorem{theorem}{Theorem}
\newenvironment{theorem*}[2][]
{\renewcommand\thetheorem{#2}\begin{theorem}[#1]}
	{\end{theorem}}
\newtheorem{proposition}{Proposition}
\newtheorem{definition}{Definition}

\newtheorem{lemma}{Lemma}[section]
\newtheorem{remark}{Remark}
\newtheorem{assumption}{Assumption}
\newenvironment{assumption*}[2][]
{\renewcommand\theassumption{#2}\begin{assumption}[#1]}
	{\end{assumption}}

\title{Bootstraps for Dynamic Panel Threshold Models}
\author{Woosik Gong\thanks{This research was supported by the BK21 FOUR (Fostering Outstanding Universities for Research) funded by the Ministry of Education(MOE, Korea) and National Research Foundation of Korea(NRF). The author is also grateful to the student travel grant award by the IAAE 2022 conference.} \\ \small University of Wisconsin-Madison  \\ \texttt{wgong28@wisc.edu}
	\and
	Myung Hwan Seo\thanks{This work was supported by the Ministry of Education of the Republic of Korea and the National Research Foundation of Korea (NRF-2018S1A5A2A01033487) and Korea Bureau of Economic Research in the Institute of Economic Research of Seoul National University.} \\ \small Seoul National University \\ \texttt{myunghseo@snu.ac.kr}}

\begin{document}
	\onehalfspacing
	
	\sloppy
	\maketitle
	\begin{abstract}
		
		This paper develops valid bootstrap inference methods for the dynamic short panel threshold regression. We show that the standard nonparametric bootstrap is inconsistent for the first-differenced generalized method of moments (GMM) estimator. The inconsistency arises from an $n^{1/4}$-consistent non-normal asymptotic distribution of the threshold estimator when the true parameter lies in the continuity region of the parameter space, which stems from the rank deficiency of the approximate Jacobian of the sample moment conditions on the continuity region. To address this, we propose a grid bootstrap to construct confidence intervals for the threshold and a residual bootstrap to construct confidence intervals for the coefficients. They are shown to be valid regardless of the model's continuity. Moreover, we establish a uniform validity for the grid bootstrap. A set of Monte Carlo experiments compares the proposed bootstraps with the standard nonparametric bootstrap. An empirical application to a firm investment model illustrates our methods.

		KEYWORDS: Dynamic Panel Threshold; Kink; Bootstrap; Endogeneity; Identification; Rank Deficiency; Uniformity.
		
		JEL: C12, C23, C24
	\end{abstract}
	
	\section{Introduction}
	\label{sec:intro}
	
	Threshold regression models are widely used in empirical research, and their usefulness has grown substantially with extenstions to the panel data settings. Estimation and inference methods for the threshold model in non-dynamic panels were developed by \cite{hansen_threshold_1999} and \cite{wang2015fixed}. 
	Dynamic panel threshold models were considered by \cite{seo_dynamic_2016}, which proposes the generalized method of moments (GMM) estimation by generalizing the \cite{arellano_tests_1991} dynamic panel estimator.  More recently, a latent group structure in the parameters of the panel threshold model was investigated by \cite{miao2020panel}.
	
	Applications of the panel threshold models cover numerous topics in economics. 
	The effect of debt on economic growth is a well-known example that has been analyzed using panel threshold models, e.g., \cite{adam_fiscal_2005}, \cite{cecchetti_real_2011}, and \cite{chudik_is_2017}. 
	Another example is the threshold effect of inflation on economic growth such as the works by \cite{khan_threshold_2001}, \cite{rousseau_inflation_2002}, \cite{bick_2010}, and \cite{kremer_inflation_2013}.
	The benefit of foreign direct investment to productivity growth that depends on the regime determined by absorptive capacity is studied by \cite{girma_absorptive_2005} using firm-level panel data.

    In empirical applications of threshold regression models, inference is usually performed after imposing an assumption about whether the model is continuous or not. Continuous threshold models that have kinks at the tipping points have received active research attention, e.g., \cite{Hansen_2017, seo_estimation_2019} and \cite{yang_panel_2020}.
	In the literature, kink threshold models are analyzed for estimators that impose the continuity restriction as in \cite{chan_tsay_1998}, \cite{Hansen_2017}, and \cite{zhang_2017}.
	On the other hand, unrestricted estimators are commonly used for discontinuous threshold models as in \cite{hansen_sample_2000}.
	However, \cite{hidalgo_robust_2019} showed that the unrestricted least squares estimator possesses a different asymptotic property in the absence of discontinuity. Specifically, while the unrestricted model is not misspecified under continuity, failing to impose the restriction results in incorrect inference without proper care. 
	
	In the empirical literature, there has been mixed use of kink/discontinuous threshold models without much consideration of a possible specification error. Among the empirical examples referred to previously, \cite{khan_threshold_2001} use a continuous threshold model and impose continuity on their estimation procedure. 
	They claim that the continuous model is desirable to prevent small changes in inflation rate from yielding different impacts around the threshold level. On the other hand, \cite{bick_2010} claims that the discontinuous threshold model is more appropriate for the same research question since overlooking a regime-dependent intercept can result in omitted variable bias. However, both of them do not provide econometric evidence that supports their choice of models.
	
	For the dynamic panel threshold model, asymptotic normality of the GMM estimator is derived by \cite{seo_dynamic_2016} under the fixed $T$ scheme. However, the asymptotic normality is valid only for the discontinuous models since it requires a full rank condition on the Jacobian of the population moment, which is violated in continuous models.	Although the continuity-restricted estimator described in \cite{seo_estimation_2019} is asymptotically normal, it may be problematic since empirical researchers often do not agree about whether their threshold models should have a kink or a jump at the threshold as in \cite{khan_threshold_2001} and \cite{bick_2010}.
	Therefore, we focus on the unrestricted GMM estimator and bootstrap inference methods which do not require any pretest on continuity or prior knowledge about continuity of true models.
	
	We first show that when the true model is continuous, the asymptotic normality of the unrestricted GMM estimator breaks down and the convergence rate of the threshold estimator becomes  $n^{1/4}$-rate, which is slower than the standard $\sqrt{n}$-rate. 
	Moreover, the standard nonparametric bootstrap is inconsistent in this case because the Jacobian from the bootstrap distribution does not degenerate fast enough due to the slow convergence rate of the threshold estimator. 
	
	We propose two different bootstrap methods to obtain confidence intervals for the parameters that are consistent regardless of whether the true model is continuous or not. One is for the threshold location, and the other is for the coefficients. The two bootstrap methods achieve consistency irrespective of the continuity of the model by adaptively setting the recentering parameter at the bootstrap for GMM introduced by \cite{hall_bootstrap_1996}. This means that our bootstrap moment function achieves zero not at the sample estimator but at the parameter values that we propose.
	In the bootstrap for the threshold location, we employ a grid bootstrap to fix the recentering parameter. The grid bootstrap was originally proposed by \cite{hansen_grid_1999} for inference on an autoregressive parameter and applies test inversion. In case of the bootstrap for the coefficients, the recentering parameter is set to adjust the unrestricted estimator by a data driven criterion on the model's continuity. 
	We also introduce a bootstrap test of model continuity.

 Furthermore, we establish the uniform validity of the grid bootstrap for the unknown continuity (or discontinuity) of the threshold model. The importance of uniform validity is well recognized in the literature, notably in the works of \cite{mikusheva_2007}, \cite{andrewsguggenberger2009}, and \cite{romanoshaikh2012}, among others, who have studied the uniformity of resampling procedures. In particular, \cite{mikusheva_2007} showed the uniform validity of the grid bootstrap for linear autoregressive models. Our work extends the advantage of the grid bootstrap to a broader class of nonstandard inference problems characterized by Jacobian degeneracy.
	
	A set of Monte Carlo simulations demonstrate that the grid bootstrap performs favorably for inference on the threshold location, not only when the model is continuous but also when it includes a jump for various jump sizes.
	However, inference on the coefficients turns out to be more challenging. Our residual bootstrap confidence intervals for the coefficients, based on the lower and upper quantiles of bootstrap distributions, tend to exhibit undercoverage, even though they generally provide higher coverage rates than the standard nonparametric bootstrap.
	
 We apply our inference methods to the dynamic firm investment model, whose static version was studied by \cite{fazzari_financing_1988} and \cite{hansen_threshold_1999} among others. It takes financial constraints into account via the threshold effect to determine a firm's investment decision.
	
	In the literature, \cite{dovonon_testing_2013} and \cite{DovononHall2018} also deal with the degeneracy of the Jacobian in the context of the common conditional heteroskedasticity testing problem. In addition, a bootstrap based test for the common conditional heteroskedasticity feature was proposed by \cite{dovonon_bootstrapping_2017}.
	However, their works do not deal with a discontinuous criterion function.
    Moreover, \cite{dovonon_testing_2013} and \cite{dovonon_bootstrapping_2017} study testing null hypothesis that always induces the degeneracy of the first-order derivative,
    while \cite{DovononHall2018} study the asymptotic distribution of an estimator when the degeneracy holds within the model. Therefore, they do not have to address the uncertainty associated with the potential degeneracy of the Jacobian. 
	
	Meanwhile, there is also a substantial body of literature on singularity-robust inference such as \cite{andrews_estimation_2012, andrews_gmm_2014} and \cite{han_estimation_2019}, among many others. 
	They are motivated by weak or non-identification problems, where models are not point identified. 
	In contrast, we focus on an inference problem that does not involve identification failure even though the Jacobian of the moment function can become singular.  
	\cite{andrews_identification-_2019} study more general singular cases than non-identification, but their approach requires differentiability of sample moments for subvector inference. 
	Since our model exhibits discontinuity, the method of \cite{andrews_identification-_2019} is not applicable.
	
	This paper is organized as follows. 
	\cref{sec:dptr} explains the dynamic panel threshold model.
	\cref{sec:asym} presents the asymptotic distribution theories of the estimators and test statistics related to the threshold location and continuity.
	\cref{sec:bootstrap} proposes the bootstrap methods.
	\cref{sec:mc} reports Monte Carlo simulation results.
	\cref{sec:empirical} contains an empirical application.
	\cref{sec:conclude} concludes.
	The mathematical proofs and technical details are left to the Appendix.

	\section{Dynamic Panel Threshold Model}
	\label{sec:dptr}
	
	We consider the dynamic panel threshold model,
	\begin{equation}
		\label{eq:model}
		y_{it}=x_{it}'\beta+(1,x_{it}')\delta1\{q_{it}>\gamma\}+\eta_{i}+\epsilon_{it},
	\end{equation}
	where $1\leq i \leq n$, $1\leq t \leq T$, and $x_{it}\in\mathbb{R}^{p}$ is a regressor vector that includes $y_{i,t-1}$ and $q_{it}$. The threshold variable $q_{it}\in\mathbb{R}$ is allowed to be endogenous and is the last element of $x_{it}$.\footnote{Our analysis still holds if researchers have two sets of regressors $x_{1it}$ and $x_{2it}$ such that $y_{it}=x_{1it}'\beta+(1,x_{2it}')\delta1\{q_{it}>\gamma\}+\eta_{i}+\epsilon_{it}$ where $q_{it}$ is an element of $x_{2it}$. However, this paper sticks to the current form to keep the exposition simple.} We partition $x_{it} $ and such that $x_{it}=(\xi_{it}',q_{it})'\in\mathbb{R}^{p}$. 

	When $x_{it}$ consists of the lagged dependent variables, the model becomes the well-known self-exciting threshold autoregressive (TAR) model popularized by \cite{chan1985use}. The static version where the lagged dependent variables are excluded from $x_{it}$ was considered by \cite{hansen_threshold_1999}, while the current dynamic model was studied by \cite{seo_dynamic_2016}. 
	
	The parameter $\gamma \in \Gamma $ denotes the threshold location, where $\Gamma$ is a compact set in $\mathbb{R}$, and $\alpha=(\beta',\delta')'\in A \subset \mathbb{R}^{2p+1}$ denotes the collection of coefficients. Let $\theta=(\alpha',\gamma)=(\beta',\delta',\gamma)'\in \Theta=A\bigtimes\Gamma $ denote the vector of all the parameters. 
	The fixed effect $\eta_{i}$ is constant across time for each individual in the panel data.  It is not identified but is eliminated after first-differencing for the GMM estimation.
    The idiosyncratic error $\epsilon_{it}$  is independent across individuals but can be dependent across time.
	
	For the estimation, we use the GMM after the first-difference transformation
	\begin{equation}
		\label{eq:fd}
		\Delta y_{it}=\Delta x_{it}'\beta+1_{it}(\gamma)'X_{it}\delta+\Delta \epsilon_{it},
	\end{equation}
	where
	\begin{equation}
		X_{it}=\begin{pmatrix}
			(1,x_{it}')
			\\
			(1,x_{it-1}')
		\end{pmatrix},\text{ and }
		1_{it}(\gamma)=\begin{pmatrix}
			1\{q_{it}>\gamma\}\\
			-1\{q_{it-1}>\gamma\}
		\end{pmatrix}.
	\end{equation}
	Let $z_{it}$ denote a set of instrumental variables at time $t$ such that $E[z_{it}\Delta\epsilon_{it}]$ becomes a zero vector,  which may include lagged dependent variables $y_{it-2},...,y_{i1}$ and certain lagged variables of covariates $x_{it}$ and/or $q_{it}$, depending on the assumptions regarding exogeneity of those variables.

	Then, we can define a vector of moment functions for the GMM estimation,
	\begin{equation}
			g_{i}(\theta)=\begin{pmatrix}
				z_{it_{0}}(\Delta y_{it_{0}}-\Delta x_{it_{0}}'\beta-1_{it_{0}}(\gamma)'X_{it_{0}}\delta)\\
				\vdots\\
				z_{iT}(\Delta y_{iT}-\Delta x_{iT}'\beta-1_{iT}(\gamma)'X_{iT}\delta)
			\end{pmatrix}\in\mathbb{R}^{k},
	\end{equation}
	where $k\geq dim(\theta)=2p+2$ and $t_{0}\geq 2$ is the earliest period that the regressor and instrument can be defined.
	For example, $k=(T-1)(T-2)/2$ when $z_{it}=(y_{it-2},...,y_{i1})'$ and $t_{0}=3$.
	Denote the population moment by $g_{0}(\theta)=E[g_{i}(\theta)]$ and the sample moment by $$\bar{g}_{n}(\theta)=\frac{1}{n}\sum_{i=1}^{n}g_{i}(\theta).$$
	We write $g_{i}$ instead of $g_{i}(\theta_{0})$ for simplicity of notations. 
	
	We consider the two-stage GMM estimation of the dynamic panel threshold model. In the first stage, we get an initial estimate by $\hat{\theta}_{(1)}=\arg\min_{\theta\in\Theta}\bar{g}_{n}(\theta)'\bar{g}_{n}(\theta)$ to compute a weight matrix
	\begin{equation*}
		W_{n}=\left(\frac{1}{n}\sum_{i=1}^{n}[g_{i}(\hat{\theta}_{(1)})g_{i}(\hat{\theta}_{(1)})']-\bar{g}_{n}(\hat{\theta}_{(1)})\bar{g}_{n}(\hat{\theta}_{(1)})'\right)^{-1},
	\end{equation*}
	and obtain the second stage estimator 
	\begin{equation*}
		\hat{\theta}=\arg\min_{\theta\in\Theta}\hat{Q}_{n}(\theta),    
	\end{equation*}
	where $\hat{Q}_{n}(\theta)=\bar{g}_{n}(\theta)'W_{n}\bar{g}_{n}(\theta)$. \cite{seo_dynamic_2016} proposed averaging of a class of GMM estimators that are constructed from randomized first stage estimators. We do not pursue the averaging since our primary goal is the bootstrap inference. 
	
	In practice, the grid search algorithm is employed to compute the estimates. Note that when $\gamma$ is given, $\hat{\alpha}(\gamma)=\arg\min_{\alpha\in A}\hat{Q}_{n}(\alpha,\gamma)$ can be easily computed because the problem becomes the estimation of a linear dynamic panel model.
	Then, $\hat{\gamma}$ minimizes the profiled criterion $\tilde{Q}_{n}(\gamma)=\hat{Q}_{n}(\hat{\alpha}(\gamma),\gamma)$ over the grid of $\Gamma$. 
    
	Let $\theta_{0}=(\alpha_{0}',\gamma_{0})'=(\beta_{0}',\delta_{0}',\gamma_{0})'$ denote the true parameter value that lies in the interior of $\Theta$.
	For the point identification of $\theta_0$, $g_{0}(\theta)=0_{k}$ should hold if and only if $\theta=\theta_{0}$, where $0_{k}=(0,...,0)'\in\mathbb{R}^{k}$.
	Let
	\begin{equation*}
		M_{1i} = -\begin{bmatrix}
			z_{it_{0}}\Delta x_{it_{0}}' \\
			\vdots \\
			z_{iT}\Delta x_{iT}'
		\end{bmatrix}\in\mathbb{R}^{k\times p},
		\quad
		M_{2i}(\gamma) = -\begin{bmatrix}
			z_{it_{0}}1_{it_{0}}(\gamma)'X_{it_{0}} \\
			\vdots \\
			z_{iT}1_{iT}(\gamma)'X_{iT}
		\end{bmatrix}\in\mathbb{R}^{k\times (p+1)},
	\end{equation*}and $M_{i}(\gamma)=\left[\begin{array}{c;{2pt/2pt}c}
		M_{1i} & M_{2i}(\gamma)
	\end{array}\right]$. Define $M_{0}(\gamma)=E[M_{i}(\gamma)]$, $M_{10}=E[M_{1i}]$, $M_{20}(\gamma)=E[M_{2i}(\gamma)]$, $\bar{M}_{n}(\gamma)=n^{-1}\sum_{i=1}^{n}M_{i}(\gamma)$, $\bar{M}_{1n}=n^{-1}\sum_{i=1}^{n}M_{1i}$, and $\bar{M}_{2n}(\gamma)=n^{-1}\sum_{i=1}^{n}M_{2i}(\gamma)$.
	We write $M_{0}$, $M_{20}$ and $\bar{M}_{n}$ instead of $M_{0}(\gamma_{0})$, $M_{20}(\gamma_{0})$ and $\bar{M}_{n}(\gamma_{0})$, respectively, for simplicity of notation. 
	The identification condition is stated in \cref{thm:id} that follows.
	
	\begin{theorem}
		\label{thm:id}
		Let the following two conditions hold:
		
		\noindent
		(\romannumeral 1) The matrix $M_{0}$ is of full column rank.
		
		\noindent	
		(\romannumeral 2) For any $\gamma\neq\gamma_{0}$, $M_{20}\delta_{0}$ is not in the column space of $M_{20}(\gamma)$. \newline
		Then,  $\theta_0$ is a unique solution to $g_{0}(\theta)=0_{k}$.
	\end{theorem}
	
	\cref{thm:id} (i) is the identification condition for the coefficients once the true threshold location is identified. 
	This means that instruments should be relevant to the first-differenced regressors appearing in $\eqref{eq:fd}$ when $\gamma=\gamma_{0}$.
	
	\cref{thm:id} (ii) is for the identification of the threshold location, which excludes the possibility of $\delta_{0}=0_{p+1}$. In the standard GMM problem, it is usually assumed that the Jacobian of $g_{0}(\theta)$ at $\theta_{0}$ is of full column rank for both the point identification and the asymptotic normality of the GMM estimator.  The condition (ii) does not require the full rank condition on the Jacobian, which is related to the presence of a jump in the threshold model, and thus it generalizes the identification conditions in \cite{seo_dynamic_2016}.
When the model is continuous and has a kink at the threshold location, the last column of the Jacobian matrix, which is the first-order derivative with respect to $\gamma$ at the true parameter, becomes a zero vector. 
The exact formula for the Jacobian is given later in this section.
This degeneracy does not violate the condition (ii), but it fails the asymptotic normality of the standard GMM estimator, which relies on the linearization of $g_{0}(\theta)$ near $\theta_{0}$ as in \cite{newey_chapter_1994}. 
	
	To define the continuity, recall that  $q_{it}$ is the last element of $x_{it}$ such that $x_{it}=(\xi_{it}',q_{it})'\in\mathbb{R}^{p}$. Accordingly, partition $\delta = (\delta_{1},\delta_{2}',\delta_{3})'$, where $\delta_{2}\in\mathbb{R}^{p-1}$ and $\delta_{1},\delta_{3}\in\mathbb{R}$,	and $\delta_{0}=(\delta_{10},\delta_{20}',\delta_{30})'$.
	Hence, $\delta_{3}$ is the change in the coefficient of the threshold variable when the threshold variable surpasses the tipping point. 	Likewise, $\delta_{2}$ and $\delta_{1}$ are the changes in the coefficients for the other regressors, $\xi_{it}$, and the intercept, respectively.
	The continuity of the dynamic panel threshold model is formally given in \cref{def:c}.
	
        \begin{definition}
		\label{def:c}
		Let $\delta \neq 0_{p+1}$. 	A dynamic panel threshold model is continuous with respect to the threshold variable if $ \theta \in \Theta_{c} = \{\theta\in\Theta : \delta \neq 0_{p+1}, \delta_{2}=0_{p-1}$ and $\delta_{1}+\delta_{3}\gamma = 0\}$. Otherwise, it is discontinuous at the threshold location. 
	\end{definition}
	
	Note that this definition of continuity requires that $\delta_3 \neq 0$; otherwise, $\delta=0_{p+1}$.

 The rank of the first-order derivative matrix, say $D_1$, of $g_{0}(\theta)$ at $\theta=\theta_{0}$ is crucial to the standard asymptotic normality of the GMM estimator. Let $G$ denote the first-order derivative of $g_{0}(\theta)$ with respect to $\gamma$ at $\theta=\theta_{0}$.
 Then,
	\begin{equation}
		\label{eq:firstderivative}
		G = \underbrace{\begin{bmatrix}
			E_{t_{0}}[z_{it_{0}}(1,x_{it_{0}}')|\gamma_{0}]f_{t_{0}}(\gamma_{0}) - E_{t_{0}-1}[z_{it_{0}}(1,x_{it_{0}-1}')|\gamma_{0}]f_{t_{0}-1}(\gamma_{0})\\
			\vdots \\
			E_{T}[z_{iT}(1,x_{iT}')|\gamma_{0}]f_{T}(\gamma_{0}) - E_{T-1}[z_{iT}(1,x_{iT-1}')|\gamma_{0}]f_{T-1}(\gamma_{0})
		\end{bmatrix}}_{\strut\textstyle G_{0}}\times\delta_{0}\in\mathbb{R}^{k},
	\end{equation} 
 where the conditional expectation $E_{t}[\cdot|q] = E[\cdot|q_{it}=q]$ and the density function $f_t(\cdot)$ of $q_{it}$ are assumed to exist. The derivation of $G$ is provided in the proof of \cref{lem:loc.conti}. Note that the first-order derivative of $g_{0}(\theta)$ with respect to $\alpha$ at $\theta=\theta_{0}$ is $M_{0}$. The linear independence of $G$ from the other columns in $D_1$ is required for the standard linear approximation 
 $$g_{0}(\theta)\approx D_1 (\theta -\theta_0 ) =  M_{0}(\alpha-\alpha_{0})+G(\gamma-\gamma_{0}).$$ 
 Recall that the vector $G$ can be written as the product of the matrix $G_0$ and the vector $\delta_0$, \eqref{eq:firstderivative}, and the first and last columns of $G_0$  are linearly dependent due to conditioning on $q_{it}=\gamma_0$ and $q_{it-1}=\gamma_{0}$.  Then, the standard rank condition on the first derivative matrix $D_1$ can follow from a more primitive rank condition on $\left[\begin{array}{c;{2pt/2pt}c}	M_{0} & G_{0,-(p+1)} \end{array}\right]$, which requires the linear independence of all columns in $M_0$ and all but the last column of $G_0$.
 Even if the primitive condition is met, however, the continuity restriction makes $G=0_{k}$  since  $E_{s}[z_{it}(1,x_{is}')\delta_{0}|\gamma_{0}]=(\delta_{10}+\delta_{30}\gamma_{0})E_{s}[z_{it}|\gamma_{0}]=0$  for $s=t-1,t$, which leads to degeneracy of $D_{1}$.

When the rank condition fails due to the continuity, the expansion becomes 
$$g_{0}(\theta)\approx M_{0}(\alpha-\alpha_{0})+H(\gamma-\gamma_{0})^{2},$$
where 
\begin{equation}
		\label{eq:secondderivative}
		H = \frac{\partial^2 g_0 (\theta_0)}{2 \partial \gamma \partial \gamma} 
  =\frac{\delta_{30}}{2}\begin{pmatrix}
			E_{t_{0}}[z_{it_{0}}|\gamma_{0}]f_{t_{0}}(\gamma_{0}) - E_{t_{0}-1}[z_{it_{0}}|\gamma_{0}]f_{t_{0}-1}(\gamma_{0})\\
			\vdots \\
			E_{T}[z_{iT}|\gamma_{0}]f_{T}(\gamma_{0}) - E_{T-1}[z_{iT}|\gamma_{0}]f_{T-1}(\gamma_{0})
		\end{pmatrix}\in\mathbb{R}^{k}.
	\end{equation}
The detailed derivation is given in the proof of \cref{lem:loc.conti}.  It is worth noting that $H$ is identical to the first column of $G_0$ up to a constant multiple.  Then, the rank condition on $\left[\begin{array}{c;{2pt/2pt}c}	M_{0} & H \end{array}\right]$ is implied by the rank condition on $\left[\begin{array}{c;{2pt/2pt}c}	M_{0} & G_{0,-(p+1)} \end{array}\right]$. Thus, the rank condition on $\left[\begin{array}{c;{2pt/2pt}c}	M_{0} & G_{0,-(p+1)} \end{array}\right]$ can be viewed as a sufficient condition for both Assumptions \ref{ass:loc} and \ref{ass:locDis} in the next section, apart from the continuity restriction on $\theta$. Next section formalizes this discussion and presents the asymptotic distribution of the GMM estimator $\hat{\theta}$  under the continuity.

	\section{Asymptotic theory}
	\label{sec:asym}
	
	This section considers the asymptotic analysis when $T$ is fixed, the data are independent and identically distributed across $i$, and $n\rightarrow\infty$. Specifically, the data for each individual $i$ is determined by the realization of $\{(z_{it},x_{it},\epsilon_{it})_{t=1}^{T},y_{i0},\eta_{i}\}$, where $y_{i0}$ denotes an initial value.
	We make the following assumptions.
	\begin{assumption*}{G}
		\label{ass:gmm}
		The parameter space $\Theta$ is compact and	$\theta_{0}\in \text{int }\Theta$. $M_{0}$ is of full column rank, and $M_{20}\delta_{0}$ is not in the column space of $M_{20}(\gamma)$ for any $\gamma\neq\gamma_{0}$. $\Omega=E[g_{i}g_{i}']$ is positive definite. $E\|z_{it}\|^{4}$, $E\|x_{it}\|^{4}$, and $E\epsilon_{it}^{4}$ are finite for all $t$.
	\end{assumption*}
	\begin{assumption*}{D}
		\label{ass:dgp} For all $t$,
		(\romannumeral 1)  $q_{it}$ has a continuous distribution and a bounded density $f_{t}(\cdot)$, which is continuously differentiable at $\gamma_{0}$ and $f_{t}(\gamma_{0})>0$.
		(\romannumeral 2) $E_{t}[z_{it}(1,x_{it}')|q]$ and $E_{t-1}[z_{it}(1,x_{it-1}')|q]$ are continuous on $q\in\Gamma$ and continuously differentiable at $q=\gamma_{0}$. 
	\end{assumption*}
	\begin{assumption*}{LK}
		\label{ass:loc}
		$D_{2}=\left[\begin{array}{c;{2pt/2pt}c}
			M_{0} & H
		\end{array}\right]\in\mathbb{R}^{k\times(2p+2)}$ has full column rank.
	\end{assumption*}
	
	Assumptions \ref{ass:gmm} and \ref{ass:dgp} are similar to Assumptions 1 and 2 in \cite{seo_dynamic_2016} except for the differentiability conditions in \cref{ass:dgp} which allow the second-order derivative of the population moment to be defined. Since the regressors include lagged dependent variables, \cref{ass:gmm} requires the individual fixed effects and initial values to have finite fourth moments, too. The assumption also includes the conditions in \cref{thm:id}. 
 \cref{ass:loc} is a rank condition for a nondegenerate asymptotic distribution when the underlying model is continuous. This condition may be viewed as less restrictive than the standard rank assumption as discussed in the previous section where $G$ and $H$ are defined.  
 For easy reference, we restate the standard full rank assumption for the asymptotic normality of the GMM estimator for the discontinuous threshold regression below.

	\begin{assumption*}{LJ}
		\label{ass:locDis}
		$D_{1}=\left[\begin{array}{c;{2pt/2pt}c}
			M_{0} & G
		\end{array}\right]\in\mathbb{R}^{k\times(2p+2)}$ has full column rank.
	\end{assumption*}

 	In a simple model, where $y_{it}=x_{it}'\beta+(\delta_{1}+\delta_{3}q_{it})1\{q_{it}>\gamma\} +\eta_{i}+\epsilon_{it}$, both Assumptions \ref{ass:loc} and \ref{ass:locDis} require $\left[\begin{array}{c;{2pt/2pt}c}
			M_{0} & G_{01}
		\end{array}\right]$ to have full rank, where $G_{01}$ is the first column of $G_{0}$ in \eqref{eq:firstderivative}, because $G=(\delta_{10}+\delta_{30}\gamma_{0})G_{01}$ while $H={\delta_{30}}G_{01}/2$.

	\cref{thm:asym} below establishes the asymptotic distribution of the GMM estimator when the dynamic panel threshold model is continuous.
	\begin{theorem}	\label{thm:asym} 
		When the true model is continuous and Assumptions \ref{ass:gmm}, \ref{ass:dgp}, and \ref{ass:loc} hold,
		\begin{equation*}
			\begin{pmatrix}
				\sqrt{n}(\hat{\alpha}-\alpha_{0}) \\
				\sqrt{n}(\hat{\gamma}-\gamma_{0})^{2}
			\end{pmatrix}\xrightarrow{d}
			\begin{pmatrix}
				U - (M_{0}'\Omega^{-1}M_{0})^{-1}M_{0}'\Omega^{-1}HV \\
				V
			\end{pmatrix},
		\end{equation*}
		where $U\sim N(0,(M_{0}'\Omega^{-1}M_{0})^{-1})$ and $V\sim \max\{0,N(0,(H'\Xi H)^{-1})\}$ are independent of each other, while $\Xi=\Omega^{-1}-\Omega^{-1}M_{0}(M_{0}'\Omega^{-1}M_{0})^{-1}M_{0}'\Omega^{-1}$. 
	\end{theorem}
	
	We observe that the convergence rate of $\hat{\gamma}$ is $n^{1/4}$, which is slower than the standard $\sqrt{n}$-rate.
	Meanwhile, \cite{seo_dynamic_2016} show the $\sqrt{n}$-convergence rate for $\hat{\gamma}$ when the model is discontinuous. Intuitively, it would be more difficult to detect the precise threshold location when there is a kink than when there is a jump at the tipping point. More technically, when the threshold model is discontinuous and the Jacobian is not singular, the limit of the GMM objective function admits a quadratic approximation with respect to $\gamma$ at the true value, while the limit admits a quartic approximation for the continuous model. Hence, the limit objective function becomes flatter in $\gamma$ at the true value resulting in the slower convergence rate. On the other hand, \cite{hidalgo_robust_2019} showed that the least squares criterion converges to a limit which is quadratic near the true $\gamma$ if the model is continuous and has a kink otherwise.
	
	Moreover, we can observe that the asymptotic distribution of $\hat{\alpha}$ is also shifting to a non-normal distribution. Hence, standard inference methods based on the asymptotic normality become invalid for the continuous dynamic panel threshold model.

	The asymptotic distribution of the GMM estimator is identical to the distribution reported in Theorem 1 (b) in \cite{DovononHall2018}, which studies a smooth GMM problem with the degeneracy of the Jacobian. 
	\cref{thm:asym} shows that even though the criterion of our threshold model is discontinuous with respect to the parameter $\gamma$, the same asymptotic distribution as that of \cite{DovononHall2018} appears. Meanwhile, \cite{dovonon_bootstrapping_2017} show that the standard nonparametric bootstrap becomes invalid when the Jacobian degenerates. To address this issue, we propose different bootstrap methods in \cref{sec:bootstrap} for inference of the parameters.
	
	The censored normal distribution also appears in \cite{andrews2002generalized} which studies the estimation of a parameter on a boundary. 
	Heuristically, because our analysis depends on the second-order derivative of $\gamma$ for the local polynomial expansion of $g_{0}(\theta)$ near $\theta_{0}$, only the asymptotic distribution of $(\hat{\gamma}-\gamma_{0})^{2}$ can be derived.
	Since $(\hat{\gamma}-\gamma_{0})^{2}$ should be nonnegative, the asymptotic censored normal distribution appears as in \cite{andrews2002generalized}. 

        The asymptotic distribution in \cref{thm:asym} can be used for parameter inference when the true model is continuous, but the estimator is obtained without imposing the continuity restriction.
		As discussed in \cite{seo_dynamic_2016}, $M_{0}$ and $\Omega$ can be consistently estimated, while $H$ can be nonparametrically estimated similarly to $G$. Then, it is straightforward to simulate the limit distribution of \cref{thm:asym} by generating random numbers for $U$ and $V$.
		However, there are several drawbacks to that approach, and hence we do not recommend it.
		First, empirical researchers might construct confidence intervals based on \cref{thm:asym} when they cannot reject the continuity.
		However, \cite{Leeb_Potscher_2005} show that confidence intervals after model selection are subject to size-distortion.
		Second, even if the true model is known to be continuous, the continuity-restricted estimator explained in \cite{seo_estimation_2019} is more efficient and asymptotically normal.
		Therefore, using the continuity-restricted estimator for estimation and inference is preferable.
		Finally, the nonparametric estimation of $H$ requires a tuning parameter and has a slower convergence rate.
		
	\cite{seo_dynamic_2016} derived the asymptotic distribution of the GMM estimator and proposed an inference method when the underlying model is discontinuous. When the true model is discontinuous and Assumptions \ref{ass:gmm}, \ref{ass:dgp}, and \ref{ass:locDis} hold,
		\begin{equation*}
			\begin{pmatrix}
				\sqrt{n}(\hat{\alpha}-\alpha_{0}) \\
				\sqrt{n}(\hat{\gamma}-\gamma_{0})
			\end{pmatrix}\xrightarrow{d} N(0,(D_{1}'\Omega^{-1}D_{1})^{-1}).
		\end{equation*}
		$\Omega$ can be estimated by $\hat{\Omega}=\frac{1}{n}\sum_{i=1}^{n}[g_{i}(\hat{\theta})g_{i}(\hat{\theta})']-\bar{g}_{n}(\hat{\theta})\bar{g}_{n}(\hat{\theta})'$. Note that $D_{1}=\left[\begin{array}{c;{2pt/2pt}c}
				M_{0} & G
			\end{array}\right]$, and $M_{0}$ can be estimated by $\bar{M}_{n}(\hat{\gamma})$, while the estimation of $G$ involves nonparametric estimation of the conditional means and densities. See section 4 of \cite{seo_dynamic_2016} for more details.
            Note that $(\hat{D}_{1}\hat{\Omega}^{-1}\hat{D}_{1})^{-1}$ diverges when the model is continuous since the last column of $\hat{D}_{1}$ converges to a zero vector when it is consistent. This paper does not study the behavior of the asymptotic confidence intervals when the true model is continuous.

	\subsection{Testing for threshold value}
	\label{sec:asym dist}
	
	Since the asymptotic distribution of the threshold estimator is not standard, we consider the GMM distance test introduced by \cite{newey_hypothesis_1987} for a hypothesis on the location of the threshold. Let the test statistic for the threshold location at $\gamma$ be
	\[\mathcal{D}_{n}(\gamma)=n(\min_{\alpha\in A}\hat{Q}_{n}(\alpha,\gamma)-\hat{Q}_{n}(\hat{\theta})),\]
	and let $\chi^{2}_{1}$ denote the chi-square distribution with 1 degree of freedom.
	
	\begin{theorem}
		\label{thm:distance test}
		(\romannumeral 1) If $\gamma=\gamma_{0}$, the true model is continuous, and Assumptions \ref{ass:gmm}, \ref{ass:dgp}, and \ref{ass:loc} hold, then
		\begin{equation*}
			\mathcal{D}_{n}(\gamma)\xrightarrow{d} Z_{0}^{2}
		\end{equation*}
		where $Z_{0}=\max(0,Z_{0}^{*})$, $Z_{0}^{*}\sim N(0,1)$.
		
		\noindent	
		(\romannumeral 2) If $\gamma=\gamma_{0}$, the true model is discontinuous, and Assumptions \ref{ass:gmm}, \ref{ass:dgp}, and \ref{ass:locDis} hold, then
		\begin{equation*}
			\mathcal{D}_{n}(\gamma)\xrightarrow{d} \chi^{2}_{1}.
		\end{equation*}
		
		\noindent	
		(\romannumeral 3) If $\gamma\neq\gamma_{0}$, then for any $M<\infty$, $\lim_{n\rightarrow\infty}P(\mathcal{D}_{n}(\gamma)<M)=0$.
	\end{theorem}
	
	\cref{thm:distance test} (i) presents the asymptotic distribution of the distance statistic under the continuity. Due to censoring, the asymptotic distribution becomes a mixture of the $\chi^{2}_{1}$ distribution with weight 1/2 and zero with weight 1/2. 

	Meanwhile, the chi-square limit in \cref{thm:distance test} (ii) extends \cite{newey_hypothesis_1987} for a discontinuous moment function. \cite{seo_dynamic_2016} did not study the distance statistic.
	
	\cref{thm:distance test} (iii) shows that the GMM distance test for the threshold location is consistent. 
	It also serves as the consistency of a bootstrap test together with \cref{thm:grid boot} since the bootstrap statistic is stochastically bounded whether or not the threshold location is true.

	Since the limit distribution depends on the continuity of the model, we introduce a bootstrap in \cref{sec:grid bootstrap}, which is valid regardless of the model continuity. Furthermore, \cref{sec:bootstrap_uniformity} establishes the uniform validity of the bootstrap inference for the threshold location under some simplifying assumptions.

	\subsection{Testing continuity}
	\label{sec:asym continuity}
	
	We propose a test for the continuity of the threshold model, similar to the approach used by \cite{Gonzalo_Wolf_TAR} or \cite{Hidalgo_et_at_2022} in the threshold regression literature. 
	While empirical researchers may employ the test to select a model, we utilize the test to modify the standard nonparametric bootstrap to make the bootstrap valid irrespective of the model continuity. Details of the use of the continuity test statistic in the bootstrap method are explained in \cref{sec:perc bootstrap}.
	
	The continuity hypothesis is a joint hypothesis. We employ the GMM distance test. Let $\tilde{\theta}=\arg\min_{\theta\in\Theta_{c}} \hat{Q}_{n}(\theta)$ be the continuity-restricted estimator. The GMM distance test statistic is 
	\[\mathcal{T}_{n}=n(\hat{Q}_{n}(\tilde{\theta})-\hat{Q}_{n}(\hat{\theta})).\]
	
	\begin{theorem}
		\label{thm:continuity test}
		\noindent	
		(\romannumeral 1)
		When the true model  is continuous and Assumptions \ref{ass:gmm}, \ref{ass:dgp}, and \ref{ass:loc} hold,
		\[\mathcal{T}_{n}\xrightarrow{d}V_{1}-V_{2}+V_{3},\]
		where $V_{1} \equiv Z'\Psi M_{20}(M_{20}'\Psi M_{20})^{-1}M_{20}'\Psi Z$, $	V_{2} \equiv Z'\Psi N_{20}(N_{20}'\Psi N_{20})^{-1}N_{20}'\Psi Z $, $V_{3} \equiv Z_{0}^{2}$,
		$Z \sim N(0,\Omega)$, $Z_{0} =\max(0, Z_{0}^{*})$, $Z_{0}^{*} \sim N(0,1)$, $Z_{0}$ and $Z$ are independent, 
		$\Psi  = \Omega^{-1}-\Omega^{-1}M_{10}(M_{10}'\Omega^{-1}M_{10})^{-1}M_{10}'\Omega^{-1}$, and
		$N_{20} = M_{20}\scalebox{0.8}{$\left(\begin{array}{ccc}
			     -\gamma_{0}& 0_{p-1}'& 1\\
			     -\delta_{30}& 0_{p-1}'& 0
			\end{array}\right)'$}$
		
		\noindent	
		(\romannumeral 2) If the model is discontinuous, then $\lim_{n\rightarrow\infty}P(n^{-m}\mathcal{T}_{n}<M)=0$ for any $m\in [0,1)$ and $M<\infty$.
	\end{theorem}
	
	While the limit distribution in \cref{thm:continuity test} (i) is non-standard, it can be  simulated to obtain critical values for the test using consistent plug-in sample analogue estimators, e.g., 
	$\hat{\Omega}=\frac{1}{n}\sum_{i=1}^{n}[g_{i}(\hat{\theta})g_{i}(\hat{\theta})']-\bar{g}_{n}(\hat{\theta})\bar{g}_{n}(\hat{\theta})'$, $\hat{M}_{1}=\bar{M}_{1n}$, $\hat{M}_{2}=\bar{M}_{2n}(\hat{\gamma})$, etc. 
	Another way to obtain the critical values is via a bootstrap method, which is introduced in \cref{sec:continuity bootstrap}.
	
	\cref{thm:continuity test} (ii) shows that the continuity test is consistent. It also implies the consistency of the bootstrap test together with \cref{thm:continuity boot}, which shows that the bootstrap test statistic is stochastically bounded even when the true model is not continuous. 
	The divergence rate of $\mathcal{T}_{n}$, which is faster than $n^{m}$ for any $0\leq m<1$, is exploited to modify the standard nonparametric bootstrap for the coefficients as detailed in \cref{sec:perc bootstrap}.

\section{Bootstrap}
\label{sec:bootstrap}
	
	As usual, the superscript ``*'' denotes the bootstrap quantities or the convergence of bootstrap statistics under the bootstrap probability law conditional on the original sample. For example,    $E^{*}$ denotes the expectation with respect to the bootstrap probability law conditional on the data. 
	``$\xrightarrow{d^{*}}$, in $P$'' denotes the distributional convergence of bootstrap statistics under the bootstrap probability law with probability approaching one. We write	``$\nu_{n}^{*}=O_{p}^{*}(1)$, in $P$'' if a sequence $\nu_{n}^{*}$ is stochastically bounded under the bootstrap probability law with probability approaching one. 
	More details are written in \cref{sec:boot prelim}.
 Let $\widehat{F}^{*-1}_{n}(\varphi;S^{*})$ denote the empirical $\varphi$ quantile of a bootstrap statistic $S^{*}$.
	
	This section introduces three different bootstrap schemes.
	The first bootstrap is for constructing bootstrap confidence interval(CI)s for the threshold, while the second bootstrap is for constructing bootstrap CIs for the coefficients. 
	Both methods aim to provide valid inferences, regardless of whether the model is continuous or not.
	The third bootstrap is for testing continuity of the threshold model.
	The three bootstrap methods can be represented by means of \cref{alg:bootstrap} with suitable choices of $\theta_0^{*}=(\beta_{0}^{*\prime},\delta_{0}^{*\prime},\gamma_{0}^{*})'$.
	
	\begin{algorithm}[htbp]
		\caption{Bootstrap with  $\theta_{0}^{*}$}
		\label{alg:bootstrap}
		\begin{algorithmic}[1]
			\State For $i=1,...,n$, let $i^{*}$ be the $i$th i.i.d. random draw from the discrete uniform distribution on $\{1,...,n\}$.
			Generate a bootstrap sample $\{(x_{it}^{*},x_{it-1}^{*},z_{it}^{*},\widehat{\Delta\epsilon}_{it}^{*})_{t=t_{0}}^{T}:i=1,...,n\}$ by setting $(x_{it}^{*},x_{it-1}^{*},z_{it}^{*},\widehat{\Delta\epsilon}_{it}^{*})_{t=t_{0}}^{T}=(x_{i^{*}t},x_{i^{*}t-1},z_{i^{*}t},\widehat{\Delta\epsilon}_{i^{*}t})_{t=t_{0}}^{T}$ for each $i$, where $\widehat{\Delta\epsilon}_{it}=\Delta y_{it}-\Delta x_{it}'\hat{\beta}-1_{it}(\hat{\gamma})'X_{it}\hat{\delta}$.
			\State Generate $\{(\Delta y_{it}^{*})_{t=t_{0}}^{T}:i=1,...,n\}$ using $\theta_{0}^{*}$ by
			\begin{equation*}
				\Delta y_{it}^{*} = \Delta x_{it}^{*\prime}\beta_{0}^{*} + 1_{it}^{*}(\gamma_{0}^{*})'X_{it}^{*}\delta_{0}^{*}+\widehat{\Delta\epsilon}_{it}^{*},
			\end{equation*}
			where $\Delta x_{it}^{*} = x_{it}^{*}-x_{it-1}^{*}$,
			\begin{equation*}
				X_{it}^{*}=\begin{pmatrix}
					(1,x_{it}^{*\prime})\\
					(1,x_{it-1}^{*\prime})
				\end{pmatrix},
				\text{ and }
				1_{it}^{*}(\gamma)=\begin{pmatrix}
					1\{q_{it}^{*}>\gamma\}\\
					-1\{q_{it-1}^{*}>\gamma\}
				\end{pmatrix}.
			\end{equation*}
			\State Define the bootstrap moment function $g_{i}^{*}(\theta)=(g_{it_{0}}^{*}(\theta)',...,g_{iT}^{*}(\theta)')'$ where $g_{it}^{*}(\theta)=z_{it}^{*}(\Delta y_{it}^{*}-\Delta x_{it}^{*\prime}\beta-1_{it}^{*}(\gamma)'X_{it}^{*}\delta)$.
			\State Define the (recentered) bootstrap sample moment  
			\[\bar{g}_{n}^{*}(\theta)=\tfrac{1}{n}{\textstyle \sum}_{i=1}^{n}(g_{i}^{*}(\theta)-\bar{g}_{n}(\hat{\theta})).\] 
			\State Compute the initial estimator $\hat{\theta}_{(1)}^{*}=\arg\min_{\theta}\bar{g}_{n}^{*}(\theta)'\bar{g}_{n}^{*}(\theta)$
			and
			the weight matrix $W_{n}^{*}=(\frac{1}{n}\sum_{i=1}^{n} g_{i}^{*}(\hat{\theta}_{(1)}^{*})g_{i}^{*}(\hat{\theta}_{(1)}^{*})'-[\frac{1}{n}\sum_{i=1}^{n}g_{i}^{*}(\hat{\theta}_{(1)}^{*})][\frac{1}{n}\sum_{i=1}^{n}g_{i}^{*}(\hat{\theta}_{(1)}^{*})]')^{-1}$.
			\State Define the bootstrap criterion function $\hat{Q}_{n}^{*}(\theta) = \bar{g}_{n}^{*}(\theta)'W_{n}^{*}\bar{g}_{n}^{*}(\theta)$, and obtain the bootstrap estimator or the test statistics.
		\end{algorithmic}
	\end{algorithm}
 
	In step 1, we resample the regressors, the instruments, and the residuals jointly to maintain the dependence among them, unlike in the usual residual bootstrap. See e.g., \cite{giannerini2024validity} for the description of the standard residual bootstrap, which resamples the residuals only, and the wild bootstrap for the testing of linearity in the threshold regression. There could be other ways of resampling not mentioned here and we do not attempt to decide which is the best here.   

	The parameter $\theta_{0}^{*}$ is used in step 2 of \cref{alg:bootstrap} to generate the dependent variables in the bootstrap samples.
	In step 4, recentering of the bootstrap sample moment is done by subtracting $\bar{g}_{n}(\hat{\theta})=(\frac{1}{n}\sum_{i=1}^{n}z_{it_{0}}'\widehat{\Delta\epsilon}_{it_{0}},...,\frac{1}{n}\sum_{i=1}^{n}z_{iT}'\widehat{\Delta\epsilon}_{iT})'$. 		Note that the expectation of $\bar{g}_{n}^{*}(\theta)$ by the bootstrap probability law conditional on the data becomes zero when $\theta=\theta_{0}^{*}$ due to the recentering, which can be easily checked from the following equations: $g_{it}^{*}(\theta_{0}^{*})=z_{it}^{*}(\Delta y_{it}^{*}-\Delta x_{it}^{*}\beta_{0}^{*}-1_{it}(\gamma_{0}^{*})'X_{it}^{*}\delta_{0}^{*})=z_{it}^{*}\widehat{\Delta\epsilon}_{it}^{*}$ and $E^{*}[g_{it}^{*}(\theta_{0}^{*})] = n^{-1}\sum_{i=1}^n z_{it} \widehat{\Delta\epsilon}_{it}$ for $t=t_{0},...,T$.

	A different choice of $\theta_{0}^{*}$ leads to a different bootstrap. 
	For example, if $\theta_{0}^{*}=\hat{\theta}$, then the bootstrap becomes the standard nonparametric bootstrap in \cite{hall_bootstrap_1996} because $\Delta y_{it}^{*}=\Delta y_{i^{*}t}$ holds true for $i=1,...,n$ and $t=t_{0},...,T$ in step 2.
        Note that, for $\theta_{0}^{*}$ not equal to $\hat{\theta}$, step 2 of \cref{alg:bootstrap} generates $\Delta y_{it}^{*}$'s that are generally different from $\Delta y_{i^{*}t}$'s.
	The following subsections detail three different choices of $\theta_{0}^{*}$ for three different inference problems.
	
\subsection{Grid bootstrap for threshold location}
\label{sec:grid bootstrap}
	
	To construct CIs for the threshold location, we propose to employ the grid bootstrap method introduced by \cite{hansen_grid_1999} for autoregressive models.
	Let $\Gamma_{n}=\{\gamma_{\ell}\in\Gamma:\ell=1,...,L\}$ be a grid of the candidate thresholds. The grid bootstrap constructs the confidence set by inverting the bootstrap threshold location tests over $\Gamma_{n}$.  Specifically, a sequence of hypothesis tests for the hypothesized threshold locations in $\Gamma_{n}$ are performed by the bootstrap that imposes the null to generate bootstrap samples. 
	
	The null imposed bootstrap at a point $\gamma_{\ell}\in \Gamma_{n}$ can be implemented by setting $\theta_{0}^{*}=(\hat{\alpha}(\gamma_{\ell})',\gamma_{\ell})'$ in \cref{alg:bootstrap},
	and the bootstrap test statistic is
	\[\mathcal{D}_{n}^{*}(\gamma_{\ell})=n(\min_{\alpha\in A}\hat{Q}_{n}^{*}(\alpha,\gamma_{\ell})-\min_{\theta\in\Theta}\hat{Q}_{n}^{*}(\theta)).\]
	The null hypothesis $\mathcal{H}_{0}:\gamma=\gamma_{\ell}$ is rejected at size $\tau$ if $\mathcal{D}_{n}(\gamma_{\ell})> \widehat{F}^{*-1}_{n}(1-\tau;\mathcal{D}_{n}^{*}(\gamma_{\ell}))$.
	Consequently, after running the null imposed bootstrap for each point in $\Gamma_{n}$, we can construct the $100(1-\tau)$\% confidence set of $\gamma$ by
	\begin{equation}
		\label{eq:grid bcs}
		CI_{n,1-\tau}^{grid}=\{\gamma\in\Gamma_{n}: \mathcal{D}_{n}(\gamma)\leq \widehat{F}^{*-1}_{n}(1-\tau;\mathcal{D}_{n}^{*}(\gamma))\}.
	\end{equation}
	Note that the confidence set is not necessarily a connected set, even though researchers can convexify the set to get a connected CI.
	The CI does not become an empty set because $\mathcal{D}_{n}(\hat{\gamma})=0$ while $\mathcal{D}_{n}^{*}(\hat{\gamma})\geq 0$.
	The consistency of the grid bootstrap method is implied by \cref{thm:grid boot} that follows.
	
	\begin{theorem}
		\label{thm:grid boot}
		For a given $\gamma\in\Gamma$, assume that $\mathcal{D}_{n}^{*}(\gamma)$ is obtained by \cref{alg:bootstrap} with $\theta_{0}^{*}=(\hat{\alpha}(\gamma)',\gamma)'$.
		
		\noindent
		(\romannumeral 1) If $\gamma=\gamma_{0}$, the true model is continuous, and Assumptions \ref{ass:gmm}, \ref{ass:dgp}, and \ref{ass:loc} hold, then
		\begin{equation*}
			\mathcal{D}_{n}^{*}(\gamma)\xrightarrow{d*} Z_{0}^{2}\quad\text{ in $P$},
		\end{equation*}
		where $Z_{0}=\max(0,Z_{0}^{*})$ and $Z_{0}^{*}\sim N(0,1)$.
		
		\noindent	
		(\romannumeral 2) If $\gamma=\gamma_{0}$, the true model is discontinuous, and Assumptions \ref{ass:gmm}, \ref{ass:dgp}, and \ref{ass:locDis} hold, then
		\begin{equation*}
			\mathcal{D}_{n}^{*}(\gamma)\xrightarrow{d*} \chi^{2}_{1}\quad\text{ in $P$}.
		\end{equation*}
		
		\noindent	
		(\romannumeral 3) If $\gamma\neq \gamma_{0}$, then $\mathcal{D}_{n}^{*}(\gamma)=O_{p}^{*}(1)$ in $P$.
	\end{theorem}
	
	\cref{thm:grid boot} (i) and (ii) show that the limit distribution of the bootstrap test statistic, conditional on the data, is identical to that of the sample test statistic regardless of the continuity of the true model.
	Therefore, the CI for the threshold location by the grid bootstrap, \eqref{eq:grid bcs}, achieves an exact coverage rate for both continuous and discontinuous models asymptotically. Specifically, $\lim_{n\rightarrow\infty}P(\gamma_{0}\in CI_{n,1-\tau}^{grid})=1-\tau$ for both cases (i) and (ii).
	\cref{thm:grid boot} (iii) says that the bootstrap test statistic is still stochastically bounded, conditionally on the data, under the alternative.
	As \cref{thm:distance test} (iii) shows that the sample test statistic is stochastically unbounded under the alternative, the grid bootstrap CI has power against fixed alternatives.

 \subsubsection{Uniform validity of grid bootstrap}
 \label{sec:main.uniform.gridB}
 We extend \cref{thm:grid boot} to the uniform validity of the grid bootstrap, which is important for good finite sample performance when the model is nearly continuous. 
 We establish the uniform validity for the following simplified specification for analytical tractability:
   \begin{equation*}
       y_{it} = x_{it}'\beta + (\delta_{1}+\delta_{3}q_{it})1\{q_{it}>\gamma\}+\eta_{i}+\epsilon_{it},
   \end{equation*}
   where $\theta=(\beta',\delta',\gamma)'$ and $\delta=(\delta_{1},\delta_{3})'$ in this subsection.

 This section briefly states the uniformity result of the grid bootstrap and gives a heuristic justification. 
Our derivation follows \cite{andrews_cheng_guggenberger_2020}. It is highly complicated and involves more technical conditions, which are stated in \cref{sec:bootstrap_uniformity}.
 
Specifically, we establish  in \cref{thm:uniform-sufficient} that 
   \begin{equation*}\liminf_{n\rightarrow\infty}\inf_{\phi_{0}\in\Phi_{0}}P_{\phi_{0}}(\gamma_{0}\in CI_{n,1-\tau}^{grid}) = \limsup_{n\rightarrow\infty}\sup_{\phi_{0}\in\Phi_{0}}P_{\phi_{0}}(\gamma_{0}\in CI_{n,1-\tau}^{grid}) = 1-\tau,
   \end{equation*}
   where $P_{\phi}$ is the probability law when the model is specified by $\phi=(\theta,F)$ and $F$ is the distribution of $\{\eta_{i},y_{i0},(z_{it},x_{it},\epsilon_{it})_{t=1}^{T}\}$.
   The collection of probabilistic models $\Phi_{0}$ includes both continuous and discontinuous threshold models.
   More detailed discussions of technical assumptions about $\Phi_{0}$ are given in \cref{sec:bootstrap_uniformity}. 
   
   For the uniformity analysis, we need to consider drifting sequences of true parameters $\phi_{0n}=(\theta_{0n},F_{0n})$ such that $\theta_{0n}\rightarrow\theta_{0,\infty}$ and $F_{0n}\rightarrow F_{0,\infty}$. Here, the distance between $F_{0n}$ and $F_{0,\infty}$ is induced by a specific choice of norm that is explained in \cref{sec:bootstrap_uniformity}.
   To show the uniform validity of the grid bootstrap CI, we need to verify that the limit distribution of $\mathcal{D}_{n}^{*}(\gamma_{0n})$ conditional on the data is identical to the limit distribution of $\mathcal{D}_{n}(\gamma_{0n})$ under all the above drifting sequences of models. 
   Our analysis finds that the limit distribution of the threshold location test statistic under the true null, i.e., the limit distribution of $\mathcal{D}_{n}(\gamma_{0n})$, is determined by $\zeta=\lim_{n\rightarrow\infty} n^{1/4}(\delta_{10n}+\delta_{30n}\gamma_{0n})$; see \cref{lem:uniform-gridboot} for details.
   When $\zeta=0$, the limit distribution of $\mathcal{D}_{n}(\gamma_{0n})$ is as described in \cref{thm:distance test} (i). In contrast, when $|\zeta|=\infty$, the limit distribution is the $\chi^{2}_{1}$-distribution as in \cref{thm:distance test} (ii). When $\zeta$ is finite and nonzero, then $\mathcal{D}_{n}(\gamma_{0n})$ has a nonstandard limit distribution that depends on $\zeta$.
   
   Therefore, if $\theta_{0n}^{*}$ comprises a sequence of true parameters for a bootstrap scheme, then $n^{1/4}(\delta_{10n}^{*}+\delta_{30n}^{*}\gamma_{0n}^{*})$ should consistently estimate $\zeta$ for the bootstrap statistics to exhibit the same asymptotic behavior as the sample statistics.
   
   Note that under the grid bootstrap scheme, the bootstrap test statistic $\mathcal{D}_{n}^{*}(\gamma_{0n})$ is drawn from the bootstrap that imposes the null threshold location $\gamma_{0n}$. 
   The true parameter of the bootstrap data generating process (dgp) is $\theta_{0n}^{*}=(\hat{\alpha}_{n}(\gamma_{0n})',\gamma_{0n})'$.
   The restricted estimator satisfies $\|\hat{\alpha}(\gamma_{0n})-\alpha_{0n}\|=O_{p}(n^{-1/2})$, as the problem becomes estimating a standard linear dynamic panel model, and hence $n^{1/4}(\hat{\delta}_{1n}(\gamma_{0n})+\hat{\delta}_{3n}(\gamma_{0n})\gamma_{0n})=\zeta+o_{p}(1)$.
   Therefore, $\mathcal{D}_{n}^{*}(\gamma_{0n})$ conditionally converges to the limit distribution of $\mathcal{D}_{n}(\gamma_{0n})$, which leads to the uniform validity of the grid bootstrap confidence intervals.
   In contrast, $\hat{\theta}$ does not satisfy this property for some $\zeta$ and the bootstrap building on $\hat{\theta}$ is not uniformly valid.

	\subsection{Residual bootstrap for coefficients}
	\label{sec:perc bootstrap}
	
	The bootstrap CIs for the coefficients can be obtained by applying \cref{alg:bootstrap} with $\theta_{0}^{*}$ set as
	\begin{equation}
		\label{eq:perc boot}
		\theta_{0}^{*}=w_{n}\hat{\theta}+(1-w_{n})\tilde{\theta},\quad w_{n}=\min\left(\frac{\mathcal{T}_{n}}{\hat{C}n^{1/4}}, 1\right),
	\end{equation}
 where $\tilde{\theta}=\arg\min_{\theta\in\Theta_{c}}\hat{Q}_{n}(\theta)$ is the continuity-restricted estimator. $\hat{C}$ is some estimated quantile, such as the $50$th percentile, of the limit distribution of the continuity test statistic $\mathcal{T}_{n}$ when the model is continuous.
	$\hat{C}$ can be obtained either by methods in \cref{sec:asym continuity} or \cref{sec:continuity bootstrap}.
    As long as $\hat{C}=O_{p}(1)$, the asymptotic validity of the residual bootstrap holds.
	Since $w_{n}=O_{p}(n^{-1/4})$ if the true model is continuous, and $w_{n}=1+o_p(1)$ if the model is discontinuous, the true parameter value for the bootstrap adapts to the model continuity. 
	
	After collecting the bootstrap estimators 
	\[\hat{\theta}^{*}=(\hat{\alpha}^{*\prime},\hat{\gamma}^{*})'=\arg\min_{\theta\in\Theta}\hat{Q}_{n}^{*}(\theta),\] 
we can construct the CIs for the coefficients using the percentiles of either $|\hat{\alpha}_j^{*}-\alpha_{j0}^{*}|$ or $(\hat{\alpha}_j^{*}-\alpha_{j0}^{*})$. 
 Here, $\hat{\alpha}^{*}_{j}$ and $\alpha^{*}_{j0}$ are the $j$th elements of $\hat{\alpha}^{*}$ and $\alpha^{*}_{0}$, respectively.
 The $100(1-\tau)$\% CI for the $j$th element of the coefficients, $\alpha_{j}$, can be constructed by
\begin{equation}
    \label{eq:rbci_asymm}
    CI^{RB}_{n,1-\tau}(\alpha_{j})=\left[ 
    \hat{\alpha}_{j}- \widehat{F}^{*-1}_{n}(1-\tfrac{\tau}{2};\hat{\alpha}^{*}_{j}-\alpha_{j0}^{*}), 
    \hat{\alpha}_{j}- \widehat{F}^{*-1}_{n}(\tfrac{\tau}{2};\hat{\alpha}^{*}_{j}-\alpha_{j0}^{*})
    \right]
\end{equation}
or
\begin{equation}
    \label{eq:rbci_symm}
    CI^{RB(S)}_{n,1-\tau}(\alpha_{j})=\left[ 
    \hat{\alpha}_{j}- \widehat{F}^{*-1}_{n}(1-\tau;|\hat{\alpha}^{*}_{j}-\alpha_{j0}^{*}|), 
    \hat{\alpha}_{j}+ \widehat{F}^{*-1}_{n}(1-\tau;|\hat{\alpha}^{*}_{j}-\alpha_{j0}^{*}|)
    \right],
\end{equation}
        which leads to a symmetric CI.

According to \cref{thm:perc boot} that follows, both CIs are asymptotically (pointwise) valid, and they should provide similar coverage rates close to the nominal rate for any fixed data generating process in large sample. 
However, our Monte Carlo experiments in \cref{sec:mc} show big differences in coverage rates between the two confidence intervals.
Specifically, \eqref{eq:rbci_asymm} shows severe undercoverage while \eqref{eq:rbci_symm} seems to provide much higher coverage rates. 
This phenomenon also appears for the nonparametric bootstrap.
We provide further numerical investigation of the phenomenon in \cref{app:MC_symm}, which suggests challenges for reliable bootstrap inference for the coefficients.
    
	\begin{theorem}
		\label{thm:perc boot}
		Let $\hat{\theta}^{*}$ be obtained by \cref{alg:bootstrap} with $\theta_{0}^{*}$ set as \eqref{eq:perc boot}. 
        
		\noindent
		(\romannumeral 1) When the true model is continuous and Assumptions \ref{ass:gmm}, \ref{ass:dgp}, and \ref{ass:loc} hold,
		\begin{equation*}
			\begin{pmatrix}
				\sqrt{n}(\hat{\alpha}^{*}-\alpha_{0}^{*}) \\
				\sqrt{n}(\hat{\gamma}^{*}-\gamma_{0}^{*})^{2}
			\end{pmatrix}\xrightarrow{d^{*}}
			\begin{pmatrix}
				U - (M_{0}'\Omega^{-1}M_{0})^{-1}M_{0}'\Omega^{-1}HV \\
				V
			\end{pmatrix}\quad \text{ in $P$},
		\end{equation*}
		where $U$ and $V$ are defined as in \cref{thm:asym}.
		
		\noindent
		(\romannumeral 2)  When the true model is discontinuous and Assumptions \ref{ass:gmm}, \ref{ass:dgp}, and \ref{ass:locDis} hold,
		\begin{equation*}
			\begin{pmatrix}
				\sqrt{n}(\hat{\alpha}^{*}-\alpha_{0}^{*}) \\
				\sqrt{n}(\hat{\gamma}^{*}-\gamma_{0}^{*})
			\end{pmatrix}\xrightarrow{d^{*}} N(0,(D_{1}'\Omega^{-1}D_{1})^{-1})\quad \text{ in $P$}.
		\end{equation*}
	\end{theorem}
	
	The asymptotic distributions of the bootstrap estimators in \cref{thm:perc boot}, conditional on the data, match those of the sample estimators for both continuous and discontinuous cases. 	Therefore, the residual bootstrap CI becomes asymptotically valid in a pointwise sense, regardless of whether the model is continuous or discontinuous.
    {We acknowledge that \cref{thm:perc boot} does not guarantee the uniform validity of the bootstrap CI. The difficulty in establishing the uniform validity lies in analyzing asymptotic behaviors of $\mathcal{T}_{n}$ and $w_{n}$ for drifting sequences of the true models. $\mathcal{T}_{n}$ already exhibits an irregular limit distribution even in the pointwise setup, as shown in \cref{thm:continuity test} (i). This paper does not provide a theoretical analysis of whether the uniformity of the residual bootstrap can be achieved. Instead, we conduct Monte Carlo experiments for nearly continuous cases in \cref{sec:mc} and leaves theoretical work on the uniformity of the bootstrap method to future research.}
    
	The key motivation for setting $\theta_{0}^{*}$, the true parameter of the bootstrap dgp, by \eqref{eq:perc boot} is to make $\delta^{*}_{10}+\delta^{*}_{30}\gamma^{*}_{0}$ degenerate fast enough when the underlying model is continuous. The $n^{1/4}$ convergence rate of the unrestricted estimator $\hat{\gamma}$ to $\gamma_0$ is not sufficiently fast. To see this, let the first-derivative of the population moment with respect to $\gamma$ at $\theta$ be 
	\begin{multline}
		\label{eq:boot firstorderderivative}
		G(\theta)=(\delta_{1}+\delta_{3}\gamma)\cdot
		\begin{bmatrix}
			E_{t_{0}}[z_{it_{0}}|\gamma]f_{t_{0}}(\gamma) - E_{t_{0}-1}[z_{it_{0}}|\gamma]f_{t_{0}-1}(\gamma)\\
			\vdots \\
			E_{T}[z_{iT}|\gamma]f_{T}(\gamma) - E_{T-1}[z_{iT}|\gamma]f_{T-1}(\gamma)
		\end{bmatrix}\\
		+
		\begin{bmatrix}
			E_{t_{0}}[z_{it_{0}}\xi_{it_{0}}'\delta_{2}|\gamma]f_{t_{0}}(\gamma) - E_{t_{0}-1}[z_{it_{0}}\xi_{it_{0}-1}'\delta_{2}|\gamma]f_{t_{0}-1}(\gamma)\\
			\vdots \\
			E_{T}[z_{iT}\xi_{iT}'\delta_{2}|\gamma]f_{T}(\gamma) - E_{T-1}[z_{iT}\xi_{iT-1}'\delta_{2}|\gamma]f_{T-1}(\gamma)
		\end{bmatrix},
	\end{multline}
	for which we recall that $x_{it}=(\xi_{it}',q_{it})'$ and that $G(\theta_{0})=0_{k}$ under continuity. 
	For the validity of a bootstrap method under continuity, the degeneracy of the Jacobian should be mimicked by the bootstrap dgp.  
	In our residual bootstrap method, the Jacobian is $G(\theta_{0}^{*})=O_p(n^{-1/2})$. 
	However, it is $G(\hat{\theta})=O_p(n^{-1/4})$ for the standard nonparametric bootstrap. 
	This fails the standard nonparametric bootstrap.
	More formal treatment of the invalidity of the standard nonparametric bootstrap is given in \cref{sec:std boot invalid}.
	
	It is not difficult to check $G(\hat{\theta})=O_{p}(n^{-1/4})$ but not $o_{p}(n^{-1/4})$ under continuity, which is directly implied by $n^{1/4}(\hat{\delta}_{1}+\hat{\delta}_{3}\hat{\gamma})=O_{p}(1)$ but not $o_{p}(1)$ due to \cref{thm:asym}.
	Meanwhile, in our residual bootstrap method, $\delta_{10}^{*}+\delta_{30}^{*}\gamma_{0}^{*}= w_{n}(\hat{\delta}_{1}+\hat{\delta}_{3}\hat{\gamma})+o_{p}(n^{-1/2})=O_{p}(n^{-1/2})$ and $\delta_{20}^{*}=w_{n}\hat{\delta}_{2}=O_{p}(n^{-3/4})$, which leads to $G(\theta_{0}^{*})=O_{p}(n^{-1/2})$. 
    The exact formula for $\delta_{10}^{*}+\delta_{30}^{*}\gamma_{0}^{*}$ is provided in the comment after \cref{lem:boot plim1}.
        
    According to the proof of \cref{thm:perc boot} in \cref{section:boot proofs}, $(\delta_{10}^{*}+\delta_{30}^{*}\gamma_{0}^{*})=O_{p}(n^{-1/2})$ is sufficient for the first-order asymptotic validity when the true model is continuous. This requirement is explicitly stated in the conditions of \cref{lem:boot plim1}. While our choice of $n^{1/4}$ decay rate for $w_{n}$ guarantees this condition, it remains an open question whether there exists a rate of decay for $w_{n}$ that ensures uniform validity.
	
	The idea of shrinking the first-order derivative in our bootstrap is closely related to other bootstrap methods developed for the case when asymptotic distributions of estimators are irregular.
	For example, \cite{chatterjee2011bootstrapping} propose a bootstrap method for the lasso estimator, and \cite{cavaliere2022bootstrap} study bootstrap inference on the boundary of a parameter space.
	Both papers set up the model where the problem appears if the true parameter value is zero, and they obtain true parameters of bootstrap dgps by thresholding unrestricted estimators, i.e.,  $\theta_{j0}^{*}=\hat{\theta}_{j}1\{|\hat{\theta}_{j}|>c_{n}\}$, where $c_{n}$ converges to zero in a proper rate.

	\subsection{Bootstrap for testing continuity}
	\label{sec:continuity bootstrap}
	
	The critical value for the continuity test introduced in \cref{sec:asym continuity} can also be obtained by bootstrapping.
	Recall that $\tilde{\theta}=\arg\min_{\theta\in\Theta_{c}} \hat{Q}_{n}(\theta)$ is the continuity-restricted estimator.
	By setting $\theta_{0}^{*}=\tilde{\theta}$ in \cref{alg:bootstrap}, and collecting the bootstrap test statistic
	\[\mathcal{T}_{n}^{*}=n\left(\min_{\theta\in\Theta_{c}}\hat{Q}_{n}^{*}(\theta)-\min_{\theta\in\Theta}\hat{Q}_{n}^{*}(\theta)\right),\]
	we can get the critical value using the empirical quantile of $\mathcal{T}_{n}^{*}$.
	To run the bootstrap continuity test at size $\tau$, reject the continuity if $\mathcal{T}_{n}>\widehat{F}^{*-1}_{n}(1-\tau;\mathcal{T}_{n}^{*})$, where $\widehat{F}^{*-1}_{n}(1-\tau;\mathcal{T}_{n}^{*})$ is the empirical $(1-\tau)$ quantile of $\mathcal{T}_{n}^{*}$.
	The consistency of the bootstrap is implied by \cref{thm:continuity boot} that follows.
	
	\begin{theorem}
		\label{thm:continuity boot}
		Assume that $\mathcal{T}_{n}^{*}$ is obtained by \cref{alg:bootstrap} with $\theta_{0}^{*}=\tilde{\theta}$.
		
		\noindent
		(\romannumeral 1) When the true model is continuous and Assumptions \ref{ass:gmm}, \ref{ass:dgp}, and \ref{ass:loc} hold,
		\begin{equation*}
			\mathcal{T}_{n}^{*}\xrightarrow{d*} V_{1}-V_{2}+V_{3}\quad \text{ in $P$},
		\end{equation*}
		where the distributions of $V_{1}$, $V_{2}$, and $V_{3}$ are specified in \cref{thm:continuity test}.
		
		\noindent
		(\romannumeral 2) When the model is discontinuous, then $\mathcal{T}_{n}^{*}=O_{p}^{*}(1)$ in $P$.
	\end{theorem}
	
	\cref{thm:continuity boot} (i) shows that the limit distribution of $\mathcal{T}_{n}^{*}$, conditional on the data, is identical to that of $\mathcal{T}_{n}$ under the null hypothesis.
	Moreover, \cref{thm:continuity boot} (ii) says that $\mathcal{T}_{n}^{*}$ is still stochastically bounded, conditionally on the data, when the true model is discontinuous.
	As $\mathcal{T}_{n}$ is shown to be stochastically unbounded under the alternative, according to \cref{thm:continuity test} (ii), the bootstrap continuity test has power against fixed alternatives. 

 \section{Monte Carlo results}
	\label{sec:mc}
	
This section presents Monte Carlo simulations to investigate finite sample performances of our bootstrap methods. The data are generated by
\begin{align}
\nonumber
y_{it} &= \beta_{2}y_{it-1}+\beta_{3}q_{it}+(\delta_{1}+\delta_{2}y_{it-1}+\delta_{3}q_{it})1\{q_{it}>\gamma\}+\sigma e_{it}\\
\nonumber
q_{it} &= \rho q_{it-1}+ u_{it}, \\
\label{eq:mc}
& \text{where } \begin{pmatrix}
    e_{it} \\ u_{it+1}
\end{pmatrix} \overset{iid}{\sim} N\left(\begin{pmatrix} 0 \\ 0 \end{pmatrix},\begin{pmatrix}
    1 & \rho_{eu}\\
    \rho_{eu} & 1
\end{pmatrix} \right),
\end{align}
with $\beta_{2}=0.6$, $\beta_{3}=1$, $\delta_{2}=0$, $\delta_{3}=2$, $\gamma=0.25$, $\sigma=0.5$, $\rho=0.7$, and $\rho_{eu}=0.5$.
Note that \eqref{eq:mc} implies that the threshold variable is weakly exogenous. That is, $E[e_{it}|q_{is}]=0$  for $s\leq t$ while $E[e_{it}|q_{is}]\neq 0$ for $s\geq t+1$.
Additional results when the threshold variable is weakly endogenous are also presented in \cref{app:MC_othdgp}.
This section focuses on comparing bootstrap methods, while results based on the asymptotic method by \cite{seo_dynamic_2016} are reported in \cref{app:MC_asym}.
	
    To investigate how coverage rates of CIs change depending on continuity, we try different values of $\delta_{1}\in\{-0.5, -0.4, -0.3, 0, 0.5\}$, which implies different degrees of (dis)continuity $\delta_{1}+\delta_{3}\gamma\in\{0,0.1,0.2,0.5,1\}$.
	If $\delta_{1}=-0.5$, then $\delta_{1}+\delta_{3}\gamma=0$ and the model is continuous.
	Otherwise, the model is discontinuous.
	As near continuous designs, we try $\delta_{1}+\delta_{3}\gamma=0.1, 0.2$ and check for any poor CI performance. 
	We generate samples of size $n\in\{400, 800, 1600\}$ and $T=6$. 
	The number of repetitions for the Monte Carlo simulations is 2000.
	We use $z_{it}=(y_{it-2},...,y_{i1},q_{it-1},...,q_{i1})'$ for $t=t_{0},\dots, T$ as instruments.
	Since $t_{0}=3$, the total number of the instruments becomes 24.
 The number of bootstrap repetitions is set at 500 for each bootstrap method.
	
	We begin with examining the finite sample coverage probabilities of bootstrap CIs for the threshold location. Specifically, the grid bootstrap CI (Grid-B) is compared with both percentile nonparametric bootstrap CI (NP-B) and symmetric percentile nonparametric bootstrap CI (NP-B(S)) that are defined as follows:
 \begin{align}
    \label{eq:npb_grid_asymm}
    CI^{NPB}_{n,1-\tau}(\gamma) &= 
    \left[ 
    \hat{\gamma}- \widehat{F}^{*-1}_{n}(1-\tfrac{\tau}{2};\hat{\gamma}^{*}-\hat{\gamma}), 
    \hat{\gamma}- \widehat{F}^{*-1}_{n}(\tfrac{\tau}{2};\hat{\gamma}^{*}-\hat{\gamma})
    \right]
    ,\\
    \label{eq:npb_grid_symm}
    CI^{NPB(S)}_{n,1-\tau}(\gamma) &= 
    \left[ 
    \hat{\gamma}- \widehat{F}^{*-1}_{n}(1-\tau;|\hat{\gamma}^{*}-\hat{\gamma}|), 
    \hat{\gamma}+ \widehat{F}^{*-1}_{n}(1-\tau;|\hat{\gamma}^{*}-\hat{\gamma}|)
    \right].
 \end{align}
	
 \cref{tab:ch1} reports the coverage rates of 95\% CIs for the threshold location. 
 First, it shows that the bootstrap CI by NP-B is subject to severe undercoverage in all cases. 
 This is the case even when $\delta_{1}+\delta_{3}\gamma=1$, despite the theoretical validity of NP-B when the model is discontinuous.
 Meanwhile, NP-B(S) exhibits extreme over-coverage in all cases. The large discrepancy between NP-B and NP-B(S) suggests that the distribution of the nonparametric bootstrap statistic $\hat{\gamma}^{*}-\hat{\gamma}$ poorly approximates that of $\hat{\gamma}-\gamma_{0}$, undermining its reliability for inference.

 In contrast, \cref{tab:ch1} shows that Grid-B provides more reasonable coverage rates.
 It seems that a larger jump yields coverage rates closer to the nominal level as a bigger jump is easier to detect. As expected from the uniform validity of Grid-B against near continuity, coverage rates remain valid for all the parameter values, if somewhat over-coveraged near continuity or under smaller sample sizes. 

 \begin{table}[htbp]
    \caption{\label{tab:ch1} Coverage rates of 95\% CIs for the threshold location. 
        Grid-B denotes the grid bootstrap CI defined as \eqref{eq:grid bcs}. 
        NP-B and NP-B(S) denote the percentile and the symmetric percentile CIs by the standard nonparametric bootstrap defined as \eqref{eq:npb_grid_asymm} and \eqref{eq:npb_grid_symm}.
    }
    \centering
    \begin{tabular}[t]{c|c|c c c c c}
        \hline
        \multicolumn{2}{c|}{ } & \multicolumn{5}{c}{$\delta_{1}+\delta_{3}\gamma$} \\
        \hline
        & n & 0 & 0.1 & 0.2 & 0.5 & 1 \\
        \hline
            & 400 & 0.992 & 0.995 & 0.993 & 0.988 & 0.966 \\
            Grid-B & 800 & 0.986 & 0.986 & 0.985 & 0.973 & 0.955 \\
            & 1600 & 0.988 & 0.987 & 0.988 & 0.979 & 0.959 \\
        \hline
            & 400 & 0.484 & 0.491 & 0.494 & 0.524 & 0.631\\
          NP-B & 800 & 0.478 & 0.472 & 0.487 & 0.518 & 0.611\\
          & 1600 & 0.471 & 0.468 & 0.476 & 0.521 & 0.642\\
        \hline
            & 400 & 1.000 & 1.000 & 1.000 & 1.000 & 0.998\\
            NP-B(S) & 800 & 1.000 & 1.000 & 1.000 & 0.999 & 0.994\\
            & 1600 & 1.000 & 1.000 & 1.000 & 1.000 & 0.994\\
          \hline
    \end{tabular}
\end{table}
 
 Compared to Grid-B, NP-B(S) exhibits higher coverage probabilities that are one or almost one for all cases. It indicates that NP-B(S) CIs are overly wide and non-informative. 
 To investigate this further, we examine some power properties as reported in \cref{tab:ch2}. It shows that the NP-B(S) based test for the threshold location is trivial for many parametrizations, specifically when the design is continuous or near-continuous. In contrast, the Grid-B test is more powerful, oftentime twice more powerful than the NP-B(S) test. We report test power instead of CI lengths because of the computational burden associated with Grid-B, which constructs CIs by test inversion.

\begin{table}[htbp]
		
		\caption{\label{tab:ch2} Rejection rates of 5\% level tests for alternative threshold locations $\gamma=\gamma_{0}+c$.
			Grid-B denotes the test using the 95\% grid bootstrap CI defined as \eqref{eq:grid bcs}. 
			NP-B(S) denotes the test using the symmetric percentile CI constructed by the standard nonparametric bootstrap defined as \eqref{eq:npb_grid_symm}.
            }
		\centering
		\begin{tabular}[t]{c|c|c c c c c||c c c c c}
			\hline
			\multicolumn{2}{c|}{ } & \multicolumn{5}{c||}{Grid-B} & \multicolumn{5}{c}{NP-B(S)} \\
			\cline{3-7} \cline{8-12}
			\multicolumn{2}{c|}{ } & \multicolumn{5}{c||}{$\delta_{1}+\delta_{3}\gamma$} & \multicolumn{5}{c}{$\delta_{1}+\delta_{3}\gamma$} \\
			\hline
			c & n & 0 & 0.1 & 0.2 & 0.5 & 1 & 0 & 0.1 & 0.2 & 0.5 & 1\\
			\hline
			& 400 & 0.015 & 0.015 & 0.015 & 0.027 & 0.096 & 0.000 & 0.000 & 0.000 & 0.004 & 0.018\\
			0.10 & 800 & 0.011 & 0.014 & 0.015 & 0.038 & 0.112 & 0.000 & 0.000 & 0.000 & 0.004 & 0.017\\
			& 1600 & 0.017 & 0.020 & 0.021 & 0.040 & 0.125 & 0.000 & 0.000 & 0.002 & 0.004 & 0.023\\
			\hline
			& 400 & 0.020 & 0.030 & 0.042 & 0.100 & 0.281 & 0.002 & 0.004 & 0.009 & 0.043 & 0.135\\
			0.25 & 800 & 0.020 & 0.034 & 0.041 & 0.112 & 0.325 & 0.002 & 0.003 & 0.007 & 0.035 & 0.154\\
			& 1600 & 0.029 & 0.034 & 0.048 & 0.126 & 0.351 & 0.002 & 0.006 & 0.007 & 0.044 & 0.152\\
			\hline
			& 400 & 0.102 & 0.137 & 0.172 & 0.314 & 0.581 & 0.062 & 0.109 & 0.142 & 0.274 & 0.298\\
			0.50 & 800 & 0.114 & 0.162 & 0.207 & 0.362 & 0.632 & 0.078 & 0.117 & 0.169 & 0.310 & 0.327\\
			& 1600 & 0.136 & 0.186 & 0.240 & 0.396 & 0.652 & 0.076 & 0.124 & 0.189 & 0.332 & 0.316\\
			\hline
		\end{tabular}
	\end{table}

Next, we examine the coverage probabilities of the regression coefficients using different bootstrap CIs.
We first report results for percentile bootstrap CIs that use the lower and upper quantiles of the bootstrap distributions.
\cref{tab:ch3} reports the coverage rates of the percentile CIs using the residual bootstrap (R-B) defined as \eqref{eq:rbci_asymm}, and the standard nonparametric bootstrap (NP-B) defined as
\begin{equation}
    \label{eq:npbci_asymm}
        CI^{NPB}_{n,1-\tau}(\alpha_{j})
    = 
    \left[ 
    \hat{\alpha}_{j}- \widehat{F}^{*-1}_{n}(1-\tfrac{\tau}{2};\hat{\alpha}_{j}^{*}-\hat{\alpha}_{j}), 
    \hat{\alpha}_{j}- \widehat{F}^{*-1}_{n}(\tfrac{\tau}{2};\hat{\alpha}_{j}^{*}-\hat{\alpha}_{j})
    \right].
\end{equation}
$\hat{C}$ in \eqref{eq:perc boot} is set as the 50th percentile of the bootstrap distribution of the test statistic $\mathcal{T}_{n}$ under the null hypothesis that the model is continuous, using the bootstrap method explained in \cref{sec:continuity bootstrap} with 500 repetitions. 

As in the threshold inference case, the percentile CIs for the coefficients constructed using NP-B exhibit undercoverage across all specifications and sample sizes. Even when $\delta_{1} + \delta_{3}\gamma = 1$, where the model is discontinuous and NP-B is theoretically valid, the undercoverage remains severe. Although R-B yields higher coverage rates than NP-B, they still fall short of the nominal 95\% level.  As reported in \cref{tab:ch4}, R-B results in wider average CI lengths compared to NP-B, partly accounting for its improved coverage. \cref{app:MC_large} presents the results with a much larger sample size, $n=10000$, and $\delta_{1}+\delta_{3}\gamma\in\{0,1\}$. When $n=10000$, the coverage rates of R-B approach the nominal level, although undercoverage persists for some coefficients.

	\begin{table}[htbp]
		\caption{\label{tab:ch3} Coverage rates of 95\% percentile CIs for the coefficients.
                R-B denotes the percentile CIs by the residual bootstrap defined as \eqref{eq:rbci_asymm}. 
			NP-B denotes the percentile CIs by the standard nonparametric bootstrap defined as \eqref{eq:npbci_asymm}.
		}
		\centering
		\begin{tabular}[t]{c|c|c c c c c||c c c c c}
			\hline
			\multicolumn{2}{c|}{ } & \multicolumn{5}{c||}{R-B} & \multicolumn{5}{c}{NP-B} \\
			\hline
			$\delta_{1}+\delta_{3}\gamma$ & n & $\beta_{2}$ & $\beta_{3}$ & $\delta_{1}$ & $\delta_{2}$ & $\delta_{3}$ & $\beta_{2}$ & $\beta_{3}$ & $\delta_{1}$ & $\delta_{2}$ & $\delta_{3}$\\
			\hline
    		& 400 & 0.839 & 0.780 & 0.746 & 0.815 & 0.801 & 0.799 & 0.691 & 0.627 & 0.712 & 0.709\\
    		0.0 & 800 & 0.837 & 0.790 & 0.721 & 0.807 & 0.806 & 0.790 & 0.723 & 0.607 & 0.725 & 0.716\\
    		& 1600 & 0.849 & 0.782 & 0.727 & 0.840 & 0.835 & 0.833 & 0.709 & 0.602 & 0.754 & 0.718\\
    		\hline
    		& 400 & 0.837 & 0.784 & 0.749 & 0.813 & 0.799 & 0.794 & 0.697 & 0.624 & 0.706 & 0.708\\
    		0.1 & 800 & 0.830 & 0.779 & 0.724 & 0.803 & 0.800 & 0.786 & 0.714 & 0.599 & 0.720 & 0.710\\
    		& 1600 & 0.853 & 0.787 & 0.727 & 0.840 & 0.829 & 0.827 & 0.700 & 0.598 & 0.760 & 0.719\\
    		\hline
    		& 400 & 0.838 & 0.786 & 0.749 & 0.819 & 0.811 & 0.794 & 0.701 & 0.623 & 0.713 & 0.716\\
    		0.2 & 800 & 0.833 & 0.776 & 0.720 & 0.803 & 0.794 & 0.784 & 0.707 & 0.585 & 0.718 & 0.712\\
    		& 1600 & 0.855 & 0.789 & 0.728 & 0.846 & 0.832 & 0.830 & 0.707 & 0.606 & 0.764 & 0.722\\
    		\hline
    		& 400 & 0.836 & 0.775 & 0.739 & 0.820 & 0.802 & 0.787 & 0.703 & 0.601 & 0.718 & 0.724\\
    		0.5 & 800 & 0.841 & 0.789 & 0.732 & 0.815 & 0.807 & 0.787 & 0.714 & 0.602 & 0.716 & 0.727\\
    		& 1600 & 0.843 & 0.799 & 0.728 & 0.826 & 0.834 & 0.815 & 0.717 & 0.595 & 0.753 & 0.737\\
    		\hline
    		& 400 & 0.858 & 0.815 & 0.745 & 0.832 & 0.805 & 0.800 & 0.741 & 0.627 & 0.741 & 0.743\\
    		1.0 & 800 & 0.858 & 0.827 & 0.749 & 0.846 & 0.820 & 0.808 & 0.731 & 0.620 & 0.741 & 0.738\\
    		& 1600 & 0.863 & 0.846 & 0.759 & 0.830 & 0.837 & 0.820 & 0.738 & 0.622 & 0.761 & 0.747\\
    		\hline
		\end{tabular}
	\end{table}
	
	\begin{table}[htbp]
		\caption{\label{tab:ch4} Ratios of the average lengths of 95\% percentile CIs for the coefficients.
			R-B denotes the percentile CIs by the residual bootstrap defined as \eqref{eq:rbci_asymm}. 
			NP-B denotes the percentile CIs by the standard nonparametric bootstrap defined as \eqref{eq:npbci_asymm}.
            }
		\centering
		\begin{tabular}[t]{c|c|c c c c c}
			\hline
			\multicolumn{2}{c|}{ } & \multicolumn{5}{c}{Ratios of average lengths of CIs:} \\
			\multicolumn{2}{c|}{ } & \multicolumn{5}{c}{R-B / NP-B} \\
			\hline
			$\delta_{1}+\delta_{3}\gamma$ & n & $\beta_{2}$ & $\beta_{3}$ & $\delta_{1}$ & $\delta_{2}$ & $\delta_{3}$ \\
			\hline
			& 400 & 1.076 & 1.091 & 1.099 & 1.074 & 1.046 \\
                0.0 & 800 & 1.081 & 1.086 & 1.093 & 1.070 & 1.046 \\
                 & 1600 & 1.088 & 1.100 & 1.111 & 1.083 & 1.057 \\
                 \hline
                 & 400 & 1.087 & 1.098 & 1.101 & 1.074 & 1.047 \\
                0.1 & 800 & 1.080 & 1.082 & 1.090 & 1.075 & 1.043 \\
                 & 1600 & 1.086 & 1.102 & 1.111 & 1.077 & 1.057 \\
                 \hline
                 & 400 & 1.080 & 1.088 & 1.097 & 1.074 & 1.047 \\
                0.2 & 800 & 1.079 & 1.089 & 1.094 & 1.075 & 1.047 \\
                 & 1600 & 1.085 & 1.100 & 1.106 & 1.077 & 1.054 \\
                 \hline
                 & 400 & 1.097 & 1.100 & 1.100 & 1.083 & 1.056 \\
                0.5 & 800 & 1.083 & 1.095 & 1.089 & 1.076 & 1.051 \\
                 & 1600 & 1.098 & 1.110 & 1.098 & 1.089 & 1.059 \\
                 \hline
                 & 400 & 1.164 & 1.159 & 1.084 & 1.114 & 1.074 \\
                1.0 & 800 & 1.158 & 1.159 & 1.079 & 1.109 & 1.076 \\
                 & 1600 & 1.158 & 1.177 & 1.084 & 1.109 & 1.079 \\
			\hline
		\end{tabular}
	\end{table}

Finally, we report the coverage rates of symmetric percentile CIs for the coefficients that are constructed using the nonparametric bootstrap (NP-B(S)) defined as
\begin{equation}
    \label{eq:npbci_symm}
    CI^{NPB(S)}_{n,1-\tau}(\alpha_{j})=
    \left[ 
    \hat{\alpha}_{j}- \widehat{F}^{*-1}_{n}(1-\tau;|\hat{\alpha}^{*}_{j}-\hat{\alpha}_{j}|), 
    \hat{\alpha}_{j}+ \widehat{F}^{*-1}_{n}(1-\tau;|\hat{\alpha}^{*}_{j}-\hat{\alpha}_{j}|)
    \right].
\end{equation}
and the residual bootstrap (R-B(S)) defined as \eqref{eq:rbci_symm}.
Tables \ref{tab:ch5} and \ref{tab:ch6} show the coverage rates and the ratios of the average lengths of CIs by the two bootstrap methods.

When the symmetric percentile CIs are used for the coefficients, \cref{tab:ch5} shows that the coverage rates increase, as also observed in \cref{tab:ch1}.
However, R-B(S) yields lower coverage rates than NP-B(S) and even produces undercoverage for the dgp reported in \cref{app:MC_othdgp}.
Nevertheless, R-B(S) tends to return wider CIs than NP-B(S) according to \cref{tab:ch6}.

NP-B(S) may appear to be the most suitable method for inference on the coefficients, given its higher coverage rates and shorter average CI lengths.
However, a more detailed numerical analysis in \cref{app:MC_symm} reveals an undesirable property of the nonparametric bootstrap: the conditional distribution of the bootstrap statistic $\sqrt{n}(\hat{\alpha}^{*}-\hat{\alpha})$ is not centered at zero. 
This misalignment also results in the unexpected relationship between the coverage and the average length of CIs in Tables \ref{tab:ch5} and \ref{tab:ch6}, which is further illustrated in \cref{app:MC_symm}. 
These findings highlight the difficulty of reliable inference for the coefficients $\beta$ and $\delta$. 
A more comprehensive theoretical and methodological investigation is needed to address these challenges in future research.

\begin{table}[htbp]
    \caption{\label{tab:ch5} Coverage rates of 95\% symmetric percentile CIs for the coefficients. 
        R-B(S) denotes the symmetric percentile CIs by the residual bootstrap defined as \eqref{eq:rbci_symm}. 
        NP-B(S) denotes the symmetric percentile CIs by the standard nonparametric bootstrap defined as \eqref{eq:npbci_symm}.
    }
    \centering
    \begin{tabular}[t]{c|c|c c c c c||c c c c c}
        \hline
            \multicolumn{2}{c|}{ } & \multicolumn{5}{c||}{R-B(S)} & \multicolumn{5}{c}{NP-B(S)}  \\
			\hline
			$\delta_{1}+\delta_{3}\gamma$ & n & $\beta_{2}$ & $\beta_{3}$ & $\delta_{1}$ & $\delta_{2}$ & $\delta_{3}$ & $\beta_{2}$ & $\beta_{3}$ & $\delta_{1}$ & $\delta_{2}$ & $\delta_{3}$\\
        \hline
        & 400 & 0.964 & 0.976 & 0.980 & 0.974 & 0.930 & 0.996 & 0.996 & 0.996 & 0.992 & 0.982\\
        0.0 & 800 & 0.951 & 0.974 & 0.971 & 0.967 & 0.931 & 0.987 & 0.992 & 0.995 & 0.988 & 0.976\\
        & 1600 & 0.955 & 0.972 & 0.964 & 0.961 & 0.923 & 0.983 & 0.994 & 0.995 & 0.980 & 0.977\\
        \hline
        & 400 & 0.964 & 0.976 & 0.979 & 0.974 & 0.933 & 0.994 & 0.993 & 0.995 & 0.991 & 0.982\\
        0.1 & 800 & 0.952 & 0.975 & 0.970 & 0.968 & 0.935 & 0.990 & 0.992 & 0.995 & 0.989 & 0.978\\
        & 1600 & 0.959 & 0.975 & 0.973 & 0.961 & 0.924 & 0.986 & 0.995 & 0.997 & 0.979 & 0.977\\
        \hline
        & 400 & 0.963 & 0.974 & 0.978 & 0.977 & 0.939 & 0.995 & 0.993 & 0.997 & 0.993 & 0.986\\
        0.2 & 800 & 0.959 & 0.972 & 0.977 & 0.974 & 0.929 & 0.992 & 0.994 & 0.996 & 0.987 & 0.978\\
        & 1600 & 0.958 & 0.972 & 0.976 & 0.964 & 0.933 & 0.986 & 0.995 & 0.996 & 0.979 & 0.980\\
        \hline
        & 400 & 0.964 & 0.971 & 0.982 & 0.978 & 0.940 & 0.992 & 0.994 & 0.998 & 0.994 & 0.989\\
        0.5 & 800 & 0.960 & 0.973 & 0.987 & 0.974 & 0.945 & 0.991 & 0.994 & 0.998 & 0.988 & 0.985\\
        & 1600 & 0.957 & 0.977 & 0.985 & 0.970 & 0.945 & 0.985 & 0.996 & 0.998 & 0.981 & 0.987\\
        \hline
        & 400 & 0.970 & 0.982 & 0.985 & 0.984 & 0.967 & 0.991 & 0.995 & 0.992 & 0.991 & 0.993\\
        1.0 & 800 & 0.968 & 0.982 & 0.988 & 0.981 & 0.967 & 0.992 & 0.993 & 0.995 & 0.989 & 0.994\\
        & 1600 & 0.960 & 0.981 & 0.987 & 0.972 & 0.963 & 0.989 & 0.995 & 0.995 & 0.988 & 0.989\\
        \hline
    \end{tabular}
\end{table}

\begin{table}[htbp]
    \caption{\label{tab:ch6} Ratios of the average lengths of 95\% symmetric percentile CIs for the coefficients.
        R-B(S) denotes the symmetric percentile CIs by the residual bootstrap defined as \eqref{eq:rbci_symm}. 
        NP-B(S) denotes the symmetric percentile CIs by the standard nonparametric bootstrap defined as \eqref{eq:npbci_symm}.
        }
    \centering
    \begin{tabular}[t]{c|c|c c c c c}
        \hline
        \multicolumn{2}{c|}{ } & \multicolumn{5}{c}{Ratios of average lengths of CIs:} \\
        \multicolumn{2}{c|}{ } & \multicolumn{5}{c}{R-B(S) / NP-B(S)} \\
        \hline
        $\delta_{1}+\delta_{3}\gamma$ & n & $\beta_{2}$ & $\beta_{3}$ & $\delta_{1}$ & $\delta_{2}$ & $\delta_{3}$ \\
        \hline
        & 400 & 1.017 & 1.035 & 1.008 & 0.996 & 1.010\\
            0.0 & 800 & 1.033 & 1.037 & 1.007 & 1.004 & 1.018\\
             & 1600 & 1.040 & 1.046 & 1.012 & 1.015 & 1.014\\
             \hline
             & 400 & 1.028 & 1.040 & 1.008 & 0.996 & 1.012\\
            0.1 & 800 & 1.032 & 1.033 & 1.000 & 1.004 & 1.015\\
             & 1600 & 1.039 & 1.047 & 1.011 & 1.020 & 1.016\\
             \hline
             & 400 & 1.022 & 1.035 & 1.003 & 0.996 & 1.012\\
            0.2 & 800 & 1.032 & 1.039 & 1.001 & 1.004 & 1.015\\
             & 1600 & 1.039 & 1.048 & 1.009 & 1.025 & 1.016\\
             \hline
             & 400 & 1.037 & 1.046 & 0.991 & 1.014 & 1.016\\
            0.5 & 800 & 1.044 & 1.045 & 0.991 & 1.008 & 1.024\\
             & 1600 & 1.052 & 1.056 & 0.996 & 1.035 & 1.022\\
             \hline
             & 400 & 1.101 & 1.107 & 0.989 & 1.042 & 1.042\\
            1.0 & 800 & 1.096 & 1.111 & 0.988 & 1.039 & 1.052\\
             & 1600 & 1.115 & 1.136 & 0.996 & 1.051 & 1.048\\
        \hline
    \end{tabular}
\end{table}

	\section{Empirical example}
	\label{sec:empirical}
	
	Our empirical example examines a firm's investment decision model that incorporates financial constraints, as in \cite{hansen_threshold_1999} and \cite{seo_dynamic_2016}.
	In a perfect financial market, firms can borrow as much money as they need to finance their investment projects, regardless of their financial conditions. 
	Therefore, the financial conditions of firms are irrelevant to their investment decisions. 
	However, in an imperfect financial market, some firms may be restricted in their access to external financing. These firms are said to be financially constrained. 
	Financially constrained firms are more sensitive to the availability of internal financing, as they cannot rely on external financing to fund their investment projects.
	
	\cite{fazzari_financing_1988} argue that firms' investments are positively related to their cash flow if they are financially constrained, where those firms are identified by low dividend payments. \cite{hansen_threshold_1999} applies the threshold panel regression more systematically to show that a more positive relationship between investment and cash flow is present for firms with higher leverage. 
	
	Since there are multiple candidate measures of the financial constraint for the threshold variable, we compare the following three dynamic panel threshold models:
	\begin{align}
		\label{eq:example1}
		I_{it} &= \eta_{i}+\xi_{it-1}'\beta+(\delta_{1}+\xi_{it-1}'\delta_{2}+LEV_{it-1}\delta_{3})1\{LEV_{it-1}>\gamma\}+\epsilon_{it} \\
		\label{eq:example2}
		I_{it} &= \eta_{i}+\xi_{it-1}'\beta+(\delta_{1}+\xi_{it-1}'\delta_{2}+TQ_{it-1}\delta_{3})1\{TQ_{it-1}>\gamma\}+\epsilon_{it} \\
		\label{eq:example3}
		I_{it} &= \eta_{i}+\xi_{it-1}'\beta+(\delta_{1}+TQ_{it-1}\delta_{3})1\{TQ_{it-1}>\gamma\}+\epsilon_{it} 
	\end{align}
	where $\xi_{it-1}=(I_{it-1},CF_{it},PPE_{it-1},ROA_{it-1})'$. 
	Here, $I_{it}$ is investment, $CF_{it}$ is cash flow, $PPE_{it}$ is property, plant and equipment, and $ROA_{it}$ is return on assets. 
	$I_{it}$, $CF_{it}$ and $PPE_{it}$ are normalized by total assets. 
	We have two candidate threshold variables, $LEV_{it}$ and $TQ_{it}$, which are leverage and Tobin's Q, respectively. 
	Choice of the regressors and threshold variables is based on previous works like \cite{hansen_threshold_1999} and \cite{lang1996leverage}.
	Note that the regression model \eqref{eq:example3} is nested within \eqref{eq:example2} and it is closer to a continuous threshold model.
	
	Unlike the previous works, we do not need to assume either continuity or discontinuity for valid inferences since the bootstrap methods in this paper are adaptive to each case. 
	With an assumption that the regressors are predetermined, we use the variables dated one period before as instruments. 
	Hence, the instruments include $I_{t-2}$, $CF_{t-1}$, $PPE_{t-2}$, $ROA_{t-2}$ added by $LEV_{t-2}$ or $TQ_{t-2}$ for each period.
	
	We construct a balanced panel of 1459 U.S. firms, excluding finance and utility firms, from 2010 to 2019 available in Compustat.
	To deal with extreme values, we drop firms if any of their non-threshold variables' values fall within the top or bottom 0.5\% tails.
	Moreover, we exclude firms whose Tobin's Q is larger than 5 for more than 5 years when the threshold variable is Tobin's Q, leaving 1222 firms in the sample.
	Meanwhile, \cite{STREBULAEV20131} claims that firms with large CEO ownership or CEO-friendly boards show persistent zero-leverage behavior.
	To prevent our threshold regression from capturing corporate governance characteristics rather than financial constraints, we exclude firms whose leverage is zero for more than half of the time periods when leverage is the threshold variable, leaving 1056 firms in the sample.

\begin{figure}[htbp]
		\centering
		\caption{
            Threshold inference results via the grid bootstrap.
			Panels (a), (b), and (c) are for the models \eqref{eq:example1}, \eqref{eq:example2}, and \eqref{eq:example3}, respectively.
			Black solid lines in each subplot denote the test statistics, red dashed lines denote the 5\% size bootstrapped critical values, and horizontal blue arrows visualize the 95\% CIs.
			The regions where the test statistics are below the bootstrapped critical values become the CIs for the threshold locations.}
		\label{figure:gridboot}
		\begin{subfigure}[b]{.3\textwidth}
			\centering
			\caption{}
			\includegraphics[width=\linewidth]{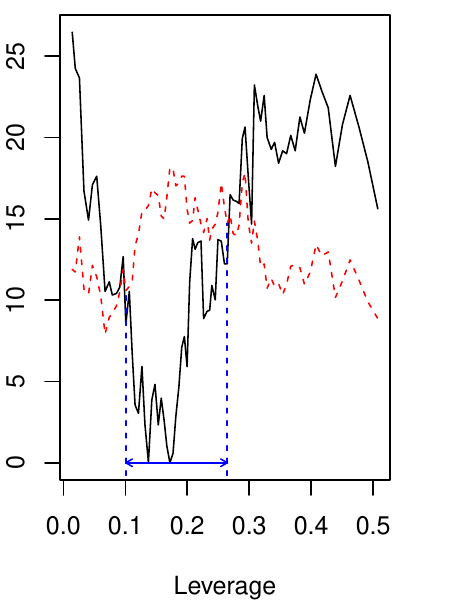}
		\end{subfigure}
		\hfill
		\begin{subfigure}[b]{.3\textwidth}
			\centering
			\caption{}
			\includegraphics[width=\linewidth]{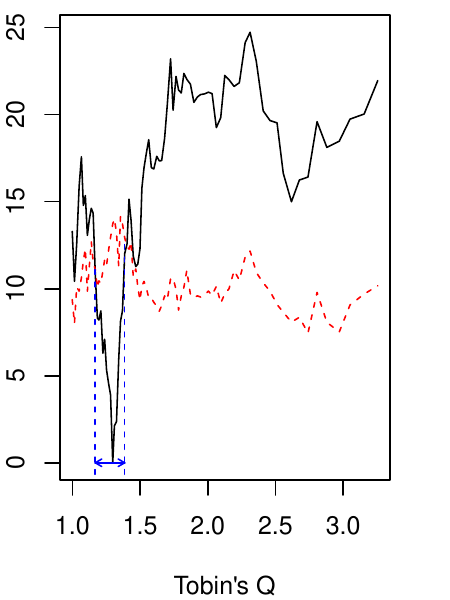}
		\end{subfigure}
		\hfill
		\begin{subfigure}[b]{.3\textwidth}
			\centering
			\caption{}
			\includegraphics[width=\linewidth]{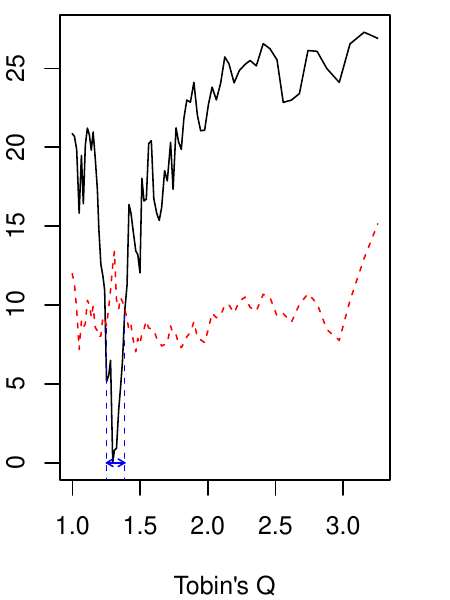}
		\end{subfigure}
	\end{figure}

	\cref{table:empirical} reports the estimates and 95\% CIs for \eqref{eq:example1} and \eqref{eq:example2}, and \cref{table:empirical1} for \eqref{eq:example3}.
	\cref{figure:gridboot} visualizes how the grid bootstrap CIs are obtained.
	The CIs for the coefficients are constructed by using the percentiles obtained from the residual bootstrap, defined as \eqref{eq:rbci_symm}\footnote{The symmetric percentile CIs via residual bootstrap that use the 0.95 quantiles of $|\hat{\alpha}_{j}^{*}-\alpha_{j0}^{*}|$'s return similar results, unlike in Monte Carlo results from \cref{sec:mc}. We report them in \cref{sec:empirical_diffboot}.}. 
	$\hat{C}$ for the precentile bootstrap is set at the 50th percentile of the bootstrap statistic for the continuity test, explained in \cref{sec:continuity bootstrap}.
	For the threshold locations, the CIs are obtained by the grid bootstrap with convexification.
	For the grid bootstrap, we make 500 bootstrap draws for each grid point. 
	The grids of the threshold locations have 81 points from the 10th percentile to the 90th percentile of the threshold variables, and there are equal number of observations between two consecutive points. 
	\cref{table:empirical} and \cref{table:empirical1} also report the bootstrap p-values for the continuity and linearity tests by the bootstrap methods explained in \cref{sec:continuity bootstrap} and \cref{sec:linearity bootstrap}, respectively. 
	The null hypothesis of the linearity test is $\mathcal{H}_{0}:\delta=(0,...,0)'$, which implies no threshold effects.

	We find supporting evidence for the presence of the threshold effect when the threshold variable is Tobin's Q, but the statistical evidence is not strong for the leverage threshold model. 
	\cref{table:empirical} and \cref{table:empirical1} report the bootstrap p-values at .135, .011, and .011, for specifications \eqref{eq:example1} - \eqref{eq:example3}, respectively.
	The statistical evidence to reject the continuity is not trivial for all specifications and gets stronger when it is the restricted model using Tobin's Q. 
	The estimated bootstrap p-values are .028 and .004 for the unrestricted and the restricted using Tobin's Q. 
	Furthermore, the confidence interval for the threshold location is narrower for the restricted model \eqref{eq:example3} than for the unrestricted model \eqref{eq:example2}.  

        \begin{table}[htbp]
            \centering
            \caption{Estimates and 95\% confidence intervals for the models \eqref{eq:example1} and \eqref{eq:example2}. Columns (a) and (b) report results of \eqref{eq:example1} and \eqref{eq:example2}, respectively. The percentile of each threshold location value is shown in parentheses below each value. The significance levels for the coefficients are given by stars: * - 10\%, ** - 5\% and *** - 1\%.}
            \label{table:empirical}
            \begin{tabular}{|llrr|llrr|}
            \hline
            \multicolumn{4}{|c|}{(a)} & \multicolumn{4}{c|}{(b)}\\
            \hline
            \hline
              & est. & \multicolumn{2}{c|}{[95\% CI]} &  & est. & \multicolumn{2}{c|}{[95\% CI]} \\ 
              \hline
              \multicolumn{4}{|l|}{\underline{Lower regime}}&\multicolumn{4}{|l|}{\underline{Lower regime}}\\
              $I_{t-1}$ & 0.778** & 0.124 & 1.154 & $I_{t-1}$ & 0.252 & -0.258 & 0.724 \\ 
		$CF_{t-1}$ & 0.047 & -0.034 & 0.145 & $CF_{t-1}$ & 0.266* & -0.003 & 0.535 \\ 
		$PPE_{t-1}$ & -0.147 & -0.385 & 0.171 & $PPE_{t-1}$ & 0.027 & -0.103 & 0.264 \\ 
		$ROA_{t-1}$ & -0.032 & -0.132 & 0.047 & $ROA_{t-1}$ & -0.017 & -0.180 & 0.090 \\ 
		$LEV_{t-1}$ & 0.231 & -0.843 & 1.849 & $TQ_{t-1}$ & 0.246* & -0.031 & 0.577 \\ 
               \multicolumn{4}{|l|}{\underline{Upper regime}}&\multicolumn{4}{|l|}{\underline{Upper regime}}\\
            $I_{t-1}$ & -0.154 & -0.717 & 0.551 & $I_{t-1}$ & 0.410 & -0.049 & 0.751 \\ 
		$CF_{t-1}$ & 0.148 & -0.015 & 0.326 & $CF_{t-1}$ & 0.081** & 0.021 & 0.200 \\ 
		$PPE_{t-1}$ & -0.291* & -0.519 & 0.015 & $PPE_{t-1}$ & 0.044 & -0.214 & 0.398 \\ 
		$ROA_{t-1}$ & 0.013 & -0.066 & 0.113 & $ROA_{t-1}$ & 0.050* & -0.019 & 0.153 \\ 
		$LEV_{t-1}$ & -0.081 & -0.234 & 0.037 & $TQ_{t-1}$ & 0.005 & -0.004 & 0.012 \\ 
               \multicolumn{4}{|l|}{\underline{Difference between regimes}}&\multicolumn{4}{|l|}{\underline{Difference between regimes}}\\
            intercept & 0.068 & -0.024 & 0.200 & intercept & 0.236* & -0.014 & 0.580 \\ 
		$I_{t-1}$ & -0.932** & -1.830 & -0.097 & $I_{t-1}$ & 0.158 & -0.559 & 0.843 \\ 
		$CF_{t-1}$ & 0.101 & -0.107 & 0.322 & $CF_{t-1}$ & -0.185 & -0.479 & 0.108 \\ 
		$PPE_{t-1}$ & -0.144 & -0.519 & 0.134 & $PPE_{t-1}$ & 0.017 & -0.227 & 0.275 \\ 
		$ROA_{t-1}$ & 0.045 & -0.111 & 0.232 & $ROA_{t-1}$ & 0.066 & -0.074 & 0.287 \\ 
		$LEV_{t-1}$ & -0.312* & -1.893 & 0.792 & $TQ_{t-1}$ & -0.242* & -0.573 & 0.038 \\ 
               \multicolumn{4}{|l|}{\underline{Threshold}} & \multicolumn{4}{|l|}{\underline{Threshold}}\\
            $LEV_{t-1}$ & 0.172 & 0.101 & 0.265 & $TQ_{t-1}$ & 1.298 & 1.169 & 1.386 \\ 
               & (38\%) & (24\%) & (58\%) &  & (30\%) & (21\%) & (36\%) \\ 
               \multicolumn{4}{|l|}{\underline{Testing (p-val)}} & \multicolumn{4}{|l|}{\underline{Testing (p-val)}}\\
            Linearity & 0.135 &  &  & Linearity & 0.011 &  &  \\ 
              Continuity & 0.033 &  &  & Continuity & 0.028 &  &  \\ 
              \hline
              \end{tabular}
        \end{table}

	\begin{table}[htbp]
            \centering
            \caption{Estimates and 95\% confidence intervals for the model \eqref{eq:example3}. The percentile of each threshold location value is shown in parentheses below each value. The significance levels for the coefficients are given by stars: * - 10\%, ** - 5\% and *** - 1\%.}
            \label{table:empirical1}
            \begin{tabular}{|llrr|}
            \hline
              & est. & \multicolumn{2}{c|}{[95\% CI]} \\ 
              \hline
			\multicolumn{4}{|l|}{\underline{Coefficients}}  \\
         $I_{t-1}$ & 0.392*** & 0.304 & 0.539 \\ 
		$CF_{t-1}$ & 0.122*** & 0.084 & 0.154 \\ 
		$PPE_{t-1}$ & 0.076 & -0.027 & 0.271 \\ 
		$ROA_{t-1}$ & 0.027*** & 0.006 & 0.046 \\ 
		$TQ_{t-1}1\{TQ_{t-1}\leq\gamma\}$ & 0.298** & 0.073 & 0.571 \\ 
		$TQ_{t-1}1\{TQ_{t-1}>\gamma\}$ & 0.008** & 0.001 & 0.015 \\ 
			\multicolumn{4}{|l|}{\underline{Difference between regimes}}  \\
            intercept & \;0.275** & 0.010 & 0.540 \\ 
		$TQ_{t-1}$ & -0.290** & -0.562 & -0.018 \\ 
			\multicolumn{4}{|l|}{\underline{Threshold}}\\
            $TQ_{t-1}$ & 1.298 & 1.253 & 1.386 \\ 
               & (30\%) & (27\%) & (36\%) \\ 
			\multicolumn{4}{|l|}{\underline{Testing (p-val)}}\\
            Linearity & 0.011 &  &  \\ 
              Continuity & 0.004 &  &  \\ 
              \hline
              \end{tabular}
        \end{table}
        
	A notable finding concerning the coefficients estimates is that the relationship between cash flow and investment is positive and has larger magnitude for the low Tobin's Q firms and the high leverage firms compared to their other respective regimes, although they are not statistically significant at 5\% level.
	Even though the sign and magnitude of the estimates align with the observations by \cite{lang1996leverage} and \cite{hansen_threshold_1999} that a firm is subject to financial constraints when its Tobin's Q is low or leverage is high, there is uncertainty in the interpretation of our results due to the lack of statistical significance.
	
	Next, the autoregressive coefficient of the lagged investment is significant at 5\% level in the low leverage regime and is larger than in the high leverage regime. 
	This lends supporting evidence for the presence of asymmetric dynamics in investment, akin to the dynamics of leverage analyzed by \cite{dang_asymmetric_2012}. 
	In the meantime, we note that the autoregressive coefficients for the low and high leverage regimes in Column (a) are 0.778 and -0.154, respectively, which appear more extreme than findings of the literature where the estimates are between 0.1 and 0.5, e.g., \cite{blundell_investment_1992}. 
	The autoregressive coefficients in the Column (b) are more in line with these estimates. Since the changes of the estimated coefficients in Column (b) are moderate, we also estimate the restricted model \eqref{eq:example3}. 
	
	Turning to \cref{table:empirical1}, we observe that the differences between the coefficients of the two regimes become significant at 5\% level, and the CI for the threshold location becomes narrower while the estimate of the threshold location remains close to the estimate under the unrestricted model.
	The autoregressive coefficient of the lagged investment and the sensitivity of investment to both cash flow and return on assets are all positive and significant. 
	The effect of Tobin's Q is both positive and significant for both high and low Tobin's Q regimes, but it almost disappears once it surpasses the threshold location.
	This suggests that low Tobin's Q is related to low investment but higher Tobin's Q does not cause higher investment once it reaches some level.

\section{Conclusion}
\label{sec:conclude}

This paper studies the asymptotic properties of the GMM estimator in dynamic panel threshold models, showing that the limiting distribution depends critically on whether the true model exhibits a kink or a jump at the threshold. We demonstrate that the standard nonparametric bootstrap is inconsistent when the true model has a kink. To address this, we propose alternative bootstrap procedures for constructing confidence intervals for the threshold location and the model coefficients, which are shown to be consistent regardless of the model's continuity. In particular, we establish that the grid bootstrap for the threshold parameter is uniformly valid. 
Monte Carlo simulations confirm that the grid bootstrap outperforms the standard bootstrap in finite samples.

Several directions remain for future research. Our simulation results reveal highly asymmetric bootstrap distributions for the coefficient estimates, which distort finite sample inference. This highlights the need for a more thorough theoretical understanding of the bootstrap's behavior. In particular, whether uniform validity of the bootstrap for coefficients is achievable remains an important open question. Extensions of our bootstrap algorithms to incorporate latent group structures, interactive fixed effects, or threshold indices, as studied in \citet{miao2020panel}, \citet{miao2020panel_interactive}, and \citet{seo2007smoothed, lee2021factor}, respectively, would also be valuable.
	\clearpage

	\bibliographystyle{apalike}
	\bibliography{RobustBootDPTR.bib}{}

	\clearpage

		\begin{appendices}
			\crefalias{section}{appendix}
			\numberwithin{equation}{section}
			\numberwithin{table}{section}
			\numberwithin{figure}{section}
			
			\paragraph{Additional Notations.}
			
			For $k,p\in\mathbb{N}$, $0_{k\times p}$ denotes $k\times p$ a matrix whose elements are all zero.
			``$\rightsquigarrow$'' denotes the weak convergence as in section 1.3 of \cite{van_der_vaart_weak_1996}.
			$\|\cdot\|$ is a norm for either vectors or matrices.
			For a vector, it is the Euclidean norm.
			For a matrix, it is the Frobenius norm, i.e., $\|M\|=\sqrt{tr(M'M)}$ for a matrix $M$.
			
			\section{Proofs for \cref{sec:asym}.}
			\label{sec:proofs}
			
			\subsection{Proof of \texorpdfstring{\cref{thm:id}}{\cref{thm:id}}.}
			Note that $E[z_{it}(\Delta y_{it}-\Delta x_{it}'\beta-1_{it}(\gamma)'X_{it}\delta)] = -E[z_{it}\Delta x_{it}'](\beta-\beta_{0})-E[z_{it}1_{it}(\gamma)'X_{it}]\delta+E[z_{it}1_{it}(\gamma_{0})'X_{it}\delta_{0}]$ due to $\Delta y_{it}=\Delta x_{it}'\beta_{0}+ 1_{it}(\gamma_{0})'X_{it}\delta_{0}+\Delta\epsilon_{it}$.
			Hence, the population moment equation is
			$g_{0}(\theta)=M_{10}(\beta-\beta_{0}) +M_{20}(\gamma)\delta -M_{20}\delta_{0}=
			\left[\begin{array}{c;{2pt/2pt}c}
				M_{0}(\gamma) & M_{20}\delta_{0} \\
			\end{array}\right]\times((\beta'-\beta_{0}',\delta'),-1)',$
			when $\gamma\neq\gamma_{0}$.
			The condition (ii) of \cref{thm:id} implies that $\left[\begin{array}{c;{2pt/2pt}c}
				M_{0}(\gamma) & M_{20}\delta_{0} \\
			\end{array}\right]$ has full column rank, and hence $g_{0}(\theta)\neq 0_{k}$ if $\gamma\neq\gamma_{0}$.
			$g_{0}(\theta)=M_{0}\times(\alpha-\alpha_{0})$, when $\gamma=\gamma_{0}$.
			The condition (i) of \cref{thm:id} implies that $M_{0}\times(\alpha-\alpha_{0})$ is not zero if  $\alpha\neq\alpha_{0}$.
			Therefore, $g_{0}(\theta)\neq 0_{k}$ if $\theta\neq \theta_{0}$, and $g_{0}(\theta)=0_{k}$ if $\theta=\theta_{0}$, which is the standard identification condition in the literature, e.g., Section 2.2.3 in \cite{newey_chapter_1994}.

			\subsection{Proof of \texorpdfstring{\cref{thm:asym}}{\cref{thm:asym}}.}
			
			To obtain limit distribution of $\hat{\theta}$, we first establish consistency of $\hat{\theta}$ to $\theta_{0}$ and rate of $\hat{\theta}$'s convergence.
			Then, we show asymptotic distribution of the estimates using rescaled versions of the parameters and criterions.
			
			\subsubsection{Consistency.} 
			Constrained estimator of the coefficients, $\hat{\alpha}(\gamma)=\arg\min_{\alpha\in A}\hat{Q}_{n}(\alpha,\gamma)$, given a fixed $\gamma$ can be expressed as
			\[
			\hat{\alpha}(\gamma)
			=
			-(\bar{M}_{n}(\gamma)'W_{n}\bar{M}_{n}(\gamma))^{-1}\bar{M}_{n}(\gamma)'W_{n}\bar{v}_{n}
			\]
			where
			\[\bar{v}_{n}=-\bar{M}_{n}\alpha_{0}+u_{n},\quad u_{n}=\frac{1}{n}\sum_{i=1}^{n}\begin{pmatrix}
				z_{it_{0}}\Delta\epsilon_{it_{0}}\\
				\vdots\\
				z_{iT}\Delta\epsilon_{iT}
			\end{pmatrix}.\]
			Therefore,
			\[
			\hat{\alpha}(\gamma)
			=-
			(\bar{M}_{n}(\gamma)'W_{n}\bar{M}_{n}(\gamma))^{-1}\bar{M}_{n}(\gamma)'W_{n}(-\bar{M}_{n}\alpha_{0}+u_{n}).
			\]
			Define profiled criterion with respect to $\gamma$ by $\tilde{g}_{n}(\gamma) =\bar{g}_{n}(\hat{\alpha}(\gamma),\gamma)$ and $\tilde{Q}_{n}(\gamma)=\tilde{g}_{n}(\gamma)'W_{n}\tilde{g}_{n}(\gamma)$. 
			The threshold location estimator is $\hat{\gamma}=\arg\min_{\gamma\in\Gamma}\tilde{g}_{n}(\gamma)'W_{n}\tilde{g}_{n}(\gamma)$.
			By the law of large numbers (LLN), $u_{n}\xrightarrow{p}0$.
			By the uniform law of large numbers (ULLN) in \cref{lem:ULLN}, $\bar{M}_{n}(\gamma)\xrightarrow{p}M_{0}(\gamma)$ uniformly with respect to $\gamma\in\Gamma$.
			Hence, $\hat{\gamma}\xrightarrow{p}\gamma_{0}$ would imply $\bar{M}_{n}(\hat{\gamma})\xrightarrow{p}M_{0}$, and then $\hat{\alpha}(\hat{\gamma})\xrightarrow{p}\alpha_{0}$, which completes the proof.
			
			To show consistency of $\hat{\gamma}$ to $\gamma_{0}$, we apply the argmin/argmax continuous mapping theorem (CMT) as in Theorem 3.2.2 in \cite{van_der_vaart_weak_1996}. 
			It is sufficient to check (i) $\tilde{Q}_{n}(\gamma)$ uniformly converges to some function $\tilde{Q}_{0}(\gamma)$ in probability, and (ii) $\tilde{Q}_{0}(\gamma_{0})<\inf_{\gamma\not\in \mathcal{O}}\tilde{Q}_{0}(\gamma)$ for any open set $\mathcal{O}$ contatining $\gamma_{0}$.
			(ii) can be shown if $\tilde{Q}_{0}(\gamma)$ is uniquely minimized at $\gamma_{0}$ and continuous as $\Gamma$ is compact.
			
			The profiled moment can be rewritten as
			\begin{equation*}
				\tilde{g}_{n}(\gamma)=[I-\bar{M}_{n}(\gamma)\left(\bar{M}_{n}(\gamma)'W_{n}\bar{M}_{n}(\gamma)\right)^{-1}\bar{M}_{n}(\gamma)'W_{n}](-\bar{M}_{n}\alpha_{0}+u_{n}).
			\end{equation*}
			Therefore,
			\begin{equation*}
				W_{n}^{1/2}\tilde{g}_{n}(\gamma)=
				[I-P_{W_{n}^{1/2}\bar{M}_{n}(\gamma)}]
				(-W_{n}^{1/2}\bar{M}_{n}
				\alpha_{0}+W_{n}^{1/2}u_{n}),
			\end{equation*}
			where $P_{W_{n}^{1/2}\bar{M}_{n}(\gamma)}=W_{n}^{1/2}\bar{M}_{n}(\gamma)\left(\bar{M}_{n}(\gamma)'W_{n}\bar{M}_{n}(\gamma)\right)^{-1}\bar{M}_{n}(\gamma)'W_{n}^{1/2}$ is a projection matrix to the column space of $W_{n}^{1/2}\bar{M}_{n}(\gamma)$. 
			The profiled objective can be written as
			\[\tilde{Q}_{n}(\gamma)=\|(I-P_{W_{n}^{1/2}\bar{M}_{n}(\gamma)})(-W_{n}^{1/2}\bar{M}_{n}\alpha_{0}+W_{n}^{1/2}u_{n})\|^{2}.\]
			By $W_{n}\xrightarrow{p}W$, $u_{n}\xrightarrow{p}0$, and $\sup_{\gamma\in\Gamma}\|\bar{M}_{n}(\gamma)-M_{0}(\gamma)\|\xrightarrow{p}0$, we can derive that
			\[\tilde{Q}_{n}(\gamma)\xrightarrow{p}\tilde{Q}_{0}(\gamma)=\|(I-P_{W^{1/2}M_{0}(\gamma)})W^{1/2}M_{0}\alpha_{0}\|^{2}\]
			uniformly with respect to $\gamma$, where $P_{W^{1/2}M_{0}(\gamma)}=W^{1/2}M_{0}(\gamma)\left(M_{0}(\gamma)'WM_{0}(\gamma)\right)^{-1}M_{0}(\gamma)'W^{1/2}$.
			Note that $W=\Omega^{-1}$ in the second stage of the two-step GMM estimation.
			$W=I$ when we consider the first stage.
			$\tilde{Q}_{0}(\gamma)$ is uniquely minimized when $\gamma=\gamma_{0}$.
			This is because $W$ is positive definite, and the conditions in \cref{thm:id} implies that $M_{0}\alpha_{0}$ does not lie in the column space of $M_{0}(\gamma)$ whenever $\gamma\neq\gamma_{0}$.
			Moreover, $\tilde{Q}_{0}(\gamma)$ is continuous as $M_{0}(\gamma)$ is continuous with respect to $\gamma$ by \cref{ass:dgp}.
			
			\subsubsection{Convergence rate.} 
			$\|W_{n}-\Omega^{-1}\|\xrightarrow{p}0$ as the consistency of $\hat{\theta}_{(1)}$ is shown.
			Our proof follows arguments similar to the proof of Theorem 3.3 by \cite{pakes_simulation_1989}.
			By the consistency of $\hat{\theta}$ and by \cref{lem:equicont},
			\[\sqrt{n}\|\bar{g}_{n}(\hat{\theta})-\bar{g}_{n}(\theta_{0})-g_{0}(\hat{\theta})\|=o_{p}(1).\]
			By $\|W_{n}-\Omega^{-1}\|\xrightarrow{p}0$, we can obtain
			\[\sqrt{n}\|W_{n}^{1/2}\bar{g}_{n}(\hat{\theta})-W_{n}^{1/2}\bar{g}_{n}(\theta_{0})-\Omega^{-1/2}g_{0}(\hat{\theta})\|=o_{p}(1).\]
			Apply triangle inequality to get 
			\[\sqrt{n}\|\Omega^{-1/2}g_{0}(\hat{\theta})\|\leq o_{p}(1)+\sqrt{n}\|W_{n}^{1/2}\bar{g}_{n}(\theta_{0})\|+\sqrt{n}\|W_{n}^{1/2}\bar{g}_{n}(\hat{\theta})\|.\]
			As $\hat{\theta}$ is the minimizer of the GMM criterion,
			$\sqrt{n}\|W_{n}^{1/2}\bar{g}_{n}(\hat{\theta})\|\leq o_{p}(1)+\sqrt{n}\|W_{n}^{1/2}\bar{g}_{n}(\theta_{0})\|=O_{p}(1)$.
			Therefore,
			\[\sqrt{n}\|\Omega^{-1/2}g_{0}(\hat{\theta})\|\leq O_{p}(1).\] 
			$\sqrt{n}\|\Omega^{-1/2}g_{0}(\hat{\theta})\|\geq \sqrt{n}\|\Omega^{-1/2}D_{2}(\hat{\alpha}'-\alpha_{0}',(\hat{\gamma}-\gamma_{0})^{2})'\|-\sqrt{n}\|\Omega^{-1/2}(g(\hat{\theta})-D_{2}(\hat{\alpha}'-\alpha_{0}',(\hat{\gamma}-\gamma_{0})^{2})')\|$, while $\sqrt{n}\|\Omega^{-1/2}(g(\hat{\theta})-D_{2}(\hat{\alpha}'-\alpha_{0}',(\hat{\gamma}-\gamma_{0})^{2})')\|\leq o_{p}(1+\sqrt{n}\|(\hat{\alpha}'-\alpha_{0}',(\hat{\gamma}-\gamma_{0})^{2})'\|)$ by \cref{lem:loc.conti}.
			Thus, 
			\[\sqrt{n}(\|\hat{\alpha}-\alpha_{0}\|+(\hat{\gamma}-\gamma_{0})^{2})\leq O_{p}(1)\]
			which implies $\|\hat{\alpha}-\alpha_{0}\|=O_{p}(n^{-1/2})$ and $(\hat{\gamma}-\gamma_{0})^{2}=O_{p}(n^{-1/2})$.
			
			\subsubsection{Asymptotic distribution.}
			
			This section derives asymptotic distribution of the estimator through the argmin/argmax continuous mapping theorem (CMT) as in Theorem 3.2.2 in \cite{van_der_vaart_weak_1996}. 
			
			Introduce a local reparametrization by $a=\sqrt{n}(\alpha-\alpha_{0})$ and $b=n^{\frac{1}{4}}(\gamma-\gamma_{0})$, and let $a$ consist of subvectors $a_{1}=\sqrt{n}(\beta-\beta_{0})$ and $a_{2}=\sqrt{n}(\delta-\delta_{0})$.
			Additionally, define $\hat{a}=\sqrt{n}(\hat{\alpha}-\alpha_{0})$ and $\hat{b}=n^{\frac{1}{4}}(\hat{\gamma}-\gamma_{0})$. 
			Note that $(\hat{a},\hat{b}^{2})$ is uniformly tight due to the convergence rate we obtained.\footnote{
				A random variable $X$ is tight if for any $\epsilon>0$, there exists a compact set $\mathbb{K}$ such that $P(X\in\mathbb{K})>1-\epsilon$, and $X_{n}$ is uniformly tight if for any $\epsilon>0$, there exists a compact set $\mathbb{K}$ such that $P(X_{n}\in\mathbb{K})>1-\epsilon$ for all $n\in\mathbb{N}$.
				Note that by the convergence rate we derived, for any $\epsilon>0$, there exists a compact $\mathbb{K}_{0}$ such that $\lim_{n\rightarrow\infty}P((\sqrt{n}(\hat{\alpha}-\alpha_{0})',\sqrt{n}(\hat{\gamma}-\gamma_{0})^{2})'\in \mathbb{K}_{0})> 1-\epsilon/2$, and $N<\infty$ such that $P((\sqrt{n}(\hat{\alpha}-\alpha_{0})',\sqrt{n}(\hat{\gamma}-\gamma_{0})^{2})'\in \mathbb{K}_{0})> 1-\epsilon$ if $n\geq N$. 
				Then, we can define a compact set $\mathbb{K}=(\cup_{j=1}^{N-1}\mathbb{K}_{j})\cup\mathbb{K}_{0}$, where $\mathbb{K}_{j}$ is a compact set such that $P((\sqrt{j}(\hat{\alpha}-\alpha_{0})',\sqrt{j}(\hat{\gamma}-\gamma_{0})^{2})'\in \mathbb{K}_{j})> 1-\epsilon$, which satisfies $P((\sqrt{n}(\hat{\alpha}-\alpha_{0})',\sqrt{n}(\hat{\gamma}-\gamma_{0})^{2})'\in \mathbb{K})> 1-\epsilon$ for all $n\in\mathbb{N}$.
			}
			Let
			\begin{equation*}
				\mathbb{S}_{n}(a,b) = n\hat{Q}_{n}(\alpha_{0}+\tfrac{a}{\sqrt{n}},\gamma_{0}+\tfrac{b}{n^{\frac{1}{4}}})=n\bar{g}_{n}(\alpha_{0}+\tfrac{a}{\sqrt{n}},\gamma_{0}+\tfrac{b}{n^{\frac{1}{4}}})'W_{n}\bar{g}_{n}(\alpha_{0}+\tfrac{a}{\sqrt{n}},\gamma_{0}+\tfrac{b}{n^{\frac{1}{4}}}).
			\end{equation*}
			We show that (i) $\mathbb{S}_{n}$ weakly converges to a stochastic process $\mathbb{S}$ in $\ell^{\infty}(\mathbb{K})$ for every compact $\mathbb{K}$ in the Euclidean space, (ii) $\mathbb{S}$ is continuous, and (iii) $\mathbb{S}$ possesses an unique optimum not in $b$ but in its square $b^2$ since $\mathbb{S}(a,b)=\mathbb{S}(a,-b)$. 
			Thus, we will establish that $(\hat{a}',\hat{b}^{2})'$ converges in distribution to $(a_{0}',b_{0}^{2})'=\arg\min_{a,b^{2}}\mathbb{S}(a,\sqrt{b^{2}})$.
			In the characterization of the minimizers, $(a_{0}',b_{0}^{2})'$ is shown to be tight.

			The rescaled and reparametrized sample moment can be written as
			\begin{multline*}
				\sqrt{n}\bar{g}_{n}(\alpha_{0}+\tfrac{a}{\sqrt{n}}, \gamma_{0}+\tfrac{b}{n^{\frac{1}{4}}})
				=
				\sqrt{n}\begin{pmatrix}
					\frac{1}{n}\sum_{i=1}^{n}z_{it_{0}}\Delta\epsilon_{it_{0}} \\
					\vdots \\
					\frac{1}{n}\sum_{i=1}^{n}z_{iT}\Delta\epsilon_{iT}
				\end{pmatrix}
				-
				\begin{pmatrix}
					\tfrac{1}{n}\sum_{i=1}^{n} z_{it_{0}}\Delta x_{it_{0}}'\\
					\vdots\\
					\tfrac{1}{n}\sum_{i=1}^{n} z_{iT}\Delta x_{iT}'
				\end{pmatrix}a_{1}
				\\
				-\begin{pmatrix}
					\tfrac{1}{n}\sum_{i=1}^{n} z_{it_{0}}1_{it_{0}}(\gamma_{0}+\tfrac{b}{n^{\frac{1}{4}}})' X_{it_{0}}\\
					\vdots\\
					\tfrac{1}{n}\sum_{i=1}^{n} z_{iT}1_{iT}(\gamma_{0}+\tfrac{b}{n^{\frac{1}{4}}})' X_{iT}
				\end{pmatrix}a_{2}
				+\begin{pmatrix}
					\tfrac{1}{\sqrt{n}}\sum_{i=1}^{n} z_{it_{0}}(1_{it_{0}}(\gamma_{0})'-1_{it_{0}}(\gamma_{0}+\tfrac{b}{n^{\frac{1}{4}}})')X_{it_{0}}\\
					\vdots\\
					\tfrac{1}{\sqrt{n}}\sum_{i=1}^{n} z_{iT}(1_{iT}(\gamma_{0})'-1_{iT}(\gamma_{0}+\tfrac{b}{n^{\frac{1}{4}}})')X_{iT}
				\end{pmatrix}\delta_{0}.
			\end{multline*}
			By the central limit theorem (CLT),
			\begin{equation*}
				\sqrt{n}\begin{pmatrix}
					\frac{1}{n}\sum_{i=1}^{n}z_{it_{0}}\Delta\epsilon_{it_{0}} \\
					\vdots \\
					\frac{1}{n}\sum_{i=1}^{n}z_{iT}\Delta\epsilon_{iT}
				\end{pmatrix}
				\xrightarrow{d}-e\sim N(0,\Omega).
			\end{equation*}
			By the LLN,
			\begin{equation*}
				\begin{pmatrix}
					\tfrac{1}{n}\sum_{i=1}^{n} z_{it_{0}}\Delta x_{it_{0}}'\\
					\vdots\\
					\tfrac{1}{n}\sum_{i=1}^{n} z_{iT}\Delta x_{iT}'
				\end{pmatrix}
				\xrightarrow{p}
				\begin{pmatrix}
					Ez_{it_{0}}\Delta x_{it_{0}}'\\
					\vdots\\
					Ez_{iT}\Delta x_{iT}'
				\end{pmatrix}
			\end{equation*}
			Let $K<\infty$ be arbitrary.
			By the ULLN in \cref{lem:ULLN},
			\begin{equation*}
				\left\|
				\begin{pmatrix}
					\tfrac{1}{n}\sum_{i=1}^{n} z_{it_{0}}1_{it_{0}}(\gamma_{0}+\tfrac{b}{n^{\frac{1}{4}}})' X_{it_{0}}\\
					\vdots\\
					\tfrac{1}{n}\sum_{i=1}^{n} z_{iT}1_{iT}(\gamma_{0}+\tfrac{b}{n^{\frac{1}{4}}})' X_{iT}
				\end{pmatrix}
				-
				\begin{pmatrix}
					Ez_{it_{0}}1_{it_{0}}(\gamma_{0}+\tfrac{b}{n^{\frac{1}{4}}})' X_{it_{0}}\\
					\vdots\\
					Ez_{iT}1_{iT}(\gamma_{0}+\tfrac{b}{n^{\frac{1}{4}}})' X_{iT}
				\end{pmatrix}
				\right\|\xrightarrow{p}0
			\end{equation*}
			uniformly with respect to $b\in[-K,K]$.
			Then, by continuity of $\kappa\mapsto E[z_{it}1_{it}(\gamma+\kappa)X_{it}]$ at $\kappa=0$,
			\begin{equation*}
				\begin{pmatrix}
					\tfrac{1}{n}\sum_{i=1}^{n} z_{it_{0}}1_{it_{0}}(\gamma_{0}+\tfrac{b}{n^{\frac{1}{4}}})' X_{it_{0}}\\
					\vdots\\
					\tfrac{1}{n}\sum_{i=1}^{n} z_{iT}1_{iT}(\gamma_{0}+\tfrac{b}{n^{\frac{1}{4}}})' X_{iT}
				\end{pmatrix}
				\xrightarrow{p}
				\begin{pmatrix}
					Ez_{it_{0}}1_{it_{0}}(\gamma_{0})' X_{it_{0}}\\
					\vdots\\
					Ez_{iT}1_{iT}(\gamma_{0})' X_{iT}
				\end{pmatrix}
			\end{equation*}
			uniformly with respect to $b\in [-K,K]$.
			By \cref{lem:plim1},
			\begin{equation*}
				\scalebox{0.8}{$\begin{pmatrix}
					\tfrac{1}{\sqrt{n}}\sum_{i=1}^{n} z_{it_{0}}(1_{it_{0}}(\gamma_{0})'-1_{it_{0}}(\gamma_{0}+\tfrac{b}{n^{\frac{1}{4}}})')X_{it_{0}}\delta_{0}\\
					\vdots\\
					\tfrac{1}{\sqrt{n}}\sum_{i=1}^{n} z_{iT}(1_{iT}(\gamma_{0})'-1_{iT}(\gamma_{0}+\tfrac{b}{n^{\frac{1}{4}}})')X_{iT}\delta_{0}
				\end{pmatrix}$} 
				\xrightarrow{p}
				\frac{\delta_{30}b^{2}}{2}
				\begin{pmatrix}
					E_{t_{0}}[z_{it_{0}}|\gamma_{0}]f_{t_{0}}(\gamma_{0})-E_{t_{0}-1}[z_{it_{0}}|\gamma_{0}]f_{t_{0}-1}(\gamma_{0})\\
					\vdots\\
					E_{T}[z_{iT}|\gamma_{0}]f_{T}(\gamma_{0})-E_{T-1}[z_{iT}|\gamma_{0}]f_{T-1}(\gamma_{0})
				\end{pmatrix}
			\end{equation*}
			uniformly with respect to $b\in [-K,K]$.
			
			Therefore, $\mathbb{S}_{n}(a,b)$ weakly converges to 
			\[\mathbb{S}(a,b)=(M_{0}a+Hb^{2}-e)'\Omega^{-1}(M_{0}a+Hb^{2}-e),\]
			in $\ell^{\infty}(\mathbb{K})$ for any compact $\mathbb{K}\subset \mathbb{R}^{2p+2}$.
			Then, by the CMT,
			\[(\hat{a},\hat{b}^{2}) \xrightarrow{d} \arg\min_{a,b^{2}} (M_{0}a+Hb^{2}-e)'\Omega^{-1}(M_{0}a+Hb^{2}-e).\]
			
			\paragraph{Characterization of the minimizers} 
			Next, we characterize the minimizers. 
			The objective function of the minimization problem is strictly convex with respect to $a$ and $b^{2}$, since $\left[\begin{array}{c;{2pt/2pt}c}
				M_{0} & H
			\end{array}\right]$ has full column rank and $\Omega^{-1}$ is positive definite. 
			Hence, a solution $(a_{0}',b_{0}^{2})'$ can be characterized by the Karush-Kuhn-Tucker (KKT) conditions.
			See Chapter 5 in \cite{boyd_vandenberghe_2004} for more details.
			
			The Lagrangian for this problem is
			\[\mathcal{L}(a,b,\lambda) = a'M_{0}'\Omega^{-1}M_{0}a+2a'M_{0}'\Omega^{-1}Hb^{2} + H'\Omega^{-1}Hb^{4}-2a'M_{0}'\Omega^{-1}e-2H'\Omega^{-1}e\cdot b^{2}+e'\Omega^{-1}e-\lambda b^{2} \]
			and the gradient of the Lagrangian with respect to $a$ and $b^{2}$ should vanish: 
			\begin{align*}
				a: &\quad M_{0}'\Omega^{-1}M_{0}a+M_{0}'\Omega^{-1}Hb^{2}-M_{0}'\Omega^{-1}e=0 \\
				b^{2}: &\quad H'\Omega^{-1}Hb^{2} + H'\Omega^{-1}M_{0}a-H'\Omega^{-1}e -\lambda=0.
			\end{align*}
			In addition, $\lambda\geq0$ and $\lambda b^{2}=0$ should hold.
			\begin{enumerate}[label=(\roman*), itemsep=0em]
				\item When $\lambda = 0$ and $b^{2} \geq 0$, we can obtain
				\[b^{2} = (H'\Omega^{-1/2}(I-P_{\Omega^{-1/2}M_{0}})\Omega^{-1/2}H)^{-1}H'\Omega^{-1/2}(I-P_{\Omega^{-1/2}M_{0}})\Omega^{-1/2}e,\]
				where $P_{\Omega^{-1/2}M_{0}}=\Omega^{-1/2}M_{0}(M_{0}'\Omega^{-1}M_{0})^{-1}M_{0}'\Omega^{-1/2}$ is the projection matrix to the column space of $\Omega^{-1/2}M_{0}$. 
				$H'\Omega^{-1/2}(I-P_{\Omega^{-1/2}M_{0}})\Omega^{-1/2}H>0$ because the matrix $\left[\begin{array}{c;{2pt/2pt}c}
					M_{0} & H
				\end{array}\right]$
				has full column rank, 
				and $\Omega^{-1/2}H$ cannot be in the column space of $\Omega^{-1/2}M_{0}$ and $(I-P_{\Omega^{-1/2}M_{0}})\Omega^{-1/2}H\neq 0$.
				Therefore,
				\[H'\Omega^{-1/2}(I-P_{\Omega^{-1/2}M_{0}})\Omega^{-1/2}e\geq 0\] 
				should hold for the feasibility condition $b^{2}\geq0$. 
				\item When $\lambda > 0$ and $b^{2} = 0$, we can obtain 
				\[a=(M_{0}'\Omega^{-1}M_{0})^{-1}M_{0}'\Omega^{-1}e.\]
				By plugging this into the equation for $b^{2}$, we get
				\[H'\Omega^{-1/2}(I-P_{\Omega^{-1/2}M_{0}})\Omega^{-1/2}e<0.\]
			\end{enumerate}
			Thus,
			\[b_{0}^{2} = \begin{cases}
				[H'\Xi H]^{-1}H'\Xi e & \text{if }H'\Xi e\geq 0\\
				\quad\quad 0 & \text{else} \\
			\end{cases}\]
			where $\Xi=\Omega^{-1/2}(I-P_{\Omega^{-1/2}M_{0}})\Omega^{-1/2}$. 
			$b_{0}^{2}$ follows a normal distribution that is left censored at 0.
			Then,
			\begin{equation*}
				a_{0}  = \begin{cases}
					(M_{0}'\Omega^{-1}M_{0})^{-1}M_{0}'\Omega^{-1}[I-H[H'\Xi H]^{-1}H'\Xi ]e & \text{if }H'\Xi e\geq 0 \\
					(M_{0}'\Omega^{-1}M_{0})^{-1}M_{0}'\Omega^{-1}e & \text{else.} 
				\end{cases}
			\end{equation*}
			Note that the two normal variables  $(M_{0}'\Omega^{-1}M_{0})^{-1}M_{0}'\Omega^{-1}H[H'\Xi H]^{-1}H'\Xi e$
			and $(M_{0}'\Omega^{-1}M_{0})^{-1}M_{0}'\Omega^{-1}e$ are independent of each other, because  $E[H'\Xi ee'\Omega^{-1}M_{0}] =  H'\Omega^{-1/2}(I-P_{\Omega^{-1/2}M_{0}})\Omega^{-1/2}M_{0}$ becomes zero.
			
			\section{Proofs for Section \ref{sec:bootstrap}}
			\label{section:boot proofs}
			
			\subsection{Preliminaries}
			\label{sec:boot prelim}
			
			The bootstrap methods we consider are \cref{alg:bootstrap} with different choices of $\theta_{0}^{*}$.
			There are three bootstrap methods this paper propose: (i) $\theta_{0}^{*}=(\hat{\alpha}(\gamma),\gamma)'$ for $\gamma\in\Gamma$, (ii) $\theta_{0}^{*}$ set as \eqref{eq:perc boot}, and (iii)  $\theta_{0}^{*}=\tilde{\theta}$ which is the continuity-restricted estimator.
			In \cref{sec:std boot invalid}, we consider the case $\theta_{0}^{*}=\hat{\theta}$ which results in the standard nonparametric bootstrap.
			
			The probability law for the bootstrap is formalized following \cite{Goncalves_white_2004}.
			Let $P$ be the probability measure for data and $P^{*}$ be the conditional probability law of bootstrap given observations.
			$Z_{n}^{*}\xrightarrow{p^{*}}0$ in $P$ ($Z_{n}^{*}=o_{p}^{*}(1)$ in $P$) if for any $\epsilon,\delta>0$, $P(P^{*}(|Z_{n}^{*}|>\epsilon)>\delta)\rightarrow 0$ as $n\rightarrow\infty$.
			$Z_{n}^{*}=O_{p}^{*}(1)$ in $P$ if for any $\epsilon>0$ and $\delta>0$, there exists $M<\infty$ such that $\limsup_{n}P(P^{*}(|Z_{n}^{*}|\geq M)>\delta)< \epsilon$.
			$Z_{n}^{*}\xrightarrow{d^{*}}Z$ in $P$ if $E^{*}f(Z_{n}^{*})\rightarrow Ef(Z)$ in $P$ for every continuous and bounded function $f$, where $E^{*}$ is the expectation by the bootstrap probability law conditional on observations.
			$Z_{n}^{*}\overset{*}{\rightsquigarrow}Z$ in $\ell^{\infty}(\mathbb{K})$ in $P$ if $\sup_{f\in BL_{1}}|E^{*}f(Z_{n}^{*})-Ef(Z_{n})|\xrightarrow{p}0$, where $BL_{1}$ is the set of all Lipschitz functions on $\ell^{\infty}(\mathbb{K})$ bounded in $[0,1]$ such that $|f(z_{1})-f(z_{2})|\leq \|z_{1}-z_{2}\|_{\ell^{\infty}(\mathbb{K})}=\sup_{x\in\mathbb{K}}|z_{1}(x)-z_{2}(x)|$. 
			
			The following lemma is useful in analyzing bootstrap stochastic orders.
			
			\begin{lemma}
				\label{lem:bootstrap order}
				\begin{enumerate}[label=(\roman*)]
					\setlength\itemsep{0mm}
					\item If $A_{n}=o_{p}(1)$ or $O_{p}(1)$, then $A_{n}=o_{p}^{*}(1)$ or $O_{p}^{*}(1)$ in $P$, respectively.
					\item Let $Z_{n}^{*}=o_{p}^{*}(1)$ in $P$ and $W_{n}^{*}=O_{p}^{*}(1)$ in $P$. Then, $Z_{n}^{*}\times W_{n}^{*}=o_{p}^{*}(1)$ in $P$.
				\end{enumerate}
			\end{lemma}
			
			\begin{proof}
				See Lemma 3 in \cite{cheng_huang_2010}.
			\end{proof}
			
			Recall that $W_{n}^{*}=\{[\frac{1}{n}\sum_{i=1}^{n}g_{i}^{*}(\hat{\theta}_{(1)}^{*})g_{i}^{*}(\hat{\theta}_{(1)}^{*})']-[\frac{1}{n}\sum_{i=1}^{n}g_{i}^{*}(\hat{\theta}_{(1)}^{*})][\frac{1}{n}\sum_{i=1}^{n}g_{i}^{*}(\hat{\theta}_{(1)}^{*})']\}^{-1}$.
			$\|W_{n}^{*}-\Omega^{-1}\|=o_{p}^{*}(1)$ in $P$ when $\hat{\theta}_{(1)}^{*}\xrightarrow{p^{*}}\theta_{0}$ in $P$.
			This would be the case when $\|\hat{\theta}_{(1)}^{*}-\theta_{0}^{*}\|\xrightarrow{p^{*}}0$ in $P$ and $\|\theta_{0}^{*}-\theta_{0}\|=o_{p}(1)$ since then $\|\hat{\theta}_{(1)}^{*}-\theta_{0}\|\leq \|\hat{\theta}_{(1)}^{*}-\theta_{0}^{*}\|+\|\theta_{0}^{*}-\theta_{0}\|=o_{p}^{*}(1)$ in $P$ by \cref{lem:bootstrap order}.
			
		\subsection{Proof of \texorpdfstring{\cref{thm:perc boot}}{\cref{thm:perc boot}}.}

            As in the proof of \cref{thm:asym}, consistency and convergence rates of the bootstrap estimator should be derived first. These results are summarized in the following proposition, with the proof provided in Online Appendix \ref{sec: online.app.boot}.
                \begin{proposition}
                \label{prop:boot.conv.rate}
                    (\romannumeral 1) Under the assumptions of the case (i) in Theorems \ref{thm:grid boot}, \ref{thm:perc boot}, or \ref{thm:continuity boot},
		          \begin{equation*}
				    \sqrt{n}(\hat{\alpha}^{*}-\alpha_{0}^{*}) = O_{p}^{*}(1)\text{ in $P$, and }
				    \sqrt{n}(\hat{\gamma}^{*}-\gamma_{0}^{*})^{2} = O_{p}^{*}(1)\text{ in $P$.}
		          \end{equation*}
		
		      \noindent
		      (\romannumeral 2) Under the assumptions of the case (ii) in Theorems \ref{thm:grid boot} or \ref{thm:perc boot}, 
		          \begin{equation*}
				    \sqrt{n}(\hat{\alpha}^{*}-\alpha_{0}^{*}) = O_{p}^{*}(1)\text{ in $P$, and }
				    \sqrt{n}(\hat{\gamma}^{*}-\gamma_{0}^{*}) = O_{p}^{*}(1)\text{ in $P$.}
		          \end{equation*}
                \end{proposition}
            Then, we derive the (conditional) weak convergence limit of the rescaled criterion and apply the CMT to obtain the asymptotic distribution of the bootstrap estimator.

			\paragraph{Asymptotic distribution under continuity.}
			
			Based on the convergence rate in \cref{prop:boot.conv.rate}, introduce the local reparametrization by $a=\sqrt{n}(\alpha-\alpha_{0}^{*})$ and $b=n^{\frac{1}{4}}(\gamma-\gamma_{0}^{*})$, and let $a$ consist of subvectors $a_{1}=\sqrt{n}(\beta-\beta_{0}^{*})$ and $a_{2}=\sqrt{n}(\delta-\delta_{0}^{*})$.
			
			The asymptotic distributions of the bootstrap estimators can be derived by using the argmin/argmax CMT as in the proof of \cref{thm:asym}.
			Let
			\begin{equation*}
				\mathbb{S}_{n}^{*}(a,b) = n\hat{Q}_{n}^{*}(\alpha_{0}^{*}+\tfrac{a}{\sqrt{n}},\gamma_{0}^{*}+\tfrac{b}{n^{\frac{1}{4}}})=n\bar{g}_{n}^{*}(\alpha_{0}^{*}+\tfrac{a}{\sqrt{n}},\gamma_{0}^{*}+\tfrac{b}{n^{\frac{1}{4}}})'W_{n}^{*}\bar{g}_{n}^{*}(\alpha_{0}^{*}+\tfrac{a}{\sqrt{n}},\gamma_{0}^{*}+\tfrac{b}{n^{\frac{1}{4}}}).
			\end{equation*}
			We show that $\mathbb{S}_{n}^{*}\overset{*}{\rightsquigarrow}\mathbb{S}$ in $\ell^{\infty}(\mathbb{K})$ in $P$ for every compact $\mathbb{K}$ in the Euclidean space.
			Recall that $\mathbb{S}(a,b)=(M_{0}a+Hb^{2}-e)'\Omega^{-1}(M_{0}a+Hb^{2}-e)$.
			
			The rescaled and reparametrized bootstrap moment can be written as
			\begin{align*}
				\sqrt{n}\bar{g}_{n}^{*}(\alpha_{0}^{*}+\tfrac{a}{\sqrt{n}}, \gamma_{0}^{*}+\tfrac{b}{n^{\frac{1}{4}}})
				=&
				\sqrt{n}
				\left\{
				\begin{pmatrix}
					\frac{1}{n}\sum_{i=1}^{n}z_{it_{0}}^{*}\widehat{\Delta\epsilon}_{it_{0}}^{*} \\
					\vdots \\
					\frac{1}{n}\sum_{i=1}^{n}z_{iT}^{*}\widehat{\Delta\epsilon}_{iT}^{*}
				\end{pmatrix}
				-
				\begin{pmatrix}
					\frac{1}{n}\sum_{i=1}^{n}z_{it_{0}}\widehat{\Delta\epsilon}_{it_{0}} \\
					\vdots \\
					\frac{1}{n}\sum_{i=1}^{n}z_{iT}\widehat{\Delta\epsilon}_{iT}
				\end{pmatrix}
				\right\}
				\\
				&
				-
				\begin{pmatrix}
					\tfrac{1}{n}\sum_{i=1}^{n} z_{it_{0}}^{*}\Delta x_{it_{0}}^{*\prime}\\
					\vdots\\
					\tfrac{1}{n}\sum_{i=1}^{n} z_{iT}^{*}\Delta x_{iT}^{*\prime}
				\end{pmatrix}a_{1}
				-\begin{pmatrix}
					\tfrac{1}{n}\sum_{i=1}^{n} z_{it_{0}}^{*}1_{it_{0}}^{*}(\gamma_{0}^{*}+\tfrac{b}{n^{\frac{1}{4}}})' X_{it_{0}}^{*}\\
					\vdots\\
					\tfrac{1}{n}\sum_{i=1}^{n} z_{iT}^{*}1_{iT}^{*}(\gamma_{0}^{*}+\tfrac{b}{n^{\frac{1}{4}}})' X_{iT}^{*}
				\end{pmatrix}a_{2}
				\\
				&
				+\sqrt{n}\begin{pmatrix}
					\tfrac{1}{n}\sum_{i=1}^{n} z_{it_{0}}^{*}(1_{it_{0}}^{*}(\gamma_{0}^{*})'-1_{it_{0}}(\gamma_{0}^{*}+\tfrac{b}{n^{\frac{1}{4}}})')X_{it_{0}}^{*}\\
					\vdots\\
					\tfrac{1}{n}\sum_{i=1}^{n} z_{iT}^{*}(1_{iT}^{*}(\gamma_{0}^{*})'-1_{iT}(\gamma_{0}^{*}+\tfrac{b}{n^{\frac{1}{4}}})')X_{iT}^{*}
				\end{pmatrix}\delta_{0}^{*}.
			\end{align*}
			By \cref{lem:boot res CLT},
			\begin{equation*}
				\sqrt{n}
				\left\{
				\begin{pmatrix}
					\frac{1}{n}\sum_{i=1}^{n}z_{it_{0}}^{*}\widehat{\Delta\epsilon}_{it_{0}}^{*} \\
					\vdots \\
					\frac{1}{n}\sum_{i=1}^{n}z_{iT}^{*}\widehat{\Delta\epsilon}_{iT}^{*}
				\end{pmatrix}
				-
				\begin{pmatrix}
					\frac{1}{n}\sum_{i=1}^{n}z_{it_{0}}\widehat{\Delta\epsilon}_{it_{0}} \\
					\vdots \\
					\frac{1}{n}\sum_{i=1}^{n}z_{iT}\widehat{\Delta\epsilon}_{iT}
				\end{pmatrix}
				\right\}
				\xrightarrow{d^{*}}-e\sim N(0,\Omega)\quad\text{ in $P$.}
			\end{equation*}
			By the bootstrap LLN,
			\begin{equation*}
				\begin{pmatrix}
					\tfrac{1}{n}\sum_{i=1}^{n} z_{it_{0}}^{*}\Delta x_{it_{0}}^{*\prime}\\
					\vdots\\
					\tfrac{1}{n}\sum_{i=1}^{n} z_{iT}^{*}\Delta x_{iT}^{*\prime}
				\end{pmatrix}
				\xrightarrow{p^{*}}
				\begin{pmatrix}
					Ez_{it_{0}}\Delta x_{it_{0}}'\\
					\vdots\\
					Ez_{iT}\Delta x_{iT}'
				\end{pmatrix}
				\quad
				\text{ in $P$.}
			\end{equation*}
			Let $K<\infty$ be arbitrary.
			By bootstrap Glivenko-Cantelli, e.g., Lemma 3.6.16 in \cite{van_der_vaart_weak_1996}, 
			\begin{equation*}
				\sup_{b:|b|\leq K,\gamma\in\Gamma}
				\left\|
				\begin{pmatrix}
					\tfrac{1}{n}\sum_{i=1}^{n} z_{it_{0}}^{*}1_{it_{0}}^{*}(\gamma+\tfrac{b}{n^{\frac{1}{4}}})' X_{it_{0}}^{*}\\
					\vdots\\
					\tfrac{1}{n}\sum_{i=1}^{n} z_{iT}^{*}1_{iT}^{*}(\gamma+\tfrac{b}{n^{\frac{1}{4}}})' X_{iT}^{*}
				\end{pmatrix}
				-
				\begin{pmatrix}
					Ez_{it_{0}}1_{it_{0}}(\gamma+\tfrac{b}{n^{\frac{1}{4}}})' X_{it_{0}}\\
					\vdots\\
					Ez_{iT}1_{iT}(\gamma+\tfrac{b}{n^{\frac{1}{4}}})' X_{iT}
				\end{pmatrix}
				\right\|\xrightarrow{p^{*}}0\quad\text{ in $P$.}
			\end{equation*}
			By continuity of $J(\gamma):=E[z_{it}1_{it}(\gamma)X_{it}]$ at $\gamma=\gamma_{0}$, for any $c>0$, there exists $h>0$ such that
			$\| J(\gamma)- J(\gamma_{0})\| < c$ if $|\gamma-\gamma_{0}|<h$.
			For any $h>0$, $P(|\gamma_{0}-\gamma_{0}^{*}-\frac{b}{n^{\frac{1}{4}}}|>h)\rightarrow 0$.
                Note that $\{\|\frac{1}{n}\sum_{i=1}^{n} z_{it}^{*}1_{it}^{*}(\gamma_{0}^{*}+\tfrac{b}{n^{\frac{1}{4}}})' X_{it}^{*}-J(\gamma_{0})\|>2c \}\subseteq \{\|\frac{1}{n}\sum_{i=1}^{n} z_{it}^{*}1_{it}^{*}(\gamma_{0}^{*}+\tfrac{b}{n^{\frac{1}{4}}})' X_{it}^{*}-J(\gamma_{0}^{*}+\tfrac{b}{n^{\frac{1}{4}}})\|>c\} \cup \{\| J(\gamma_{0}^{*}+\tfrac{b}{n^{\frac{1}{4}}})- J(\gamma_{0})\|>c\}$, and hence $P^{*}(\|\frac{1}{n}\sum_{i=1}^{n} z_{it}^{*}1_{it}^{*}(\gamma_{0}^{*}+\tfrac{b}{n^{\frac{1}{4}}})' X_{it}^{*}-J(\gamma_{0})\|>2c)\leq P^{*}(\|\frac{1}{n}\sum_{i=1}^{n} z_{it}^{*}1_{it}^{*}(\gamma_{0}^{*}+\tfrac{b}{n^{\frac{1}{4}}})' X_{it}^{*}-J(\gamma_{0}^{*}+\tfrac{b}{n^{\frac{1}{4}}})\|>c)$ with probability approaching 1, while $P^{*}(\|\frac{1}{n}\sum_{i=1}^{n} z_{it}^{*}1_{it}^{*}(\gamma_{0}^{*}+\frac{b}{n^{\frac{1}{4}}})' X_{it}^{*}-J(\gamma_{0}^{*}+\frac{b}{n^{\frac{1}{4}}})\|>c)\xrightarrow{p}0$ uniformly with respect to $b\in[-K,K]$. Thus,
			\begin{equation*}
				\begin{pmatrix}
					\tfrac{1}{n}\sum_{i=1}^{n} z_{it_{0}}^{*}1_{it_{0}}^{*}(\gamma_{0}^{*}+\tfrac{b}{n^{\frac{1}{4}}})' X_{it_{0}}^{*}\\
					\vdots\\
					\tfrac{1}{n}\sum_{i=1}^{n} z_{iT}^{*}1_{iT}^{*}(\gamma_{0}^{*}+\tfrac{b}{n^{\frac{1}{4}}})' X_{iT}^{*}
				\end{pmatrix}
				\xrightarrow{p^{*}}
				\begin{pmatrix}
					Ez_{it_{0}}1_{it_{0}}(\gamma_{0})' X_{it_{0}}\\
					\vdots\\
					Ez_{iT}1_{iT}(\gamma_{0})' X_{iT}
				\end{pmatrix}\text{ in $P$,}
			\end{equation*}
			both uniformly with respect to $b\in [-K,K]$.
			By \cref{lem:boot plim1},
			\begin{multline*}
				\begin{pmatrix}
					\tfrac{1}{\sqrt{n}}\sum_{i=1}^{n} z_{it_{0}}^{*}(1_{it_{0}}^{*}(\gamma_{0}^{*})'-1_{it_{0}}(\gamma_{0}^{*}+\tfrac{b}{n^{\frac{1}{4}}})')X_{it_{0}}^{*}\\
					\vdots\\
					\tfrac{1}{\sqrt{n}}\sum_{i=1}^{n} z_{iT}^{*}(1_{iT}^{*}(\gamma_{0}^{*})'-1_{iT}(\gamma_{0}^{*}+\tfrac{b}{n^{\frac{1}{4}}})')X_{iT}^{*}
				\end{pmatrix}\delta_{0}^{*}
				\\
				\xrightarrow{p^{*}}
				\frac{\delta_{30}}{2}
				\begin{pmatrix}
					E_{t_{0}}[z_{it_{0}}|\gamma_{0}]f_{t_{0}}(\gamma_{0})-E_{t_{0}-1}[z_{it_{0}}|\gamma_{0}]f_{t_{0}-1}(\gamma_{0})\\
					\vdots\\
					E_{T}[z_{iT}|\gamma_{0}]f_{T}(\gamma_{0})-E_{T-1}[z_{iT}|\gamma_{0}]f_{T-1}(\gamma_{0})
				\end{pmatrix}b^{2}\quad\text{ in $P$}
			\end{multline*}
			uniformly with respect to $b\in [-K,K]$.
			
			Therefore, $\mathbb{S}_{n}^{*}(a,b)\overset{*}{\rightsquigarrow}\mathbb{S}(a,b)$ in $\ell^{\infty}(\mathbb{K})$ in $P$ for any compact $\mathbb{K}\subset \mathbb{R}^{2p+2}$.
			Then, by applying the argmin CMT as in the proof of \cref{thm:asym}, we can obtain the limit distribution of the bootstrap estimates conditional on the data.

			\paragraph{Asymptotic distribution under discontinuity.}
			
			The proof for the discontinuous model only requires a slight change to the proof for the continuous model.
			As the convergence rate for the discontinuous model is $\sqrt{n}$ for both coefficients and threshold location estimators,
			let $a$ be unchanged and $b=\sqrt{n}(\gamma-\gamma_{0}^{*})$ for the local reparametrization.
			Let
			\begin{equation*}
				\mathbb{S}_{n}^{*}(a,b) = n\hat{Q}_{n}^{*}(\alpha_{0}^{*}+\tfrac{a}{\sqrt{n}},\gamma_{0}^{*}+\tfrac{b}{\sqrt{n}})=n\bar{g}_{n}^{*}(\alpha_{0}^{*}+\tfrac{a}{\sqrt{n}},\gamma_{0}^{*}+\tfrac{b}{\sqrt{n}})'W_{n}^{*}\bar{g}_{n}^{*}(\alpha_{0}^{*}+\tfrac{a}{\sqrt{n}},\gamma_{0}^{*}+\tfrac{b}{\sqrt{n}}).
			\end{equation*}
			We can write the rescaled and reparametrized moment as follows:
			\begin{multline*}
				\sqrt{n}\bar{g}_{n}^{*}(\alpha_{0}^{*}+\tfrac{a}{\sqrt{n}}, \gamma_{0}^{*}+\tfrac{b}{\sqrt{n}})
				= \\
				\sqrt{n}
				\left\{
				\begin{pmatrix}
					\frac{1}{n}\sum_{i=1}^{n}z_{it_{0}}^{*}\widehat{\Delta\epsilon}_{it_{0}}^{*} \\
					\vdots \\
					\frac{1}{n}\sum_{i=1}^{n}z_{iT}^{*}\widehat{\Delta\epsilon}_{iT}^{*}
				\end{pmatrix}
				-
				\begin{pmatrix}
					\frac{1}{n}\sum_{i=1}^{n}z_{it_{0}}\widehat{\Delta\epsilon}_{it_{0}} \\
					\vdots \\
					\frac{1}{n}\sum_{i=1}^{n}z_{iT}\widehat{\Delta\epsilon}_{iT}
				\end{pmatrix}
				\right\}
				-
				\begin{pmatrix}
					\tfrac{1}{n}\sum_{i=1}^{n} z_{it_{0}}^{*}\Delta x_{it_{0}}^{*\prime}\\
					\vdots\\
					\tfrac{1}{n}\sum_{i=1}^{n} z_{iT}^{*}\Delta x_{iT}^{*\prime}
				\end{pmatrix}a_{1}
				\\
				-\begin{pmatrix}
					\tfrac{1}{n}\sum_{i=1}^{n} z_{it_{0}}^{*}1_{it_{0}}^{*}(\gamma_{0}^{*}+\tfrac{b}{\sqrt{n}})' X_{it_{0}}^{*}\\
					\vdots\\
					\tfrac{1}{n}\sum_{i=1}^{n} z_{iT}^{*}1_{iT}^{*}(\gamma_{0}^{*}+\tfrac{b}{\sqrt{n}})' X_{iT}^{*}
				\end{pmatrix}a_{2}
				+\sqrt{n}\begin{pmatrix}
					\tfrac{1}{n}\sum_{i=1}^{n} z_{it_{0}}^{*}(1_{it_{0}}^{*}(\gamma_{0}^{*})'-1_{it_{0}}(\gamma_{0}^{*}+\tfrac{b}{\sqrt{n}})')X_{it_{0}}^{*}\\
					\vdots\\
					\tfrac{1}{n}\sum_{i=1}^{n} z_{iT}^{*}(1_{iT}^{*}(\gamma_{0}^{*})'-1_{iT}(\gamma_{0}^{*}+\tfrac{b}{\sqrt{n}})')X_{iT}^{*}
				\end{pmatrix}\delta_{0}^{*}.
			\end{multline*}
			The limit of $\sqrt{n}\bar{g}_{n}^{*}(\alpha_{0}^{*}+\tfrac{a}{\sqrt{n}}, \gamma_{0}^{*}+\tfrac{b}{\sqrt{n}})$ can be obtained similarly to the continuous model case, except that we use \cref{lem:boot plim2} instead of \cref{lem:boot plim1} to get
			\begin{multline*}
				\sqrt{n}
				\begin{pmatrix}
					\tfrac{1}{n}\sum_{i=1}^{n} z_{it_{0}}^{*}(1_{it_{0}}^{*}(\gamma_{0}^{*})'-1_{it_{0}}(\gamma_{0}^{*}+\tfrac{b}{n^{\frac{1}{4}}})')X_{it_{0}}^{*}\\
					\vdots\\
					\tfrac{1}{n}\sum_{i=1}^{n} z_{iT}^{*}(1_{iT}^{*}(\gamma_{0}^{*})'-1_{iT}(\gamma_{0}^{*}+\tfrac{b}{n^{\frac{1}{4}}})')X_{iT}^{*}
				\end{pmatrix}\delta_{0}^{*}
				\\
				\xrightarrow{p^{*}}
				\begin{pmatrix}
					E_{t_{0}}[z_{it_{0}}(1,x_{it_{0}}')\delta_{0}|\gamma_{0}]f_{t_{0}}(\gamma_{0})-E_{t_{0}-1}[z_{it_{0}}(1,x_{it_{0}}')\delta_{0}|\gamma_{0}]f_{t_{0}-1}(\gamma_{0})\\
					\vdots\\
					E_{T}[z_{iT}(1,x_{iT}')\delta_{0}|\gamma_{0}]f_{T}(\gamma_{0})-E_{T-1}[z_{iT}(1,x_{iT-1}')\delta_{0}|\gamma_{0}]f_{T-1}(\gamma_{0})
				\end{pmatrix}b\quad\text{ in $P$}
			\end{multline*}
			uniformly with respect to $b\in [-K,K]$. 
			
			Then, $\mathbb{S}_{n}^{*}(a,b)$ conditonally weakly converges to $\mathbb{S}_{J}(a,b)=(M_{0}a+Gb-e)'\Omega^{-1}(M_{0}a+Gb-e)$ in $\ell^{\infty}(\mathbb{K})$ in $P$ for any compact $\mathbb{K}\subset \mathbb{R}^{2p+2}$.
			And the argmin CMT yields the asymptotic distribution of the bootstrap estimators. 
			The limit distributions of the bootstrap estimators are normal because $(a_{0}',b_{0})'=\arg\min_{a,b}\mathbb{S}_{J}(a,b)=(D_{1}'\Omega^{-1}D_{1})^{-1}D_{1}'\Omega^{-1}e$. \qed

\clearpage
\renewcommand{\thepage}{S-\arabic{page}}
\setcounter{page}{1}
\setcounter{section}{2}

\section*{Online Supplements for ``Bootstraps for Dynamic Panel Threshold Models'' (Not for Publication)}

{\centering \subsection*{Woosik Gong and Myung Hwan Seo}}

\begin{quote}
    
    This part of the appendix is only for online supplements. It contains supplementary results for the Monte Carlo simulations, the remaining proofs for \cref{thm:distance test}, \cref{thm:continuity test}, \cref{prop:boot.conv.rate}, \cref{thm:grid boot}, \cref{thm:continuity boot}, as well as additional lemmas with proofs. It also presents invalidity of the standard nonparametric bootstrap, percentile bootstrap confidence intervals for empirical application, explanation of bootstrap for linearity test, and the uniform validity of the grid bootstrap. 
    
\end{quote}

\section{Supplementary Results for Monte Carlo Simulation}
\label{app:MC}

In this section, we present supplementary results for the Monte Carlo simulations in \cref{sec:mc}.

\subsection{Percentile Confidence Intervals with Large Sample}
\label{app:MC_large}

\cref{sec:mc} shows that the coverage rates by percentile CIs for the coefficients are very low for all specifications in finite sample; see \cref{tab:ch3}.
This is inconsistent to the (pointwise) validity of the residual bootstrap derived in \cref{sec:asym}.
It is the case even when the model is discontinuous where the nonparametric bootstrap is consistent.
To investigate large sample behavior, \cref{tab:large} reports the coverage rates of 95\% percentile CIs when $n=10000$ and $\delta_{1}+\delta_{3}\gamma\in\{0,1\}$.
The number of Monte Carlo repetitions is 1000.

\cref{tab:large} shows that the coverage rates for both R-B and NP-B get closer to the nominal 95\% level than those in \cref{tab:ch3}, although the undercoverage still remains for some coefficients.
The undercoverage by NP-B is especially severe for $\beta_{3}$, $\delta_{1}$, and $\delta_{3}$ when the model is continuous, i.e., when $\delta_{1}+\delta_{3}\gamma=0$, which suggests that the nonparametric bootstrap performs poorly when the true model is continuous.

\begin{table}[htbp]
		\caption{\label{tab:large} Coverage rates of 95\% percentile CIs for the coefficients when $n=10000$. 
                R-B denotes the percentile CIs by the residual bootstrap defined as \eqref{eq:rbci_asymm}. 
			NP-B denotes the percentile CIs by the standard nonparametric bootstrap defined as \eqref{eq:npbci_asymm}.
		}
		\centering
		\begin{tabular}[t]{c|c c c c c||c c c c c}
			\hline
			$n=10000$ & \multicolumn{5}{c||}{R-B} & \multicolumn{5}{c}{NP-B} \\
			\hline
			$\delta_{1}+\delta_{3}\gamma$ & $\beta_{2}$ & $\beta_{3}$ & $\delta_{1}$ & $\delta_{2}$ & $\delta_{3}$ & $\beta_{2}$ & $\beta_{3}$ & $\delta_{1}$ & $\delta_{2}$ & $\delta_{3}$\\
			\hline
    		0.0 & 0.922 & 0.879 & 0.836 & 0.934 & 0.887 & 0.940 & 0.761 & 0.763 & 0.905 & 0.741\\
    		1.0 & 0.982 & 0.941 & 0.856 & 0.964 & 0.827 & 0.962 & 0.893 & 0.817 & 0.933 & 0.835\\
    		\hline
		\end{tabular}
	\end{table}

\subsection{Percentile and Symmetric Percentile Confidence Intervals for Coefficients}
\label{app:MC_symm}

\begin{figure}[htbp]
    \begin{center}
    \caption{Scatter plot of sample statistic and bootstrap quantiles}
    \label{figure:bootquantiles}
    \begin{subfigure}{0.45\textwidth}
    \caption{NP-B}
    \includegraphics[width=\textwidth]{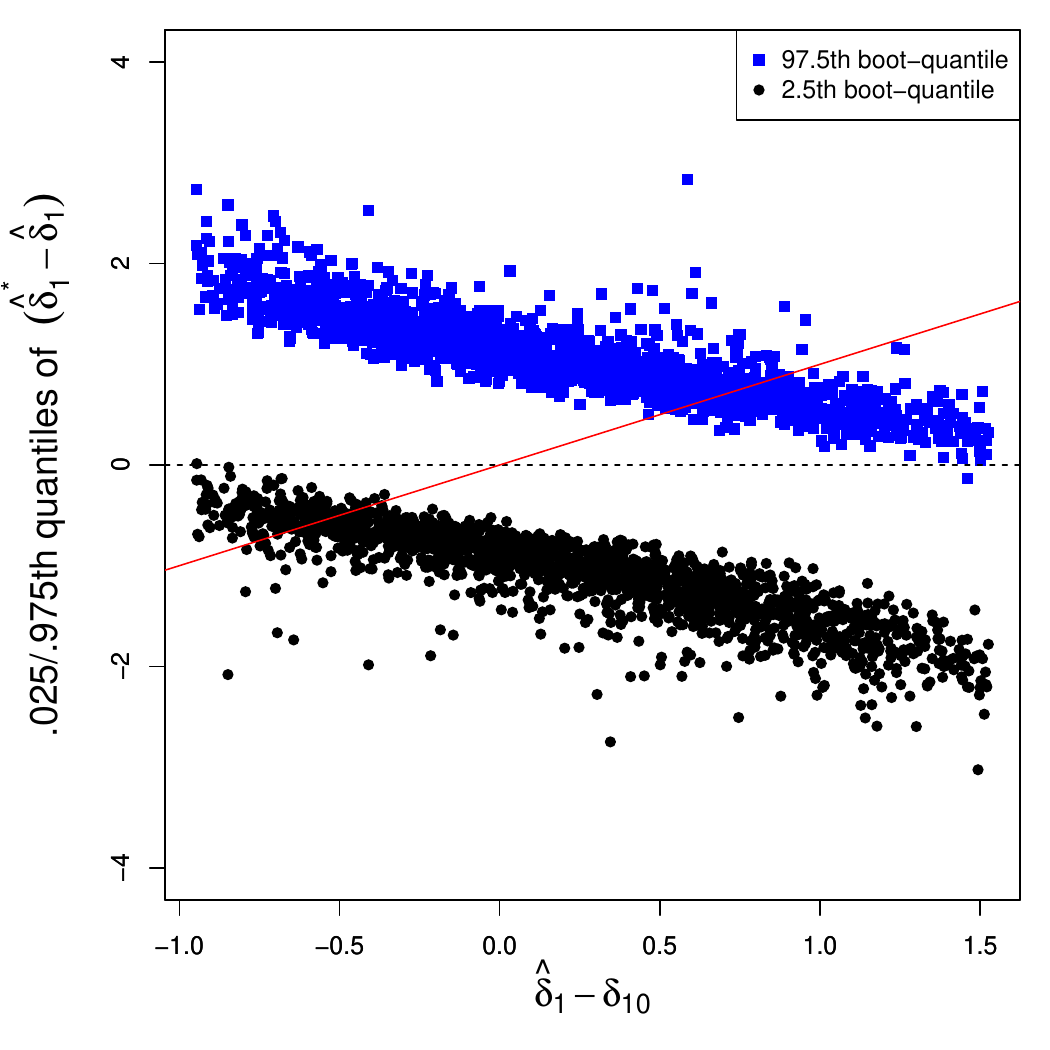}
    \end{subfigure}
    \hfill
    \begin{subfigure}{0.45\textwidth}
    \caption{R-B}
    \includegraphics[width=\textwidth]{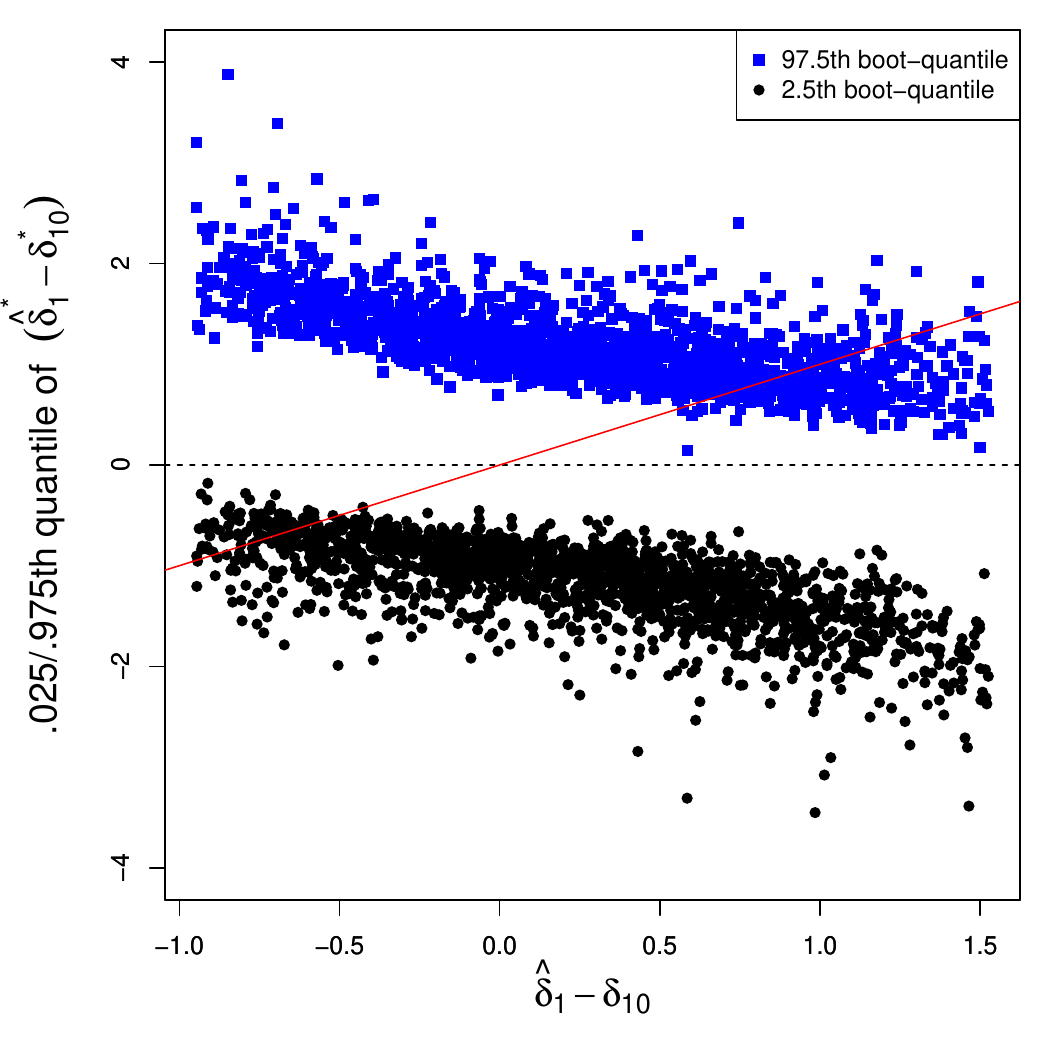}
    \end{subfigure}
    \\
    \begin{subfigure}{0.45\textwidth}
    \caption{NP-B(S)}
    \includegraphics[width=\linewidth]{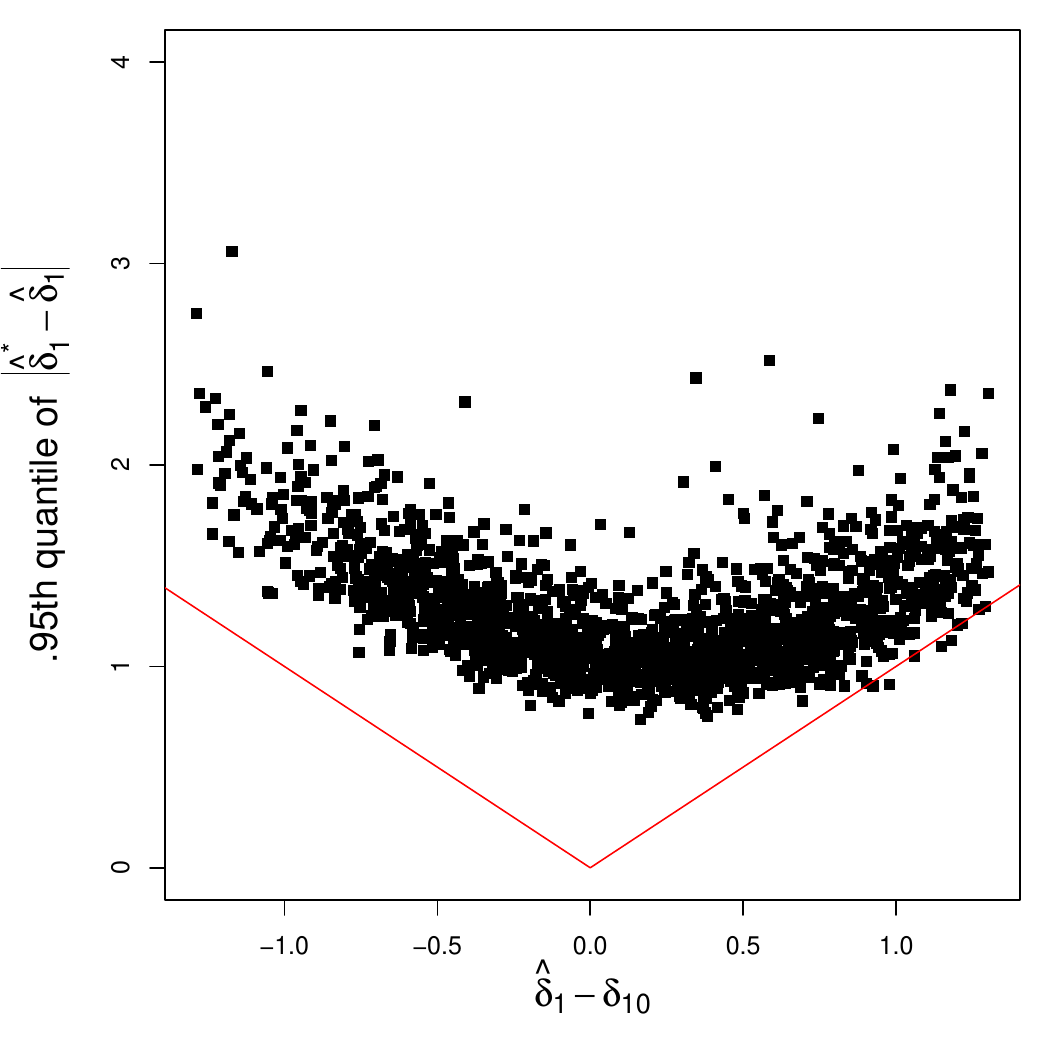}
    \end{subfigure}
    \hfill
    \begin{subfigure}{0.45\textwidth}
    \caption{R-B(S)}
    \includegraphics[width=\linewidth]{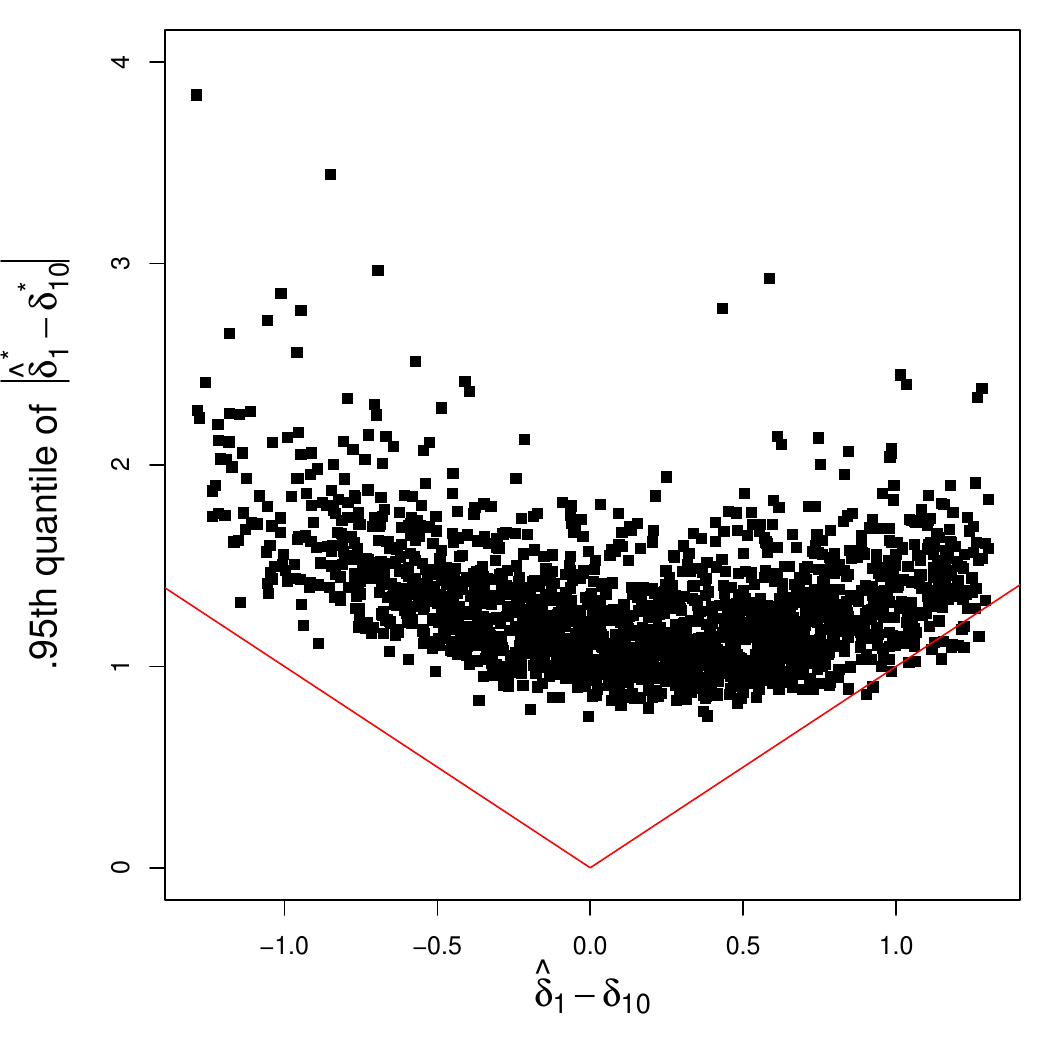}
    \end{subfigure}
    \end{center}
    {\scriptsize Notes: 
    The figures plot the sample statistic $(\hat{\delta}_{1}-\delta_{10})$ and the quantiles of the bootstrap test statistics relevant for confidence intervals for each simulated dataset from the continuous dgp where $\delta_{1}+\delta_{3}\gamma=0$ with $n=1600$.
    Panels (a) and (b) show the 0.025 and 0.975 bootstrap quantiles of $(\hat{\delta}_{1}^{*}-\hat{\delta}_{1})$ (used for NP-B) and $(\hat{\delta}_{1}^{*}-\delta_{10}^{*})$ (for R-B), respectively.
    Panels (c) and (d) show the 0.95 bootstrap quantiles of $|\hat{\delta}_{1}^{*}-\hat{\delta}_{1}|$ (for NP-B(S)) and $|\hat{\delta}_{1}^{*}-\delta_{10}^{*}|$ (for R-B(S)), respectively.
    Red line represents a linear line with 45 degree in Panels (a) and (b), and the line $y=|x|$ in Panels (c) and (d).
    In Panels (a) and (b), the coverage probability is the frequency that the upper and lower bootstrap quantiles (dots) include the red line (45 degree line) between them.
    In Panels (c) and (d), the coverage probability is the frequency  with which the bootstrap quantile (dot) lies above the red line.}
\end{figure}

To investigate the cause of the large difference in coverage rates between symmetric and non-symmetric CIs in \cref{sec:mc}, we present \cref{figure:bootquantiles}, which displays the sample statistic $(\hat{\delta}_{1}-\delta_{10})$ and the quantiles of the bootstrap test statistics that is used for confidence intervals for each simulated dataset.
\cref{figure:bootquantiles} collects results under the specification $\delta_{1}+\delta_{3}\gamma=0$, where the model is continuous, with the sample size 1600.
Results for other coefficients and other specifications are almost identical and are therefore omitted.

Panels (a) and (b) show the 0.025 and 0.975 bootstrap quantiles of $(\hat{\delta}_{1}^{*}-\hat{\delta}_{1})$ (used for NP-B) and $(\hat{\delta}_{1}^{*}-\delta_{10}^{*})$ (for R-B), respectively.
The coverage probability is the frequency that the upper and lower bootstrap quantiles (dots) include the red line (45 degree line) between them. 
We observe that R-B method improves upon NP-B, as the distance between the two bootstrap quantiles tends to be wider.
However, the improvement is not sufficiently large to resolve the undercoverage; see \cref{tab:ch3}.

Note that the bootstrap quantiles (dots of each color) would be horizontally flat if they are asymptotically independent to the sample statistic. 
The nonparametric bootstrap CIs are asymptotically valid if 
$$\sqrt{n}(\hat{\theta}^{*}-\hat{\theta})\xrightarrow{d^{*}}\mathcal{Z}^{*}\text{ in $P$ when }\sqrt{n}(\hat{\theta}-\theta_{0})\xrightarrow{d}\mathcal{Z},$$
where $\mathcal{Z}^{*}$ is an independent copy of $\mathcal{Z}$.
Therefore, the empirical 95\% percentile of $\sqrt{n}(\hat{\delta}_{1}^{*}-\hat{\delta}_{1})$ should be asymptotically independent to $\sqrt{n}(\hat{\delta}_{1}-\delta_{10})$ for the nonparametric bootstrap CI to be valid.

However, as shown in Panel (a), the bootstrap quantiles are negatively correlated with the sample statistic.
Specifically, the correlations between the sample statistic $(\hat{\delta}_{1}-\delta_{10})$ and the 0.975 and 0.025 bootstrap quantiles from NP-B are -0.9037 and -0.8892, respectively.
Our residual bootstrap (R-B) mitigates this issue. The bootstrap quantiles in Panel (b) appear flatter compared to those in Panel (a). 
The corresponding correlations from R-B are -0.7083 and -0.7003 for the 0.975 and 0.025 quantiles, respectively.
While the correlations have decreased, they remain far from zero. Further investigation is warranted, although we leave this for future research.

Panels (c) and (d) show the 0.95 bootstrap quantiles of $|\hat{\delta}_{1}^{*}-\hat{\delta}_{1}|$ (for NP-B(S)) and $|\hat{\delta}_{1}^{*}-\delta_{10}^{*}|$ (for R-B(S)), respectively.
The coverage probability is the frequency of the dots that lie above the red line.
Contrary to Panels (a) and (b), there is no rejection if $\hat{\delta}_{1}-\delta_{10}<0$. Although this brings the coverage probabilities of both bootstraps closer to the nominal level, it is not desirable and misleading.

\subsection{Weakly Endogenous Threshold Variable}

\label{app:MC_othdgp}

We additionally report Monte Carlo results when the threshold variable is not weakly exogenous but weakly endogenous, that is, when the variable is predetermined.
We consider the dgp same with the one in \cref{sec:mc} with an exception that \eqref{eq:mc} is replaced by
\begin{equation}
\label{eq:mc2}
    \begin{pmatrix}
    e_{it} \\ u_{it}
\end{pmatrix} \overset{iid}{\sim} N\left(\begin{pmatrix} 0 \\ 0 \end{pmatrix},\begin{pmatrix}
    1 & \rho_{eu}\\
    \rho_{eu} & 1
\end{pmatrix} \right),
\end{equation}
where $\rho_{eu} = 0.5$.
Other parameters such as $\theta=(\beta',\delta',\gamma)'$ and $\sigma$ remain the same as in \cref{sec:mc}.
Note that under \eqref{eq:mc}, $E[q_{is}\Delta e_{it}] = 0$ if $s\leq t-1$. 
On the other hand, $E[q_{is}\Delta e_{it}] = 0$ if $s\leq t-2$ but $E[q_{it-1}\Delta e_{it}] \neq 0$ under \eqref{eq:mc2}.
Therefore, we need to exclude $q_{it-1}$ from the instrument such that $z_{it}=(y_{it-2},\dots,y_{i1},q_{it-2},\dots,q_{i1})'$.

We consider the specifications where $\delta_{1}+\delta_{3}\gamma=0,0.5,1$ and repeat Monte Carlo iterations 1,000 times.
We report coverage rates of 95\% CIs constructed by different bootstrap methods.
Tables \ref{tab:ch7} and \ref{tab:ch8} show the coverage rates of the threshold location and the coefficients, respectively.

\cref{tab:ch7} shows that Grid-B achieves the most reasonable coverage rates, similar to the results in \cref{tab:ch1} in \cref{sec:mc}.
\cref{tab:ch8} shows that both R-B and NP-B are subject to undercoverage for the coefficients, although R-B offers higher coverage rates than NP-B.
R-B(S) and NP-B(S) return higher coverage rates compared to R-B and NP-B, while NP-B(S) provides higher coverage rates than R-B(S).

\begin{table}[htbp]
    \caption{\label{tab:ch7} Coverage rates of 95\% CIs for the threshold location. 
    Grid-B denotes the grid bootstrap CI defined as \eqref{eq:grid bcs}. 
    NP-B and NP-B(S) denote the percentile and the symmetric percentile CIs by the standard nonparametric bootstrap defined as \eqref{eq:npb_grid_asymm} and \eqref{eq:npb_grid_symm}.
    }
    \centering
    \begin{tabular}[t]{c|c|c|c|c}
    \hline
    \multicolumn{2}{c|}{ } & \multicolumn{3}{c}{$\delta_{1}+\delta_{3}\gamma$} \\
    \hline
     & n & 0 & 0.5 & 1\\
    \hline
           & 400 & 0.990 & 0.983 & 0.975\\
    Grid-B & 800 & 0.986 & 0.983 & 0.965\\
           & 1600 & 0.981 & 0.975 & 0.959\\
           \hline
         & 400 & 0.508 & 0.519 & 0.634\\
    NP-B & 800 & 0.443 & 0.496 & 0.612\\
         & 1600 & 0.468 & 0.501 & 0.610\\
         \hline
            & 400 & 1.000 & 0.998 & 0.994\\
    NP-B(S) & 800 & 1.000 & 1.000 & 0.996\\
            & 1600 & 1.000 & 0.999 & 0.999\\
    \hline
    \end{tabular}
\end{table}

\begin{table}[htbp]
    \caption{\label{tab:ch8}  Coverage rates of 95\% percentile CIs for the coefficients are shown. 
        R-B denotes the percentile CIs by the residual bootstrap defined as \eqref{eq:rbci_asymm}. 
        NP-B denotes the percentile CIs by the standard nonparametric bootstrap defined as \eqref{eq:npbci_asymm}.
        R-B(S) denotes the symmetric percentile CIs by the residual bootstrap defined as \eqref{eq:rbci_symm}. 
        NP-B(S) denotes the symmetric percentile CIs by the standard nonparametric bootstrap defined as \eqref{eq:npbci_symm}.
    }
    \centering
    \begin{tabular}[t]{c|c|c|c|c|c|c|c|c|c|c|c}
    \hline
    \multicolumn{2}{c|}{ } & \multicolumn{5}{c|}{R-B} & \multicolumn{5}{c}{NP-B} \\
    \hline
    $\delta_{1}+\delta_{3}\gamma$ & n & $\beta_{2}$ & $\beta_{3}$ & $\delta_{1}$ & $\delta_{2}$ & $\delta_{3}$ & $\beta_{2}$ & $\beta_{3}$ & $\delta_{1}$ & $\delta_{2}$ & $\delta_{3}$\\\hline
        & 400 & 0.753 & 0.739 & 0.781 & 0.796 & 0.765 & 0.726 & 0.658 & 0.636 & 0.706 & 0.691\\
    0.0 & 800 & 0.795 & 0.729 & 0.783 & 0.786 & 0.756 & 0.764 & 0.629 & 0.640 & 0.709 & 0.669\\
        & 1600 & 0.832 & 0.746 & 0.803 & 0.787 & 0.755 & 0.800 & 0.647 & 0.640 & 0.720 & 0.674\\
        \hline
        & 400 & 0.773 & 0.756 & 0.757 & 0.806 & 0.750 & 0.740 & 0.672 & 0.601 & 0.725 & 0.670\\
    0.5 & 800 & 0.816 & 0.736 & 0.755 & 0.802 & 0.770 & 0.778 & 0.661 & 0.580 & 0.717 & 0.675\\
        & 1600 & 0.835 & 0.746 & 0.776 & 0.791 & 0.770 & 0.811 & 0.660 & 0.605 & 0.720 & 0.660\\
        \hline
        & 400 & 0.805 & 0.777 & 0.743 & 0.822 & 0.754 & 0.765 & 0.712 & 0.618 & 0.731 & 0.701\\
    1.0 & 800 & 0.829 & 0.770 & 0.725 & 0.798 & 0.742 & 0.784 & 0.685 & 0.582 & 0.727 & 0.683\\
        & 1600 & 0.867 & 0.799 & 0.751 & 0.815 & 0.762 & 0.822 & 0.697 & 0.576 & 0.747 & 0.673\\
    \hline
    \multicolumn{2}{c|}{ } & \multicolumn{5}{c|}{R-B(S)} & \multicolumn{5}{c}{NP-B(S)} \\
    \hline
        & 400 & 0.817 & 0.865 & 0.969 & 0.918 & 0.940 & 0.826 & 0.890 & 1.000 & 0.952 & 0.995\\
    0.0 & 800 & 0.843 & 0.878 & 0.973 & 0.913 & 0.943 & 0.868 & 0.901 & 1.000 & 0.942 & 0.996\\
        & 1600 & 0.896 & 0.881 & 0.973 & 0.920 & 0.923 & 0.932 & 0.947 & 1.000 & 0.952 & 0.995\\
        \hline
        & 400 & 0.843 & 0.885 & 0.973 & 0.930 & 0.952 & 0.869 & 0.919 & 1.000 & 0.960 & 0.998\\
    0.5 & 800 & 0.880 & 0.894 & 0.982 & 0.937 & 0.952 & 0.883 & 0.939 & 0.998 & 0.947 & 0.997\\
        & 1600 & 0.907 & 0.906 & 0.980 & 0.942 & 0.940 & 0.930 & 0.970 & 1.000 & 0.964 & 0.995\\
        \hline
        & 400 & 0.880 & 0.911 & 0.966 & 0.945 & 0.966 & 0.875 & 0.951 & 0.999 & 0.965 & 0.999\\
    1.0 & 800 & 0.900 & 0.918 & 0.965 & 0.951 & 0.974 & 0.894 & 0.969 & 0.994 & 0.960 & 0.999\\
        & 1600 & 0.940 & 0.932 & 0.967 & 0.954 & 0.963 & 0.948 & 0.987 & 1.000 & 0.974 & 0.993\\
    \hline
    \end{tabular}
\end{table}

\subsection{Coverage Rates by Asymptotic Confidence Intervals}

\label{app:MC_asym}

We additionally report coverage rates of CIs based on the asymptotic method described in \cite{seo_dynamic_2016}.
The dgp remains the same as in \cref{sec:mc}.
Tables \ref{tab:ch9} and \ref{tab:ch10} show the results for the threshold and the coefficients, respectively.

For the threshold inference, \cref{tab:ch9} shows that the asymptotic method suffers undercoverage for all specifications we consider and does not improve as the sample size grows.
This remains true even when $\delta_{1} + \delta_{3} \gamma = 1$, a case in which the model is discontinuous and the asymptotic CIs are theoretically valid, as shown in \cite{seo_dynamic_2016}.
This especially highlights the desirability of our grid bootstrap method for inference of the threshold location, which achieves good coverage rates in finite samples.

On the other hand, in \cref{tab:ch10}, the coverage rates of the coefficients by the asymptotic method are much closer to the nominal level compared to those obtained from the nonparametric bootstrap or our residual bootstrap for both continuous and discontinuous models; see \cref{tab:ch3}.
We ask readers to be cautious, as it is unclear how the coverage rates of the asymptotic CIs behave when the true model is continuous, as explained in the last paragraph of \cref{sec:asym}.

\begin{table}
\caption{\label{tab:ch9}Coverage rates of 95\% CIs for the threshold location by the asymptotic method described in \cite{seo_dynamic_2016}. The method is based on the asymptotic normality, which holds only when the true model is discontinuous.}
\centering
\begin{tabular}[t]{c|c|c|c|c|c}
\hline
& \multicolumn{5}{c}{$\delta_{1}+\delta_{3}\gamma$} \\
\cline{2-6}
 n & 0 & 0.1 & 0.2 & 0.5 & 1\\
\hline
400 & 0.881 & 0.881 & 0.885 & 0.884 & 0.899\\
800 & 0.864 & 0.862 & 0.860 & 0.846 & 0.869\\
1600 & 0.837 & 0.836 & 0.837 & 0.836 & 0.864\\
\hline
\end{tabular}
\end{table}

\begin{table}
\caption{\label{tab:ch10}Coverage rates of 95\% CIs for the coefficients by the asymptotic method described in \cite{seo_dynamic_2016}. The method is based on the asymptotic normality, which holds only when the true model is discontinuous.}
\centering
\begin{tabular}[t]{c|c|c|c|c|c|c}
\hline
$\delta_{1}+\delta_{3}\gamma$ & n & $\beta_{2}$ & $\beta_{3}$ & $\delta_{1}$ & $\delta_{2}$ & $\delta_{3}$ \\
\hline
    & 400 & 0.950 & 0.923 & 0.951 & 0.916 & 0.970 \\
0.0 & 800 & 0.956 & 0.921 & 0.952 & 0.921 & 0.973 \\
    & 1600 & 0.960 & 0.927 & 0.956 & 0.931 & 0.979 \\
    \hline
    & 400 & 0.947 & 0.922 & 0.947 & 0.917 & 0.972 \\
0.1 & 800 & 0.961 & 0.923 & 0.952 & 0.928 & 0.973 \\
    & 1600 & 0.960 & 0.929 & 0.956 & 0.933 & 0.983 \\
    \hline
    & 400 & 0.942 & 0.919 & 0.947 & 0.915 & 0.974 \\
0.2 & 800 & 0.959 & 0.926 & 0.952 & 0.926 & 0.971 \\
    & 1600 & 0.957 & 0.923 & 0.954 & 0.933 & 0.982 \\
    \hline
    & 400 & 0.943 & 0.922 & 0.944 & 0.914 & 0.977 \\
0.5 & 800 & 0.959 & 0.934 & 0.953 & 0.937 & 0.977 \\
    & 1600 & 0.953 & 0.934 & 0.953 & 0.930 & 0.983 \\
    \hline
    & 400 & 0.949 & 0.937 & 0.950 & 0.925 & 0.987 \\
1.0 & 800 & 0.958 & 0.952 & 0.952 & 0.945 & 0.985 \\
    & 1600 & 0.958 & 0.949 & 0.955 & 0.936 & 0.981 \\
\hline
\end{tabular}
\end{table}

			\section{Proofs of Theorems in Section \ref{sec:asym} and Auxiliary Lemmas}

                \paragraph{Additional notations}
			We introduce additional notations as lemmas in this online appendix involve more empirical process theory.
			Suppose that $(\mathcal{X},\mathcal{A})$ is a measurable space and $\omega_{1},\omega_{2},...$ are i.i.d. random elements in $(\mathcal{X},\mathcal{A})$ with probability law $P$.
			For a point $\omega\in\mathcal{X}$, let $\delta_{\omega}$ be a dirac measure at $\omega$\footnote{Although we already use $\delta$ as the subvector of the parameter $\theta=(\beta',\delta',\gamma)'$, we still use $\delta$ to represent dirac measure as it is strong convention in the literature. We explicitly mention if $\delta$ is used as dirac measure to avoid confusion.}.
			The empirical measure of a sample $\omega_{1},...,\omega_{n}$ is $\mathbb{P}_{n}=\frac{1}{n}\sum_{i=1}^{n}\delta_{\omega_{i}}$, and the empirical process is $\mathbb{G}_{n}=\sqrt{n}(\mathbb{P}_{n}-P)$.
			Let $\mathcal{F}$ be a functional class, elements of which are measurable functions from $\mathcal{X}$ to $\mathbb{R}$.
			We call a function $F:\mathcal{X}\rightarrow\mathbb{R}$ an envelope of $\mathcal{F}$ if $|f|\leq F$ for all $f\in\mathcal{F}$.
			For a stochastic process $\mathbb{G}$ and a functional class $\mathcal{F}$, define $\|\mathbb{G}\|_{\mathcal{F}}:=\sup_{f\in\mathcal{F}}|\mathbb{G}f|$.
            
			\subsection{Proof of \texorpdfstring{\cref{thm:distance test}}{\cref{thm:distance test}}.}
			
			\subsubsection{Continuous Model.} 
			
			\paragraph{When $\gamma=\gamma_{0}$.}
			Note that the constrained estimator $\hat{\alpha}(\gamma_{0})=\arg\min_{\alpha\in A}\hat{Q}_{n}(\alpha,\gamma_{0})$ is $\sqrt{n}$-consistent to $\alpha_{0}$, which is identical to the convergence rate of $\hat{\alpha}$, since the problem becomes a standard linear dynamic panel estimation.
			Let $a=\sqrt{n}(\alpha-\alpha_{0})$ and $b=n^{1/4}(\gamma-\gamma_{0})$. The distance test statistic can be rewritten as follows:
			\begin{eqnarray*}
				\mathcal{D}_{n}(\gamma_{0}) &=& \inf_{a}\mathbb{S}_{n}(a,0)-\inf_{a,b}\mathbb{S}_{n}(a,b) + o_p (1) \\ 
				&\xrightarrow{d}& \inf_{a}\mathbb{S}(a,0)-\inf_{a,b}\mathbb{S}(a,b) \\
				&=& \min_{a} (M_{0}a-e)'\Omega^{-1}(M_{0}a-e) -\min_{a,b^{2}}(M_{0}a+Hb^{2}-e)'\Omega^{-1}(M_{0}a+Hb^{2}-e),
			\end{eqnarray*}
			where we apply the CMT. \cite{lee_testing_2011} showed that the difference between the constrained and unconstrained infima is a continuous operator on $\ell^{\infty}(\mathbb{K})$. 
			
			Note that $\min_{a} (M_{0}a-e)'\Omega^{-1}(M_{0}a-e)=e'(\Omega^{-1}-\Omega^{-1}M_{0}(M_{0}'\Omega^{-1}M_{0})^{-1}M_{0}'\Omega^{-1})e$, 
			while
			\begin{align*}
				&\min_{a,b^{2}}(M_{0}a+Hb^{2}-e)'\Omega^{-1}(M_{0}a+Hb^{2}-e)\\
				&=(M_{0}a_{0}+Hb_{0}^{2}-e)'\Omega^{-1}(M_{0}a_{0}+Hb_{0}^{2}-e) \\
				&=(M_{0}'\Omega^{-1}M_{0}a_{0}+M_{0}'\Omega^{-1}Hb_{0}^{2})'(M_{0}'\Omega^{-1}M_{0})^{-1}(M_{0}'\Omega^{-1}M_{0}a_{0}+M_{0}'\Omega^{-1}Hb_{0}^{2}) \\
				&\quad +b_{0}^{2}H'\Omega^{-1/2}(I-P_{\Omega^{-1/2}M_{0}})\Omega^{-1/2}Hb_{0}^{2}-2e'\Omega^{-1}M_{0}(M_{0}'\Omega^{-1}M_{0})^{-1}(M_{0}'\Omega^{-1}M_{0}a_{0}+M_{0}'\Omega^{-1}Hb_{0}^{2}) \\
				&\quad -2e'\Omega^{-1/2}(I-P_{\Omega^{-1/2}M_{0}})\Omega^{-1/2}Hb_{0}^{2} + e'\Omega^{-1}e,
			\end{align*}
			where $(a_{0},b_{0}^{2})$ is the argmin, whose formula is derived in the proof of \cref{thm:asym}.
			By plugging in one of the first order conditions, $M_{0}'\Omega^{-1}M_{0}a_{0}+M_{0}'\Omega^{-1}Hb_{0}^{2}=M_{0}'\Omega^{-1}e$, and the formula for $b_{0}$, we can get
			\begin{align*}
				&\min_{a,b^{2}}(M_{0}a+Hb^{2}-e)'\Omega^{-1}(M_{0}a+Hb^{2}-e)\\
				&\quad =\begin{cases}
					-e'\Omega^{-1}M_{0}(M_{0}'\Omega^{-1}M_{0})^{-1}M_{0}'\Omega^{-1}e-e'\Xi H(H'\Xi H)^{-1}H'\Xi e+e'\Omega^{-1}e & \text{if }H'\Xi e\geq 0 \\
					-e'\Omega^{-1}M_{0}(M_{0}'\Omega^{-1}M_{0})^{-1}M_{0}'\Omega^{-1}e+e'\Omega^{-1}e & \text{else}.
				\end{cases}
			\end{align*}
			Therefore, the limit distribution of the test statistic is identical to
			\[
			\begin{cases}
				e'\Xi H(H'\Xi H)^{-1}H'\Xi e & \text{if }H'\Xi e\geq 0 \\
				0 & \text{else}.
			\end{cases}
			\]
			Note that $e'\Xi H(H'\Xi H)^{-1}H'\Xi e\sim\chi^{2}_{1}$ as $H'\Xi e \sim N(0,H'\Xi \Omega\Xi H)$, and $H'\Xi \Omega\Xi H= H'\Xi H$.
			
			\paragraph{When $\gamma\neq\gamma_{0}$.}
			We show that $\mathcal{D}_{n}(\gamma)$ diverges to infinity in probability.
			There is a constant $C_{1}\in(0,+\infty)$ such that $\inf_{\alpha\in A}\|g_{0}(\alpha,\gamma)\| \geq C_{1}$. This is because $g_{0}(\theta)$ is zero if and only if $\theta=\theta_{0}$, by \cref{ass:gmm} and \cref{thm:id}, and continuous on $\Theta$, by \cref{ass:dgp}, while the restricted parameter set $\{\theta=(\beta',\delta',\gamma)'\in\Theta:\gamma=c\}$ is closed for all $c\in\Gamma$.
			$\mathcal{G}=\{g(\omega_{i},\theta):\theta\in\Theta\}$ is shown to satisfy the uniform entropy condition in the proof of \cref{lem:equicont}, and hence $\sup_{\theta\in\Theta}\|\bar{g}_{n}(\theta)-g_{0}(\theta)\|=o_{p}(1)$ by Glivenko-Cantelli theorem. 
			By triangle inequality, $ C_{1} \leq \|g_{0}(\hat{\alpha}(\gamma),\gamma)\|\leq \|\bar{g}_{n}(\hat{\alpha}(\gamma),\gamma)\|+ o_{p}(1)$.
			Meanwhile, $\|\bar{g}_{n}(\hat{\theta})\|= O_{p}(n^{-1/2})$ because $\|\bar{g}_{n}(\hat{\theta})\|\leq \|\bar{g}_{n}(\theta_{0})\|=O_{p}(n^{-1/2})$.
			Therefore, there exists $C_{2}\in(0,+\infty)$ such that $\hat{Q}_{n}(\hat{\alpha}(\gamma),\gamma)-\hat{Q}_{n}(\hat{\theta})\geq C_{2}+O_{p}(n^{-1})$, which implies that $P(\mathcal{D}_{n}(\gamma)>M)=P(\hat{Q}_{n}(\hat{\alpha}(\gamma),\gamma)-\hat{Q}_{n}(\hat{\theta})>M/n)\rightarrow1$ for any $M<\infty$.

			\subsubsection{Discontinuous Model.} 
			
			\paragraph{When $\gamma=\gamma_{0}$.}
			As in the proof for the continuous model, we apply the CMT to the test statistic.
			Let $a=\sqrt{n}(\alpha-\alpha_{0})$ and $b=\sqrt{n}(\gamma-\gamma_{0})$.
			First, we will show that when the model is discontinuous and Assumptions \ref{ass:gmm}, \ref{ass:dgp}, and \ref{ass:locDis} are true, $\mathbb{S}_{n}(a,b) \rightsquigarrow \mathbb{S}_{J}(a,b)=(M_{0}a+Gb-e)'\Omega^{-1}(M_{0}a+Gb-e)$ in $\ell^{\infty}(\mathbb{K})$ for any compact $\mathbb{K}\subset\mathbb{R}^{2p+2}$.
			Note that
			\begin{align}
				\label{eq:distant discont re-crit 1}
				\sqrt{n}\bar{g}_{n}(\alpha_{0}+\tfrac{a}{\sqrt{n}}, \gamma_{0}+\tfrac{b}{\sqrt{n}})
				=&
				\sqrt{n}\begin{pmatrix}
					\frac{1}{n}\sum_{i=1}^{n}z_{it_{0}}\Delta\epsilon_{it_{0}} \\
					\vdots \\
					\frac{1}{n}\sum_{i=1}^{n}z_{iT}\Delta\epsilon_{iT}
				\end{pmatrix}
				-
				\begin{pmatrix}
					\tfrac{1}{n}\sum_{i=1}^{n} z_{it_{0}}\Delta x_{it_{0}}'\\
					\vdots\\
					\tfrac{1}{n}\sum_{i=1}^{n} z_{iT}\Delta x_{iT}'
				\end{pmatrix}a_{1}
				\\
				\label{eq:distant discont re-crit 2}
				&
				-\begin{pmatrix}
					\tfrac{1}{n}\sum_{i=1}^{n} z_{it_{0}}1_{it_{0}}(\gamma_{0}+\tfrac{b}{\sqrt{n}})' X_{it_{0}}\\
					\vdots\\
					\tfrac{1}{n}\sum_{i=1}^{n} z_{iT}1_{iT}(\gamma_{0}+\tfrac{b}{\sqrt{n}})' X_{iT}
				\end{pmatrix}a_{2}
				\\
				\label{eq:distant discont re-crit 3}
				&
				+\sqrt{n}\begin{pmatrix}
					\tfrac{1}{n}\sum_{i=1}^{n} z_{it_{0}}(1_{it_{0}}(\gamma_{0})'-1_{it_{0}}(\gamma_{0}+\tfrac{b}{\sqrt{n}})')X_{it_{0}}\\
					\vdots\\
					\tfrac{1}{n}\sum_{i=1}^{n} z_{iT}(1_{iT}(\gamma_{0})'-1_{iT}(\gamma_{0}+\tfrac{b}{\sqrt{n}})')X_{iT}
				\end{pmatrix}\delta_{0}.
			\end{align}
			The terms in the first two lines of the right hand side \eqref{eq:distant discont re-crit 1} and \eqref{eq:distant discont re-crit 2} converge in distribution to $(M_{0}a-e)$ uniformly with respect to $b\in [-K,K]$.
			Since $\sup_{b:|b|\leq K}\sqrt{n}\|\bar{g}_{n}(\alpha_{0},\gamma_{0}+\frac{b}{\sqrt{n}})-\bar{g}_{n}(\alpha_{0},\gamma_{0})-g_{0}(\alpha_{0},\gamma_{0}+\frac{b}{\sqrt{n}})+g_{0}(\alpha_{0},\gamma_{0})\|=o_{p}(1)$ by \cref{lem:equicont},
			\begin{equation*}
				\sqrt{n}
				\left\|
				\begin{pmatrix}
					\tfrac{1}{n}\sum_{i=1}^{n} z_{it_{0}}(1_{it_{0}}(\gamma_{0})'-1_{it_{0}}(\gamma_{0}+\tfrac{b}{\sqrt{n}})')X_{it_{0}}\delta_{0}\\
					\vdots\\
					\tfrac{1}{n}\sum_{i=1}^{n} z_{iT}(1_{iT}(\gamma_{0})'-1_{iT}(\gamma_{0}+\tfrac{b}{\sqrt{n}})')X_{iT}\delta_{0}
				\end{pmatrix}
				-
				\begin{pmatrix}
					E [z_{it_{0}}(1_{it_{0}}(\gamma_{0})'-1_{it_{0}}(\gamma_{0}+\tfrac{b}{\sqrt{n}})')X_{it_{0}}\delta_{0}]\\
					\vdots\\
					E [z_{iT}(1_{iT}(\gamma_{0})'-1_{iT}(\gamma_{0}+\tfrac{b}{\sqrt{n}})')X_{iT}\delta_{0}]
				\end{pmatrix}
				\right\|
			\end{equation*}
			converges in probability to zero uniformly with respect to $b\in[-K,K]$.
			Suppose $b>0$.
			The result for $b<0$ is similar.
			By application of Talyor expansion,
			\begin{equation*}
				\sqrt{n}E[z_{it}(1,x_{it}')\delta_{0}1\{\gamma_{0}+\tfrac{b}{\sqrt{n}}\geq q_{it}>\gamma_{0}\}] \rightarrow E_{t}[z_{it}(1,x_{it}')\delta_{0}|\gamma_{0}]f_{t}(\gamma_{0}) b,
			\end{equation*}
			uniformly with respect to $b\in[-K,K]$, and similar limit result can be derived for $\sqrt{n}E[z_{it}((1,x_{it-1}')\delta_{0}1\{\gamma_{0}+\tfrac{b}{\sqrt{n}}\geq q_{it-1}>\gamma_{0}\}]$.
			Hence, we can derive that the term \eqref{eq:distant discont re-crit 3} converges in probability to $Gb$ uniformly with respect to $b\in [-K,K]$.
			
			By the CMT, the test statistic converges in distribution to
			\begin{equation*}
				\min_{a} (M_{0}a-e)'\Omega^{-1}(M_{0}a-e) -\min_{a,b}(M_{0}a+Gb-e)'\Omega^{-1}(M_{0}a+Gb-e).
			\end{equation*}
			Note that $\min_{a} (M_{0}a-e)'\Omega^{-1}(M_{0}a-e)=e'(\Omega^{-1}-\Omega^{-1}M_{0}(M_{0}'\Omega^{-1}M_{0})^{-1}M_{0}'\Omega^{-1})e$, and
			$\min_{a,b}(M_{0}a+Gb-e)'\Omega^{-1}(M_{0}a+Gb-e)=e'(\Omega^{-1}-\Omega^{-1}D_{1}(D_{1}'\Omega^{-1}D_{1})^{-1}D_{1}'\Omega^{-1})e$. 
			Therefore, the limit distribution of the test statistic is identical to the distribution of
			\[
			e'\Omega^{-1/2}[\Omega^{-1/2}D_{1}(D_{1}'\Omega^{-1}D_{1})^{-1}D_{1}'\Omega^{-1/2}-\Omega^{-1/2}M_{0}(M_{0}'\Omega^{-1}M_{0})^{-1}M_{0}'\Omega^{-1/2}]\Omega^{-1/2}e.
			\]
			The matrix $\Omega^{-1/2}D_{1}(D_{1}'\Omega^{-1}D_{1})^{-1}D_{1}'\Omega^{-1/2}-\Omega^{-1/2}M_{0}(M_{0}'\Omega^{-1}M_{0})^{-1}M_{0}'\Omega^{-1/2}$ is idempotent since the column space of $\Omega^{-1/2}M_{0}$ lies in the column space of $\Omega^{-1/2}D_{1}$.
			The rank of the matrix is 1. 
			Since $\Omega^{-1/2}e\sim N(0,I)$, the chi-square distribution with 1 degree of freedom is the limit distribution.
			
			\paragraph{When $\gamma\neq\gamma_{0}$.}
			The proof showing that $\mathcal{D}_{n}(\gamma)$ diverges when $\gamma\neq\gamma_{0}$ for the discontinuous model is identical to the proof written for the continuous model.

			\subsection{Proof of \texorpdfstring{\cref{thm:continuity test}}{\cref{thm:continuity test}}.}
			\paragraph{Under the null hypothesis.}
			Define a map $T$ such that $T(\psi)=(\beta',-\gamma\delta_{3},0,...,0,\delta_{3},\gamma)'\in\mathbb{R}^{2p+2}$ if $\psi=(\beta',\delta_{3},\gamma)'\in\mathbb{R}^{p+2}$.
			Let $\psi_{0}=(\beta_{0}',\delta_{30},\gamma_{0})'$.
			Note that
			\[g_{i}(T(\psi))=\begin{pmatrix}
				z_{it_{0}}\{\Delta y_{it_{0}}-\Delta x_{it_{0}}'\beta-[(q_{it_{0}}-\gamma)1_{\{q_{it_{0}}>\gamma\}}-(q_{it_{0}-1}-\gamma)1_{\{q_{it_{0}-1}>\gamma\}}]\delta_{3}\} \\
				\vdots \\
				z_{iT}\{\Delta y_{iT}-\Delta x_{iT}'\beta-[(q_{iT}-\gamma)1_{\{q_{iT}>\gamma\}}-(q_{iT-1}-\gamma)1_{\{q_{iT-1}>\gamma\}}]\delta_{3}\}
			\end{pmatrix}.\]
			The first-order derivative of $g_{0}(T(\psi))$ with respect to $\psi$ is
			\begin{multline*}
				D_{\psi}=\\
				E\begin{bmatrix}
					-z_{it_{0}}\Delta x_{it_{0}}',& -z_{it_{0}}[(q_{it_{0}}-\gamma_{0})1_{\{q_{it_{0}}>\gamma_{0}\}}-(q_{it_{0}-1}-\gamma_{0})1_{\{q_{it_{0}-1}>\gamma_{0}\}}],& z_{it_{0}}[1_{\{q_{it_{0}}>\gamma_{0}\}}-1_{\{q_{it_{0}-1}>\gamma_{0}\}}]\delta_{30}\\
					\vdots & \vdots & \vdots \\
					-z_{iT}\Delta x_{iT}',& -z_{iT}[(q_{iT}-\gamma_{0})1_{\{q_{iT}>\gamma_{0}\}}-(q_{iT-1}-\gamma_{0})1_{\{q_{iT-1}>\gamma_{0}\}}],& z_{iT}[1_{\{q_{iT}>\gamma_{0}\}}-1_{\{q_{iT-1}>\gamma_{0}\}}]\delta_{30}
				\end{bmatrix}.
			\end{multline*}
			$D_{\psi}$ is a matrix that is identical to a binding of the columns of $M_{10}$ and $N_{20}$.
			If $\hat{\psi}=\arg\min_{\psi}\hat{Q}_{n}(T(\psi))$, then $\sqrt{n}(\hat{\psi}-\psi_{0})\xrightarrow{d}N(0,(D_{\psi}'\Omega D_{\psi})^{-1})$ (see \cite{seo_estimation_2019}).
			The continuity test statistic $\mathcal{T}_{n}=n(\hat{Q}_{n}(\tilde{\theta})-\hat{Q}_{n}(\hat{\theta}))$ can be rewritten as 
			\begin{equation*}
				n(\hat{Q}_{n}(T(\hat{\psi}))-\hat{Q}_{n}(\hat{\theta}))=n\left(\min_{(\theta',\psi')':\theta=\theta_{0}}(\hat{Q}_{n}(T(\psi))-\hat{Q}_{n}(\theta))-\min_{(\theta',\psi')':\psi=\psi_{0}}(-\hat{Q}_{n}(T(\psi))+\hat{Q}_{n}(\theta))\right).
			\end{equation*}
			Reparametrize such that $a=\sqrt{n}(\alpha-\alpha_{0})$, $b=n^{1/4}(\gamma-\gamma_{0})$, and $r=\sqrt{n}(\psi-\psi_{0})$.
			Define a centered criterion by
			\begin{equation*}
				\mathbb{M}_{n}(a,b,r)=n(\hat{Q}_{n}(T(\psi_{0}+\tfrac{r}{\sqrt{n}}))-\hat{Q}_{n}(\alpha_{0}+\tfrac{a}{\sqrt{n}},\gamma_{0}+\tfrac{b}{n^\frac{1}{4}})).
			\end{equation*}
			We will show that $\mathbb{M}_{n}$ weakly converges to a process $\mathbb{M}$ in $\ell^{\infty}(\mathbb{K})$ for every compact $\mathbb{K}\subset \mathbb{R}^{3p+4}$. 
			Then, by the CMT, the continuity test statistic converges in distribution to
			\begin{equation*}
				\min_{(a',b,r')':(a',b)'=0}\mathbb{M}(a,b,r)-\min_{(a',b,r')':r=0}(-\mathbb{M}(a,b,r)).
			\end{equation*}
			In the proof of \cref{thm:asym}, it is shown that $\sqrt{n}\bar{g}_{n}(\alpha_{0}+\tfrac{a}{\sqrt{n}},\gamma_{0}+\tfrac{b}{n^\frac{1}{4}})\rightsquigarrow (M_{0}a+Hb^{2}-e)$ and
			\begin{equation*}
				n\hat{Q}_{n}(\alpha_{0}+\tfrac{a}{\sqrt{n}},\gamma_{0}+\tfrac{b}{n^\frac{1}{4}})\rightsquigarrow (M_{0}a+Hb^{2}-e)'\Omega^{-1}(M_{0}a+Hb^{2}-e).
			\end{equation*}
			Let $r_{1}=\sqrt{n}(\beta-\beta_{0})$, $r_{2}=(r_{21},r_{22})'$, $r_{21}=\sqrt{n}(\delta_{3}-\delta_{30})$, and $r_{22}=\sqrt{n}(\gamma-\gamma_{0})$.
			Then,
			\begin{align*}
				&\sqrt{n}\bar{g}_{n}(T(\psi_{0}+\tfrac{r}{\sqrt{n}})) \\
				&=\begin{pmatrix}
					\frac{1}{\sqrt{n}}\sum_{i=1}^{n}z_{it_{0}}\Delta\epsilon_{it_{0}} \\
					\vdots \\
					\frac{1}{\sqrt{n}}\sum_{i=1}^{n}z_{iT}\Delta\epsilon_{iT} 
				\end{pmatrix}
				-
				\begin{pmatrix}
					\frac{1}{n}\sum_{i=1}^{n}z_{it_{0}}\Delta x_{it_{0}} \\
					\vdots \\
					\frac{1}{n}\sum_{i=1}^{n}z_{iT}\Delta x_{iT} 
				\end{pmatrix}r_{1}\\
				&-
				\begin{pmatrix}
					\frac{1}{n}\sum_{i=1}^{n}z_{it_{0}}[(q_{it_{0}}-\gamma_{0}-\frac{r_{22}}{\sqrt{n}})1\{q_{it_{0}}>\gamma_{0}+\frac{r_{22}}{\sqrt{n}}\}-(q_{it_{0}-1}-\gamma_{0}-\frac{r_{22}}{\sqrt{n}})1\{q_{it_{0}-1}>\gamma_{0}+\frac{r_{22}}{\sqrt{n}}\}] \\
					\vdots \\
					\frac{1}{n}\sum_{i=1}^{n}z_{iT}[(q_{iT}-\gamma_{0}-\frac{r_{22}}{\sqrt{n}})1\{q_{iT}>\gamma_{0}+\frac{r_{22}}{\sqrt{n}}\}-(q_{iT-1}-\gamma_{0}-\frac{r_{22}}{\sqrt{n}})1\{q_{iT-1}>\gamma_{0}+\frac{r_{22}}{\sqrt{n}}\}]
				\end{pmatrix}r_{21}\\
				&+\sqrt{n}\left\{
				\begin{pmatrix}
					\frac{1}{n}\sum_{i=1}^{n}z_{it_{0}}[(q_{it_{0}}-\gamma_{0})1\{q_{it_{0}}>\gamma_{0}\}-(q_{it_{0}}-\gamma_{0}-\frac{r_{22}}{\sqrt{n}})1\{q_{it_{0}}>\gamma_{0}+\frac{r_{22}}{\sqrt{n}}\}] \\
					\vdots \\
					\frac{1}{n}\sum_{i=1}^{n}z_{iT}[(q_{iT}-\gamma_{0})1\{q_{iT}>\gamma_{0}\}-(q_{iT}-\gamma_{0}-\frac{r_{22}}{\sqrt{n}})1\{q_{iT}>\gamma_{0}+\frac{r_{22}}{\sqrt{n}}\}]
				\end{pmatrix}
				\right.\\
				&
				\left.
				\qquad -
				\begin{pmatrix}
					\frac{1}{n}\sum_{i=1}^{n}z_{it_{0}}[(q_{it_{0}-1}-\gamma_{0})1\{q_{it_{0}-1}>\gamma_{0}\}-(q_{it_{0}-1}-\gamma_{0}-\frac{r_{22}}{\sqrt{n}})1\{q_{it_{0}-1}>\gamma_{0}+\frac{r_{22}}{\sqrt{n}}\}] \\
					\vdots \\
					\frac{1}{n}\sum_{i=1}^{n}z_{iT}[(q_{iT-1}-\gamma_{0})1\{q_{iT-1}>\gamma_{0}\}-(q_{iT-1}-\gamma_{0}-\frac{r_{22}}{\sqrt{n}})1\{q_{iT-1}>\gamma_{0}+\frac{r_{22}}{\sqrt{n}}\}]
				\end{pmatrix}
				\right\}\delta_{30}.
			\end{align*}
			By the CLT and LLN,
			\begin{equation*}
				\begin{pmatrix}
					\frac{1}{\sqrt{n}}\sum_{i=1}^{n}z_{it_{0}}\Delta\epsilon_{it_{0}} \\
					\vdots \\
					\frac{1}{\sqrt{n}}\sum_{i=1}^{n}z_{iT}\Delta\epsilon_{iT} 
				\end{pmatrix}
				-
				\begin{pmatrix}
					\frac{1}{n}\sum_{i=1}^{n}z_{it_{0}}\Delta x_{it_{0}} \\
					\vdots \\
					\frac{1}{n}\sum_{i=1}^{n}z_{iT}\Delta x_{iT} 
				\end{pmatrix}r_{1}
				\xrightarrow{d} (M_{10}r_{1}-e).
			\end{equation*}
			By the ULLN (application of \cref{lem:ULLN}) and continuity of $\kappa\mapsto E[z_{it}(1,q_{it})1\{q_{it}>\gamma_{0}+\kappa\}]$ and $\kappa\mapsto E[z_{it}(1,q_{it-1})1\{q_{it-1}>\gamma_{0}+\kappa\}]$ at $\kappa=0$,
			\begin{multline*}
				\begin{pmatrix}
					\frac{1}{n}\sum_{i=1}^{n}z_{it_{0}}[(q_{it_{0}}-\gamma_{0}-\frac{r_{22}}{\sqrt{n}})1\{q_{it_{0}}>\gamma_{0}+\frac{r_{22}}{\sqrt{n}}\}-(q_{it_{0}-1}-\gamma_{0}-\frac{r_{22}}{\sqrt{n}})1\{q_{it_{0}-1}>\gamma_{0}+\frac{r_{22}}{\sqrt{n}}\}] \\
					\vdots \\
					\frac{1}{n}\sum_{i=1}^{n}z_{iT}[(q_{iT}-\gamma_{0}-\frac{r_{22}}{\sqrt{n}})1\{q_{iT}>\gamma_{0}+\frac{r_{22}}{\sqrt{n}}\}-(q_{iT-1}-\gamma_{0}-\frac{r_{22}}{\sqrt{n}})1\{q_{iT-1}>\gamma_{0}+\frac{r_{22}}{\sqrt{n}}\}]
				\end{pmatrix}r_{21}\\
				\xrightarrow{p} 
				\begin{pmatrix}
					Ez_{it_{0}}[(q_{it_{0}}-\gamma_{0})1\{q_{it_{0}}>\gamma_{0}\}-(q_{it_{0}-1}-\gamma_{0})1\{q_{it_{0}-1}>\gamma_{0}\}] \\
					\vdots \\
					Ez_{iT}[(q_{iT}-\gamma_{0})1\{q_{iT}>\gamma_{0}\}-(q_{iT-1}-\gamma_{0})1\{q_{iT-1}>\gamma_{0}\}]
				\end{pmatrix}r_{21}
			\end{multline*}
			uniformly with respect to $r_{22}\in [-K,K]$.
			Finally,
			\begin{multline*}
				\sqrt{n}\left\{
				\begin{pmatrix}
					\frac{1}{n}\sum_{i=1}^{n}z_{it_{0}}[(q_{it_{0}}-\gamma_{0})1\{q_{it_{0}}>\gamma_{0}\}-(q_{it_{0}}-\gamma_{0}-\frac{r_{22}}{\sqrt{n}})1\{q_{it_{0}}>\gamma_{0}+\frac{r_{22}}{\sqrt{n}}\}] \\
					\vdots \\
					\frac{1}{n}\sum_{i=1}^{n}z_{iT}[(q_{iT}-\gamma_{0})1\{q_{iT}>\gamma_{0}\}-(q_{iT}-\gamma_{0}-\frac{r_{22}}{\sqrt{n}})1\{q_{iT}>\gamma_{0}+\frac{r_{22}}{\sqrt{n}}\}]
				\end{pmatrix}
				\right.\\
				\left.
				\qquad -
				\begin{pmatrix}
					\frac{1}{n}\sum_{i=1}^{n}z_{it_{0}}[(q_{it_{0}-1}-\gamma_{0})1\{q_{it_{0}-1}>\gamma_{0}\}-(q_{it_{0}-1}-\gamma_{0}-\frac{r_{22}}{\sqrt{n}})1\{q_{it_{0}-1}>\gamma_{0}+\frac{r_{22}}{\sqrt{n}}\}] \\
					\vdots \\
					\frac{1}{n}\sum_{i=1}^{n}z_{iT}[(q_{iT-1}-\gamma_{0})1\{q_{iT-1}>\gamma_{0}\}-(q_{iT-1}-\gamma_{0}-\frac{r_{22}}{\sqrt{n}})1\{q_{iT-1}>\gamma_{0}+\frac{r_{22}}{\sqrt{n}}\}]
				\end{pmatrix}
				\right\}
				\\
				\xrightarrow{p}
				\begin{pmatrix}
					Ez_{it_{0}}[1\{q_{it_{0}}>\gamma_{0}\}-1\{q_{it_{0}-1}>\gamma_{0}\}] \\
					\vdots \\
					Ez_{iT}[1\{q_{iT}>\gamma_{0}\}-1\{q_{iT-1}>\gamma_{0}\}]
				\end{pmatrix}r_{22}
			\end{multline*}
			uniformly with respect to $r_{22}\in[-K,K]$.
			Suppose that $r_{22}>0$.
			The case for $r_{22}<0$ follows similarly.
			The last uniform convergence holds because \cref{lem:equicont} yields $\sqrt{n}\|\bar{g}_{n}(T(\beta_{0},\delta_{30},\gamma_{0}+\frac{r_{22}}{\sqrt{n}}))-\bar{g}_{n}(T(\beta_{0},\delta_{30},\gamma_{0}))-g_{0}(T(\beta_{0},\delta_{30},\gamma_{0}+\frac{r_{22}}{\sqrt{n}}))+g_{0}(T(\beta_{0},\delta_{30},\gamma_{0}))\|=o_{p}(1)$ uniformly with respect to $r_{22}\in [-K,K]$
			and the following application of Taylor expansion:
			\begin{align*}
				&\sqrt{n}E[z_{it}((q_{it}-\gamma_{0})1\{q_{it}>\gamma_{0}\}-(q_{it}-\gamma_{0}-\frac{r_{22}}{\sqrt{n}})1\{q_{it}>\gamma_{0}+\frac{r_{22}}{\sqrt{n}}\})]\\
				&=\sqrt{n}E[z_{it}(q_{it}-\gamma_{0})1\{\gamma_{0}+\frac{r_{22}}{\sqrt{n}}\geq q_{it}>\gamma_{0}\}] +r_{22}E[z_{it}1\{q_{it}>\gamma_{0}+\frac{r_{22}}{\sqrt{n}}\}]\\
				&\rightarrow E_{t}[z_{it}(q_{it}-\gamma_{0})|\gamma_{0}]f_{t}(\gamma_{0}) r_{22} + E[z_{it}1\{q_{it}>\gamma_{0}\}]r_{22}=E[z_{it}1\{q_{it}>\gamma_{0}\}]r_{22}
			\end{align*}
			uniformly with respect to $r_{22}\in[-K,K]$ as $n\rightarrow\infty$.
			
			In conclusion, $\sqrt{n}\bar{g}_{n}(T(\psi_{0}+\tfrac{r}{\sqrt{n}}))\rightsquigarrow (D_{\psi}r-e)$, and
			\begin{align*}
				\mathbb{M}(a,b,r)
				&=(D_{\psi}r-e)'\Omega^{-1}(D_{\psi}r-e)-(M_{0}a+Hb^{2}-e)'\Omega^{-1}(M_{0}a+Hb^{2}-e) \\
				&=(M_{10}r_{1}+N_{20}r_{2}-e)'\Omega^{-1}(M_{10}r_{1}+N_{20}r_{2}-e)\\
				&\qquad -(M_{10}a_{1}+M_{20}a_{2}+Hb^{2}-e)'\Omega^{-1}(M_{10}a_{1}+M_{20}a_{2}+Hb^{2}-e),
			\end{align*}
			where $a_{1}=\sqrt{n}(\beta-\beta_{0})$ and $a_{2}=\sqrt{n}(\delta-\delta_{0})$.
			By applying the CMT, the continuity test statistic converges in distribution to 
			\begin{align*}
				&\min_{r}(M_{10}r_{1}+N_{20}r_{2}-e)'\Omega^{-1}(M_{10}r_{1}+N_{20}r_{2}-e)\\
				&\qquad -\min_{a,b^{2}}(M_{10}a_{1}+M_{20}a_{2}+Hb^{2}-e)'\Omega^{-1}(M_{10}a_{1}+M_{20}a_{2}+Hb^{2}-e).
			\end{align*}
			By similar computations to the proof of \cref{thm:distance test},
			\begin{align*}
				&\min_{r_{1},r_{2}}(M_{10}r_{1}+N_{20}r_{2}-e)'\Omega^{-1}(M_{10}r_{1}+N_{20}r_{2}-e)\\
				&\quad = e'\Omega^{-1}e-e'\Omega^{-1}M_{10}(M_{10}'\Omega^{-1}M_{10})^{-1}M_{10}'\Omega^{-1}e -e'\Xi _{1}N_{20}(N_{20}'\Xi _{1}N_{20})^{-1}N_{20}'\Xi _{1}e,\\
				&\min_{a_{1},a_{2},b^{2}}(M_{10}a_{1}+M_{20}a_{2}+Hb^{2}-e)'\Omega^{-1}(M_{10}a_{1}+M_{20}a_{2}+Hb^{2}-e) \\
				&\quad= \begin{cases}\begin{array}{l} 
						-e'\Omega^{-1}M_{10}(M_{10}'\Omega^{-1}M_{10})^{-1}M_{10}'\Omega^{-1}e -e'\Xi _{1}M_{20}(M_{20}'\Xi _{1}M_{20})^{-1}M_{20}'\Xi _{1}e\\
						-e'\Xi _{12}H(H'\Xi _{12}H)^{-1}H'\Xi _{12}e +e'\Omega^{-1}e
					\end{array}
					& \hspace*{-9mm}\text{if }H'\Xi _{12}e\geq 0  \\
					& \\
					-e'\Omega^{-1}M_{10}(M_{10}'\Omega^{-1}M_{10})^{-1}M_{10}'\Omega^{-1}e -e'\Xi _{1}M_{20}(M_{20}'\Xi _{1}M_{20})^{-1}M_{20}'\Xi _{1}e +e'\Omega^{-1}e & \text{ else}
				\end{cases}
			\end{align*}
			where $\Xi_{1}= \Omega^{-1/2}(I-\Omega^{-1/2}M_{10}(M_{10}'\Omega^{-1}M_{10})^{-1}M_{10}'\Omega^{-1/2})\Omega^{-1/2}$ and $\Xi _{12}= \Xi_{1}^{1/2}(I-\Xi_{1}^{1/2}M_{20}(M_{20}'\Xi _{1}M_{20})^{-1}M_{20}'\Xi_{1}^{1/2})\Xi_{1}^{1/2}$. 
			As $\Xi_{1}\Omega\Xi_{1}=\Xi_{1}$, we can derive $\Xi_{12}\Omega\Xi_{12}=(\Xi_{1}-\Xi_{1}M_{20}(M_{20}'\Xi _{1}M_{20})^{-1}M_{20}'\Xi_{1})\Omega(\Xi_{1}-\Xi_{1}M_{20}(M_{20}'\Xi _{1}M_{20})^{-1}M_{20}'\Xi_{1})=\Xi_{1}-\Xi_{1}M_{20}(M_{20}'\Xi _{1}M_{20})^{-1}M_{20}'\Xi_{1}=\Xi_{12}$, and hence $e'\Xi_{12}H(H'\Xi_{12}H)^{-1}H'\Xi_{12}e\sim\chi^{2}_{1}$.
			Since $E[H'\Xi_{12}ee'\Xi_{1}M_{20}]$ is zero,
			$(e'\Xi_{1}M_{20}(M_{20}'\Xi_{1}M_{20})^{-1}M_{20}'\Xi_{1}e, e'\Xi_{1}N_{20}(N_{20}'\Xi_{1}N_{20})^{-1}N_{20}'\Xi_{1}e)$ is independent to
			$e'\Xi_{12}H(H'\Xi_{12}H)^{-1}H'\Xi_{12}e$. 
			
			\paragraph{Under the alternative hypothesis.}
			There is a constant $C_{1}\in(0,+\infty)$ such that $\inf_{\theta\in\Theta:\delta_{1}+\delta_{3}\gamma=0,\delta_{2}=0_{p-1}}\|g_{0}(\theta)\| \geq C_{1}$. 
			This is because $g_{0}(\theta)$ is zero if and only if $\theta=\theta_{0}$, by \cref{ass:gmm} and \cref{thm:id}, and continuous on $\Theta$,  by \cref{ass:dgp}, while the restricted parameter set $\{\theta=(\beta',\delta',\gamma)':\delta_{2}=0_{p-1},\delta_{1}+\delta_{3}\gamma=0\}$ is closed.
			$\mathcal{G}=\{g(\omega_{i},\theta):\theta\in\Theta\}$ is shown to satisfy the uniform entropy condition in the proof of \cref{lem:equicont}, and hence $\sup_{\theta\in\Theta}\|\bar{g}_{n}(\theta)-g_{0}(\theta)\|=o_{p}(1)$ by Glivenko-Cantelli theorem. 
			By triangle inequality, $ C_{1} \leq \|g_{0}(\tilde{\theta})\|\leq \|\bar{g}_{n}(\tilde{\theta})\|+ o_{p}(1)$.
			Recall that $\tilde{\theta}$ is the continuity-restricted estimator.
			Meanwhile, $\|\bar{g}_{n}(\hat{\theta})\|= O_{p}(n^{-1/2})$ because $\|\bar{g}_{n}(\hat{\theta})\|\leq \|\bar{g}_{n}(\theta_{0})\|=O_{p}(n^{-1/2})$.
			Therefore, there exists $C_{2}\in(0,+\infty)$ such that $\hat{Q}_{n}(\tilde{\theta})-\hat{Q}_{n}(\hat{\theta})\geq C_{2}+O_{p}(n^{-1})$, which implies that
			$P(n^{-m}\mathcal{T}_{n}>M)=P(\hat{Q}_{n}(\tilde{\theta})-\hat{Q}_{n}(\hat{\theta})>M/(n^{1-m}))\rightarrow1$, for any $m\in[0,1)$ and $M<\infty$.
			
			\subsection{Auxiliary Lemmas}
			\begin{lemma}
				\label{lem:loc.conti}
				Suppose that the true model is continuous and Assumptions \ref{ass:gmm}, \ref{ass:dgp}, and \ref{ass:loc} are true.
				For any $\eta>0$, there is a neighborhood $\mathcal{O}$ of $\theta_{0}$ such that the population moment function $g_{0}(\theta)$ satisfies
				\[\lim_{n\rightarrow\infty}\sup_{\theta\in \mathcal{O}}
				\frac{\sqrt{n}\|g_{0}(\theta)-D_{2}\left(\alpha'-\alpha_{0}',(\gamma-\gamma_{0})^{2}\right)'\|}
				{1+\sqrt{n}\|\left(\alpha'-\alpha_{0}',(\gamma-\gamma_{0})^{2}\right)'\|}
				<\eta.\]
			\end{lemma}
			\begin{proof}
				
				Recall that $G$, whose formula is \eqref{eq:firstderivative}, is the first-order derivative of $g_{0}(\theta)$ with respect to $\gamma$ at $\theta=\theta_{0}$, and $H$, whose formula is \eqref{eq:secondderivative}, is a half of the second-order derivative.
				$G$ can be obtained by applying the Leibniz rule as follows:
				\begin{align*}
					\left. \frac{d}{d\gamma}E[-z_{it}(1,x_{it}')\delta_{0}1\{q_{it}>\gamma\}] \right|_{\gamma=\gamma_{0}} 
					& = \left. \frac{d}{d\gamma} \int_{\gamma}^{\infty} -E_{t}[z_{it}(1,x_{it}')\delta_{0}|q]f_{t}(q)dq \right|_{\gamma=\gamma_{0}} \\
					& = E_{t}[z_{it}(1,x_{it}')\delta_{0}|\gamma_{0}]f_{t}(\gamma_{0}).
				\end{align*}
				Similarly, we can get
				\[\left. \frac{d}{d\gamma}E[z_{it}(1,x_{it-1}')\delta_{0}1\{q_{it-1}>\gamma\}] \right|_{\gamma=\gamma_{0}} = -E_{t-1}[z_{it}(1,x_{it-1}')\delta_{0}|\gamma_{0}]f_{t-1}(\gamma_{0}).\]
				This implies the formula \eqref{eq:firstderivative} for $G$.
				$H$ can also be obtained by the Leibniz rule as follows:
				\begin{align*}
					\left. \frac{d}{d\gamma}E_{t}[z_{it}(1,x_{it}')\delta_{0}|\gamma]f_{t}(\gamma)\right|_{\gamma=\gamma_{0}} 
					& = \left. \frac{d}{d\gamma} E_{t}[z_{it}(\delta_{10}+\delta_{30}\gamma)|\gamma]f_{t}(\gamma) \right|_{\gamma=\gamma_{0}} \\
					& = \left. \frac{d}{d\gamma} (\delta_{10}+\delta_{30}\gamma)\cdot E_{t}[z_{it}|\gamma]f_{t}(\gamma) \right|_{\gamma=\gamma_{0}} \\
					& = \delta_{30}E_{t}[{z_{it}}|\gamma_{0}]f_{t}(\gamma_{0})+(\delta_{10}+\delta_{30}\gamma_{0})
					\left.\frac{d}{d\gamma}E_{t}[z_{it}|\gamma]f_{t}(\gamma) \right|_{\gamma=\gamma_{0}}\\
					& = \delta_{30}E_{t}[{z_{it}}|\gamma_{0}]f_{t}(\gamma_{0}).
				\end{align*}
				Similarly, we can get
				\[\left. \frac{d}{d\gamma}\{-E_{t-1}[z_{it}(1,x_{it-1}')\delta_{0}|\gamma]f_{t-1}(\gamma)\}\right|_{\gamma=\gamma_{0}}  = -\delta_{30}E_{t-1}[{z_{it}}|\gamma_{0}]f_{t-1}(\gamma_{0}).\]
				This implies the formula \eqref{eq:secondderivative} for $H$.
				
				The population moment can be expressed as,
				\begin{equation*}
					g_{0}(\alpha,\gamma) =
					M_{0}(\gamma)(\alpha-\alpha_{0}) + H(\gamma-\gamma_{0})^{2} + o((\gamma-\gamma_{0})^{2}).
				\end{equation*}
				Define
				$M_{0,G}=\left[
				\begin{array}{c;{2pt/2pt}c}
					0_{k\times p} & M_{G}
				\end{array}
				\right]\in\mathbb{R}^{k\times(2p+1)}$ where
				\begin{equation*}
					M_{G}=\begin{bmatrix}
						E_{t_{0}}[z_{it_{0}}(1,x_{it_{0}}')|\gamma_{0}]f_{t_{0}}(\gamma_{0})-E_{t_{0}-1}[z_{it_{0}}(1,x_{it_{0}-1}')|\gamma_{0}]f_{t_{0}-1}(\gamma_{0})\\
						\vdots\\
						E_{T}[z_{iT}(1,x_{iT}')|\gamma_{0}]f_{T}(\gamma_{0})-E_{T-1}[z_{iT}(1,x_{iT-1}')|\gamma_{0}]f_{T-1}(\gamma_{0})
					\end{bmatrix}\in\mathbb{R}^{k\times(p+1)}.
				\end{equation*}
				The polynomial expansion $M_{0}(\gamma) = M_{0} + M_{0,G}(\gamma-\gamma_{0})+o(|\gamma-\gamma_{0}|)$ implies
				\begin{equation*}
					g_{0}(\alpha,\gamma) = M_{0}(\alpha-\alpha_{0})+H(\gamma-\gamma_{0})^{2}+o(\|\alpha-\alpha_{0}\|+(\gamma-\gamma_{0})^{2}).
				\end{equation*}
				Thus, $\sqrt{n}\|g_{0}(\theta)-D_{2}\left(\alpha'-\alpha_{0}',(\gamma-\gamma_{0})^{2}\right)'\| = o(\sqrt{n}(\|\alpha-\alpha_{0}\|+(\gamma-\gamma_{0})^{2}))$, which completes the proof.
			\end{proof}

			\begin{lemma}
				\label{lem:ULLN}
				If \cref{ass:gmm} is true, then
				\begin{equation*}
					\sup_{\gamma\in\Gamma}\|\bar{M}_{n}(\gamma)-M_{0}(\gamma)\|\xrightarrow{p}0.
				\end{equation*}
			\end{lemma}
			
			\begin{proof}
				We show that the classes $\{z_{it}(1,x_{it}')1\{q_{it}>\gamma\}:\gamma\in\Gamma\}$ and $\{z_{it}(1,x_{it-1}')1\{q_{it-1}>\gamma\}:\gamma\in\Gamma\}$ are P-Glivenko-Cantelli. 
				We focus on the former class since the verification for the latter class is exactly identical.
				Let $\omega_{i}=\{(z_{it},y_{it},x_{it},\epsilon_{it})_{t=1}^{T}\}$ be a random element in a measurable space $(\mathcal{X},\mathcal{A})$.
				A collection of measurable index functions $\mathcal{G}_{index}=\{1\{q_{it}>\gamma\}:\gamma\in\Gamma\}$ on $\mathcal{X}$ is a VC class with a VC index 2.
				If $m_{ij}$ is the $(i,j)$th element of $z_{it}(1,x_{it}')$, then $\mathcal{G}_{index}\cdot m_{ij}=\{g_{index}\cdot m_{ij}:g_{index}\in\mathcal{G}_{index}\}$ is also a VC class as discussed by Lemma 2.6.18 in \cite{van_der_vaart_weak_1996}.
				The envelope for $\mathcal{G}_{index}\cdot m_{ij}$ would be $|m_{ij}|$ since an index function is always bounded by 1.
				The expectation of the envelope is bounded since $E\|z_{it}(1,x_{it}')\|\leq \sqrt{E\|z_{it}\|^{2}E\|(1,x_{it}')'\|^{2}}<\infty$.
				In conclusion, $\mathcal{G}_{index}\cdot m_{ij}$ is a $P$-Glivenko-Cantelli for each $(i,j)$, and thus the ULLN for $\{z_{it}(1,x_{it}')1\{q_{it}>\gamma\}:\gamma\in\Gamma\}$ holds.
			\end{proof}
			
			\begin{lemma}
				\label{lem:equicont}
				Let \cref{ass:gmm} hold. 
				If $h_{n}\rightarrow 0$, then
				\begin{equation*}
					\sup_{\|\theta_{1}-\theta_{2}\|<h_{n}}\sqrt{n}\|\bar{g}_{n}(\theta_{1})-\bar{g}_{n}(\theta_{2})-g_{0}(\theta_{1})+g_{0}(\theta_{2})\|=o_{p}(1).
				\end{equation*}
			\end{lemma}
			
			\begin{proof}
				Let $\omega_{i}=\{(z_{it},y_{it},x_{it},\epsilon_{it})_{t=1}^{T}\}$ be a random element in a measurable space $(\mathcal{X},\mathcal{A})$, and $P$ is the probability measure for $\omega_{i}$.
				Define a functional class $\mathcal{G}=\{g(\omega_{i},\theta):\theta\in\Theta\}$ on $\mathcal{X}$ such that
				\begin{align}
					\label{eq:momtclass}
					g(\omega_{i},\theta) &=(g_{t_{0}}(\omega_{i},\theta)', ..., g_{T}(\omega_{i},\theta)')', \\
					\nonumber
					g_{t}(\omega_{i},\theta) & = z_{it}\Delta y_{it} - z_{it}\Delta x_{it}'\beta - z_{it}1_{it}(\gamma)'X_{it}\delta \\
					\nonumber
					& = z_{it}\Delta \epsilon_{it} - z_{it}\Delta x_{it}'(\beta-\beta_{0}) - z_{it}1_{it}(\gamma)'X_{it}(\delta-\delta_{0}) + z_{it}(1_{it}(\gamma_{0})'-1_{it}(\gamma)')X_{it}\delta_{0}.
				\end{align}
				and $\mathcal{G}_{h}=\{g(\omega_{i},\theta_{1})-g(\omega_{i},\theta_{2}):\|\theta_{1}-\theta_{2}\|<h,\theta_{1},\theta_{2}\in\Theta\}$.
				We need to show that
				$P(\|\mathbb{G}_{n}\|_{\mathcal{G}_{h}}>x)\rightarrow 0$ if $h\rightarrow 0$ as $n\rightarrow\infty$, which is the asymptotic equicontinuity.
				To show the asymptotic equicontinuity, it is sufficient to show that each element of $\mathcal{G}$ is P-Donsker, e.g., 2.3.11 Lemma and its corollary in \cite{van_der_vaart_weak_1996}, which is implied by the uniform entropy condtion:
				\begin{equation*}
					\int_{0}^{\infty}\sup_{Q}\sqrt{\log N(\varepsilon\|G\|_{Q,2},\mathcal{G},L_{2}(Q))}d\varepsilon<\infty,
				\end{equation*}
				where supremum is taken over all probability measures $Q$ on $(\mathcal{X},\mathcal{A})$ such that $QG^{2}<\infty$, and $G$ is an envelope for $\mathcal{G}$.
				For more details, see section 2.1 in \cite{van_der_vaart_weak_1996}.
				As we only need to consider each scalar element of $\mathcal{G}$, it is sufficient to consider the following functional class
				\begin{multline*}
					\widetilde{\mathcal{G}}^{(t)}=\{z_{it}\Delta\epsilon_{it}-z_{it}\Delta x_{it}\bar{\beta}-z_{it}1_{it}(\gamma_{1})'X_{it}\delta_{1}+z_{it}1_{it}(\gamma_{2})'X_{it}\delta_{2}\\
					:\|\bar{\beta}\|\leq K, \|\delta_{1}\|\leq K, \|\delta_{2}\|\leq K, \gamma_{1},\gamma_{2}\in\Gamma\},
				\end{multline*}
				where $K<\infty$ is a constant such that $\|\theta\|\leq K/2$ if $\theta\in\Theta$.
				Assume that $z_{it}$ is a scalar without losing of generality.
				Note that $g_{t}(\omega_{i},\theta)=z_{it}(\Delta y_{it}-\Delta x_{it}'\beta-1_{it}(\gamma)'X_{it}\delta)=z_{it}\Delta\epsilon_{it}-z_{it}\Delta x_{it}(\beta-\beta_{0n})-z_{it}1_{it}(\gamma)'X_{it}\delta+z_{it}1_{it}(\gamma_{0})'X_{it}\delta_{0}$ is an element of $\widetilde{\mathcal{G}}^{(t)}$.
				So it is sufficient to show $\widetilde{\mathcal{G}}^{(t)}$ satisfies the uniform entropy condition.
				
				Let $\mathcal{G}_{1}=\{z_{it}\Delta x_{it}'\bar{\beta}:\|\bar{\beta}\|\leq K\}$.
				$\mathcal{G}_{1}$ is a $p$-dimensional vector space and is a VC class by 2.6.15 Lemma in \cite{van_der_vaart_weak_1996}, with an envelope function $G_{1}(\omega_{i})=C\|z_{it}\Delta x_{it}'\|$ for some constant $C<\infty$, and $E G_{1}^{2}<\infty$.
				Let $\mathcal{G}_{2}=\{z_{it}(1,x_{it}')'\delta1\{q_{it}>\gamma\}:\|\delta\|\leq K,\gamma\in\Gamma\}$, $\mathcal{G}_{2a}=\{z_{it}(1,x_{it}')'\delta:\|\delta\|\leq K\}$, and $\mathcal{G}_{2b}=\{1\{q_{it}>\gamma\}:\gamma\in\Gamma\}$.
				$G_{2a}=C\|z_{it}(1,x_{it}')\|$ for some $C<\infty$ and $G_{2b}=1$ are envelopes for $\mathcal{G}_{2a}$ and $\mathcal{G}_{2b}$, respectively.
				Note that $\mathcal{G}_{2}=\mathcal{G}_{2a}\mathcal{G}_{2b}$, i.e., $\mathcal{G}_{2}$ is a collection of $g_{2a}\cdot g_{2b}$ where $g_{2a}\in\mathcal{G}_{2a}$ and $g_{2b}\in\mathcal{G}_{2b}$. 
				$\mathcal{G}_{2}$ satisfies the uniform entropy condition as pairwise sum or product of functional classes preserve the uniform entropy condition, e.g., Theorem 2.10.20 in \cite{van_der_vaart_weak_1996}.
				Note that for every $d>0$,
				\begin{multline*}
					\int_{0}^{d}\sup_{Q}\sqrt{\log N(\varepsilon\|(2G_{2a}^{2}G_{2b}^{2})^{1/2} \|_{Q,2},\mathcal{G}_{2}, L_{2}(Q))}d\varepsilon \\
					\leq \int_{0}^{d}\sup_{Q}\sqrt{\log N(\varepsilon\|G_{2a} \|_{Q,2},\mathcal{G}_{2a}, L_{2}(Q))}d\varepsilon 
					+ \int_{0}^{d}\sup_{Q}\sqrt{\log N(\varepsilon\|G_{2b} \|_{Q,2},\mathcal{G}_{2b}, L_{2}(Q))}d\varepsilon,
				\end{multline*}
				while $G_{2a}G_{2b}$ is an envelope of $\mathcal{G}_{2}$.
				So the uniform entropy condition for $\mathcal{G}_{2}$ holds.
				Similarly, we can show that $\mathcal{G}_{3}=\{z_{it}(1,x_{it-1}')'\delta1\{q_{it-1}>\gamma\}:\|\delta\|\leq K,\gamma\in\Gamma\}$ satisfies the uniform entropy condition.
				Hence, the functional class $(\mathcal{G}_{2}-\mathcal{G}_{3})$ defined by pairwise sum, which is a set of functions $g_{2}-g_{3}$ for all $g_{2}\in\mathcal{G}_{2}$ and $g_{3}\in\mathcal{G}_{3}$, also satisfies the uniform entropy condition, e.g., Theorem 2.10.20 in \cite{van_der_vaart_weak_1996}.
				As $(\mathcal{G}_{2}-\mathcal{G}_{3})$ is a superset of $\{z_{it}1_{it}(\gamma)'X_{it}\delta:\|\delta\|\leq K,\gamma\in\Gamma\}$, the functional class $\{z_{it}1_{it}(\gamma)'X_{it}\delta:\|\delta\|\leq K,\gamma\in\Gamma\}$ also satisfies the uniform entropy condition .
				Thus, $\{z_{it}\Delta\epsilon_{it}\}-\mathcal{G}_{1}-(\mathcal{G}_{2}-\mathcal{G}_{3})+(\mathcal{G}_{2}-\mathcal{G}_{3})$, which is a superset of $\widetilde{\mathcal{G}}^{(t)}$, satisfies the uniform entropy condition by repetitively applying Theorem 2.10.20 in \cite{van_der_vaart_weak_1996}, and hence $\widetilde{\mathcal{G}}^{(t)}$ also satisfies the condition.
				
				Note that for some constant $C<\infty$,
				\begin{equation*}
					\widetilde{G}=C(\|z_{it}\Delta x_{it}'\|+\|z_{it}(1,x_{it}')\|+\|z_{it}(1,x_{it-1}')\|)+\|z_{it}\Delta \epsilon_{it}\|
				\end{equation*}
				is an envelope for $\widetilde{\mathcal{G}}^{(t)}$, and $E\widetilde{G}^{2}<\infty$ by \cref{ass:gmm}.
				
			\end{proof}
			
			\begin{lemma}
				\label{lem:plim1}
				When the true model is continuous and Assumptions \ref{ass:gmm}, \ref{ass:dgp}, and \ref{ass:loc} are true,
				\begin{equation*}
					\frac{1}{\sqrt{n}}\sum_{i=1}^{n} z_{it}(1_{it}(\gamma_{0})'-1_{it}(\gamma_{0}+\tfrac{b}{n^{\frac{1}{4}}})')X_{it}\delta_{0}
					\xrightarrow{p}
					\frac{\delta_{30}}{2}\left\{E_{t}[z_{it}|\gamma_{0}]f_{t}(\gamma_{0})-E_{t-1}[z_{it}|\gamma_{0}]f_{t-1}(\gamma_{0})\right\}b^{2}
				\end{equation*}
				uniformly over $b\in[-K,K]$ for any $K<\infty$.
			\end{lemma}
			
			\begin{proof}
				Note that
				\begin{align}
					\nonumber
					& \frac{1}{\sqrt{n}}\sum_{i=1}^{n} z_{it}(1_{it}(\gamma_{0})'-1_{it}(\gamma_{0}+\tfrac{b}{n^{\frac{1}{4}}})')X_{it}\delta_{0}\\
					\label{eq:CLTlimit}
					&=\frac{1}{\sqrt{n}}\sum_{i=1}^{n} 
					\left\{
					z_{it}(1_{it}(\gamma_{0})'-1_{it}(\gamma_{0}+\tfrac{b}{n^{\frac{1}{4}}})')X_{it}\delta_{0}-
					E[z_{it}(1_{it}(\gamma_{0})'-1_{it}(\gamma_{0}+\tfrac{b}{n^{\frac{1}{4}}})')X_{it}\delta_{0}]
					\right\} \\
					\label{eq:deter.limit}
					&\quad+
					\sqrt{n}
					E[z_{it}(1_{it}(\gamma_{0})'-1_{it}(\gamma_{0}+\tfrac{b}{n^{\frac{1}{4}}})')X_{it}\delta_{0}].
				\end{align}
				The stochastic term \eqref{eq:CLTlimit} converges in probability to zero uniformly with respect to $b\in[-K,K]$.
				This is because \cref{lem:equicont} shows that when $h_{n}\downarrow0$, then
				\begin{equation*}
					\sup_{|\gamma-\gamma_{0}|<h_{n}}\sqrt{n}\left\{\frac{1}{n}\sum_{i=1}^{n}z_{it}(1_{it}(\gamma_{0})-1_{it}(\gamma))'X_{it}\delta_{0}-E[z_{it}(1_{it}(\gamma_{0})-1_{it}(\gamma))'X_{it}\delta_{0}]\right\}=o_{p}(1)
				\end{equation*}
				as it can be expressed as $\sup_{|\gamma-\gamma_{0}|<h_{n}}\| \bar{g}_{n}(\alpha_{0},\gamma)- \bar{g}_{n}(\alpha_{0},\gamma_{0})-g_{0}(\alpha_{0},\gamma)+g_{0}(\alpha_{0},\gamma_{0})\|$.
				
				Suppose $b>0$.
				The case for $b<0$ follows similarly.
				As $n\rightarrow\infty$, the deterministic term \eqref{eq:deter.limit} converges as follows:
				\begin{multline*}
					\sqrt{n}E
					z_{it}(1_{it}(\gamma_{0})'-1_{it}(\gamma_{0}+\tfrac{b}{n^{\frac{1}{4}}})')X_{it}\delta_{0} \\
					=\sqrt{n}\left\{
					E[z_{it}(\delta_{10}+\delta_{30}q_{it})1\{\gamma_{0}+\tfrac{b}{n^\frac{1}{4}}\geq q_{it}>\gamma_{0}\}]
					-E[z_{it}(\delta_{10}+\delta_{30}q_{it-1})1\{\gamma_{0}+\tfrac{b}{n^\frac{1}{4}}\geq q_{it-1}>\gamma_{0}\}]
					\right\}
					\\
					\rightarrow 
					\frac{\delta_{30}}{2}\left\{E_{t}[z_{it}|\gamma_{0}]f_{t}(\gamma_{0})-E_{t-1}[z_{it}|\gamma_{0}]f_{t-1}(\gamma_{0})\right\}b^{2},
				\end{multline*}
				uniformly with respect to $b\in[-K,K]$.
				To show that, use the (second-order) derivative of $\kappa\mapsto E[z_{it}(\delta_{10}+\delta_{30}q_{it})1\{\gamma_{0}+\kappa\geq q_{it}>\gamma_{0}\}]$ and derive the Taylor expansion
				\begin{align*}
					&\sqrt{n}E[z_{it}(\delta_{10}+\delta_{30}q_{it})1\{\gamma_{0}+\tfrac{b}{n^\frac{1}{4}}\geq q_{it}>\gamma_{0}\}]\\
					&=\frac{b^{2}}{2}\left(\delta_{30}E_{t}[z_{it}|\gamma_{n,b}]f_{t}(\gamma_{n,b})+(\delta_{10}+\delta_{30}\gamma_{n,b})\frac{d}{d\gamma}E_{t}[z_{it}|\gamma]f_{t}(\gamma)|_{\gamma=\gamma_{n,b}}\right),
				\end{align*}
				where $\gamma_{n,b}\in[\gamma_{0},\gamma_{0}+\frac{b}{n^{1/4}}]$.
				Note that $|\gamma_{n,b}-\gamma_{0}|\rightarrow 0$ unifromly with respect to $b\in[-K,K]$.
				Since $E_{t}[z_{it}|\gamma]$ and $f_{t}(\gamma)$ are continuously differentiable at $\gamma_{0}$ by \cref{ass:dgp}, both $\frac{d}{d\gamma}E_{t}[z_{it}|\gamma]f_{t}(\gamma)|_{\gamma=\gamma_{n,b}}\rightarrow \frac{d}{d\gamma}E_{t}[z_{it}|\gamma]f_{t}(\gamma)|_{\gamma=\gamma_{0}}$ and $(\delta_{10}+\delta_{30}\gamma_{n,b})\rightarrow 0$ hold uniformly with respect to $b\in[-K,K]$.
				On the other hand, $E_{t}[z_{it}|\gamma_{n,b}]f_{t}(\gamma_{n,b})\rightarrow E_{t}[z_{it}|\gamma_{0}]f_{t}(\gamma_{0})$ uniformly with respect to $b\in[-K,K]$.
				Hence, $\sqrt{n}E[z_{it}(\delta_{10}+\delta_{30}q_{it})1\{\gamma_{0}+\tfrac{b}{n^\frac{1}{4}}\geq q_{it}>\gamma_{0}\}$ converges to $\frac{\delta_{30}}{2}E_{t}[z_{it}|\gamma_{0}]f_{t}(\gamma_{0})b^{2}$ uniformly with respect to $b\in[-K,K]$ as $n\rightarrow\infty$.
				We can derive the similar result for $\sqrt{n}E[z_{it}(\delta_{10}+\delta_{30}q_{it-1})1\{\gamma_{0}+\tfrac{b}{n^\frac{1}{4}}\geq q_{it-1}>\gamma_{0}\}]$.
				
			\end{proof}

			\section{Proofs of Theorems in Section \ref{sec:bootstrap} and Auxiliary Lemmas}
			\label{sec: online.app.boot}
   
			\subsection{Preliminaries}
			\label{sec: online.app.boot1}
			
			Proofs in this section are regarding bootstrap results, and hence we explain empirical process framework for our bootstrap analysis.
			Let $\omega_{1}^{*},...,\omega_{n}^{*}$ be i.i.d. resampling draws from a given sample $\{\omega_{i}:1\leq i\leq n\}$.
			We set $\omega_{i}=\{(z_{it},y_{it},x_{it},\epsilon_{it})_{t=1}^{T}\}$ as in the proofs of Lemmas \ref{lem:ULLN} and \ref{lem:equicont}.
			An important functional class for our bootstrap analysis is $\mathcal{G}=\{g(\omega_{i},\theta):\theta\in\Theta\}$ where $g(\omega_{i},\theta)$ is defined as in \eqref{eq:momtclass}.
			
			Be mindful that $g_{i}^{*}(\theta)$ that appears in \cref{sec:bootstrap} is different from $g(\omega_{i}^{*},\theta)$.
			This is because $g_{i}^{*}(\theta) =(g_{it_{0}}^{*}(\theta)',...,g_{iT}^{*}(\theta)')'$ where
			\begin{align}
				\nonumber
				g_{it}^{*}(\theta) & = z_{it}^{*}(\Delta y_{it}^{*}-\Delta x_{it}^{*\prime}\beta-1_{it}^{*}(\gamma)'X_{it}^{*}\delta)\\
				\label{eq:boot momt}
				&=\underbrace{-z_{it}^{*}\Delta x_{it}^{*\prime}(\beta-\beta_{0}^{*}) - z_{it}^{*}1_{it}^{*}(\gamma)'X_{it}^{*}(\delta-\delta_{0}^{*})+z_{it}^{*}(1_{it}^{*}(\gamma_{0}^{*})-1_{it}^{*}(\gamma))'X_{it}^{*}\delta_{0}^{*}}
				_{\textstyle (I)}
				+
				\underbrace{z_{it}^{*}\widehat{\Delta\epsilon}_{it}^{*}}
				_{\textstyle (II)}.
			\end{align}
			Recall that $\Delta y_{it}^{*}$ is not an i.i.d. resampling draw from $\{\Delta y_{it}:1\leq i \leq n\}$ but is generated using resampled regressors and residuals with regression equation using $\theta_{0}^{*}$. The formula for $\Delta y_{it}^{*}$ is used to derive the equality in \eqref{eq:boot momt} (see Step 2 in \cref{alg:bootstrap}).
			Instead, $g_{it}^{*}(\theta)=g_{t}(\omega_{i}^{*},\theta)-g_{t}(\omega_{i}^{*},\theta_{0}^{*})+g_{t}(\omega_{i}^{*},\hat{\theta})$.
			To be more precise, $(I)$ in \eqref{eq:boot momt} is $g_{t}(\omega_{i}^{*},\theta)-g_{t}(\omega_{i}^{*},\theta_{0}^{*})$, and $(II)$ in \eqref{eq:boot momt} is $g_{t}(\omega_{i}^{*},\hat{\theta})$.

                \subsection{Proof of \texorpdfstring{\cref{prop:boot.conv.rate}}{\cref{prop:boot.conv.rate}}}

                \paragraph{Consistency of the bootstrap estimator.}
			
			The bootstrap sample moment can be rewritten by
			\begin{align*}
				\bar{g}_{n}^{*}(\theta) &= \frac{1}{n}\sum_{i=1}^{n}(g_{i}^{*}(\theta)-\bar{g}_{n}(\hat{\theta}))
				\\
				&=
				\begin{pmatrix}
					\frac{1}{n}\sum_{i=1}^{n}z_{it_{0}}^{*}\widehat{\Delta\epsilon}_{it_{0}}^{*}\\
					\vdots\\
					\frac{1}{n}\sum_{i=1}^{n}z_{iT}^{*}\widehat{\Delta\epsilon}_{iT}^{*}
				\end{pmatrix}
				-
				\begin{pmatrix}
					\frac{1}{n}\sum_{i=1}^{n}z_{it_{0}}\widehat{\Delta\epsilon}_{it_{0}}\\
					\vdots\\
					\frac{1}{n}\sum_{i=1}^{n}z_{iT}\widehat{\Delta\epsilon}_{iT}
				\end{pmatrix}-\begin{pmatrix}
					\frac{1}{n}\sum_{i=1}^{n}z_{it_{0}}^{*}\Delta x_{it_{0}}^{*\prime} \\
					\vdots \\
					\frac{1}{n}\sum_{i=1}^{n}z_{iT}^{*}\Delta x_{iT}^{*\prime}
				\end{pmatrix}(\beta-\beta_{0}^{*})\\
				& \quad -\begin{pmatrix}
					\frac{1}{n}\sum_{i=1}^{n}z_{it_{0}}^{*}1_{it_{0}}^{*}(\gamma)'X_{it_{0}}^{*}\\
					\vdots\\
					\frac{1}{n}\sum_{i=1}^{n}z_{iT}^{*}1_{iT}^{*}(\gamma)'X_{iT}^{*}
				\end{pmatrix}(\delta-\delta_{0}^{*}) +\begin{pmatrix}
					\frac{1}{n}\sum_{i=1}^{n}z_{it_{0}}^{*}(1_{it_{0}}^{*}(\gamma_{0}^{*})-1_{it_{0}}^{*}(\gamma))'X_{it_{0}}^{*} \\
					\vdots \\
					\frac{1}{n}\sum_{i=1}^{n}z_{iT}^{*}(1_{iT}^{*}(\gamma_{0}^{*})-1_{iT}^{*}(\gamma))'X_{iT}^{*}
				\end{pmatrix}\delta_{0}^{*}.
			\end{align*}
			We additionally define
			\begin{alignat*}{3}
				v_{i}^{*}=& \begin{pmatrix}
					z_{it_{0}}^{*}\Delta y_{it_{0}}^{*} \\
					\vdots \\
					z_{iT}^{*}\Delta y_{iT}^{*}
				\end{pmatrix}-\begin{pmatrix}
					\frac{1}{n}\sum_{i=1}^{n}z_{it_{0}}\widehat{\Delta\epsilon}_{it_{0}}\\
					\vdots\\
					\frac{1}{n}\sum_{i=1}^{n}z_{iT}\widehat{\Delta\epsilon}_{iT}
				\end{pmatrix}&,\quad 
				M_{i}^{*}(\gamma) =& -\begin{bmatrix}
					z_{it_{0}^{*}}(\Delta x_{it_{0}}^{*\prime},1_{it_{0}}^{*}(\gamma)'X_{it_{0}}^{*}) \\
					\vdots \\
					z_{iT}^{*}(\Delta x_{iT}^{*\prime},1_{iT}^{*}(\gamma)'X_{iT}^{*})
				\end{bmatrix},
			\end{alignat*}
			$\bar{v}_{n}^{*} =\frac{1}{n}\sum_{i=1}^{n} v_{i}^{*}$, and
			$\bar{M}_{n}^{*}(\gamma) = \frac{1}{n}\sum_{i=1}^{n} M_{i}^{*}(\gamma)$.
			Then, $\bar{g}_{n}^{*}(\theta) = \bar{v}_{n}^{*}+\bar{M}_{n}^{*}(\gamma)\alpha$.
			Given $\gamma$, we can obtain the constrained optimizer
			\begin{equation*}
				\hat{\alpha}^{*}(\gamma)=-
				(\bar{M}_{n}^{*\prime}(\gamma)W_{n}^{*}\bar{M}_{n}^{*}(\gamma))^{-1}\bar{M}_{n}^{*\prime}(\gamma)W_{n}^{*}\bar{v}_{n}^{*}
			\end{equation*}
			where
			\begin{equation*}
				\bar{v}_{n}^{*} = -\bar{M}_{n}^{*}(\gamma_{0}^{*})\alpha_{0}^{*}+\hat{u}_{n}^{*};\quad
				\hat{u}_{n}^{*}=\begin{pmatrix}
					\frac{1}{n}\sum_{i=1}^{n}z_{it_{0}}^{*}\widehat{\Delta\epsilon}_{it_{0}}^{*}\\
					\vdots\\
					\frac{1}{n}\sum_{i=1}^{n}z_{iT}^{*}\widehat{\Delta\epsilon}_{iT}^{*}
				\end{pmatrix}
				-
				\begin{pmatrix}
					\frac{1}{n}\sum_{i=1}^{n}z_{it_{0}}\widehat{\Delta\epsilon}_{it_{0}}\\
					\vdots\\
					\frac{1}{n}\sum_{i=1}^{n}z_{iT}\widehat{\Delta\epsilon}_{iT}
				\end{pmatrix}.
			\end{equation*}
			Let $\tilde{Q}_{n}^{*}(\gamma)=\hat{Q}_{n}^{*}(\hat{\alpha}^{*}(\gamma),\gamma)$ be a profiled criterion and $\hat{\gamma}^{*}=\arg\min_{\gamma\in\Gamma}\tilde{Q}_{n}^{*}(\gamma)$.
			$\hat{u}_{n}^{*}=o_{p}^{*}(1)$ in $P$ by \cref{lem:boot res LLN}.
			By \cref{lem:boot.ULLN}, $\sup_{\gamma\in\Gamma}\|\bar{M}_{n}^{*}(\gamma)-M_{0}(\gamma)\|=o_{p}^{*}(1)$ in $P$.
			Therefore, if $|\hat{\gamma}^{*}-\gamma_{0}^{*}|\xrightarrow{p^{*}}0$ in $P$, then $\|\hat{\alpha}^{*}(\hat{\gamma}^{*})-\alpha_{0}^{*}\|\xrightarrow{p^{*}}0$ in $P$, which completes the proof.

			Let $\tilde{g}_{n}^{*}(\gamma)=\bar{g}_{n}^{*}(\hat{\alpha}^{*}(\gamma),\gamma)$ which can be expressed as
			\begin{equation*}
				\tilde{g}_{n}^{*}(\gamma)=\left[I-\bar{M}_{n}^{*}(\gamma)(\bar{M}_{n}^{*\prime}(\gamma)W_{n}^{*}\bar{M}_{n}^{*}(\gamma))^{-1}\bar{M}_{n}^{*\prime}(\gamma)W_{n}^{*} \right]\left(-\bar{M}_{n}^{*}(\gamma_{0}^{*})
				\alpha_{0}^{*}+\hat{u}_{n}^{*}\right).
			\end{equation*}
			Therefore,
			\begin{equation*}
				W_{n}^{*1/2}\tilde{g}_{n}^{*}(\gamma) = \left[I-P_{W_{n}^{*1/2}\bar{M}_{n}^{*}(\gamma)}\right]\left(-W_{n}^{*1/2}\bar{M}_{n}^{*}(\gamma_{0}^{*})
				\alpha_{0}^{*}+W_{n}^{*1/2}\hat{u}_{n}^{*}\right),
			\end{equation*}
			and
			\begin{equation*}
				\sup_{\gamma\in\Gamma}\left|
				\tilde{Q}_{n}^{*}(\gamma)
				-\left\| \left[I-P_{W^{1/2}M_{0}(\gamma)}\right]\left(-W^{1/2}M_{0}(\gamma_{0})
				\alpha_{0}\right)\right\|^{2}\right| =o_{p}^{*}(1) \text{ in $P$}
			\end{equation*}
			when $\|W_{n}^{*}-W\|=o_{p}^{*}(1)$ in $P$ and $\theta_{0}^{*}\xrightarrow{p}\theta_{0}$. 
			Note that $W$ is the identity matrix if it is for the first step estimation and $\Omega^{-1}$ if it is for the second step estimation and the first step estimator is consistent.
			Since the uniform probability limit of $\tilde{Q}_{n}^{*}(\gamma)$ conditional on the data is minimized when $\gamma=\gamma_{0}$, the argmin CMT implies $\hat{\gamma}^{*}-\gamma_{0}=o_{p}^{*}(1)$ in $P$.
                Recall that $\theta_{0}^{*}$ is set as $(\hat{\alpha}(\gamma_{0}),\gamma_{0})'$ in \cref{thm:grid boot}, \eqref{eq:perc boot} in \cref{thm:perc boot}, and $\tilde{\theta}$ in \cref{thm:continuity boot}.
			For both cases (i) and (ii) of the proposition, $\gamma_{0}^{*}\xrightarrow{p}\gamma_{0}$ which implies $\gamma_{0}^{*}-\gamma_{0}=o_{p}^{*}(1)$ in $P$ by \cref{lem:bootstrap order}. Therefore, we can derive that $\hat{\gamma}^{*}-\gamma_{0}^{*}=(\hat{\gamma}^{*}-\gamma_{0})-(\gamma_{0}^{*}-\gamma_{0})=o_{p}^{*}(1)$ in $P$.
			
			\paragraph{Convergence rate under continuity.}
			
			By bootstrap equicontinuity, \cref{lem:boot.equicont}, and the consistency of $\hat{\theta}^{*}$ to $\theta_{0}^{*}$,
			\begin{equation*}
				\sqrt{n}\|\bar{g}_{n}^{*}(\hat{\theta}^{*})-\bar{g}_{n}^{*}(\theta_{0}^{*})-\bar{g}_{n}(\hat{\theta}^{*})+\bar{g}_{n}(\theta_{0}^{*})\|=o_{p}^{*}(1)\text{ in $P$.}
			\end{equation*}
			$\|W_{n}^{*}-W_{n}\|\xrightarrow{p^{*}}0$ in $P$ since $\|W_{n}-\Omega^{-1}\|=o_{p}^{*}(1)$ in $P$ and $\|W_{n}^{*}-\Omega^{-1}\|=o_{p}^{*}(1)$ in $P$. 
			The condition $\|W_{n}^{*}-\Omega^{-1}\|=o_{p}^{*}(1)$ in $P$ is implied by $\hat{\theta}_{(1)}\xrightarrow{p^{*}}\theta_{0}$ in $P$, as $|\hat{\theta}_{(1)}^{*}-\theta_{0}^{*}|\xrightarrow{p^{*}}0$ and $\theta_{0}^{*}\xrightarrow{p^{*}}\theta_{0}$ in $P$.
			Thus,
			\begin{equation*}
				\sqrt{n}\|W_{n}^{*1/2}\bar{g}_{n}^{*}(\hat{\theta}^{*})-W_{n}^{*1/2}\bar{g}_{n}^{*}(\theta_{0}^{*})-W_{n}^{1/2}\bar{g}_{n}(\hat{\theta}^{*})+W_{n}^{1/2}\bar{g}_{n}(\theta_{0}^{*})\|
				=o_{p}^{*}(1)\text{ in $P$}.
			\end{equation*}
			Apply triangle inequality to get 
			\begin{equation*}
				\sqrt{n}\|W_{n}^{1/2}\bar{g}_{n}(\hat{\theta}^{*})-W_{n}^{1/2}\bar{g}_{n}(\theta_{0}^{*})\|
				\leq o_{p}^{*}(1)+\sqrt{n}\|W_{n}^{*1/2}\bar{g}_{n}^{*}(\theta_{0}^{*})\|+\sqrt{n}\|W_{n}^{*1/2}\bar{g}_{n}^{*}(\hat{\theta}^{*})\| 
			\end{equation*}
			where $o_{p}^{*}(1)$ holds in $P$.
			As $\hat{\theta}^{*}$ is the minimizer of the bootstrap criterion,
			$\sqrt{n}\|W_{n}^{*1/2}\bar{g}_{n}^{*}(\hat{\theta}^{*})\|\leq \sqrt{n}\|W_{n}^{*1/2}\bar{g}_{n}^{*}(\theta_{0}^{*})\|=O_{p}^{*}(1)$ in $P$ where the last equality is implied by \cref{lem:boot res CLT}.
			Therefore,
			\[\sqrt{n}\|W_{n}^{1/2}\bar{g}_{n}(\hat{\theta}^{*})-W_{n}^{1/2}\bar{g}_{n}(\theta_{0}^{*})\|\leq O_{p}^{*}(1)\text{ in $P$}.\]
			By \cref{lem:equicont}, $\sqrt{n}\|W_{n}^{1/2}\bar{g}_{n}(\hat{\theta}^{*})-W_{n}^{1/2}\bar{g}_{n}(\theta_{0}^{*})-\Omega^{-1/2}g_{0}(\hat{\theta}^{*})+\Omega^{-1/2}g_{0}(\theta_{0}^{*})\|=o_{p}(1)$, so it is $o_{p}^{*}(1)$ in $P$ by \cref{lem:bootstrap order}.
			Hence,
			\[\sqrt{n}\|\Omega^{-1/2}g_{0}(\hat{\theta}^{*})-\Omega^{-1/2}g_{0}(\theta_{0}^{*})\|\leq O_{p}^{*}(1)\text{ in $P$}.\]
			By \cref{lem:loc.conti}, $\sqrt{n}\|\Omega^{-1/2}g_{0}(\hat{\theta}^{*})-\Omega^{-1/2}g_{0}(\theta_{0}^{*})\|\geq  \sqrt{n}\|\Omega^{-1/2}M_{0}(\hat{\alpha}^{*}-\alpha_{0}^{*})+\Omega^{-1/2}H\{(\hat{\gamma}^{*}-\gamma_{0})^{2}-(\gamma_{0}^{*}-\gamma_{0})^{2}\}\|+o_{p}^{*}(1+\sqrt{n}\{\|\hat{\alpha}^{*}-\alpha_{0}^{*}\|+(\hat{\gamma}^{*}-\gamma_{0})^{2}+(\gamma_{0}^{*}-\gamma_{0})^{2}\})$ in $P$. 
			Therefore, $\sqrt{n}\|\hat{\alpha}^{*}-\alpha_{0}^{*}\|=O_{p}^{*}(1)$ in $P$ and $\sqrt{n}(\hat{\gamma}^{*}-\gamma_{0})^{2}=O_{p}^{*}(1)$ in $P$.
			Suppose that $\sqrt{n}(\gamma_{0}^{*}-\gamma_{0})^{2}=O_{p}^{*}(1)$ in $P$. 
                Then, $\sqrt{n}(\hat{\gamma}^{*}-\gamma_{0}^{*})^{2}=O_{p}^{*}(1)$ in $P$ since $\sqrt{n}(\hat{\gamma}^{*}-\gamma_{0}^{*})^{2}\leq 2\sqrt{n}[(\hat{\gamma}^{*}-\gamma_{0})^{2}+(\gamma_{0}^{*}-\gamma_{0})^{2}]=O_{p}^{*}(1)$ in $P$.
   
                The condition, $\sqrt{n}(\gamma_{0}^{*}-\gamma_{0})^{2}=O_{p}^{*}(1)$ in $P$, is true if $\sqrt{n}(\gamma_{0}^{*}-\gamma_{0})^{2}=O_{p}(1)$ by \cref{lem:bootstrap order}.
			This is true for $\gamma_{0}^{*}=\gamma_{0}$ (\cref{thm:grid boot} (i)), $\gamma_{0}^{*}=w_{n}\hat{\gamma}+(1-w_{n})\tilde{\gamma}$ (\cref{thm:perc boot} (i)), or $\gamma_{0}^{*}=\tilde{\gamma}$ (\cref{thm:continuity boot} (i)).
                It is also the case for the standard nonparametric bootstrap as $\sqrt{n}(\hat{\gamma}-\gamma_{0})^{2}=O_{p}(1)$ by \cref{thm:asym}.

                \paragraph{Convergence rate under discontinuity.}
			
			Identically to the proof for the continuous model, we can get
			\[\sqrt{n}\|\Omega^{-1/2}g_{0}(\hat{\theta}^{*})-\Omega^{-1/2}g_{0}(\theta_{0}^{*})\|\leq O_{p}^{*}(1)\text{ in $P$}.\]
			Meanwhile, $\sqrt{n}\|\Omega^{-1/2}g_{0}(\hat{\theta}^{*})-\Omega^{-1/2}g_{0}(\theta_{0}^{*})\|\geq C\sqrt{n}\|\hat{\theta}^{*}-\theta_{0}^{*}\|+o_{p}^{*}(1+\sqrt{n}\|\hat{\theta}^{*}-\theta_{0}^{*}\|)$ for some $C<\infty$ in $P$ when the true model is discontinuous and \cref{ass:locDis} holds.
                This is because $g_{0}(\theta)=D_{1}(\theta-\theta_{0})+o(\|\theta-\theta_{0}\|)$ by \cref{ass:locDis} and
                \begin{equation*}
                    o(1) = \frac{\|g_{0}(\theta)-D_{1}(\theta-\theta_{0})\|}{\|\theta-\theta_{0}\|} \geq \frac{\sqrt{n}\|g_{0}(\theta)-D_{1}(\theta-\theta_{0})\|}{1+\sqrt{n}\|\theta-\theta_{0}\|}.
                \end{equation*}
			Therefore, $\sqrt{n}\|\hat{\theta}^{*}-\theta_{0}^{*}\|\leq O_{p}^{*}(1)$ in $P$.

			\subsection{Proof of \texorpdfstring{\cref{thm:grid boot}}{\cref{thm:grid boot}}.}
			In the grid bootstrap at $\gamma$, $\theta_{0}^{*}=(\hat{\alpha}(\gamma)',\gamma)'$.
			
			\paragraph{When $\gamma=\gamma_{0}$.}
			The proof of \cref{thm:perc boot} still holds, and $\mathbb{S}_{n}^{*}(a,b)$ conditionally weakly converges to either $\mathbb{S}$ or $\mathbb{S}_{J}$ in $\ell^{\infty}(\mathbb{K})$ in $P$ for every compact $\mathbb{K}$.
			The limit is $\mathbb{S}$ for the \cref{thm:grid boot} (i) case, and $\mathbb{S}_{J}$ for the \cref{thm:grid boot} (ii) case.
			By following the similar steps to the proof of \cref{thm:distance test}, we can derive the asymptotic distributions of $\mathcal{D}_{n}^{*}(\gamma)$.
			
			\paragraph{When $\gamma\neq\gamma_{0}$.}
			Note that $\bar{g}_{n}^{*}(\hat{\alpha}(\gamma),\gamma)=O_{p}^{*}(n^{-1/2})$.
			It will be shown that $\|W_{n}^{*}\|=O_{p}^{*}(1)$ in $P$.
			Then, $\min_{\alpha}\hat{Q}_{n}^{*}(\alpha,\gamma) \leq \hat{Q}_{n}^{*}(\hat{\alpha}(\gamma),\gamma)= \bar{g}_{n}^{*}(\hat{\alpha}(\gamma),\gamma)'W_{n}^{*}\bar{g}_{n}^{*}(\hat{\alpha}(\gamma),\gamma)=O_{p}^{*}(n^{-1})$, and $\mathcal{D}_{n}^{*}(\gamma)\leq n\min_{\alpha}\hat{Q}_{n}^{*}(\alpha,\gamma) = O_{p}^{*}(1)$ in $P$, which completes the proof.
			
			Recall that
			\begin{equation*}
				W_{n}^{*}= \left\{\frac{1}{n}\sum_{i=1}^{n}[g_{i}^{*}(\hat{\theta}_{(1)}^{*})g_{i}^{*}(\hat{\theta}_{(1)}^{*})']-\frac{1}{n}\sum_{i=1}^{n}g_{i}^{*}(\hat{\theta}_{(1)}^{*})\frac{1}{n}\sum_{i=1}^{n}g_{i}^{*}(\hat{\theta}_{(1)}^{*})'\right\}^{-1},
			\end{equation*}
			while $g_{i}^{*}(\theta)=g(\omega_{i}^{*},\theta)-g(\omega_{i}^{*},\theta_{0}^{*})+g(\omega_{i}^{*},\hat{\theta})$ as explained in Online Appendix \ref{sec: online.app.boot1}.
			The functional class $\mathcal{G}=\{g(\omega_{i},\theta):\theta\in\Theta\}$ is shown to satisfy the uniform entropy condition in the proof of \cref{lem:equicont}, and pairwise sum or product of functional classes preserve the uniform entropy condition by Theorem 2.10.20 in \cite{van_der_vaart_weak_1996}.
			Hence, by applying the bootstrap Glivenko-Cantelli theorem, e.g., Lemma 3.6.16 in \cite{van_der_vaart_weak_1996},
			\begin{multline*}
				\textstyle
				\sup_{\theta\in\Theta}\Biggl\|
				\frac{1}{n}\sum_{i=1}^{n}[g_{i}^{*}(\theta)g_{i}^{*}(\theta)']-\frac{1}{n}\sum_{i=1}^{n}g_{i}^{*}(\theta)\frac{1}{n}\sum_{i=1}^{n}g_{i}^{*}(\theta)' \\
				\textstyle
				-
				\Bigl(
				\frac{1}{n}\sum_{i=1}^{n}\left[\{g_{i}(\theta)-g_{i}(\theta_{0}^{*})+g_{i}(\hat{\theta})\}\{g_{i}(\theta)-g_{i}(\theta_{0}^{*})+g_{i}(\hat{\theta})\}'\right]\\
				\textstyle
				-\frac{1}{n}\sum_{i=1}^{n}\{g_{i}(\theta)-g_{i}(\theta_{0}^{*})+g_{i}(\hat{\theta})\}\frac{1}{n}\sum_{i=1}^{n}\{g_{i}(\theta)-g_{i}(\theta_{0}^{*})+g_{i}(\hat{\theta})\}'
				\Bigr)\Biggl\|
			\end{multline*}
			is $o_{p}^{*}(1)$ in $P$.
			Furthermore,
			\begin{multline*}
				\frac{1}{n}\sum_{i=1}^{n}\left[\{g_{i}(\theta)-g_{i}(\theta_{1})+g_{i}(\theta_{2})\}\{g_{i}(\theta)-g_{i}(\theta_{1})+g_{i}(\theta_{2})\}'\right]\\
				-\frac{1}{n}\sum_{i=1}^{n}\{g_{i}(\theta)-g_{i}(\theta_{1})+g_{i}(\theta_{2})\}\frac{1}{n}\sum_{i=1}^{n}\{g_{i}(\theta)-g_{i}(\theta_{1})+g_{i}(\theta_{2})\}'\\
				\xrightarrow{p}
				E\left[\{g_{i}(\theta)-g_{i}(\theta_{1})+g_{i}(\theta_{2})\}\{g_{i}(\theta)-g_{i}(\theta_{1})+g_{i}(\theta_{2})\}'\right]\\
				-E[g_{i}(\theta)-g_{i}(\theta_{1})+g_{i}(\theta_{2})]E[g_{i}(\theta)-g_{i}(\theta_{1})+g_{i}(\theta_{2})]'
			\end{multline*}
			uniformly with respect to $\theta$, $\theta_{1}$, and $\theta_{2}$.
			As $\hat{\theta}$ and $\hat{\theta}_{0}^{*}$ are consistent to $\theta_{0}$, 
			\begin{equation*}
				\frac{1}{n}\sum_{i=1}^{n}[g_{i}^{*}(\theta)g_{i}^{*}(\theta)']-\frac{1}{n}\sum_{i=1}^{n}g_{i}^{*}(\theta)\frac{1}{n}\sum_{i=1}^{n}g_{i}^{*}(\theta)' \xrightarrow{p^{*}} E\left[g_{i}(\theta)g_{i}(\theta)'\right]-E[g_{i}(\theta)]E[g_{i}(\theta)]'
			\end{equation*} uniformly with respect to $\theta$.
			By the compactness of $\Theta$, the minimum eigenvalue of $\{E\left[g_{i}(\theta)g_{i}(\theta)'\right]-E[g_{i}(\theta)]E[g_{i}(\theta)]'\}$ is bounded below by some constant $c>0$.
			Therefore, $\sup_{\theta\in\Theta}\|W_{n}^{*}(\theta)\|=O_{p}^{*}(1)$ in $P$ where
			\begin{equation*}
				W_{n}^{*}(\theta)= \left\{\frac{1}{n}\sum_{i=1}^{n}[g_{i}^{*}(\theta)g_{i}^{*}(\theta)']-\frac{1}{n}\sum_{i=1}^{n}g_{i}^{*}(\theta)\frac{1}{n}\sum_{i=1}^{n}g_{i}^{*}(\theta)'\right\}^{-1}.
			\end{equation*}
			As $W_{n}^{*}=W_{n}^{*}(\hat{\theta}_{(1)}^{*})$, we can conclude that $\|W_{n}^{*}\|=O_{p}^{*}(1)$.

			\subsection{Proof of \texorpdfstring{\cref{thm:continuity boot}}{\cref{thm:continuity boot}}.}
			In the bootstrap for continuity test, $\theta_{0}^{*}=\tilde{\theta}$, where $\tilde{\theta}$ is the continuity-restricted estimator.
			
			\paragraph{Under the null hypothesis.}
			When the true model is continuous, the proof of \cref{thm:perc boot} still holds.
			$\mathbb{S}_{n}^{*}(a,b)$ conditionally weakly converges to $\mathbb{S}$ in $\ell^{\infty}(\mathbb{K})$ in $P$ for every compact $\mathbb{K}$.
			By following the similar steps to the proof of \cref{thm:continuity test}, we can derive the asymptotic distribution of $\mathcal{T}_{n}^{*}$.
			
			\paragraph{Under the alternative hypothesis.}
			Let the true model be discontinuous.
			Note that $\bar{g}_{n}^{*}(\tilde{\theta})=O_{p}^{*}(n^{-1/2})$.
			Meanwhile, $\|W_{n}^{*}\|=O_{p}^{*}(1)$ in $P$, by the same logic used in the proof of \cref{thm:grid boot} when $\gamma\neq\gamma_{0}$.
			Then, $\min_{\theta\in\Theta: \delta_{2}=0_{p-1}, \delta_{1}=-\delta_{3}\gamma}\hat{Q}_{n}^{*}(\theta) \leq \hat{Q}_{n}^{*}(\tilde{\theta})= \bar{g}_{n}^{*}(\tilde{\theta})'W_{n}^{*}\bar{g}_{n}^{*}(\tilde{\theta})=O_{p}^{*}(n^{-1})$.
			Therefore, $\mathcal{T}_{n}^{*}\leq n\min_{\theta\in\Theta: \delta_{2}=0_{p-1}, \delta_{1}=-\delta_{3}\gamma}\hat{Q}_{n}^{*}(\theta) = O_{p}^{*}(1)$ in $P$, which completes the proof.
			
			\subsection{Lemmas}
			\begin{lemma}
				\label{lem:boot res LLN}
				If \cref{ass:gmm} holds,
				\begin{equation*}
					\hat{u}_{n}^{*}=\begin{pmatrix}
						\frac{1}{n}\sum_{i=1}^{n}z_{it_{0}}^{*}\widehat{\Delta\epsilon}_{it_{0}}^{*}\\
						\vdots\\
						\frac{1}{n}\sum_{i=1}^{n}z_{iT}^{*}\widehat{\Delta\epsilon}_{iT}^{*}
					\end{pmatrix}
					-
					\begin{pmatrix}
						\frac{1}{n}\sum_{i=1}^{n}z_{it_{0}}\widehat{\Delta\epsilon}_{it_{0}}\\
						\vdots\\
						\frac{1}{n}\sum_{i=1}^{n}z_{iT}\widehat{\Delta\epsilon}_{iT}
					\end{pmatrix}
					\xrightarrow{p^{*}}0 \text{ in $P$.}
				\end{equation*}
			\end{lemma}
			
			\begin{proof}
				Let $u_{n}^{*}(\theta) = \frac{1}{n}\sum_{i=1}^{n}[g(\omega_{i}^{*},\theta)-\frac{1}{n}\sum_{i=1}^{n}g(\omega_{i},\theta)]$ where $g(\omega_{i},\theta)$ is defined as \eqref{eq:momtclass}, and $\omega_{i}^{*}$ is a resampling draw from $\{\omega_{i}:i=1,...,n\}$.
				See Online Appendix \ref{sec: online.app.boot1} for more explanation.
				$\mathcal{G}=\{g(\omega_{i},\theta):\theta\in\Theta\}$ is shown to satisfy the uniform entropy condition in the proof of \cref{lem:equicont}.
				Therefore, by bootstrap Glivenko-Cantelli theorem, e.g., Lemma 3.6.16 in \cite{van_der_vaart_weak_1996}, $\sup_{\theta\in\Theta}\|u_{n}^{*}(\theta)\|=o_{p}^{*}(1)$ in $P$.
				Note that $\hat{u}_{n}^{*}=u_{n}^{*}(\hat{\theta})$ which completes the proof.
			\end{proof}
			
			\begin{lemma}
				\label{lem:boot res CLT}
				If \cref{ass:gmm} holds and $\hat{\theta}\xrightarrow{p}\theta_{0}$, then
				\begin{equation*}
					\sqrt{n}\left\{\begin{pmatrix}
						\frac{1}{n}\sum_{i=1}^{n}z_{it_{0}}^{*}\widehat{\Delta\epsilon}_{it_{0}}^{*}\\
						\vdots\\
						\frac{1}{n}\sum_{i=1}^{n}z_{iT}^{*}\widehat{\Delta\epsilon}_{iT}^{*}
					\end{pmatrix}
					-
					\begin{pmatrix}
						\frac{1}{n}\sum_{i=1}^{n}z_{it_{0}}\widehat{\Delta\epsilon}_{it_{0}}\\
						\vdots\\
						\frac{1}{n}\sum_{i=1}^{n}z_{iT}\widehat{\Delta\epsilon}_{iT}
					\end{pmatrix}\right\}
					\xrightarrow{d^{*}}N(0,\Omega)\text{ in $P$.}
				\end{equation*}
			\end{lemma}
			
			\begin{proof}
				Note that $g_{i}^{*}(\theta_{1})-g_{i}^{*}(\theta_{2})=g(\omega_{i}^{*},\theta_{1})-g(\omega_{i}^{*},\theta_{2})$ for any $\theta_{1}$ and $\theta_{2}$ where $g(\omega_{i},\theta)$ is defined as \eqref{eq:momtclass}, and $\omega_{i}^{*}$ is a resampling draw from $\{\omega_{i}:i=1,...,n\}$.
				See Online Appendix \ref{sec: online.app.boot1} for more explanation.
				Hence, $\bar{g}_{n}^{*}(\theta)-\bar{g}_{n}^{*}(\theta_{0})-\bar{g}_{n}(\theta)+\bar{g}_{n}(\theta_{0})=\frac{1}{n}\sum_{i=1}^{n}\left[g(\omega_{i}^{*},\theta)-\frac{1}{n}\sum_{i=1}^{n}g(\omega_{i},\theta)\right]-\frac{1}{n}\sum_{i=1}^{n}\left[g(\omega_{i}^{*},\theta_{0})-\frac{1}{n}\sum_{i=1}^{n}g(\omega_{i},\theta_{0})\right]$.
				Furthermore,
				\begin{equation*}
					\frac{1}{\sqrt{n}}\sum_{i=1}^{n}\left[g(\omega_{i}^{*},\hat{\theta})-\frac{1}{n}\sum_{i=1}^{n}g(\omega_{i},\hat{\theta})\right]
					=
					\sqrt{n}\left\{\begin{pmatrix}
						\frac{1}{n}\sum_{i=1}^{n}z_{it_{0}}^{*}\widehat{\Delta\epsilon}_{it_{0}}^{*}\\
						\vdots\\
						\frac{1}{n}\sum_{i=1}^{n}z_{iT}^{*}\widehat{\Delta\epsilon}_{iT}^{*}
					\end{pmatrix}
					-
					\begin{pmatrix}
						\frac{1}{n}\sum_{i=1}^{n}z_{it_{0}}\widehat{\Delta\epsilon}_{it_{0}}\\
						\vdots\\
						\frac{1}{n}\sum_{i=1}^{n}z_{iT}\widehat{\Delta\epsilon}_{iT}
					\end{pmatrix}\right\}.
				\end{equation*}
				By \cref{lem:boot.equicont}, $\sqrt{n}\|\bar{g}_{n}^{*}(\hat{\theta})-\bar{g}_{n}^{*}(\theta_{0})-\bar{g}_{n}(\hat{\theta})+\bar{g}_{n}(\theta_{0})\|=\sqrt{n}\|\frac{1}{n}\sum_{i=1}^{n}\left[g(\omega_{i}^{*},\hat{\theta})-\frac{1}{n}\sum_{i=1}^{n}g(\omega_{i},\hat{\theta})\right]-\frac{1}{n}\sum_{i=1}^{n}\left[g(\omega_{i}^{*},\theta_{0})-\frac{1}{n}\sum_{i=1}^{n}g(\omega_{i},\theta_{0})\right]\|=o_{p}^{*}(1)$ in $P$.
				By the bootstrap CLT (e.g., \cite{gine_zinn_1990}),
				\begin{equation*}
					\frac{1}{\sqrt{n}}\sum_{i=1}^{n}\left[g(\omega_{i}^{*},\theta_{0})-\frac{1}{n}\sum_{i=1}^{n}g(\omega_{i},\theta_{0})\right] \xrightarrow{d^{*}} N(0,\Omega)\text{ in $P$.}
				\end{equation*}
				By applying the Slutsky theorem, we can derive $\frac{1}{\sqrt{n}}\sum_{i=1}^{n}\left[g(\omega_{i}^{*},\hat{\theta})-\frac{1}{n}\sum_{i=1}^{n}g(\omega_{i},\hat{\theta})\right]\xrightarrow{d^{*}} N(0,\Omega)$ in $P$.
			\end{proof}
			
			Recall that $\bar{M}_{n}^{*}(\gamma) = \frac{1}{n}\sum_{i=1}^{n} M_{i}^{*}(\gamma)$ where
			\begin{equation*}
				M_{i}^{*}(\gamma) = -\begin{bmatrix}
					z_{it_{0}^{*}}(\Delta x_{it_{0}}^{*\prime},1_{it_{0}}^{*}(\gamma)'X_{it_{0}}^{*}) \\
					\vdots \\
					z_{iT}^{*}(\Delta x_{iT}^{*\prime},1_{iT}^{*}(\gamma)'X_{iT}^{*})
				\end{bmatrix}.
			\end{equation*}
			
			\begin{lemma}
				\label{lem:boot.ULLN}
				If \cref{ass:gmm} is true, then
				\begin{equation*}
					\sup_{\gamma\in\Gamma}\|\bar{M}_{n}^{*}(\gamma)-M_{0}(\gamma)\|\xrightarrow{p^{*}}0 \text{ in $P$}.
				\end{equation*}
			\end{lemma}
			
			\begin{proof}
				It is shown that the classes $\{z_{it}(1,x_{it}')1\{q_{it}>\gamma\}:\gamma\in\Gamma\}$ and $\{z_{it}(1,x_{it-1}')1\{q_{it-1}>\gamma\}:\gamma\in\Gamma\}$ are P-Glivenko-Cantelli in the proof of \cref{lem:ULLN}. 
				Then, by bootstrap Glivenko-Cantelli theorem, e.g., Lemma 3.6.16 in \cite{van_der_vaart_weak_1996}, the result of this lemma holds.
				
			\end{proof}
			
			\begin{lemma}
				\label{lem:boot.equicont}
				Let \cref{ass:gmm} hold.
				If $h_{n}\rightarrow 0$, then
				\begin{equation*}
					\sup_{\|\theta_{1}-\theta_{2}\|<h_{n}}\sqrt{n}\|\bar{g}_{n}^{*}(\theta_{1})-\bar{g}_{n}^{*}(\theta_{2})-\bar{g}_{n}(\theta_{1})+\bar{g}_{n}(\theta_{2})\|=o_{p}^{*}(1)\text{ in $P$.}
				\end{equation*}
				
			\end{lemma}
			
			\begin{proof}
				Note that $g_{i}^{*}(\theta_{1})-g_{i}^{*}(\theta_{2})=g(\omega_{i}^{*},\theta_{1})-g(\omega_{i}^{*},\theta_{2})$ for any $\theta_{1}$ and $\theta_{2}$ where $g(\omega_{i},\theta)$ is defined as \eqref{eq:momtclass}, and $\omega_{i}^{*}$ is a resampling from $\{\omega_{i}:i=1,...,n\}$.
				Hence, $\bar{g}_{n}^{*}(\theta_{1})-\bar{g}_{n}^{*}(\theta_{2})-\bar{g}_{n}(\theta_{1})+\bar{g}_{n}(\theta_{2})=\frac{1}{n}\sum_{i=1}^{n}\left[g(\omega_{i}^{*},\theta_{1})-\frac{1}{n}\sum_{i=1}^{n}g(\omega_{i},\theta_{1})\right]-\frac{1}{n}\sum_{i=1}^{n}\left[g(\omega_{i}^{*},\theta_{2})-\frac{1}{n}\sum_{i=1}^{n}g(\omega_{i},\theta_{2})\right]$.
				By bootstrap version of stochastic equicontinuity, e.g., C2 in the proof of Theorem 2.1 in \cite{praestgaard_wellner_1993}, the result of this lemma holds if $\{g(\omega_{i},\theta):\theta\in\Theta\}$ satisfies the uniform entropy condition and has a square integrable envelope function, which are verified in the proof of \cref{lem:equicont}.
			\end{proof}

			\begin{lemma}
				\label{lem:boot plim1}
				Suppose that Assumptions \ref{ass:gmm}, \ref{ass:dgp}, and \ref{ass:loc} hold, and the true model is continuous.
				If $\delta_{20}^{*}=O_{p}(n^{-1/2})$, $\delta_{30}^{*}-\delta_{30}=O_{p}(n^{-1/2})$, $\gamma_{0}^{*}-\gamma_{0}=O_{p}(n^{-1/4})$, and $\delta_{10}^{*}+\delta_{30}^{*}\gamma_{0}^{*}=O_{p}(n^{-1/2})$,
				then
				\begin{equation*}
					\frac{1}{\sqrt{n}}\sum_{i=1}^{n} z_{it}^{*}(1_{it}^{*}(\gamma_{0}^{*})'-1_{it}^{*}(\gamma_{0}^{*}+\tfrac{b}{n^{\frac{1}{4}}})')X_{it}^{*}\delta_{0}^{*}
					\xrightarrow{p^{*}}
					\frac{\delta_{30}}{2}\left\{E_{t}[z_{it}|\gamma_{0}]f_{t}(\gamma_{0})-E_{t-1}[z_{it}|\gamma_{0}]f_{t-1}(\gamma_{0})\right\}b^{2},
				\end{equation*}
				in $P$ uniformly with respect to $b\in [-K,K]$ for any $K<\infty$.
			\end{lemma}
			
			The conditions for $\delta_{0}^{*}$ and $\gamma_{0}^{*}$ hold if (i) $\theta_{0}^{*}=(\hat{\alpha}(\gamma_{0})',\gamma_{0})'$, (ii) $\theta_{0}^{*}$ is set as \eqref{eq:perc boot}, and (iii) $\theta_{0}^{*}=\tilde{\theta}$, which is the continuity-restricted estimator in \cref{sec:asym continuity}, under the assumptions of this lemma.
			For (i), $\sqrt{n}(\hat{\alpha}(\gamma_{0})-\alpha_{0})$ is asymptotically normal, and $\hat{\delta}_{1}(\gamma_{0})-\delta_{10}+(\hat{\delta}_{3}(\gamma_{0})-\delta_{30})\cdot\gamma_{0}=O_{p}(n^{-1/2})$.
			For (ii), note that $w_{n}=O_{p}(n^{-1/4})$. $\delta_{10}^{*}+\delta_{30}^{*}\gamma_{0}^{*}=w_{n}(\hat{\delta}_{1}+\hat{\delta}_{3}\hat{\gamma})+w_{n}(1-w_{n})(\hat{\delta}_{3}-\tilde{\delta}_{3})(\tilde{\gamma}-\hat{\gamma})+(1-w_{n})(\tilde{\delta}_{1}+\tilde{\delta}_{3}\tilde{\gamma})$, while $w_{n}(\hat{\delta}_{1}+\hat{\delta}_{3}\hat{\gamma})=O_{p}(n^{-1/2})$, $(1-w_{n})(\tilde{\delta}_{1}+\tilde{\delta}_{3}\tilde{\gamma})=0$, and $(1-w_{n})w_{n}(\hat{\delta}_{3}-\tilde{\delta}_{3})(\tilde{\gamma}-\hat{\gamma})=O_{p}(n^{-1/4})O_{p}(n^{-1/2})O_{p}(n^{-1/4})$. $\delta_{20}^{*}=w_{n}\hat{\delta}_{2}=O_{p}(n^{-3/4})$, and $\delta_{30}^{*}-\delta_{30}=w_{n}(\hat{\delta}_{3}-\delta_{30})+(1-w_{n})(\tilde{\delta}_{3}-\delta_{30})=O_{p}(n^{-3/4})+O_{p}(n^{-1/2})$. $\gamma_{0}^{*}-\gamma_{0}=w_{n}(\hat{\gamma}-\gamma_{0})+(1-w_{n})(\tilde{\gamma}-\gamma_{0})=O_{p}(n^{-1/4})O_{p}(n^{-1/4})+O_{p}(n^{-1/2})=O_{p}(n^{-1/2})$ also holds.
			For (iii), \cite{seo_estimation_2019} showed that $\tilde{\theta}-\theta_{0}=O_{p}(n^{-1/2})$, while $\tilde{\delta}_{1}+\tilde{\delta}_{3}\tilde{\gamma}=0$ and $\tilde{\delta}_{2}=0_{p-1}$ by definition.

			\begin{proof}
				
				Note that
				\begin{align}
					\nonumber
					& \frac{1}{\sqrt{n}}\sum_{i=1}^{n} z_{it}^{*}(1_{it}(\gamma_{0}^{*})'-1_{it}^{*}(\gamma_{0}^{*}+\tfrac{b}{n^{\frac{1}{4}}})')X_{it}^{*}\delta_{0}^{*}\\
					\label{eq:bootCLTlimit}
					&=\frac{1}{\sqrt{n}}\sum_{i=1}^{n} 
					z_{it}^{*}(1_{it}^{*}(\gamma_{0}^{*})'-1_{it}^{*}(\gamma_{0}^{*}+\tfrac{b}{n^{\frac{1}{4}}})')X_{it}^{*}\delta_{0}^{*}-
					\frac{1}{\sqrt{n}}\sum_{i=1}^{n}z_{it}(1_{it}(\gamma_{0}^{*})'-1_{it}(\gamma_{0}^{*}+\tfrac{b}{n^{\frac{1}{4}}})')X_{it}\delta_{0}^{*}
					\\
					\label{eq:bootdeter.limit}
					&\quad+
					\frac{1}{\sqrt{n}}\sum_{i=1}^{n}z_{it}(1_{it}(\gamma_{0}^{*})'-1_{it}(\gamma_{0}^{*}+\tfrac{b}{n^{\frac{1}{4}}})')X_{it}\delta_{0}^{*}.
				\end{align}
				First, we show that the stochastic term \eqref{eq:bootCLTlimit} is $o_{p}^{*}(1)$ in $P$ uniformly with respect to $b\in[-K,K]$.
				Note that $\{z_{it}(1_{it}(\gamma)'-1_{it}(\gamma+\kappa)')X_{it}\delta:\theta\in\Theta,|\kappa| \leq K\}=\{g(\omega_{i},(\alpha',\gamma)')-g(\omega_{i},(\alpha',\gamma+\kappa)'):\theta\in\Theta, |\kappa| \leq K\}$ while $\mathcal{G}=\{g(\omega_{i},\theta):\theta\in\Theta\}$ is shown to satisfy the uniform entropy condition and to have a square integrable envelope in the proof of \cref{lem:equicont}. 
				Then, by C2 in the proof of Theorem 2.1 in \cite{praestgaard_wellner_1993}, the following bootstrap asymptotic equicontinuity can be derived:
				\begin{equation*}
					\sup_{\substack{b\in[-K,K],\\ \theta\in\Theta}}\frac{1}{\sqrt{n}}\sum_{i=1}^{n} 
					\left\{
					z_{it}^{*}(1_{it}^{*}(\gamma)'-1_{it}^{*}(\gamma+\tfrac{b}{n^{\frac{1}{4}}})')X_{it}^{*}\delta-
					\frac{1}{n}\sum_{i=1}^{n}z_{it}(1_{it}(\gamma)'-1_{it}(\gamma+\tfrac{b}{n^{\frac{1}{4}}})')X_{it}\delta
					\right\}
				\end{equation*}
				is $o_{p}^{*}(1)$ in $P$.
				Hence, by plugging in $\theta_{0}^{*}$ to the place of $\theta$ in the last display, we can derive that \eqref{eq:bootCLTlimit} is $o_{p}^{*}(1)$ in $P$ uniformly with respect to $b\in[-K,K]$.
				
				Next, we show that \eqref{eq:bootdeter.limit} term converges to a deterministic limit.
				As $\{z_{it}1_{it}(\gamma)'X_{it}\delta:\theta\in\Theta,|\kappa| \leq K\}$ satisfies the uniform entropy condition and has a square integrable envelope function, we can derive the following asymptotic equicontinuity:
				\begin{equation*}
					\sup_{\substack{b\in[-K,K],\\ \theta\in\Theta}}\left\|
					\frac{1}{\sqrt{n}}\sum_{i=1}^{n}z_{it}(1_{it}(\gamma)'-1_{it}(\gamma+\tfrac{b}{n^{\frac{1}{4}}})')X_{it}\delta
					-
					\sqrt{n}E[z_{it}(1_{it}(\gamma)'-1_{it}(\gamma+\tfrac{b}{n^{\frac{1}{4}}})')X_{it}\delta]
					\right\|
				\end{equation*}
				is $o_{p}(1)$, and hence $o_{p}^{*}(1)$ in $P$ by \cref{lem:bootstrap order}.
				Therefore,
				\begin{equation*}
					\sup_{\substack{b\in[-K,K],\\ \theta\in\Theta}}\left\|
					\frac{1}{\sqrt{n}}\sum_{i=1}^{n}z_{it}^{*}(1_{it}^{*}(\gamma)'-1_{it}^{*}(\gamma+\tfrac{b}{n^{\frac{1}{4}}})')X_{it}^{*}\delta
					-
					\sqrt{n}E[z_{it}(1_{it}(\gamma)'-1_{it}(\gamma+\tfrac{b}{n^{\frac{1}{4}}})')X_{it}\delta]
					\right\|
				\end{equation*}
				is $o_{p}^{*}(1)$ in $P$.
				
				Let $J_{n}(\delta,\gamma,b)=\sqrt{n}E[z_{it}(1_{it}(\gamma)'-1_{it}(\gamma+\tfrac{b}{n^{\frac{1}{4}}})')X_{it}\delta]$.
				By assumption, we can reparametrize such that $\delta_{20}^{*}=\frac{r_{\delta_{2}}}{\sqrt{n}}$, $\delta_{30}^{*}=\delta_{30}+\frac{r_{\delta_{3}}}{\sqrt{n}}$, $\gamma_{0}^{*}=\gamma_{0}+\frac{r_{\gamma}}{n^{1/4}}$, and $\delta_{10}^{*}=-\delta_{30}^{*}\gamma_{0}^{*}+\frac{r_{\delta_{1}+\delta_{3}\gamma}}{\sqrt{n}}=\delta_{10}-\delta_{30}\frac{r_{\gamma}}{n^{1/4}}-\gamma_{0}\frac{r_{\delta_{3}}}{\sqrt{n}}-\tfrac{r_{\gamma}r_{\delta_{3}}}{n^{3/4}}+\frac{r_{\delta_{1}+\delta_{3}\gamma}}{\sqrt{n}}$.
				Then, we can reparametrize the function $J_{n}$ such that 
				\begin{equation}
					\label{eq:boot_reparam_limit}
					\widetilde{J}_{n}(r_{\delta_{1}+\delta_{3}\gamma},r_{\delta_{2}},r_{\delta_{3}},r_{\gamma},b)
					=
					J_{n}(\delta_{10}-\delta_{30}\tfrac{r_{\gamma}}{n^{1/4}}-\gamma_{0}\tfrac{r_{\delta_{3}}}{\sqrt{n}}-\tfrac{r_{\gamma}r_{\delta_{3}}}{n^{3/4}}+\tfrac{r_{\delta_{1}+\delta_{3}\gamma}}{\sqrt{n}},
					\tfrac{r_{\delta_{2}}}{\sqrt{n}},
					\delta_{30}+\tfrac{r_{\delta_{3}}}{\sqrt{n}}, 
					\gamma_{0}+\tfrac{r_{\gamma}}{n^{1/4}},
					b).
				\end{equation}
				Let $r=(r_{\delta_{1}+\delta_{3}\gamma},r_{\delta_{2}},r_{\delta_{3}},r_{\gamma})$ which lies in a compact set $\mathcal{R}=\{r\in\mathbb{R}^{p+2}:\|r\|\leq \overline{K}\}$ for an aribtrary  $\overline{K}<\infty$.
				
				To prove the lemma, it will be shown below that 
                $$\widetilde{J}_{n}(r_{\delta_{1}+\delta_{3}\gamma},r_{\delta_{2}},r_{\delta_{3}},r_{\gamma},b) \rightarrow \frac{\delta_{30}}{2}\left\{E_{t}[z_{it}|\gamma_{0}]f_{t}(\gamma_{0})-E_{t-1}[z_{it}|\gamma_{0}]f_{t-1}(\gamma_{0})\right\}b^{2}$$ uniformly with respect to $r\in\mathcal{R}$ and $b\in[-K,K]$, which in turn implies     
                $$\frac{1}{\sqrt{n}}\sum_{i=1}^{n}z_{it}^{*}(1_{it}^{*}(\gamma_{0}^{*})'-1_{it}^{*}(\gamma_{0}^{*}+\tfrac{b}{n^{\frac{1}{4}}})')X_{it}^{*}\delta_{0}^{*}\xrightarrow{p^{*}}\frac{\delta_{30}}{2}\left\{E_{t}[z_{it}|\gamma_{0}]f_{t}(\gamma_{0})-E_{t-1}[z_{it}|\gamma_{0}]f_{t-1}(\gamma_{0})\right\}b^{2}\text{ in $P$}$$  
                uniformly with respect to $b\in[-K,K]$ since 
                \begin{equation*}
					\sup_{b\in[-K,K]}\left\|
					\frac{1}{\sqrt{n}}\sum_{i=1}^{n}z_{it}^{*}(1_{it}^{*}(\gamma_{0}^{*})'-1_{it}^{*}(\gamma_{0}^{*}+\tfrac{b}{n^{\frac{1}{4}}})')X_{it}^{*}\delta_{0}^{*}
					-
					J_{n}(\delta_{0}^{*},\gamma_{0}^{*},b)
					\right\|=o_{p}^{*}(1)\text{ in $P$.}
			\end{equation*}
				
            Suppose $b>0$. The case for $b<0$ follows similarly.
				Note that
				\begin{multline*}
					\sqrt{n}E[z_{it}(1_{it}(\gamma)'-1_{it}(\gamma+\tfrac{b}{n^{\frac{1}{4}}})')X_{it}\delta]
					=
					\sqrt{n}E[z_{it}(1,x_{it}')\delta 1\{\gamma+\tfrac{b}{n^{\frac{1}{4}}}\geq q_{it}>\gamma\}]
					\\
					-\sqrt{n}E[z_{it}(1,x_{it-1}')\delta 1\{\gamma+\tfrac{b}{n^{\frac{1}{4}}}\geq q_{it-1}>\gamma\}].
				\end{multline*}
				We focus on the first term on the right hand side $\sqrt{n}E[z_{it}(1,x_{it}')\delta1\{\gamma+\tfrac{b}{n^{\frac{1}{4}}}\geq q_{it}>\gamma\}]$ since the limit of the second term can be analyzed similarly, and redefine $J_{n}(\delta,\gamma,b)=\sqrt{n}E[z_{it}(1,x_{it}')\delta1\{\gamma+\tfrac{b}{n^{\frac{1}{4}}}\geq q_{it}>\gamma\}]$ and $\widetilde{J}_{n}$, accordingly.
				Let $x_{it}=(\xi_{it}',q_{it})'$ where $\xi_{it}\in\mathbb{R}^{p-1}$.
				Then, $J_{n}(\delta,\gamma,b)  =  J_{1n}(\delta,\gamma,b) + J_{2n}(\delta,\gamma,b)$ where
				\begin{align*}
					J_{1n}(\delta,\gamma,b) &
					=\sqrt{n}E[z_{it}\xi_{it}'\delta_{2}1\{\gamma+\tfrac{b}{n^{\frac{1}{4}}}\geq q_{it}>\gamma\}],\text{ and}\\
					J_{2n}(\delta,\gamma,b) &
					=\sqrt{n}E[z_{it}(\delta_{1}+\delta_{3}q_{it})1\{\gamma+\tfrac{b}{n^{\frac{1}{4}}}\geq q_{it}>\gamma\}].
				\end{align*}
				Similarly to $\widetilde{J}_{n}$ in \eqref{eq:boot_reparam_limit}, we define reparametrized function $\widetilde{J}_{1n}$ and $\widetilde{J}_{2n}$.
				
				\paragraph{Limit of $\widetilde{J}_{1n}$:}
				
				We can derive the Taylor expansion 
				\begin{equation*}
					\widetilde{J}_{1n}(r,b)=E[z_{it}\xi_{it}'r_{\delta_{2}}1\{\gamma_{0}+\tfrac{b+r_{\gamma}}{n^{\frac{1}{4}}}\geq q_{it}> \gamma_{0}+\tfrac{r_{\gamma}}{n^{\frac{1}{4}}}\}]=E_{t}[z_{it}\xi_{it}'\tfrac{r_{\delta_{2}}}{n^{1/4}}|\gamma_{n,b}]f_{t}(\gamma_{n,b})b,
				\end{equation*}
				where $\gamma_{n,b}\in[\gamma_{0}+\frac{r_{\gamma}}{n^{1/4}},\gamma_{0}+\frac{b+r_{\gamma}}{n^{1/4}}]$.
				As both $r_{\gamma}$ and $b$ are in compact spaces, $\gamma_{n,b}\rightarrow \gamma_{0}$ uniformly with respect to $r_{\gamma}$ and $b$.
				By \cref{ass:dgp}, $E_{t}[z_{it}\xi_{it}'|\gamma]f_{t}(\gamma)$ is bounded and continuous on a neighborhood $\mathcal{O}$ of $\gamma_{0}$.
				Therefore, $E_{t}[z_{it}\xi_{it}'|\gamma_{n,b}]f_{t}(\gamma_{n,b})\rightarrow E_{t}[z_{it}\xi_{it}'|\gamma_{0}]f_{t}(\gamma_{0})$.
				Since $\frac{r_{\delta_{2}}}{n^{1/4}}\rightarrow 0$, we can derive $\widetilde{J}_{1n}(r,b)\rightarrow 0$ uniformly in $r$ and $b$.
				
				\paragraph{Limit of $\widetilde{J}_{2n}$:}
				
				We can derive the Taylor expansion
				\begin{align}
					\nonumber
					& \widetilde{J}_{2n}(r,b) \\
					\nonumber
					&=\sqrt{n}E[z_{it}(\delta_{10}-\delta_{30}\tfrac{r_{\gamma}}{n^{1/4}}-\gamma_{0}\tfrac{r_{\delta_{3}}}{\sqrt{n}}-\tfrac{r_{\gamma}r_{\delta_{3}}}{n^{3/4}}+\tfrac{r_{\delta_{1}+\delta_{3}\gamma}}{\sqrt{n}}+(\delta_{30}+\tfrac{r_{\delta_{3}}}{\sqrt{n}})q_{it})1\{\gamma_{0}+\tfrac{b+r_{\gamma}}{n^{\frac{1}{4}}}\geq q_{it}> \gamma_{0}+\tfrac{r_{\gamma}}{n^{\frac{1}{4}}}\}]
					\\
					\label{eq:bootdeter.limit.3}
					&=
					\tfrac{r_{\delta_{1}+\delta_{3}\gamma}}{n^{1/4}}E_{t}[z_{it}|\gamma_{0}+\tfrac{r_{\gamma}}{n^{\frac{1}{4}}}]f_{t}(\gamma_{0}+\tfrac{r_{\gamma}}{n^{\frac{1}{4}}})b \\
					\label{eq:bootdeter.limit.4}
					&\quad
					+\frac{b^{2}}{2}
					(\tfrac{r_{\delta_{1}+\delta_{3}\gamma}}{\sqrt{n}}+(\delta_{30}+\tfrac{r_{\delta_{3}}}{\sqrt{n}})(\gamma_{n,b}-\gamma_{0}-\tfrac{b}{n^{\frac{1}{4}}}))\frac{d}{d\gamma}\left\{E_{t}[z_{it}|\gamma]f_{t}(\gamma)\right\}|_{\gamma=\gamma_{n,b}}
					\\
					\label{eq:bootdeter.limit.5}
					&\quad 
					+\frac{b^{2}}{2}
					(\delta_{30}+\tfrac{r_{\delta_{3}}}{\sqrt{n}})E_{t}[z_{it}|\gamma_{n,b}]f_{t}(\gamma_{n,b}),
				\end{align}
				where $\gamma_{n,b}\in[\gamma_{0}+\frac{r_{\gamma}}{n^{1/4}},\gamma_{0}+\frac{b+r_{\gamma}}{n^{1/4}}]$.
				
				First, we can observe that \eqref{eq:bootdeter.limit.3} converges to zero uniformly with respect to $r_{\delta_{1}+\delta_{3}\gamma}$, $r_{\gamma}$, and $b$.
				This is because $\gamma_{n,b}\rightarrow\gamma_{0}$ uniformly with respect to $r_{\gamma}$ and $b$, which implies $E_{t}[z_{it}|\gamma_{0}+\tfrac{r_{\gamma}}{n^{\frac{1}{4}}}]f_{t}(\gamma_{0}+\tfrac{r_{\gamma}}{n^{\frac{1}{4}}})\rightarrow E_{t}[z_{it}|\gamma_{0}]f_{t}(\gamma_{0})$, while $\frac{r_{\delta_{1}+\delta_{3}\gamma}}{n^{1/4}}b\rightarrow 0$.
				
				Next, we check that \eqref{eq:bootdeter.limit.4} converges to zero uniformly with respect to $r_{\delta_{1}+\delta_{3}\gamma}$, $r_{\gamma}$, and $b$.
				By \cref{ass:dgp}, $\frac{d}{d\gamma}(E_{t}[z_{it}|\gamma]f_{t}(\gamma))$ is bounded and continuous on a neighborhood $\mathcal{O}$ of $\gamma_{0}$.
				As $\gamma_{n,b}\rightarrow \gamma_{0}$ uniformly with respect to $r_{\gamma}$ and $b$, $\frac{d}{d\gamma}(E_{t}[z_{it}|\gamma]f_{t}(\gamma))|_{\gamma=\gamma_{n,b}}\rightarrow \frac{d}{d\gamma}(E_{t}[z_{it}|\gamma]f_{t}(\gamma))|_{\gamma=\gamma_{0}}$ and $(\tfrac{r_{\delta_{1}+\delta_{3}\gamma}}{\sqrt{n}}+(\delta_{30}+\tfrac{r_{\delta_{3}}}{\sqrt{n}})(\gamma_{n,b}-\gamma_{0}-\tfrac{b}{n^{\frac{1}{4}}}))\rightarrow 0$, which implies the convergence of \eqref{eq:bootdeter.limit.4} to zero.
				
				Finally, we obtain the limit of \eqref{eq:bootdeter.limit.5}. 
				Since $E_{t}[z_{it}|\gamma_{n,b}]f_{t}(\gamma_{n,b})\rightarrow E_{t}[z_{it}|\gamma_{0}]f_{t}(\gamma_{0})$ and $\frac{r_{\delta_{3}}}{\sqrt{n}}\rightarrow 0$, \eqref{eq:bootdeter.limit.5} converges to $\frac{\delta_{30}}{2}E_{t}[z_{it}|\gamma_{0}]f_{t}(\gamma_{0})b^{2}$ uniformly with respect to $r\in\mathcal{R}$ and $b\in[-K,K]$.\\

				In conclusion,
				\begin{equation*}
					\widetilde{J}_{n}(r,b)\rightarrow
					\frac{\delta_{30}}{2}E_{t}[z_{it}|\gamma_{0}]f_{t}(\gamma_{0})b^{2}
				\end{equation*}
				uniformly with respect to $r\in\mathcal{R}$ and $b\in[-K,K]$, and hence
				\begin{equation*}
					\frac{1}{\sqrt{n}}\sum_{i=1}^{n}z_{it}^{*}(1,x_{it}^{*\prime})\delta_{0}^{*}1\{\gamma_{0}^{*}+\tfrac{b}{n^{\frac{1}{4}}}\geq q_{it}^{*}>\gamma_{0}^{*}\}\xrightarrow{p^{*}}\frac{\delta_{30}}{2}E_{t}[z_{it}|\gamma_{0}]f_{t}(\gamma_{0})b^{2} \quad\text{ in $P$}
				\end{equation*}
				uniformly with respect to $b\in[-K,K]$.
				Similarly, we can show that
				\begin{equation*}
					\frac{1}{\sqrt{n}}z_{it}^{*}(1,x_{it-1}^{*\prime})\delta_{0}^{*}1\{\gamma_{0}^{*}+\tfrac{b}{n^{\frac{1}{4}}}\geq q_{it-1}^{*}> \gamma_{0}^{*}\}
					\xrightarrow{p^{*}}
					\frac{\delta_{30}}{2}E_{t-1}[z_{it}|\gamma_{0}]f_{t-1}(\gamma_{0})b^{2}\quad\text{ in $P$}
				\end{equation*}
				uniformly with respect to $b\in [-K,K]$.
			\end{proof}

			\begin{lemma}
				\label{lem:boot plim2}
				Suppose that Assumptions \ref{ass:gmm}, \ref{ass:dgp}, and \ref{ass:locDis} hold, and the true model is discontinuous.
				If $\delta_{0}^{*}-\delta_{0}=O_{p}(n^{-1/2})$ and $\gamma_{0}^{*}-\gamma_{0}=O_{p}(n^{-1/2})$, then
				\begin{multline*}
					\frac{1}{\sqrt{n}}\sum_{i=1}^{n} z_{it}^{*}(1_{it}^{*}(\gamma_{0}^{*})'-1_{it}^{*}(\gamma_{0}^{*}+\tfrac{b}{\sqrt{n}})')X_{it}^{*}\delta_{0}^{*}
					\\
					\xrightarrow{p^{*}}
					\left\{E_{t}[z_{it}(1,x_{it}')\delta_{0}|\gamma_{0}]f_{t}(\gamma_{0})-E_{t-1}[z_{it}(1,x_{it-1}')\delta_{0}|\gamma_{0}]f_{t-1}(\gamma_{0})\right\}b,
				\end{multline*}
				in $P$ uniformly with respect to $b\in [-K,K]$ for any $K<\infty$.
			\end{lemma}
			The conditions for $\delta_{0}^{*}$ and $\gamma_{0}^{*}$ hold if (i) $\theta_{0}^{*}=(\hat{\alpha}(\gamma_{0})',\gamma_{0})'$ or (ii) $\theta_{0}^{*}$ is set as \eqref{eq:perc boot} under the assumptions of this lemma. Note that $\delta_{0}^{*}=w_{n}\hat{\delta}+(1-w_{n})\tilde{\delta}=\delta_{0}+O_{p}(n^{-1/2})$ since $w_{n}\xrightarrow{p} 1$, $\hat{\delta}=\delta_{0}+O_{p}(n^{-1/2})$, and $\tilde{\delta}=O_{p}(1)$.

			\begin{proof}
				By similar arguments used in the proof of \cref{lem:boot plim1}, we can derive that
				\begin{equation*}
					\sup_{\substack{b\in[-K,K],\\ \theta\in\Theta}}\left\|
					\frac{1}{\sqrt{n}}\sum_{i=1}^{n}z_{it}^{*}(1_{it}^{*}(\gamma)'-1_{it}^{*}(\gamma+\tfrac{b}{\sqrt{n}})')X_{it}^{*}\delta
					-
					\sqrt{n}E[z_{it}(1_{it}(\gamma)'-1_{it}(\gamma+\tfrac{b}{\sqrt{n}})')X_{it}\delta]
					\right\|
				\end{equation*}
				is $o_{p}^{*}(1)$ in $P$.
				
				Let $J_{n}(\delta,\gamma,b)=\sqrt{n}E[z_{it}(1_{it}(\gamma)'-1_{it}(\gamma+\tfrac{b}{\sqrt{n}})')X_{it}\delta]$.
				By assumption, we can reparametrize such that $\delta_{0}^{*}=\delta_{0}+\frac{r_{\delta}}{\sqrt{n}}$ and $\gamma_{0}^{*}=\gamma_{0}+\frac{r_{\gamma}}{\sqrt{n}}$.
				Then, we can reparametrize the function $J_{n}$ such that $\widetilde{J}_{n}(r_{\delta},r_{\gamma},b)=J_{n}(\delta_{0}+\frac{r_{\delta}}{\sqrt{n}},\gamma_{0}+\frac{r_{\gamma}}{\sqrt{n}},b)$.
				Let $r=(r_{\delta},r_{\gamma})$ which lies in a compact set $\mathcal{R}=\{r\in\mathbb{R}^{p+2}:\|r\|\leq \overline{K}\}$ for an aribtrary  $\overline{K}<\infty$.

				To prove the lemma, it will be shown that 
                $$\widetilde{J}_{n}(r_{\delta},r_{\gamma},b)\rightarrow\{E_{t}[z_{it}(1,x_{it}')\delta_{0}|\gamma_{0}]f_{t}(\gamma_{0})-E_{t-1}[z_{it}(1,x_{it-1}')\delta_{0}|\gamma_{0}]f_{t-1}(\gamma_{0})\}b$$ uniformly with respect to $r\in\mathcal{R}$ and $b\in[-K,K]$, which in turn implies     
                 \begin{multline*}
                    \frac{1}{\sqrt{n}}\sum_{i=1}^{n}z_{it}^{*}(1_{it}^{*}(\gamma_{0}^{*})'-1_{it}^{*}(\gamma_{0}^{*}+\tfrac{b}{\sqrt{n}})')X_{it}^{*}\delta_{0}^{*}\\
                    \xrightarrow{p^{*}}
                    \{E_{t}[z_{it}(1,x_{it}')\delta_{0}|\gamma_{0}]f_{t}(\gamma_{0})-E_{t-1}[z_{it}(1,x_{it-1}')\delta_{0}|\gamma_{0}]f_{t-1}(\gamma_{0})\}b\text{ in $P$}
                \end{multline*}
                uniformly with respect to $b\in[-K,K]$ since 
                \begin{equation*}
					\sup_{b\in[-K,K]}\left\|
					\frac{1}{\sqrt{n}}\sum_{i=1}^{n}z_{it}^{*}(1_{it}^{*}(\gamma_{0}^{*})'-1_{it}^{*}(\gamma_{0}^{*}+\tfrac{b}{\sqrt{n}})')X_{it}^{*}\delta_{0}^{*}
					-
					J_{n}(\delta_{0}^{*},\gamma_{0}^{*},b)
					\right\|=o_{p}^{*}(1)\text{ in $P$.}
			\end{equation*}
                
				Suppose $b>0$.
				The case for $b<0$ follows similarly.
				Then,
				\begin{multline*}
					\sqrt{n}E[z_{it}(1_{it}(\gamma)'-1_{it}(\gamma+\tfrac{b}{\sqrt{n}})')X_{it}\delta]
					=
					\sqrt{n}E[z_{it}(1,x_{it}')\delta1\{\gamma+\tfrac{b}{\sqrt{n}}\geq q_{it}> \gamma\}]
					\\
					-\sqrt{n}E[z_{it}(1,x_{it-1}')\delta1\{\gamma+\tfrac{b}{\sqrt{n}}\geq q_{it}> \gamma\}].
				\end{multline*}
				We focus on the first term of the right hand side $\sqrt{n}E[z_{it}(1,x_{it}')\delta1\{\gamma+\tfrac{b}{\sqrt{n}}\geq q_{it}> \gamma\}]$ as the limit of the second term can be derived identically, and redefine $J_{n}(\delta,\gamma,b)=\sqrt{n}E[z_{it}(1,x_{it}')\delta1\{\gamma+\tfrac{b}{\sqrt{n}}\geq q_{it}> \gamma\}]$ and $\widetilde{J}_{n}$, accordingly.
				
				We can derive the following Taylor expansion:
				\begin{equation*}
					\widetilde{J}_{n}(r,b)=\sqrt{n}E[z_{it}(1,x_{it}')(\delta_{0}+\tfrac{r_{\delta}}{\sqrt{n}})1\{\gamma_{0}+\tfrac{b+r_{\gamma}}{\sqrt{n}}> q_{it}\geq \gamma_{0}+\tfrac{r_{\gamma}}{\sqrt{n}}\}]
					=E_{t}[z_{it}(1,x_{it}')(\delta_{0}+\tfrac{r_{\delta}}{\sqrt{n}})|\gamma_{n,b}]f_{t}(\gamma_{n,b})b,
				\end{equation*}
				where $\gamma_{n,b}\in[\gamma_{0}+\frac{r_{\gamma}}{\sqrt{n}},\gamma_{0}+\frac{b+r_{\gamma}}{\sqrt{n}}]$.
				As $\gamma_{n,b}\rightarrow\gamma_{0}$ uniformly with respect to $r\in\mathcal{R}$ and $b\in[-K,K]$, $E_{t}[z_{it}(1,x_{it}')(\delta_{0}+\tfrac{r_{\delta}}{\sqrt{n}})|\gamma_{n,b}]f_{t}(\gamma_{n,b})b\rightarrow E_{t}[z_{it}(1,x_{it}')\delta_{0}|\gamma_{0}]f_{t}(\gamma_{0})b$ uniformly, and hence $\widetilde{J}_{n}(r,b) \rightarrow E_{t}[z_{it}(1,x_{it}')\delta_{0}|\gamma_{0}]f_{t}(\gamma_{0})b$ uniformly.
				
				In conclusion,
				\begin{equation*}
					\frac{1}{\sqrt{n}}\sum_{i=1}^{n}z_{it}^{*}(1,x_{it}^{*\prime})\delta_{0}^{*}1\{\gamma_{0}^{*}+\tfrac{b}{\sqrt{n}}\geq q_{it}^{*}> \gamma_{0}^{*}\}
					\xrightarrow{p^{*}}
					E_{t}[z_{it}(1,x_{it}')\delta_{0}|\gamma_{0}]f_{t}(\gamma_{0})b\quad\text{in $P$}
				\end{equation*}
				uniformly with respect to $b\in[-K,K]$.
				Similarly, we can show that
				\begin{equation*}
					\frac{1}{\sqrt{n}}\sum_{i=1}^{n}z_{it}^{*}(1,x_{it-1}^{*\prime})\delta_{0}^{*}1\{\gamma_{0}^{*}+\tfrac{b}{\sqrt{n}}\geq q_{it-1}^{*}> \gamma_{0}^{*}\}
					\xrightarrow{p^{*}}
					E_{t-1}[z_{it}(1,x_{it-1}')\delta_{0}|\gamma_{0}]f_{t-1}(\gamma_{0})b\quad\text{in $P$}
				\end{equation*}
				uniformly with respect to $b\in [-K,K]$.
			\end{proof}

			\section{Invalidity of standard nonparametric bootstrap}
			\label{sec:std boot invalid}
			
			In this section, we explain why the bootstrap estimators of the standard bootstrap does not have the asymptotic distribution in \cref{thm:asym} when the true model is continuous.
			Note that the bootstrap explained by \cref{alg:bootstrap} becomes the standard nonparametric bootstrap when $\theta_{0}^{*}=\hat{\theta}$.
			The consistency and convergence rate derivations in the proof of \cref{prop:boot.conv.rate} can still be followed, and hence $\sqrt{n}(\hat{\alpha}^{*}-\hat{\alpha})=O_{p}^{*}(1)$ and $\sqrt{n}(\hat{\gamma}^{*}-\hat{\gamma})^{2}=O_{p}^{*}(1)$ both in $P$.
			However, the conditions for \cref{lem:boot plim1} do not hold for the standard nonparametric bootstrap as $n^{1/4}(\hat{\delta}_{1}+\hat{\delta}_{3}\hat{\gamma})\neq o_{p}(1)$ as explained in \cref{sec:perc bootstrap}.
			Therefore, the rescaled versions of the criterion converges to a different limit. 
			Specifically,
			\begin{equation*}
				\sqrt{n}\bar{g}_{n}^{*}(\hat{\alpha}+\tfrac{a}{\sqrt{n}},\hat{\gamma}+\tfrac{b}{n^{1/4}})-n^{1/4}G(\hat{\theta})b\overset{*}{\rightsquigarrow} M_{0}a+Hb^{2}-e
			\end{equation*}
			in $\ell^{\infty}(\mathbb{K})$ in $P$ for every compcat $\mathbb{K}$ in the Euclidean space, where $G(\theta)$ is defined as \eqref{eq:boot firstorderderivative}.
			Recall that $n^{1/4}G(\hat{\theta})\neq o_{p}(1)$ as shown in \cref{sec:perc bootstrap}.
			The conditional weak convergence, $\overset{*}{\rightsquigarrow}$, in the last display comes from applying the following \cref{lem:boot plim3} in the place of \cref{lem:boot plim1} used in the proof of \cref{thm:perc boot}.

			\begin{lemma}
				\label{lem:boot plim3}
				Suppose that Assumptions \ref{ass:gmm}, \ref{ass:dgp}, \ref{ass:loc} are true and that the true model is continuous. Then,
				\begin{multline*}
					\frac{1}{\sqrt{n}}\sum_{i=1}^{n} z_{it}^{*}(1_{it}^{*}(\hat{\gamma})'-1_{it}^{*}(\hat{\gamma}+\tfrac{b}{n^{\frac{1}{4}}})')X_{it}^{*}\hat{\delta}
					-\left\{E_{t}[z_{it}|\gamma_{0}]f_{t}(\gamma_{0})-E_{t-1}[z_{it}|\gamma_{0}]f_{t-1}(\gamma_{0})\right\}n^{1/4}(\hat{\delta}_{1}+\hat{\delta}_{3}\hat{\gamma})b
					\\
					\xrightarrow{p^{*}}
					\frac{\delta_{30}}{2}\left\{E_{t}[z_{it}|\gamma_{0}]f_{t}(\gamma_{0})-E_{t-1}[z_{it}|\gamma_{0}]f_{t-1}(\gamma_{0})\right\}b^{2}
				\end{multline*}
				in $P$ uniformly with respect to $b\in [-K,K]$ for any $K<\infty$. 
			\end{lemma}

			\begin{proof}
				By similar arguments used in the proof of \cref{lem:boot plim1}, we can derive that
				\begin{equation*}
					\sup_{\substack{b\in[-K,K],\\ \theta\in\Theta}}\left\|
					\frac{1}{\sqrt{n}}\sum_{i=1}^{n}z_{it}^{*}(1_{it}^{*}(\gamma)'-1_{it}^{*}(\gamma+\tfrac{b}{n^{1/4}})')X_{it}^{*}\delta
					-
					\sqrt{n}E[z_{it}(1_{it}(\gamma)'-1_{it}(\gamma+\tfrac{b}{n^{1/4}})')X_{it}\delta]
					\right\|
				\end{equation*}
				is $o_{p}^{*}(1)$ in $P$.
				
				Suppose that $b>0$.
				The $b<0$ case can be analyzed similarly.
				Let $J_{n}(\delta,\gamma,b)=\sqrt{n}E[z_{it}(1,x_{it}')\delta1\{\gamma+\frac{b}{n^{1/4}}\geq q_{it}>\gamma\}]-n^{1/4}(\delta_{1}+\delta_{3}\gamma) E_{t}[z_{it}|\gamma_{0}]f_{t}(\gamma_{0})b$.
				Reparametrize such that $\hat{\gamma}=\gamma_{0}+\frac{r_{\gamma}}{n^{1/4}}$ and $\hat{\delta}=\delta_{0}+\frac{r_{\delta}}{\sqrt{n}}$.
				Let the set of $r=(r_{\delta},r_{\gamma})$ be $\mathcal{R}=\{r\in\mathbb{R}^{p+2}:\|r\|\leq \overline{K}\}$ for arbitrary $\overline{K}<\infty$. 
				Let $\widetilde{J}_{n}(r,b) = J_{n}(\delta_{0}+\frac{r_{\delta}}{\sqrt{n}},\gamma_{0}+\frac{r_{\gamma}}{n^{1/4}},b)$.
				
				We will show that $\widetilde{J}_{n}(r,b)\rightarrow \frac{\delta_{30}}{2}E_{t}[z_{it}|\gamma_{0}]f_{t}(\gamma_{0})b^{2}$ uniformly with respect to $r\in\mathcal{R}$ and $b\in[-K,K]$, which implies 
				\begin{equation*}
					\frac{1}{\sqrt{n}}\sum_{i=1}^{n}z_{it}^{*}(1,x_{it}^{*\prime})\hat{\delta}1\{\hat{\gamma}+\tfrac{b}{n^{1/4}}\geq q_{it}^{*}>\hat{\gamma}\}-n^{1/4}(\hat{\delta}_{1}+\hat{\delta}_{3}\hat{\gamma}) E_{t}[z_{it}|\gamma_{0}]f_{t}(\gamma_{0})b\xrightarrow{p^{*}}\frac{\delta_{30}}{2}E_{t}[z_{it}|\gamma_{0}]f_{t}(\gamma_{0})b^{2}
				\end{equation*} in $P$ uniformly with respect to $b\in[-K,K]$, because
                    \begin{multline*}
					\sup_{b\in[-K,K]}\left\|
					\frac{1}{\sqrt{n}}\sum_{i=1}^{n}z_{it}(1,x_{it}')\hat{\delta}1\{\hat{\gamma}+\tfrac{b}{n^{1/4}}\geq q_{it}>\hat{\gamma}\}-n^{1/4}(\hat{\delta}_{1}+\hat{\delta}_{3}\hat{\gamma}) E_{t}[z_{it}|\gamma_{0}]f_{t}(\gamma_{0})b
					-
					J_{n}(\hat{\delta},\hat{\gamma},b)
					\right\|\\
                    =o_{p}^{*}(1)\text{ in $P$.}
			      \end{multline*}
				
				Note that $J_{n}(\delta,\gamma,b)  =  J_{1n}(\delta,\gamma,b) + J_{2n}(\delta,\gamma,b)$ where
				\begin{align*}
					J_{1n}(\delta,\gamma,b) &
					=\sqrt{n}E[z_{it}\xi_{it}'\delta_{2}1\{\gamma+\tfrac{b}{n^{\frac{1}{4}}}\geq q_{it}>\gamma\}],\text{ and}\\
					J_{2n}(\delta,\gamma,b) &
					=\sqrt{n}E[z_{it}(\delta_{1}+\delta_{3}q_{it})1\{\gamma+\tfrac{b}{n^{\frac{1}{4}}}\geq q_{it}>\gamma\}]-n^{1/4}(\delta_{1}+\delta_{3}\gamma)E_{t}[z_{it}|\gamma_{0}]f_{t}(\gamma_{0})b.
				\end{align*}
				Let $\widetilde{J}_{1n}$ and $\widetilde{J}_{2n}$ denote the reparametrized version of $J_{1n}$ and $J_{2n}$, respectively.
				
				$\widetilde{J}_{1n}(r,b)$ converges to zero uniformly, for which we recall that it is identical to $\widetilde{J}_{1n}$ that appears in the proof of \cref{lem:boot plim1}.
				
				$\widetilde{J}_{2n}(r,b)=\widetilde{J}_{2an}(r,b)+\widetilde{J}_{2bn}(r,b)$ where
				\begin{align*}
					\widetilde{J}_{2an}(r,b) &
					=E[z_{it}(r_{\delta_{1}}+r_{\delta_{3}}q_{it})1\{\gamma_{0}+\tfrac{b+r_{\gamma}}{n^{\frac{1}{4}}}\geq q_{it}>\gamma_{0}+\tfrac{r_{\gamma}}{n^{\frac{1}{4}}}\}],\text{ and}\\
					\widetilde{J}_{2bn}(r,b) &
					=\sqrt{n}E[z_{it}(\delta_{10}+\delta_{30}q_{it})1\{\gamma_{0}+\tfrac{b+r_{\gamma}}{n^{\frac{1}{4}}}\geq q_{it}>\gamma_{0}+\tfrac{r_{\gamma}}{n^{\frac{1}{4}}}\}]\\
					&\quad -(\delta_{30}r_{\gamma}+\tfrac{r_{\delta_{1}}+r_{\delta_{3}}\gamma_{0}}{n^{1/4}}+\tfrac{r_{\delta_{3}}r_{\gamma}}{\sqrt{n}})E_{t}[z_{it}|\gamma_{0}]f_{t}(\gamma_{0})b.
				\end{align*}
				It can be easily checked that $\widetilde{J}_{2an}(r,b)$ converges to zero uniformly.
                It will be shown in the next paragraph that $\widetilde{J}_{2bn}(r,b)\rightarrow \frac{\delta_{30}}{2}E_{t}[z_{it}|\gamma_{0}]f_{t}(\gamma_{0})b^{2}$ uniformly, which implies $\widetilde{J}_{n}(r,b)\rightarrow \frac{\delta_{30}}{2}E_{t}[z_{it}|\gamma_{0}]f_{t}(\gamma_{0})b^{2}$ uniformly.
    
				By Taylor expansion,
				\begin{align}
                        \nonumber
                        & \widetilde{J}_{2bn}(r,b)
                        \\
					\nonumber
					& 
					=\sqrt{n}E[z_{it}(\delta_{10}+\delta_{30}q_{it})1\{\gamma_{0}+\tfrac{b+r_{\gamma}}{n^{\frac{1}{4}}}\geq q_{it}>\gamma_{0}+\tfrac{r_{\gamma}}{n^{\frac{1}{4}}}\}]-(\delta_{30}r_{\gamma}+\tfrac{r_{\delta_{1}}+r_{\delta_{3}}\gamma_{0}}{n^{1/4}}+\tfrac{r_{\delta_{3}}r_{\gamma}}{\sqrt{n}})E_{t}[z_{it}|\gamma_{0}]f_{t}(\gamma_{0})b \\
					\label{eq:bootdeter.limit3.1}
					&  =
					\delta_{30}r_{\gamma}E_{t}[z_{it}|\gamma_{0}+\tfrac{r_{\gamma}}{n^{1/4}}]f_{t}(\gamma_{0}+\tfrac{r_{\gamma}}{n^{1/4}})b-(\delta_{30}r_{\gamma}+\tfrac{r_{\delta_{1}}+r_{\delta_{3}}\gamma_{0}}{n^{1/4}}+\tfrac{r_{\delta_{3}}r_{\gamma}}{\sqrt{n}})E_{t}[z_{it}|\gamma_{0}]f_{t}(\gamma_{0})b \\
					\label{eq:bootdeter.limit3.2}
					& \quad +
					\frac{b^{2}}{2}\left((\delta_{10}+\delta_{30}\gamma_{n,b})\frac{d}{d\gamma}\{E_{t}[z_{it}|\gamma]f_{t}(\gamma)\}|_{\gamma=\gamma_{n,b}}+\delta_{30}E_{t}[z_{it}|\gamma_{n,b}]f_{t}(\gamma_{n,b})\right),
				\end{align}
				where $\gamma_{n,b}\in[\gamma_{0}+\frac{r_{\gamma}}{n^{1/4}},\gamma_{0}+\frac{b+r_{\gamma}}{n^{1/4}}]$.
				By continuity of $E_{t}[z_{it}|\gamma]f_{t}(\gamma)$ at $\gamma=\gamma_{0}$, \eqref{eq:bootdeter.limit3.1} converges to 0 uniformly with respect to $r\in\mathcal{R}$ and $b\in[-K,K]$.
				As $\gamma_{n,b}\rightarrow\gamma_{0}$ uniformly, we can derive that \eqref{eq:bootdeter.limit3.2} converges to $\frac{\delta_{30}}{2}E_{t}[z_{it}|\gamma_{0}]f_{t}(\gamma_{0})b^{2}$ uniformly.

				By similar manner, we can derive
				\begin{multline*}
					\frac{1}{\sqrt{n}}\sum_{i=1}^{n}z_{it}^{*}(1,x_{it-1}^{*\prime})\hat{\delta}1\{\hat{\gamma}+\tfrac{b}{n^{1/4}}\geq q_{it-1}^{*}>\hat{\gamma}\}-n^{1/4}(\hat{\delta}_{1}+\hat{\delta}_{3}\hat{\gamma}) E_{t-1}[z_{it}|\gamma_{0}]f_{t-1}(\gamma_{0})b\\
					\xrightarrow{p^{*}}\frac{\delta_{30}}{2}E_{t-1}[z_{it}|\gamma_{0}]f_{t-1}(\gamma_{0})b^{2}
				\end{multline*} 
				in $P$ uniformly with respect to $b\in[-K,K]$.
			\end{proof}

			\section{Symmetric percentile bootstrap confidence intervals for empirical application}
			\label{sec:empirical_diffboot}

In this section, we report the symmetric percentile residual-bootstrap confidence intervals for the coefficients for the empirical application. \cref{table:empirical_a} and \cref{table:empirical_a1} correspond to \cref{table:empirical} and \cref{table:empirical1} in \cref{sec:empirical}, respectively.

\begin{table}[htbp]
            \centering
            \caption{The 95\% symmetric percentile bootstrap confidence intervals that use the 0.95 quantile of $|\hat{\alpha}_{j}^{*}-\alpha_{j0}^{*}|$ are reported.
            Columns (a) and (b) report results of the models \eqref{eq:example1} and \eqref{eq:example2}, respectively. The percentile of each threshold location value is shown in parentheses below each value. The significance levels for the coefficients are given by stars: * - 10\%, ** - 5\% and *** - 1\%.}
            \label{table:empirical_a}
            \begin{tabular}{|llrr|llrr|}
            \hline
            \multicolumn{4}{|c|}{(a)} & \multicolumn{4}{c|}{(b)}\\
            \hline
            \hline
              & est. & \multicolumn{2}{c|}{[95\% CI]} &  & est. & \multicolumn{2}{c|}{[95\% CI]} \\ 
              \hline
              \multicolumn{4}{|l|}{\underline{Lower regime}} & \multicolumn{4}{|l|}{\underline{Lower regime}}\\
              $I_{t-1}$ & \;0.778** & 0.319 & 1.237 & $I_{t-1}$ & \;0.252 & -0.242 & 0.746 \\ 
		$CF_{t-1}$ & \;0.047 & -0.041 & 0.135 & $CF_{t-1}$ & \;0.266* & -0.004 & 0.535 \\ 
		$PPE_{t-1}$ & -0.147 & -0.428 & 0.134 & $PPE_{t-1}$ & \;0.027 & -0.175 & 0.229 \\ 
		$ROA_{t-1}$ & -0.032 & -0.128 & 0.065 & $ROA_{t-1}$ & -0.017 & -0.157 & 0.123 \\ 
		$LEV_{t-1}$ & \;0.231 & -1.219 & 1.682 & $TQ_{t-1}$ & \;0.246 & -0.071 & 0.564 \\ 
               \multicolumn{4}{|l|}{\underline{Upper regime}} & \multicolumn{4}{|l|}{\underline{Upper regime}}\\
            $I_{t-1}$ & -0.154 & -0.769 & 0.462 & $I_{t-1}$ & \;0.410** & 0.007 & 0.813 \\ 
		$CF_{t-1}$ & \;0.148* & -0.026 & 0.322 & $CF_{t-1}$ & \;0.081* & -0.023 & 0.184 \\ 
		$PPE_{t-1}$ & -0.291** & -0.566 & -0.015 & $PPE_{t-1}$ & \;0.044 & -0.251 & 0.340 \\ 
		$ROA_{t-1}$ & \;0.013 & -0.076 & 0.102 & $ROA_{t-1}$ & \;0.050 & -0.038 & 0.137 \\ 
		$LEV_{t-1}$ & -0.081 & -0.216 & 0.054 & $TQ_{t-1}$ & \;0.005 & -0.004 & 0.013 \\ 
               \multicolumn{4}{|l|}{\underline{Difference between regimes}} & \multicolumn{4}{|l|}{\underline{Difference between regimes}}\\
            intercept & \;0.068 & -0.045 & 0.181 & intercept & \;0.236 & -0.083 & 0.554 \\ 
		$I_{t-1}$ & -0.932** & -1.803 & -0.061 & $I_{t-1}$ & \;0.158 & -0.542 & 0.857 \\ 
		$CF_{t-1}$ & \;0.101 & -0.117 & 0.319 & $CF_{t-1}$ & -0.185 & -0.479 & 0.109 \\ 
		$PPE_{t-1}$ & -0.144 & -0.463 & 0.176 & $PPE_{t-1}$ & \;0.017 & -0.233 & 0.267 \\ 
		$ROA_{t-1}$ & \;0.045 & -0.129 & 0.218 & $ROA_{t-1}$ & \;0.066 & -0.128 & 0.261 \\ 
		$LEV_{t-1}$ & -0.312 & -1.754 & 1.130 & $TQ_{t-1}$ & -0.242 & -0.557 & 0.074 \\ 
		\hline
              \end{tabular}
        \end{table}

	\begin{table}[htbp]
            \centering
            \caption{The 95\% symmetric percentile bootstrap confidence intervals that use the 0.05 quantile of $|\hat{\alpha}_{j}^{*}-\alpha_{j0}^{*}|$ are reported.
            Results of the model \eqref{eq:example3} are reported. The percentile of each threshold location value is shown in parentheses below each value. The significance levels for the coefficients are given by stars: * - 10\%, ** - 5\% and *** - 1\%.}
            \label{table:empirical_a1}
            \begin{tabular}{|llrr|}
            \hline
              & est. & \multicolumn{2}{c|}{[95\% CI]} \\ 
              \hline
			\multicolumn{4}{|l|}{\underline{Coefficients}}  \\
             $I_{t-1}$ & \;0.392*** & 0.269 & 0.514 \\ 
		$CF_{t-1}$ & \;0.122*** & 0.087 & 0.156 \\ 
		$PPE_{t-1}$ & \;0.076 & -0.095 & 0.247 \\ 
		$ROA_{t-1}$ & \;0.027*** & 0.007 & 0.047 \\ 
		$TQ_{t-1}1\{TQ_{t-1}\leq\gamma\}$ & \;0.298** & 0.028 & 0.567 \\ 
		$TQ_{t-1}1\{TQ_{t-1}>\gamma\}$ & \;0.008** & 0.000 & 0.015 \\ 
			\multicolumn{4}{|l|}{\underline{Difference between regimes}}  \\
            intercept & 0.275** & 0.074 & 0.566 \\ 
		$TQ_{t-1}$ & -0.290** & -0.566 & -0.061 \\ 
		\hline
			
              \end{tabular}
        \end{table}

			\section{Bootstrap for linearity test}
			\label{sec:linearity bootstrap}
			
			We explain the bootstrap for linearity test based on sup-Wald statistic, explained in \cite{seo_dynamic_2016}.
			Null hypothesis of the test is $\delta=0_{p+1}$.
			The sup-Wald test statistic is 
			\begin{equation}
				\label{eq:sup-Wald}
				\sup_{\gamma\in\Gamma}\{n\hat{\delta}(\gamma)'[B'(\bar{M}_{n}(\gamma)'W_{n}(\gamma)\bar{M}_{n}(\gamma))^{-1}\bar{M}_{n}(\gamma)'W_{n}(\gamma)\hat{\Omega}(\gamma)W_{n}(\gamma)\bar{M}_{n}(\gamma)(\bar{M}_{n}(\gamma)'W_{n}(\gamma)\bar{M}_{n}(\gamma))^{-1}B]^{-1}\hat{\delta}(\gamma)\},
			\end{equation}
			where $B=\left[\begin{array}{c;{2pt/2pt}c}
				0_{(p+1)\times p} & I_{p+1}
			\end{array}\right]\in\mathbb{R}^{(p+1)\times(2p+1)}$, $W_{n}(\gamma)$ is the weight matrix obtained by the initial estimator with the restriction that the threshold location is $\gamma$, $\hat{\delta}(\gamma)$ is a subvector of the restricted estimator $\hat{\alpha}(\gamma)=(\hat{\beta}(\gamma)',\hat{\delta}(\gamma)')'$, and $\hat{\Omega}(\gamma)=(\frac{1}{n}\sum_{i=1}^{n}[g_{i}(\hat{\alpha}(\gamma),\gamma)g_{i}(\hat{\alpha}(\gamma),\gamma)']-[\frac{1}{n}\sum_{i=1}^{n}g_{i}(\hat{\alpha}(\gamma),\gamma)][\frac{1}{n}\sum_{i=1}^{n}g_{i}(\hat{\alpha}(\gamma),\gamma)]')$.
			
			The bootstrap for the linearity test can be implemented by setting
			\begin{equation*}
				\beta_{0}^{*}=\hat{\beta},\quad
				\delta_{0}^{*}=0_{p+1}
			\end{equation*}
			in \cref{alg:bootstrap}.
			Note that $\gamma_{0}^{*}$ does not matter in this case as $\delta_{0}^{*}=0_{p+1}$.
			The critical value for $\tau$-size test is obtained by using the $(1-\tau)$ quantile of the bootstrapped sup-Wald test statistics, defined analogously to \eqref{eq:sup-Wald}.

			\section{Uniform validity of the grid bootstrap}
			\label{sec:bootstrap_uniformity}
			
			In this section, we show the uniform validity of the grid bootstrap given in \cref{sec:grid bootstrap}. As discussed in \cref{sec:main.uniform.gridB}, the following simplified specification is analyzed for the clarity of exposition:
			\begin{equation*}
				y_{it}=x_{it}'\beta+(\delta_{1}+\delta_{3}q_{it})1\{q_{it}>\gamma\}+\eta_{i}+\epsilon_{it},\quad t=1,...,T,
			\end{equation*}
			where $\theta=(\alpha',\gamma)'=(\beta',\delta',\gamma)$, $\alpha=(\beta',\delta')'$, and $\delta=(\delta_{1},\delta_{3})'\in\mathbb{R}^{2}$. 	$x_{it}=(\xi_{it}',q_{it})'$ still includes the threshold variable. 
			The goal here is to show the uniform validity of the grid bootstrap near parameter values that make threshold models continuous. 
			Let $\Theta, \Gamma, g_{0}(\cdot), M_{0}, M_{10}, M_{20}(\gamma), M_{20}, \Omega, f_{t}(\cdot)$, and $E_{t}[\cdot|q]$ be defined as in \cref{sec:dptr}, while
			\begin{equation*}
				\widetilde{H}=\begin{pmatrix}
					E_{t_{0}}[z_{it_{0}}|\gamma_{0}]f_{t_{0}}(\gamma_{0})-E_{t_{0}-1}[z_{it_{0}}|\gamma_{0}]f_{t_{0}-1}(\gamma_{0})\\
					\vdots \\
					E_{T}[z_{iT}|\gamma_{0}]f_{T}(\gamma_{0})-E_{T-1}[z_{iT}|\gamma_{0}]f_{T-1}(\gamma_{0})
				\end{pmatrix}.
			\end{equation*}
			
			Let $\phi=(\theta,F)$ index the dgp while $F$ is an infinite dimensional index that determines the distribution of the random variables $\{\eta_{i},y_{i0},(z_{it},x_{it},\epsilon_{it})_{t=1}^{T}\}$ . 
			This section restricts $F$ to admit continuous density function.
			Let the space of the distributions be $\Phi_{F}$ which is compact and equipped with sup-norm over the space of density functions\footnote{
				That means $d(F_{1},F_{2})=\sup_{x\in\mathbb{R}^{d_{x}}}|f_{1}(x)-f_{2}(x)|$, where $f_{1}$ and $f_{2}$ are densities of the distribution functions $F_{1}$ and $F_{2}$, and $d_{x}$ is a dimension of the random vectors whose distributions are $F_{1}$ or $F_{2}$.
				It is a stronger norm than the sup-norm over the space of distribution functions as $\sup_{x\in\mathbb{R}^{d_{x}}}|f_{n}(x)-f_{0}(x)|\rightarrow 0$ implies $\sup_{x\in\mathbb{R}^{d_{x}}}|F_{n}(x)-F_{0}(x)|\rightarrow 0$.
			}, and the space of $\phi$ be $\Phi=\Theta\bigtimes\Phi_{F}$ which is compact since $\Theta$ and $\Phi_{F}$ are compact.
			
			Following the general framework explained in \cite{andrews_cheng_guggenberger_2020}, we consider a sequence of true parameters $\phi_{0n}=(\theta_{0n},F_{0n})=((\beta_{0n}',\delta_{10n},\delta_{30n},\gamma_{0n})',F_{0n})$.
			Let $\sigma_{\min}(A)$ and $\sigma_{\max}(A)$ be the square root of the minimum and maximum eigenvalues of $A'A$, respectively.
			Let the parameter space for $\phi_{0n}$ be 
			\begin{equation*}
				\begin{array}{rl}\Phi_{0} = \Bigl\{\phi_{0}\in\Phi: & 
                        (\delta_{10}+\delta_{30}\gamma_{0})^{2} + \delta_{30}^{2} \geq  c_{1}, \\
					&  c_{2} \leq  \sigma_{\min}(\Omega) \leq \sigma_{\max}(\Omega) \leq  c_{3},  \\
					& c_{4} \leq  E\|z_{it}\|^{4+r}\leq c_{5},\ c_{4}\leq E\|x_{it}\|^{4+r}\leq c_{5},\ c_{4}\leq E\|\epsilon_{it}\|^{4+r} \leq  c_{5}, \\
					& f_{t}(\cdot) \text{ is continuously differentiable at }[\gamma_{0}-c_{6},\gamma_{0}+c_{6}], \\
					& c_{7} \leq  \min_{q\in[\gamma_{0}-c_{6},\gamma_{0}+c_{6}]}f_{t}(q) \leq  \max_{q\in[\gamma_{0}-c_{6},\gamma_{0}+c_{6}]}f_{t}(q) \leq  c_{8},\\ 
					& \min_{q\in[\gamma_{0}-c_{6},\gamma_{0}+c_{6}]}|f_{t}'(q)| \leq  c_{9},\\
					& E_{t}[z_{it}|q]\text{ and }E_{t-1}[z_{it}|q]\text{ are continuously differentiable at }[\gamma_{0}-c_{10},\gamma_{0}+c_{10}], \\ 
					& \max_{q\in[\gamma_{0}-c_{10},\gamma_{0}+c_{10}]}\|E_{t}[z_{it}|q]\|\leq c_{11}, \\        
                        & \max_{q\in[\gamma_{0}-c_{10},\gamma_{0}+c_{10}]}\|E_{t-1}[z_{it}|q]\|\leq c_{11}, \\
					& \max_{q\in[\gamma_{0}-c_{10},\gamma_{0}+c_{10}]}\|\frac{d}{d\gamma}\left(E_{t}[z_{it}|\gamma]\right)_{\gamma=q}\| \leq  c_{11},\\ 
                        & \max_{q\in[\gamma_{0}-c_{10},\gamma_{0}+c_{10}]}\|\frac{d}{d\gamma}\left(E_{t-1}[z_{it}|\gamma]\right)_{\gamma=q}\| \leq  c_{11},\\ 
                       & c_{12} \leq  \sigma_{\min}\left(\textstyle{\left[\begin{array}{c;{2pt/2pt}c}M_{0} & \widetilde{H}\end{array}\right]}\right) \leq \sigma_{\max}\left(\textstyle{\left[\begin{array}{c;{2pt/2pt}c}M_{0} & \widetilde{H}\end{array}\right]}\right) \leq  c_{13} \\
                        &E_{t}[\|z_{it}\|^{1+r}|\gamma_{0}]\leq c_{14},\ E_{t-1}[\|z_{it}\|^{1+r}|\gamma_{0}]\leq c_{14},\hfill \text{ for }t=1,...,T\Bigr\},
				\end{array}
			\end{equation*}
			where $c_{1},...,c_{14},\text{ and }r$ are some positive constants.
			Note that $(\delta_{10}+\delta_{30}\gamma_{0})^{2} + \delta_{30}^{2} \geq c_{1}$ is to prevent $(\delta_{10n}+\delta_{30n}\gamma_{0n},\delta_{30n})'$ from (having a subsequence) converging to zero.\footnote{
				This implies that our threshold model has a strong threshold effect which excludes the diminishing or small threshold effect as in \cite{hansen_sample_2000}.
			} The remaining conditions for $\Phi_{0}$ other than $E_{t}[\|z_{it}\|^{1+r}|\gamma_{0}]\leq c_{14},\ E_{t-1}[\|z_{it}\|^{1+r}|\gamma_{0}]\leq c_{14}$ imply that Assumptions \ref{ass:dgp}, \ref{ass:gmm}, and \ref{ass:loc}/\ref{ass:locDis} hold uniformly.
                The condition $E_{t}[\|z_{it}\|^{1+r}|\gamma_{0}]\leq c_{14},\ E_{t-1}[\|z_{it}\|^{1+r}|\gamma_{0}]\leq c_{14}$ is a uniform integrability condition for the distribution of $z_{it}$ conditional on $q_{it}$ or $q_{it-1}$. Its role will be explained after introducing the drifting sequence framework.

			Because of the nonlinearity and discontinuity of our dynamic model, it is not trivial to answer what primitive conditions for the parameter and distributions of random variables, such as initial value $y_{i0}$ or individual fixed effect $\eta_{i}$, are sufficient for $\Phi_{0}$.
			This paper does not investigate this issue so that we can focus on uniformity analysis with respect to degeneracy of the Jacobian of nonlinear GMM.
			
			For $n=1,2,...$, let $\{\eta_{in},y_{i0n},(z_{itn},x_{itn},\epsilon_{itn})_{t=1}^{T}\}$ be drawn from distribution $F_{0n}$.
			For a function or random variable $u$, e.g., $u=z,x$ or $\Delta\epsilon$, we often write $u_{it,n}$ and $u_{it-1,n}$ to indicate more explicitly that indices in subscript are $((i,t),n)$ or $((i,t-1),n)$, while $n$ is the new index introduced in this section.
                Suppose that
			\begin{align*}
				y_{itn} =& x_{itn}'\beta_{0n}+(\delta_{10n}+\delta_{30n}q_{itn})1\{q_{itn}>\gamma_{0n}\}+\eta_{in}+\epsilon_{itn}, \text{ for }t=1,...,T, \\
				& E[z_{itn}\Delta\epsilon_{itn}]=0,\quad \text{where }\Delta \epsilon_{itn}= \epsilon_{it,n}-\epsilon_{it-1,n}.
			\end{align*}
			As in \cref{sec:dptr}, we define
			\begin{equation*}
				M_{1in} = -\begin{bmatrix}
					z_{it_{0}n}\Delta x_{it_{0}n}' \\
					\vdots \\
					z_{iTn}\Delta x_{iTn}'
				\end{bmatrix}\in\mathbb{R}^{k\times p},
				\quad
				M_{2in}(\gamma) = -\begin{bmatrix}
					z_{it_{0}n}1_{it_{0}n}(\gamma)'X_{it_{0}n} \\
					\vdots \\
					z_{iTn}1_{iTn}(\gamma)'X_{iTn}
				\end{bmatrix}\in\mathbb{R}^{k\times 2},
			\end{equation*}
			where $\Delta y_{itn}= y_{it,n}-y_{it-1,n}$, $\Delta x_{itn}= x_{it,n}-x_{it-1,n}$,
			\begin{equation*}
				X_{itn} = \begin{pmatrix}
					(1,q_{it,n})\\
					(1,q_{it-1,n})
				\end{pmatrix},\text{ and}
				\quad
				1_{itn}(\gamma) =\begin{pmatrix}
					1\{q_{it,n}>\gamma\}\\
					-1\{q_{it-1,n}>\gamma\}
				\end{pmatrix}.
			\end{equation*}
			Let $M_{in}(\gamma)=\left[\begin{array}{c;{2pt/2pt}c}
				M_{1in} & M_{2in}(\gamma)
			\end{array}\right]$, and $M_{0n}(\gamma)=E[M_{in}(\gamma)]$, $M_{10n}=E[M_{1in}]$, $M_{20n}(\gamma)=E[M_{2in}(\gamma)]$, $\bar{M}_{n}(\gamma)=\frac{1}{n}\sum_{i=1}^{n}M_{in}(\gamma)$, $\bar{M}_{1n}=\frac{1}{n}\sum_{i=1}^{n}M_{1in}$, and $\bar{M}_{2n}(\gamma)=\frac{1}{n}\sum_{i=1}^{n}M_{2in}(\gamma)$. We write $M_{0n}$, $M_{20n}$ and $\bar{M}_{n}$ instead of $M_{0n}(\gamma_{0n})$, $M_{20n}(\gamma_{0n})$ and $\bar{M}_{n}(\gamma_{0n})$.
			Define
			\begin{equation*}
				\widetilde{H}_{n}=
				\begin{pmatrix}
					E_{t_{0}n}[z_{it_{0}n}|\gamma_{0n}]f_{t_{0}n}(\gamma_{0n})-E_{t_{0}-1,n}[z_{it_{0},n}|\gamma_{0n}]f_{t_{0}-1,n}(\gamma_{0n})\\
					\vdots \\
					E_{Tn}[z_{iTn}|\gamma_{0n}]f_{Tn}(\gamma_{0n})-E_{T-1,n}[z_{iT,n}|\gamma_{0n}]f_{T-1,n}(\gamma_{0n})
				\end{pmatrix},
			\end{equation*}
			where $E_{tn}[\cdot|q]$ and $f_{tn}(\cdot)$ are the conditional expectation $E[\cdot|q_{itn}=q]$ and the density of $q_{itn}$, respectively.
			
			Suppose that a sequence $\{\phi_{0n}\}$ (or its subsequence $\{\phi_{0p_{n}}\}$) converges so that $\theta_{0n}\rightarrow\theta_{0,\infty}=(\alpha_{0,\infty}',\gamma_{0,\infty})'=(\beta_{0,\infty}',\delta_{10,\infty},\delta_{30,\infty},\gamma_{0,\infty})'$ and $F_{0n}\rightarrow F_{0,\infty}$, i.e., $\phi_{0n}\text{ (or $\phi_{0p_{n}}$)}\rightarrow\phi_{0,\infty}$.
			Note that the density of the distribution $F_{0n}$ converges to the density of $F_{0,\infty}$  uniformly by our choice of norm in $\Phi_{F}$, and $\sup_{\upsilon}\|F_{0n}(\upsilon)-F_{0,\infty}(\upsilon)\|\rightarrow 0$.
			
			Note that $M_{0,\infty}(\gamma) = E[M_{i,\infty}(\gamma)]=\lim_{n\rightarrow\infty}M_{0n}(\gamma)$ as each element of $M_{in}(\gamma)$ is uniformly integrable by $\max\{E\|z_{itn}\|^{4+r},E\|x_{itn}\|^{4+r},E\|\epsilon_{itn}\|^{4+r}\}\leq c_{5}<\infty$ for all $n$ while $F_{0n}$ converges to $F_{0,\infty}$.
			Hence, $M_{10,\infty} = E[M_{1i,\infty}] = \lim_{n\rightarrow\infty}M_{10n}$ and $M_{20,\infty}(\gamma) = E[M_{2i,\infty}(\gamma)]=\lim_{n\rightarrow\infty}M_{20n}(\gamma)$ also hold.
			Furthermore, $\widetilde{H}_{\infty}=\lim_{n\rightarrow\infty}\widetilde{H}_{n}$, where
                \begin{equation*}
				\widetilde{H}_{\infty}=
				\begin{pmatrix}
					E_{t_{0},\infty}[z_{it_{0},\infty}|\gamma_{0,\infty}]f_{t_{0},\infty}(\gamma_{0,\infty})-E_{t_{0},\infty}[z_{it_{0}-1,\infty}|\gamma_{0,\infty}]f_{t_{0}-1,\infty}(\gamma_{0,\infty})\\
					\vdots \\
					E_{T,\infty}[z_{iT,\infty}|\gamma_{0,\infty}]f_{T,\infty}(\gamma_{0,\infty})-E_{T,\infty}[z_{iT-1,\infty}|\gamma_{0,\infty}]f_{T-1,\infty}(\gamma_{0,\infty})
				\end{pmatrix}.
			\end{equation*}
                This is because $f_{tn}\rightarrow f_{t,\infty}$ uniformly by our definition of norm in $\Phi_{F}$, and it is straightforward to derive $z_{itn}|q_{isn}=\gamma_{0n}\xrightarrow{d} z_{it,\infty}|q_{is,\infty}=\gamma_{0,\infty}$ for $s=t,t-1$, which implies $E_{s}[z_{itn}|\gamma_{0n}]\rightarrow E_{s}[z_{it,\infty}|\gamma_{0,\infty}]$ due to the uniform integrability $E_{s}[\|z_{it}\|^{1+r}|\gamma_{0}]\leq c_{14}$ for $s=t,t-1$.
			Furthermore, $\|M_{0n}-M_{0,\infty}\|\rightarrow 0$ as $n\rightarrow\infty$ because $\|M_{0n}(\gamma_{0,\infty})-M_{0,\infty}(\gamma_{0,\infty})\|\rightarrow 0$,
			and $\|M_{0n}-M_{0n}(\gamma_{0,\infty})\|=\|M_{20n}-M_{20n}(\gamma_{0,\infty})\|\leq \|\mathfrak{H}_{n}(\bar{\gamma}_{n})\|(\gamma_{0n}-\gamma_{0,\infty})$, where 
			\begin{equation*}
				\mathfrak{H}_{n}(\gamma)=\begin{pmatrix}
					E_{t_{0}n}[z_{it_{0}n}(1,\gamma)|\gamma]f_{t_{0}n}(\gamma) - E_{t_{0}-1,n}[z_{it_{0}n}(1,\gamma)|\gamma]f_{t_{0}-1,n}(\gamma)\\
					\vdots\\
					E_{Tn}[z_{iTn}(1,\gamma)|\gamma]f_{Tn}(\gamma) - E_{T-1,n}[z_{iTn}(1,\gamma)|\gamma]f_{T-1,n}(\gamma)
				\end{pmatrix},
			\end{equation*}
			and $\bar{\gamma}_{n}$ is between $\gamma_{0n}$ and $\gamma_{0,\infty}$.
			Note that $\|\mathfrak{H}_{n}(\bar{\gamma}_{n})\|<C$ for some nonnegative $C<\infty$ for sufficiently large $n$ as $(\theta_{0n},F_{0n})\in\Phi_{0}$.

                Let $\omega_{in}=\{(z_{itn},y_{itn},x_{itn},\epsilon_{itn})_{t=1}^{T}\}$ and $g(\omega_{in},\theta) = (g_{t_{0}}(\omega_{in},\theta)',\dots,g_{T}(\omega_{in},\theta)')'$, where $g_{t}(\omega_{in},\theta)=z_{itn}(\Delta y_{itn}-\Delta x_{itn}'\beta-1_{itn}(\gamma)'X_{itn}\delta)$.
			Let $\Omega_{n}=E[g(\omega_{in},\theta_{0n})g(\omega_{in},\theta_{0n})']$, and $\Omega_{\infty}=E[g(\omega_{i,\infty},\theta_{0,\infty})g(\omega_{i,\infty},\theta_{0,\infty})']=\lim_{n\rightarrow\infty}\Omega_{n}$.
			Let $\bar{g}_{n}(\theta)=\frac{1}{n}\sum_{i=1}^{n}g(\omega_{in},\theta)$, $\hat{Q}_{n}(\theta)=\bar{g}_{n}(\theta)'W_{n}\bar{g}_{n}(\theta)$, and $g_{0n}(\theta)=E[g(\omega_{in},\theta)]$, while $W_{n}=\{\frac{1}{n}\sum_{i=1}^{n}[g(\omega_{in},\hat{\theta}_{(1)n})g(\omega_{in},\hat{\theta}_{(1)n})']-\bar{g}_{n}(\hat{\theta}_{(1)n})\bar{g}_{n}(\hat{\theta}_{(1)n})'\}^{-1}$ and $\hat{\theta}_{(1)n}=\arg\min_{\theta}\bar{g}_{n}(\theta)'\bar{g}_{n}(\theta)$ is the initial estimator.
			$\hat{\theta}_{n}=(\hat{\alpha}_{n}',\hat{\gamma}_{n})'=\arg\min_{\theta}\hat{Q}_{n}(\theta)$ and $\mathcal{D}_{n}(\gamma)=n(\min_{\alpha\in A}\hat{Q}_{n}(\alpha,\gamma)-\hat{Q}_{n}(\hat{\theta}_{n}))$.
			
		Let $\omega_{in}^{*}$ be an i.i.d. draw along the index $i$ from $\{\omega_{in}:i=1,...,n\}$.
			Let 
			\begin{align}
				\nonumber
				g_{in}^{*}(\theta) & = (g_{it_{0}n}^{*}(\theta)',...,g_{iTn}^{*}(\theta)')' \\
				\label{eq:unif.boot.momt}
				g_{itn}^{*}(\theta) & = g_{t}(\omega_{in}^{*},\theta) - g_{t}(\omega_{in}^{*},\theta_{0n}^{*}) + g_{t}(\omega_{in}^{*}, \hat{\theta}_{n}) \\
				\nonumber
				& = -z_{itn}^{*}\Delta x_{itn}^{*\prime}(\beta-\beta_{0n}^{*}) -z_{itn}^{*}1_{itn}^{*}(\gamma)'X_{itn}^{*}(\delta-\delta_{0n}^{*}) \\
				\nonumber
				& \quad +z_{itn}^{*}(1_{itn}^{*}(\gamma_{0n}^{*})'-1_{itn}^{*}(\gamma)')X_{itn}^{*}\delta_{0n}^{*} + z_{itn}^{*}\widehat{\Delta\epsilon}_{itn}^{*},
			\end{align}
			where $\theta_{0}^{*}=(\hat{\alpha}_{n}(\gamma_{0n})',\gamma_{0n})'$ and $\hat{\alpha}_{n}(\gamma)=\arg\min_{\alpha}\hat{Q}_{n}(\alpha,\gamma)$.
                For the justification of the representation \eqref{eq:unif.boot.momt}, please refer to \eqref{eq:boot momt} and description in \cref{sec: online.app.boot1} .
			Note that $\bar{g}_{n}^{*}(\theta) = \frac{1}{n}\sum_{i=1}^{n}[g_{in}^{*}(\theta)-\bar{g}_{n}(\hat{\theta}_{n})]$ becomes the bootstrap sample moment from the grid bootstrap.
			Then, let $\hat{Q}_{n}^{*}(\theta)=\bar{g}_{n}^{*}(\theta)'W_{n}^{*}\bar{g}_{n}^{*}(\theta)$, $W_{n}^{*}=[\frac{1}{n}\sum_{i=1}^{n}\{g_{in}^{*}(\hat{\theta}_{(1)n}^{*})g_{in}^{*}(\hat{\theta}_{(1)n}^{*})'\}-\{\frac{1}{n}\sum_{i=1}^{n}g_{in}^{*}(\hat{\theta}_{(1)n}^{*})\}\{\frac{1}{n}\sum_{i=1}^{n}g_{in}^{*}(\hat{\theta}_{(1)n}^{*})\}']^{-1}$, $\hat{\theta}_{(1)n}^{*}=\arg\min_{\theta}\bar{g}_{n}^{*}(\theta)'\bar{g}_{n}^{*}(\theta)$, $\hat{\theta}_{n}^{*}=\arg\min_{\theta}\hat{Q}_{n}^{*}(\theta)$, and $\mathcal{D}_{n}^{*}(\gamma)=n(\min_{\alpha}\hat{Q}_{n}^{*}(\alpha,\gamma)-\hat{Q}_{n}^{*}(\hat{\theta}_{n}^{*}))$.
			Recall that in \cref{sec:grid bootstrap}
			the $100(1-\tau)$\% grid bootstrap confidence set was defined as		
			\begin{equation*}
				CI_{n,1-\tau}^{grid} = \{\gamma\in\Gamma: \mathcal{D}_{n}(\gamma) \leq \widehat{F}^{*-1}_{n}(1-\tau;\mathcal{D}_{n}^{*}(\gamma))\}.
			\end{equation*}
			
			Define a mapping $\pi_{n}:\Phi_{0}\rightarrow\Pi$, where $\Pi=[-\infty,\infty]\bigtimes\mathbb{R}\bigtimes\Phi_{0}$ such that 
			\begin{equation*}
				\pi_{n}(\phi)=\begin{pmatrix}
					n^{1/4}(\delta_{1}+\delta_{3}\gamma)
					\\
					(\delta_{1}+\delta_{3}\gamma)
					\\
					\phi
				\end{pmatrix}.
			\end{equation*} 
			This is because the limits of $n^{1/4}(\delta_{1}+\delta_{3}\gamma)$ and $(\delta_{1}+\delta_{3}\gamma)$ characterize the asymptotic behaviors of the test statistic used in the grid bootstrap. 
			
			\begin{theorem*}{I.1}
				\label{thm:uniform-sufficient}
				For any subsequence $\{p_{n}\}$ of $\{n:n\in\mathbb{N}\}$ and any sequence $\{\phi_{0p_{n}}\in\Phi_{0}:n\geq 1 \}$  s.t. $\pi_{p_{n}}(\phi_{0p_{n}})\rightarrow(\zeta_{1},\zeta_{2},\phi_{0,\infty})\in\Pi$,  $$P_{\phi_{0p_{n}}}(\gamma_{0p_{n}}\in CI_{p_{n},1-\tau}^{grid})\rightarrow 1-\tau,$$ where $P_{\phi_{0p_{n}}}(\cdot)$ is the probability law under $\phi_{0p_{n}}=(\theta_{0p_{n}},F_{0p_{n}})$.
                Moreover,
                $$\liminf_{n\rightarrow\infty}\inf_{\phi_{0}\in\Phi_{0}}P_{\phi_{0}}(\gamma_{0}\in CI_{n,1-\tau}^{grid})=\limsup_{n\rightarrow\infty}\sup_{\phi_{0}\in\Phi_{0}}P_{\phi_{0}}(\gamma_{0}\in CI_{n,1-\tau}^{grid})=1-\tau,$$ 
                which establishes the uniform validity of the grid bootstrap confidence interval.
			\end{theorem*}
			
			Note that the last statement of \cref{thm:uniform-sufficient} follows from the theorem's preceding statement, as the latter verifies Assumption B* from \cite{andrews_cheng_guggenberger_2020}.
			Let $\{\pm\infty\}=\{-\infty,+\infty\}$. To show \cref{thm:uniform-sufficient}, we consider  the following four cases:
			\begin{itemize}[label=]
				\setlength\itemsep{0mm}
				\item (i) continuous: $\zeta_{1}=0$ and $\zeta_{2}=0$.
				\item (ii) semi-continuous: $\zeta_{1}\in\mathbb{R}\setminus\{0\}$ and $\zeta_{2}=0$.
				\item (iii) semi-discontinuous: $\zeta_{1}\in\{\pm\infty\}$ and $\zeta_{2}=0$.
				\item (vi) discontinuous: $\zeta_{1}\in\{\pm\infty\}$ and $\zeta_{2}\neq0$.
			\end{itemize}
			The following lemma implies \cref{thm:uniform-sufficient}.
			
			\begin{lemma}
				\label{lem:uniform-gridboot}
				For all sequences $\{\phi_{0p_{n}}\in\Phi_{0}:n\geq 1 \}$ for which $\pi_{p_{n}}(\phi_{0p_{n}})\rightarrow(\zeta_{1},\zeta_{2},\phi_{0,\infty})\in\Pi$, the following convergences hold ($P$ in ``$\xrightarrow{d^{*}}$ in $P$'' denotes the probability of $\{\omega_{ip_{n}}:1\leq i \leq p_{n}, n=1,2,...\}$): 
				
				\noindent
				\hspace{0mm}(i) For continuous case, $\mathcal{D}_{p_{n}}(\gamma_{0p_{n}})\xrightarrow{d} Z_{0}^{2}$, and $\mathcal{D}_{p_{n}}^{*}(\gamma_{0p_{n}})\xrightarrow{d^{*}} Z_{0}^{2}$ in $P$, where $Z_{0}=\max\{Z_{0}^{*},0\}$ and $Z_{0}^{*}\sim N(0,1)$.\\
				\hspace*{0mm}(ii) For semi-continuous case, $\mathcal{D}_{p_{n}}(\gamma_{0p_{n}})\xrightarrow{d} \mathcal{D}_{\infty}$, and $\mathcal{D}_{p_{n}}^{*}(\gamma_{0p_{n}})\xrightarrow{d^{*}} \mathcal{D}_{\infty}$ in $P$, where
				\begin{equation*}
					\mathcal{D}_{\infty}=
					\begin{cases}
						(\frac{U}{\sqrt{\widetilde{H}_{\infty}'\Xi_{\infty}\widetilde{H}_{\infty}}})^{2} & \text{if }U\geq\frac{-\zeta_{1}^{2}}{2|\delta_{30,\infty}|}\widetilde{H}_{\infty}'\Xi_{\infty}\widetilde{H}_{\infty} \\
						-(\frac{-\zeta_{1}^{2}}{2|\delta_{30,\infty}|})^{2}\widetilde{H}_{\infty}'\Xi_{\infty}\widetilde{H}_{\infty}+2\frac{-\zeta_{1}^{2}}{2|\delta_{30,\infty}|}U&
						\text{if }U<\frac{-\zeta_{1}^{2}}{2|\delta_{30,\infty}|}\widetilde{H}_{\infty}'\Xi_{\infty}\widetilde{H}_{\infty} 
					\end{cases},
				\end{equation*}
				$U\sim N(0,\widetilde{H}_{\infty}'\Xi_{\infty} \widetilde{H}_{\infty})$, and $\Xi_{\infty}=\Omega_{\infty}^{-1}-\Omega_{\infty}^{-1}M_{0,\infty}(M_{0,\infty}'\Omega_{\infty}^{-1} M_{0,\infty})M_{0,\infty}'\Omega_{\infty}^{-1}$.\\
			\noindent	(iii) For semi-discontinuous and discontinuous cases, $\mathcal{D}_{p_{n}}(\gamma_{0p_{n}})\xrightarrow{d} \chi^{2}_{1}$, and $\mathcal{D}_{p_{n}}^{*}(\gamma_{0p_{n}})\xrightarrow{d^{*}} \chi^{2}_{1}$ in $P$.
				
			\end{lemma}

			\begin{remark}
				Note that the distribution of $\mathcal{D}_{\infty}$ is (first-order) stochastically dominated by the $\chi^{2}_{1}$ distribution.
				This is because $f_{1}(Z_{0}):= (\frac{Z_{0}}{\sqrt{\widetilde{H}_{\infty}'\Xi_{\infty}\widetilde{H}_{\infty}}})^{2}= -(\frac{-\zeta_{1}^{2}}{2|\delta_{30,\infty}|})^{2}\widetilde{H}_{\infty}'\Xi_{\infty}\widetilde{H}_{\infty}+2\frac{-\zeta_{1}^{2}}{2|\delta_{30,\infty}|}Z_{0} =:f_{2}(Z_{0})$ when $Z_{0}=\frac{-\zeta_{1}^{2}}{2|\delta_{30,\infty}|}\widetilde{H}_{\infty}'\Xi_{\infty}\widetilde{H}_{\infty}<0$, and $f_{1}'(Z_{0}) < f_{2}'(Z_{0})$ when $Z_{0}<\frac{-\zeta_{1}^{2}}{2|\delta_{30,\infty}|}\widetilde{H}_{\infty}'\Xi_{\infty}\widetilde{H}_{\infty}$, which implies $f_{1}(Z_{0})>f_{2}(Z_{0})$ for $Z_{0}<\frac{-\zeta_{1}^{2}}{2|\delta_{30,\infty}|}\widetilde{H}_{\infty}'\Xi_{\infty}\widetilde{H}_{\infty}$.
			\end{remark}
			
			\begin{proof}[Proof of \cref{lem:uniform-gridboot}]
				We prove the result for sequence $\{n\}$ rather than $\{p_{n}\}$ to ease notation.
				Then, we can replace $\{n\}$ by $\{p_{n}\}$ to complete the proof.
				
				First, we derive the consistency, convergence rates, and asymptotic distributions of $\hat{\theta}_{n}$, and then we derive the asymptotic distributions of $\mathcal{D}_{n}(\gamma_{0n})$, depending on the regimes determined by $\zeta_{1}$ and $\zeta_{2}$.
				Then, the same results are derived for bootstrap estimator and test statistic for each case. 
				
				\paragraph{Consistency of estimator}
				Define $\hat{\alpha}_{n}(\gamma)=\arg\min_{\alpha\in A}\hat{Q}_{n}(\alpha,\gamma)$, which is
				\begin{gather*}
					\hat{\alpha}_{n}(\gamma)= -(\bar{M}_{n}(\gamma)'W_{n}\bar{M}_{n}(\gamma))^{-1}\bar{M}_{n}(\gamma)'W_{n}\bar{v}_{n} \\
					\bar{v}_{n} = -\bar{M}_{n}\alpha_{0n}+u_{n}, \quad
					u_{n}=\frac{1}{n}\sum_{i=1}^{n}\begin{pmatrix}
						z_{it_{0}n}\Delta\epsilon_{it_{0}n}\\
						\vdots \\
						z_{iTn}\Delta\epsilon_{iTn}
					\end{pmatrix}.
				\end{gather*}
				Therefore, $\hat{\alpha}_{n}(\gamma) = -(\bar{M}_{n}(\gamma)'W_{n}\bar{M}_{n}(\gamma))^{-1}\bar{M}_{n}(\gamma)'W_{n}(-\bar{M}_{n}\alpha_{0n}+u_{n})$.
				
				Note that $u_{n}\xrightarrow{p}0$ by the WLLN for triangular array which holds as $\sup_{n\in\mathbb{N}}E\|z_{itn}\Delta\epsilon_{itn}\|^{2}\leq \sup_{n\in\mathbb{N}}(E\|z_{itn}\|^{4})^{1/2}(E\|\Delta\epsilon_{itn}\|^{4})^{1/2} < \infty$.
				Furthermore, $\sup_{\gamma\in\Gamma}\|\bar{M}_{n}(\gamma)-M_{0n}(\gamma)\|\xrightarrow{p}0$ by \cref{lem:unif.ULLN}.
				Thus, $\sup_{\gamma\in\Gamma}\|\hat{\alpha}_{n}(\gamma)-(M_{0n}(\gamma)'W M_{0n}(\gamma))^{-1}M_{0n}(\gamma)'WM_{0n}\alpha_{0n}\|\xrightarrow{p}0$ so that $\|\hat{\alpha}_{n}(\hat{\gamma}_{n})-\alpha_{0n}\|\xrightarrow{p}0$ if $\hat{\gamma}_{n}=\arg\min_{\gamma\in\Gamma}\tilde{Q}_{n}(\gamma)$, where $\tilde{Q}_{n}(\gamma)=\hat{Q}_{n}(\hat{\alpha}_{n}(\gamma),\gamma)$, is consistent such that $|\hat{\gamma}_{n}-\gamma_{0n}|\xrightarrow{p}0$.
				
				 If $\hat{\theta}_{(1)n}$ is consistent, then $\|W_{n}-\Omega_{n}^{-1}\|\rightarrow 0$ by \cref{lem:unif.Wn}.
				Then,
				\begin{equation*}
					\sup_{\gamma\in\Gamma}\left|\tilde{Q}_{n}(\gamma)-\|(I-P_{\Omega_{n}^{-1/2}M_{0n}(\gamma)})(\Omega_{n}^{-1/2}M_{0n}\alpha_{0n})\|^{2}\right|\rightarrow 0.
				\end{equation*}
                    Since $\sigma_{\min}\left(\left[\begin{array}{c;{2pt/2pt}c}
                        M_{20n} & \widetilde{H}_{n}
                    \end{array}\right]\right)\geq c_{12}$ for all $n$, $M_{20n}\delta_{0n}$ is not in the column space of $M_{20n}(\gamma)$, and $\gamma_{0n}$ is the unique minimizer of $\|(I-P_{\Omega_{n}^{-1/2}M_{0n}(\gamma)})(\Omega_{n}^{-1/2}M_{0n}\alpha_{0n})\|$.
				By applying the argmin CMT as in the proof of \cref{thm:asym}, $|\hat{\gamma}_{n}-\gamma_{0n}|\xrightarrow{p} 0$ can be derived.
                    Derivation of the consistency of $\hat{\theta}_{(1)n}$ is straightforward if we replace $\Omega_{n}^{-1/2}$ by the identity matrix.
				
				\paragraph{Convergence rate of estimator}
				
				By \cref{lem:unif.equicont} and $\|\hat{\theta}_{n}-\theta_{0n}\|\xrightarrow{p}0$, $\sqrt{n}\|\bar{g}_{n}(\hat{\theta}_{n})-\bar{g}_{n}(\theta_{0n})-g_{0n}(\hat{\theta}_{n})\|=o_{p}(1)$.
				As $\|W_{n}-\Omega_{n}^{-1}\|\xrightarrow{p}0$,
				\[\sqrt{n}\|W_{n}^{1/2}\bar{g}_{n}(\hat{\theta}_{n})-W_{n}^{1/2}\bar{g}_{n}(\theta_{0n})-\Omega_{n}^{-1/2}g_{0n}(\hat{\theta}_{n})\|=o_{p}(1).\]
				By triangle inequality, 
				$\sqrt{n}\|\Omega_{n}^{-1/2}g_{0n}(\hat{\theta}_{n})\|\leq \sqrt{n}\|W_{n}^{1/2}\bar{g}_{n}(\hat{\theta}_{n})\|+\sqrt{n}\|W_{n}^{1/2}\bar{g}_{n}(\theta_{0n})\|+o_{p}(1)$.
				As $\hat{\theta}_{n}$ minimizes $\|W_{n}^{1/2}\bar{g}_{n}(\theta)\|$, 
				$\sqrt{n}\|W_{n}^{1/2}\bar{g}_{n}(\hat{\theta}_{n})\|\leq \sqrt{n}\|W_{n}^{1/2}\bar{g}_{n}(\theta_{0n})\|$.
				Note that $\sqrt{n}\|W_{n}^{1/2}\bar{g}_{n}(\theta_{0n})\|=O_{p}(1)$ because $\|W_{n}\|=O_{p}(1)$, while the CLT for triangular array implies
				$\frac{1}{\sqrt{n}}\sum_{i=1}^{n} z_{itn}\Delta\epsilon_{itn}\xrightarrow{d} N(0,\lim_{n\rightarrow\infty}E[z_{itn}z_{itn}'\Delta\epsilon_{itn}^{2}])$.
				The CLT holds by combination of Lyapunov condition and Cram\'er-Wold if $\lim_{n\rightarrow\infty}\frac{E[(\lambda'z_{itn})^{2+r}\Delta\epsilon_{itn}^{2+r}]}{n^{r/2}\{E[(\lambda'z_{itn})^{2}\Delta\epsilon_{itn}^{2}]\}^{1+r/2}}=0$ for some $r>0$ and for any $\lambda\in\mathbb{R}^{dim(z_{it})}$, which holds as $\inf_{n\in\mathbb{N}}\sigma_{\min}(\Omega_{n})>0$ and $\sup_{n\in\mathbb{N}}\max\{(E\|z_{itn}\|^{4+2r})^{1/2},(E\Delta\epsilon_{itn}^{4+2r})^{1/2}\}<\infty$ for some $r>0$.
				Therefore,
				\[
				\begin{array}{rcl}
					\sqrt{n}\|\Omega_{n}^{-1/2}g_{0n}(\hat{\theta}_{n})\| &\leq& \sqrt{n}\|W_{n}^{1/2}\bar{g}_{n}(\hat{\theta}_{n})\|+\sqrt{n}\|W_{n}^{1/2}\bar{g}_{n}(\theta_{0n})\|+o_{p}(1)\\
					&\leq& 2\sqrt{n}\|W_{n}^{1/2}\bar{g}_{n}(\theta_{0n})\|+o_{p}(1) \\
					& = & O_{p}(1),
				\end{array}
				\]
				while $\sqrt{n}\|\Omega_{n}^{-1/2}g_{0n}(\hat{\theta}_{n})\|\geq \sqrt{n}\|\Omega_{n}^{-1/2}M_{0n}(\hat{\alpha}_{n}-\alpha_{0n})+\Omega_{n}^{-1/2}\widetilde{H}_{n}[(\delta_{10n}+\delta_{30n}\gamma_{0n})(\hat{\gamma}_{n}-\gamma_{0n})+\frac{\delta_{30n}}{2}(\hat{\gamma}_{n}-\gamma_{0n})^{2}]\|+o(\sqrt{n}(\|\hat{\alpha}_{n}-\alpha_{0n}\|+|(\delta_{10n}+\delta_{30n}\gamma_{0n})(\hat{\gamma}_{n}-\gamma_{0n})|+(\hat{\gamma}_{n}-\gamma_{0n})^{2}))$ by \cref{lem:unif.loc}.
				
				In conclusion, 
				\[\sqrt{n}(\|\hat{\alpha}_{n}-\alpha_{0n}\|+|(\delta_{10n}+\delta_{30n}\gamma_{0n})(\hat{\gamma}_{n}-\gamma_{0n})|+(\hat{\gamma}_{n}-\gamma_{0n})^{2})\leq O_{p}(1).\]
				It implies that $\sqrt{n}\|\hat{\alpha}_{n}-\alpha_{0n}\|=O_{p}(1)$ for any values of $\zeta_{1}=\lim_{n}n^{1/4}(\delta_{10n}+\delta_{30n}\gamma_{0n})$ and $\zeta_{2}=\lim_{n}(\delta_{10n}+\delta_{30n}\gamma_{0n})$, while for $\hat{\gamma}_{n}$,
				\begin{itemize}[label=]
					\setlength\itemsep{0mm}
					\item (i) $n^{1/4}(\hat{\gamma}_{n}-\gamma_{0n})=O_{p}(1)$ if $\zeta_1=\zeta_2 =0$ 
					\item (ii) $n^{1/4}(\hat{\gamma}_{n}-\gamma_{0n})=O_{p}(1)$ if $\zeta_1 \in \mathbb{R}\setminus \{0\}, \zeta_2 =0$ 
					\item (iii) $\sqrt{n}(\delta_{10n}+\delta_{30n}\gamma_{0n})(\hat{\gamma}_{n}-\gamma_{0n})=O_{p}(1)$ if $|\zeta_1|=\infty, \zeta_2 =0$ 
					\item (vi) $\sqrt{n}(\hat{\gamma}_{n}-\gamma_{0n})=O_{p}(1)$ if $|\zeta_1|=\infty, \zeta_2 \neq 0$.
				\end{itemize}
				
				\paragraph{Asymptotic distribution of estimator and test statistic}
				
				We only	consider (ii) semi-continuous and (iii) semi-discontinuous cases since the proofs for (i) continuous and (iv) discontinuous cases are almost identical to the proof of continuous and discontinuous cases in \cref{thm:distance test}.
				
				\textbf{Case (ii): }
				Let $a=\sqrt{n}(\alpha-\alpha_{0n})$ and $b=n^{1/4}(\gamma-\gamma_{0n})$. 
				Additionally, define $\hat{a}_{n}=\sqrt{n}(\hat{\alpha}-\alpha_{0n})$ and $\hat{b}_{n}=n^{\frac{1}{4}}(\hat{\gamma}-\gamma_{0n})$. 
				Let
				\begin{equation*}
					\mathbb{S}_{n}(a,b) = n\hat{Q}_{n}(\alpha_{0n}+\tfrac{a}{\sqrt{n}},\gamma_{0n}+\tfrac{b}{n^{\frac{1}{4}}})=n\bar{g}_{n}(\alpha_{0n}+\tfrac{a}{\sqrt{n}},\gamma_{0n}+\tfrac{b}{n^{\frac{1}{4}}})'W_{n}\bar{g}_{n}(\alpha_{0n}+\tfrac{a}{\sqrt{n}},\gamma_{0n}+\tfrac{b}{n^{\frac{1}{4}}}).
				\end{equation*}
				The rescaled and reparametrized sample moment can be written as
				\begin{align*}
					\sqrt{n}\bar{g}_{n}(\alpha_{0n}+\tfrac{a}{\sqrt{n}}, \gamma_{0n}+\tfrac{b}{n^{\frac{1}{4}}})
					=&
					\sqrt{n}\begin{pmatrix}
						\frac{1}{n}\sum_{i=1}^{n}z_{it_{0}n}\Delta\epsilon_{it_{0}n} \\
						\vdots \\
						\frac{1}{n}\sum_{i=1}^{n}z_{iTn}\Delta\epsilon_{iTn}
					\end{pmatrix}
					-
					\begin{pmatrix}
						\tfrac{1}{n}\sum_{i=1}^{n} z_{it_{0}n}\Delta x_{it_{0}n}'\\
						\vdots\\
						\tfrac{1}{n}\sum_{i=1}^{n} z_{iTn}\Delta x_{iTn}'
					\end{pmatrix}a_{1}
					\\
					&
					-\begin{pmatrix}
						\tfrac{1}{n}\sum_{i=1}^{n} z_{it_{0}n}1_{it_{0}n}(\gamma_{0n}+\tfrac{b}{n^{\frac{1}{4}}})' X_{it_{0}n}\\
						\vdots\\
						\tfrac{1}{n}\sum_{i=1}^{n} z_{iTn}1_{iTn}(\gamma_{0n}+\tfrac{b}{n^{\frac{1}{4}}})' X_{iTn}
					\end{pmatrix}a_{2}
					\\
					&
					+\sqrt{n}\begin{pmatrix}
						\tfrac{1}{n}\sum_{i=1}^{n} z_{it_{0}n}(1_{it_{0}n}(\gamma_{0n})'-1_{it_{0}n}(\gamma_{0n}+\tfrac{b}{n^{\frac{1}{4}}})')X_{it_{0}n}\\
						\vdots\\
						\tfrac{1}{n}\sum_{i=1}^{n} z_{iTn}(1_{iTn}(\gamma_{0n})'-1_{iT}(\gamma_{0n}+\tfrac{b}{n^{\frac{1}{4}}})')X_{iTn}
					\end{pmatrix}\delta_{0n}.
				\end{align*}
				By the CLT for triangular array,
				\begin{equation*}
					\sqrt{n}\begin{pmatrix}
						\frac{1}{n}\sum_{i=1}^{n}z_{it_{0}n}\Delta\epsilon_{it_{0}n} \\
						\vdots \\
						\frac{1}{n}\sum_{i=1}^{n}z_{iTn}\Delta\epsilon_{iTn}
					\end{pmatrix}
					\xrightarrow{d}-e\sim N(0,\Omega_{\infty}).
				\end{equation*}
				Note that the CLT holds by combination of Lyapunov condition and Cram\'er-Wold device if $\lim_{n\rightarrow\infty}\frac{E[(\lambda'z_{itn})^{2+r}\Delta\epsilon_{itn}^{2+r}]}{n^{r/2}\{E[(\lambda'z_{itn})^{2}\Delta\epsilon_{itn}^{2}]\}^{1+r/2}}=0$ for some $r>0$ for any $\lambda\in\mathbb{R}^{k}$, which holds as $\inf_{n\in\mathbb{N}}\sigma_{\min}(\Omega_{n})>0$ and $\sup_{n\in\mathbb{N}}\max\{(E\|z_{itn}\|^{4+2r})^{1/2},(E\Delta\epsilon_{itn}^{4+2r})^{1/2}\}<\infty$ for some $r>0$.
				By the WLLN for triangular array,
				\begin{equation*}
					\begin{pmatrix}
						\tfrac{1}{n}\sum_{i=1}^{n} z_{it_{0}n}\Delta x_{it_{0}n}'\\
						\vdots\\
						\tfrac{1}{n}\sum_{i=1}^{n} z_{iTn}\Delta x_{iTn}'
					\end{pmatrix}
					\xrightarrow{p}
					\begin{pmatrix}
						Ez_{it_{0},\infty}\Delta x_{it_{0},\infty}'\\
						\vdots\\
						Ez_{iT,\infty}\Delta x_{iT,\infty}'
					\end{pmatrix},
				\end{equation*}
				which holds as $\sup_{n\in\mathbb{N}}E\|z_{itn}\Delta x_{itn}\|^{2}\leq \sup_{n\in\mathbb{N}}(E\|z_{itn}\|^{4})^{1/2}(E\|\Delta x_{itn}\|^{4})^{1/2} < \infty$.
				Let $K<\infty$ be some constant.
				By the ULLN in \cref{lem:unif.ULLN},
				\begin{equation*}
					\left\|
					\begin{pmatrix}
						\tfrac{1}{n}\sum_{i=1}^{n} z_{it_{0}n}1_{it_{0}n}(\gamma_{0n}+\tfrac{b}{n^{\frac{1}{4}}})' X_{it_{0}n}\\
						\vdots\\
						\tfrac{1}{n}\sum_{i=1}^{n} z_{iTn}1_{iTn}(\gamma_{0n}+\tfrac{b}{n^{\frac{1}{4}}})' X_{iTn}
					\end{pmatrix}
					-
					\begin{pmatrix}
						Ez_{it_{0},\infty}1_{it_{0},\infty}(\gamma_{0,\infty}+\tfrac{b}{n^{\frac{1}{4}}})' X_{it_{0},\infty}\\
						\vdots\\
						Ez_{iT,\infty}1_{iT,\infty}(\gamma_{0,\infty}+\tfrac{b}{n^{\frac{1}{4}}})' X_{iT,\infty}
					\end{pmatrix}
					\right\|\xrightarrow{p}0
				\end{equation*}
				uniformly with respect to $b\in[-K,K]$.
				Then, by the continuity of $\kappa\mapsto E[z_{it,\infty}1_{it,\infty}(\gamma_{0,\infty}+\kappa)X_{it,\infty}]$ at $\kappa=0$,
				\begin{equation*}
					\begin{pmatrix}
						\tfrac{1}{n}\sum_{i=1}^{n} z_{it_{0},\infty}1_{it_{0},\infty}(\gamma_{0,\infty}+\tfrac{b}{n^{\frac{1}{4}}})' X_{it_{0},\infty}\\
						\vdots\\
						\tfrac{1}{n}\sum_{i=1}^{n} z_{iT,\infty}1_{iT,\infty}(\gamma_{0,\infty}+\tfrac{b}{n^{\frac{1}{4}}})' X_{iT,\infty}
					\end{pmatrix}
					\xrightarrow{p}
					\begin{pmatrix}
						Ez_{it_{0},\infty}1_{it_{0,\infty}}(\gamma_{0,\infty})' X_{it_{0},\infty}\\
						\vdots\\
						Ez_{iT,\infty}1_{iT,\infty}(\gamma_{0,\infty})' X_{iT,\infty}
					\end{pmatrix}
				\end{equation*}
				uniformly with respect to $b\in [-K,K]$.
				By \cref{lem:unif.plim1},
				\begin{multline*}
					\sqrt{n}\begin{pmatrix}
						\tfrac{1}{n}\sum_{i=1}^{n} z_{it_{0}n}(1_{it_{0}n}(\gamma_{0n})'-1_{it_{0}n}(\gamma_{0n}+\tfrac{b}{n^{\frac{1}{4}}})')X_{it_{0}n}\delta_{0n}\\
						\vdots\\
						\tfrac{1}{n}\sum_{i=1}^{n} z_{iTn}(1_{iTn}(\gamma_{0n})'-1_{iTn}(\gamma_{0n}+\tfrac{b}{n^{\frac{1}{4}}})')X_{iTn}\delta_{0n}
					\end{pmatrix} \\
					\xrightarrow{p}
					\begin{pmatrix}
						E_{t_{0},\infty}[z_{it_{0},\infty}|\gamma_{0,\infty}]f_{t_{0},\infty}(\gamma_{0,\infty})-E_{t_{0}-1,\infty}[z_{iT,\infty}|\gamma_{0,\infty}]f_{t_{0}-1,\infty}(\gamma_{0,\infty})\\
						\vdots\\
						E_{T,\infty}[z_{iT,\infty}|\gamma_{0,\infty}]f_{T,\infty}(\gamma_{0,\infty})-E_{T-1,\infty}[z_{iT,\infty}|\gamma_{0,\infty}]f_{T-1,\infty}(\gamma_{0,\infty})
					\end{pmatrix}
					\left\{\zeta_{1}b+\frac{\delta_{30,\infty}}{2}b^{2}\right\}
				\end{multline*}
				uniformly with respect to $b\in [-K,K]$.
				Therefore, $\mathbb{S}_{n}(a,b)$ weakly converges to 
				\[\mathbb{S}(a,b)=(M_{0,\infty}a+\widetilde{H}_{\infty}(\zeta_{1}b+\frac{\delta_{30,\infty}}{2}b^{2})-e)'\Omega_{\infty}^{-1}(M_{0,\infty}a+\widetilde{H}_{\infty}(\zeta_{1}b+\frac{\delta_{30,\infty}}{2}b^{2})-e),\]
				in $\ell^{\infty}(\mathbb{K})$ for any compact $\mathbb{K}\subset \mathbb{R}^{2p+2}$.
				
				Let $\tilde{b}=\zeta_{1}b+\frac{\delta_{30,\infty}}{2}b^{2}$ and $\hat{\tilde{b}}_{n}=\zeta_{1}\hat{b}_{n}+\frac{\delta_{30,\infty}}{2}\hat{b}_{n}^{2}$.
				We consider $\delta_{30,\infty}>0$ so that $\tilde{b}\geq -\frac{\zeta_{1}^{2}}{2\delta_{30,\infty}}$.
				When $\delta_{30,\infty}<0$, derivations are almost identical and lead to the same limit distribution of the test statistic.
				Let $\underline{b}=-\frac{\zeta_{1}^{2}}{2\delta_{30,\infty}}$.
				Then, by the CMT,
				\[(\hat{a}_{n},\hat{\tilde{b}}_{n}) \xrightarrow{d} 
				(a_{0},\tilde{b}_{0})= \arg\min_{a,\tilde{b}\geq \underline{b}} (M_{0,\infty}a+\widetilde{H}_{\infty}\tilde{b}-e)'\Omega_{\infty}^{-1}(M_{0,\infty}a+\widetilde{H}_{\infty}\tilde{b}-e).\]
				KKT conditions, as in the proof of \cref{thm:asym}, imply
				\begin{align*}
					&M_{0,\infty}'\Omega_{\infty}^{-1}M_{0,\infty}a_{0}+M_{0,\infty}'\Omega_{\infty}^{-1}\widetilde{H}_{\infty}\tilde{b}_{0}-M_{0,\infty}'\Omega_{\infty}^{-1}e=0, \\
					&\widetilde{H}_{\infty}'\Omega_{\infty}^{-1}\widetilde{H}_{\infty}\tilde{b}_{0} + \widetilde{H}_{\infty}'\Omega_{\infty}^{-1}M_{0,\infty}a_{0}-\widetilde{H}_{\infty}'\Omega_{\infty}^{-1}e -\lambda=0,
				\end{align*}
				$\lambda\geq0$, $\tilde{b}_{0}\geq\underline{b}$, and $\lambda (\tilde{b}_{0}-\underline{b})=0$ should hold.
				Then, we can get
				\[\tilde{b}_{0} = \begin{cases}
					[\widetilde{H}_{\infty}'\Xi_{\infty} \widetilde{H}_{\infty}]^{-1}\widetilde{H}_{\infty}'\Xi_{\infty} e & \text{if }[\widetilde{H}_{\infty}'\Xi_{\infty} \widetilde{H}_{\infty}]^{-1}\widetilde{H}_{\infty}'\Xi_{\infty} e\geq \underline{b}\\
					\quad\quad \underline{b} & \text{else} \\
				\end{cases}\]
				where $\Xi_{\infty}=\Omega_{\infty}^{-1/2}(I-P_{\Omega_{\infty}^{-1/2}M_{0,\infty}})\Omega_{\infty}^{-1/2}$. 
				$\tilde{b}_{0}$ follows a normal distribution that is left censored at $\underline{b}$.
				Then,
				\begin{equation*}
					a_{0}  = \begin{cases}
						(M_{0,\infty}'\Omega_{\infty}^{-1}M_{0,\infty})^{-1}M_{0,\infty}'\Omega_{\infty}^{-1}[I-\widetilde{H}_{\infty}[\widetilde{H}_{\infty}'\Xi_{\infty} \widetilde{H}_{\infty}]^{-1}\widetilde{H}_{\infty}'\Xi_{\infty} ]e & \text{if }[\widetilde{H}_{\infty}'\Xi_{\infty} \widetilde{H}_{\infty}]^{-1}\widetilde{H}_{\infty}'\Xi_{\infty} e\geq \underline{b} \\
						(M_{0,\infty}'\Omega_{\infty}^{-1}M_{0,\infty})^{-1}M_{0,\infty}'\Omega_{\infty}^{-1}(e-\widetilde{H}_{\infty}\underline{b}) & \text{else.} 
					\end{cases}
				\end{equation*}
				
				Asymptotic distribution of the test statistic $\mathcal{D}_{n}(\gamma_{0n})$ can be derived by
				\begin{align*}
					\mathcal{D}_{n}(\gamma_{0n}) 
					\xrightarrow{d}
					&
					\min_{a} (M_{0,\infty}a-e)'\Omega_{\infty}^{-1}(M_{0,\infty}a-e)
					\\ 
					&
					-\min_{a,\tilde{b}\geq\underline{b}}(M_{0,\infty}a+\widetilde{H}_{\infty}\tilde{b}-e)'\Omega_{\infty}^{-1}(M_{0,\infty}a+\widetilde{H}_{\infty}\tilde{b}-e),
				\end{align*}
				where we apply the CMT.
				Note that $\min_{a} (M_{0,\infty}a-e)'\Omega_{\infty}^{-1}(M_{0,\infty}a-e)=e'\Omega_{\infty}^{-1/2}(I-P_{\Omega_{\infty}^{-1/2}M_{0,\infty}})\Omega_{\infty}^{-1/2}e$, 
				while
				\begin{align*}
					&\min_{a,\tilde{b}\geq\underline{b}}(M_{0,\infty}a+\widetilde{H}_{\infty}\tilde{b}-e)'\Omega_{\infty}^{-1}(M_{0,\infty}a+\widetilde{H}_{\infty}\tilde{b}-e)\\
					&=(M_{0,\infty}a_{0}+\widetilde{H}_{\infty}\tilde{b}_{0}-e)'\Omega_{\infty}^{-1}(M_{0,\infty}a_{0}+\widetilde{H}_{\infty}\tilde{b}_{0}-e) \\
					&=(M_{0,\infty}'\Omega_{\infty}^{-1}M_{0,\infty}a_{0}+M_{0,\infty}'\Omega_{\infty}^{-1}\widetilde{H}_{\infty}\tilde{b}_{0})'(M_{0,\infty}'\Omega_{\infty}^{-1}M_{0,\infty})^{-1}(M_{0,\infty}'\Omega_{\infty}^{-1}M_{0,\infty}a_{0}+M_{0,\infty}'\Omega_{\infty}^{-1}\widetilde{H}_{\infty}\tilde{b}_{0}) \\
					&\quad +\tilde{b}_{0}\widetilde{H}_{\infty}'\Omega_{\infty}^{-1/2}(I-P_{\Omega_{\infty}^{-1/2}M_{0,\infty}})\Omega_{\infty}^{-1/2}\widetilde{H}_{\infty}\tilde{b}_{0}\\
					&\quad -2e'\Omega_{\infty}^{-1}M_{0,\infty}(M_{0,\infty}'\Omega_{\infty}^{-1}M_{0,\infty})^{-1}(M_{0,\infty}'\Omega_{\infty}^{-1}M_{0,\infty}a_{0}+M_{0,\infty}'\Omega_{\infty}^{-1}\widetilde{H}_{\infty}\tilde{b}_{0}) \\
					&\quad -2e'\Omega_{\infty}^{-1/2}(I-P_{\Omega_{\infty}^{-1/2}M_{0,\infty}})\Omega_{\infty}^{-1/2}\widetilde{H}_{\infty}\tilde{b}_{0} + e'\Omega_{\infty}^{-1}e.
				\end{align*}
				By plugging in the formula for $(a_{0},\tilde{b}_{0})$ (note that $M_{0,\infty}'\Omega_{\infty}^{-1}M_{0,\infty}a_{0}+M_{0,\infty}'\Omega_{\infty}^{-1}\widetilde{H}_{\infty}\tilde{b}_{0}=M_{0,\infty}'\Omega_{\infty}^{-1}e$) we can get
				\begin{align*}
					&\min_{a,\tilde{b}\geq\underline{b}}(M_{0,\infty}a+\widetilde{H}_{\infty}\tilde{b}-e)'\Omega_{\infty}^{-1}(M_{0,\infty}a+\widetilde{H}_{\infty}\tilde{b}-e)\\
					&\quad =\begin{cases}
						e'\Omega_{\infty}^{-1/2}(I-P_{\Omega_{\infty}^{-1/2}M_{0,\infty}})\Omega_{\infty}^{-1/2}e  -e'\Xi_{\infty}\widetilde{H}_{\infty}(\widetilde{H}_{\infty}'\Xi_{\infty}\widetilde{H}_{\infty})^{-1}\widetilde{H}_{\infty}\Xi_{\infty}e & \text{if }[\widetilde{H}_{\infty}'\Xi_{\infty} \widetilde{H}_{\infty}]^{-1}\widetilde{H}_{\infty}'\Xi_{\infty} e\geq \underline{b}\\
						e'\Omega_{\infty}^{-1/2}(I-P_{\Omega_{\infty}^{-1/2}M_{0,\infty}})\Omega_{\infty}^{-1/2}e +(\widetilde{H}_{\infty}'\Xi_{\infty} \widetilde{H}_{\infty})\underline{b}^{2} - 2(e'\Xi_{\infty}\widetilde{H}_{\infty})\underline{b} & \text{else}.
					\end{cases}
				\end{align*}
				Therefore, the limit distribution of the test statistic is identical to
				\[
				\begin{cases}
					e'\Xi_{\infty}\widetilde{H}_{\infty}(\widetilde{H}_{\infty}'\Xi_{\infty}\widetilde{H}_{\infty})^{-1}\widetilde{H}_{\infty}\Xi_{\infty}e & \text{if }[\widetilde{H}_{\infty}'\Xi_{\infty} \widetilde{H}_{\infty}]^{-1}\widetilde{H}_{\infty}'\Xi_{\infty} e\geq \underline{b} \\
					-(\widetilde{H}_{\infty}'\Xi_{\infty} \widetilde{H}_{\infty})\underline{b}^{2} + 2(e'\Xi_{\infty}\widetilde{H}_{\infty})\underline{b} & \text{else}.
				\end{cases}
				\]
				
				\textbf{Case (iii): }
				Let $a=\sqrt{n}(\alpha-\alpha_{0n})$ and $b=\sqrt{n}(\delta_{10n}+\delta_{30n}\gamma_{0n})(\gamma-\gamma_{0n})$.
				The rescaled and reparametrized sample moment can be written as
				\begin{align*}
					&\sqrt{n}\bar{g}_{n}(\alpha_{0n}+\tfrac{a}{\sqrt{n}}, \gamma_{0n}+\tfrac{b}{\sqrt{n}(\delta_{10n}+\delta_{30n}\gamma_{0n})})
					=\\
					&\hspace{2cm}
					\sqrt{n}\begin{pmatrix}
						\frac{1}{n}\sum_{i=1}^{n}z_{it_{0}n}\Delta\epsilon_{it_{0}n} \\
						\vdots \\
						\frac{1}{n}\sum_{i=1}^{n}z_{iTn}\Delta\epsilon_{iTn}
					\end{pmatrix}
					-
					\begin{pmatrix}
						\tfrac{1}{n}\sum_{i=1}^{n} z_{it_{0}n}\Delta x_{it_{0}n}'\\
						\vdots\\
						\tfrac{1}{n}\sum_{i=1}^{n} z_{iTn}\Delta x_{iTn}'
					\end{pmatrix}a_{1}
					\\
					&\hspace{2cm}
					-\begin{pmatrix}
						\tfrac{1}{n}\sum_{i=1}^{n} z_{it_{0}n}1_{it_{0}n}(\gamma_{0n}+\tfrac{b}{\sqrt{n}(\delta_{10n}+\delta_{30n}\gamma_{0n})})' X_{it_{0}n}\\
						\vdots\\
						\tfrac{1}{n}\sum_{i=1}^{n} z_{iTn}1_{iTn}(\gamma_{0n}+\tfrac{b}{\sqrt{n}(\delta_{10n}+\delta_{30n}\gamma_{0n})})' X_{iTn}
					\end{pmatrix}a_{2}
					\\
					&\hspace{2cm}
					+\sqrt{n}\begin{pmatrix}
						\tfrac{1}{n}\sum_{i=1}^{n} z_{it_{0}n}(1_{it_{0}n}(\gamma_{0n})'-1_{it_{0}n}(\gamma_{0n}+\tfrac{b}{\sqrt{n}(\delta_{10n}+\delta_{30n}\gamma_{0n})})')X_{it_{0}n}\\
						\vdots\\
						\tfrac{1}{n}\sum_{i=1}^{n} z_{iTn}(1_{iTn}(\gamma_{0n})'-1_{iT}(\gamma_{0n}+\tfrac{b}{\sqrt{n}(\delta_{10n}+\delta_{30n}\gamma_{0n})})')X_{iTn}
					\end{pmatrix}\delta_{0n}.
				\end{align*}
				By the CLT for triangular array,
				\begin{equation*}
					\sqrt{n}\begin{pmatrix}
						\frac{1}{n}\sum_{i=1}^{n}z_{it_{0}n}\Delta\epsilon_{it_{0}n} \\
						\vdots \\
						\frac{1}{n}\sum_{i=1}^{n}z_{iTn}\Delta\epsilon_{iTn}
					\end{pmatrix}
					\xrightarrow{d}-e\sim N(0,\Omega_{\infty}).
				\end{equation*}
				By the WLLN for triangular array,
				\begin{equation*}
					\begin{pmatrix}
						\tfrac{1}{n}\sum_{i=1}^{n} z_{it_{0}n}\Delta x_{it_{0}n}'\\
						\vdots\\
						\tfrac{1}{n}\sum_{i=1}^{n} z_{iTn}\Delta x_{iTn}'
					\end{pmatrix}
					\xrightarrow{p}
					\begin{pmatrix}
						Ez_{it_{0},\infty}\Delta x_{it_{0},\infty}'\\
						\vdots\\
						Ez_{iT,\infty}\Delta x_{iT,\infty}'
					\end{pmatrix}.
				\end{equation*}
				By the ULLN in \cref{lem:unif.ULLN},
				\begin{multline*}
					\left\|
					\begin{pmatrix}
						\tfrac{1}{n}\sum_{i=1}^{n} z_{it_{0}n}1_{it_{0}n}(\gamma_{0n}+\tfrac{b}{\sqrt{n}(\delta_{10n}+\delta_{30n}\gamma_{0n})})' X_{it_{0}n}\\
						\vdots\\
						\tfrac{1}{n}\sum_{i=1}^{n} z_{iTn}1_{iTn}(\gamma_{0n}+\tfrac{b}{\sqrt{n}(\delta_{10n}+\delta_{30n}\gamma_{0n})})' X_{iTn}
					\end{pmatrix}
					\right.
					-\\
					\left.
					\begin{pmatrix}
						Ez_{it_{0},\infty}1_{it_{0},\infty}(\gamma_{0,\infty}+\tfrac{b}{\sqrt{n}(\delta_{10n}+\delta_{30n}\gamma_{0n})})' X_{it_{0},\infty}\\
						\vdots\\
						Ez_{iT,\infty}1_{iT,\infty}(\gamma_{0,\infty}+\tfrac{b}{\sqrt{n}(\delta_{10n}+\delta_{30n}\gamma_{0n})})' X_{iT,\infty}
					\end{pmatrix}
					\right\|\xrightarrow{p}0
				\end{multline*}
				uniformly with respect to $b\in[-K,K]$, which implies
				\begin{equation*}
					\begin{pmatrix}
						\tfrac{1}{n}\sum_{i=1}^{n} z_{it_{0},\infty}1_{it_{0},\infty}(\gamma_{0,\infty}+\tfrac{b}{\sqrt{n}(\delta_{10n}+\delta_{30n}\gamma_{0n})})' X_{it_{0},\infty}\\
						\vdots\\
						\tfrac{1}{n}\sum_{i=1}^{n} z_{iT,\infty}1_{iT,\infty}(\gamma_{0,\infty}+\tfrac{b}{\sqrt{n}(\delta_{10n}+\delta_{30n}\gamma_{0n})})' X_{iT,\infty}
					\end{pmatrix}
					\xrightarrow{p}
					\begin{pmatrix}
						Ez_{it_{0},\infty}1_{it_{0,\infty}}(\gamma_{0,\infty})' X_{it_{0},\infty}\\
						\vdots\\
						Ez_{iT,\infty}1_{iT,\infty}(\gamma_{0,\infty})' X_{iT,\infty}
					\end{pmatrix}
				\end{equation*}
				uniformly with respect to $b\in [-K,K]$.
				By \cref{lem:unif.plim2},
				\begin{multline*}
					\sqrt{n}\begin{pmatrix}
						\tfrac{1}{n}\sum_{i=1}^{n} z_{it_{0}n}(1_{it_{0}n}(\gamma_{0n})'-1_{it_{0}n}(\gamma_{0n}+\tfrac{b}{\sqrt{n}(\delta_{10n}+\delta_{30n}\gamma_{0n})})')X_{it_{0}n}\delta_{0n}\\
						\vdots\\
						\tfrac{1}{n}\sum_{i=1}^{n} z_{iTn}(1_{iTn}(\gamma_{0n})'-1_{iTn}(\gamma_{0n}+\tfrac{b}{\sqrt{n}(\delta_{10n}+\delta_{30n}\gamma_{0n})})')X_{iTn}\delta_{0n}
					\end{pmatrix} \\
					\xrightarrow{p}
					\begin{pmatrix}
						E_{t_{0},\infty}[z_{it_{0},\infty}|\gamma_{0,\infty}]f_{t_{0},\infty}(\gamma_{0,\infty})-E_{t_{0}-1,\infty}[z_{iT,\infty}|\gamma_{0,\infty}]f_{t_{0}-1,\infty}(\gamma_{0,\infty})\\
						\vdots\\
						E_{T,\infty}[z_{iT,\infty}|\gamma_{0,\infty}]f_{T,\infty}(\gamma_{0,\infty})-E_{T-1,\infty}[z_{iT,\infty}|\gamma_{0,\infty}]f_{T-1,\infty}(\gamma_{0,\infty})
					\end{pmatrix}
					b
				\end{multline*}
				uniformly with respect to $b\in [-K,K]$.
				Therefore, $\mathbb{S}_{n}(a,b) = n\hat{Q}_{n}(\alpha_{0n}+\tfrac{a}{\sqrt{n}},\gamma_{0n}+\frac{b}{\sqrt{n}(\delta_{10n}+\delta_{30n}\gamma_{0n})})$ weakly converges to 
				\[\mathbb{S}(a,b)=(M_{0,\infty}a+\widetilde{H}_{\infty}b-e)'\Omega_{\infty}^{-1}(M_{0,\infty}a+\widetilde{H}_{\infty}b-e),\]
				in $\ell^{\infty}(\mathbb{K})$ for any compact $\mathbb{K}\subset \mathbb{R}^{2p+2}$.
				Then, $\hat{a}_{n}=\sqrt{n}(\hat{\alpha}_{n}-\alpha_{0n})$ and $\hat{b}_{n}=\sqrt{n}(\delta_{10n}+\delta_{30n}\gamma_{0n})(\hat{\gamma}_{n}-\gamma_{0n})$ converges in distribution to
				\begin{equation*}
					(a_{0},b_{0})=\arg\min_{a,b}(M_{0,\infty}a+\widetilde{H}_{\infty}b-e)'\Omega_{\infty}^{-1}(M_{0,\infty}a+\widetilde{H}_{\infty}b-e).
				\end{equation*}
				by the argmin CMT.
				KKT conditions, as in the proof of \cref{thm:asym}, imply
				\begin{align*}
					&M_{0,\infty}'\Omega_{\infty}^{-1}M_{0,\infty}a_{0}+M_{0,\infty}'\Omega_{\infty}^{-1}\widetilde{H}_{\infty}b_{0}-M_{0,\infty}'\Omega_{\infty}^{-1}e=0 \\
					&\widetilde{H}_{\infty}'\Omega_{\infty}^{-1}\widetilde{H}_{\infty}b_{0} + \widetilde{H}_{\infty}'\Omega_{\infty}^{-1}M_{0,\infty}a_{0}-\widetilde{H}_{\infty}'\Omega_{\infty}^{-1}e=0.
				\end{align*}
				
				Then, we can get
				\[b_{0} = 
				[\widetilde{H}_{\infty}'\Xi_{\infty} \widetilde{H}_{\infty}]^{-1}\widetilde{H}_{\infty}'\Xi_{\infty} e,
				\]
				where $\Xi_{\infty}=\Omega_{\infty}^{-1/2}(I-P_{\Omega_{\infty}^{-1/2}M_{0,\infty}})\Omega_{\infty}^{-1/2}$, and
				\begin{equation*}
					a_{0}  = 
					(M_{0,\infty}'\Omega_{\infty}^{-1}M_{0,\infty})^{-1}M_{0,\infty}'\Omega_{\infty}^{-1}[I-\widetilde{H}_{\infty}[\widetilde{H}_{\infty}'\Xi_{\infty} \widetilde{H}_{\infty}]^{-1}\widetilde{H}_{\infty}'\Xi_{\infty} ]e 
				\end{equation*}
				
				Asymptotic distribution of the test statistic $\mathcal{D}_{n}(\gamma_{0n})$ can be derived by
				\begin{align*}
					\mathcal{D}_{n}(\gamma_{0n}) 
					\xrightarrow{d}
					&
					\min_{a} (M_{0,\infty}a-e)'\Omega_{\infty}^{-1}(M_{0,\infty}a-e)
					\\ 
					&
					-\min_{a,b}(M_{0,\infty}a+\widetilde{H}_{\infty}b-e)'\Omega_{\infty}^{-1}(M_{0,\infty}a+\widetilde{H}_{\infty}b-e),
				\end{align*}
				where we apply the CMT.
				Note that $\min_{a} (M_{0,\infty}a-e)'\Omega_{\infty}^{-1}(M_{0,\infty}a-e)=e'\Omega_{\infty}^{-1/2}(I-P_{\Omega_{\infty}^{-1/2}M_{0,\infty}})\Omega_{\infty}^{-1/2}e$, 
				while
				\begin{align*}
					&\min_{a,b}(M_{0,\infty}a+\widetilde{H}_{\infty}b-e)'\Omega_{\infty}^{-1}(M_{0,\infty}a+\widetilde{H}_{\infty}b-e)\\
					&=(M_{0,\infty}a_{0}+\widetilde{H}_{\infty}b_{0}-e)'\Omega_{\infty}^{-1}(M_{0,\infty}a_{0}+\widetilde{H}_{\infty}b_{0}-e) \\
					&=(M_{0,\infty}'\Omega_{\infty}^{-1}M_{0,\infty}a_{0}+M_{0,\infty}'\Omega_{\infty}^{-1}\widetilde{H}_{\infty}b_{0})'(M_{0,\infty}'\Omega_{\infty}^{-1}M_{0,\infty})^{-1}(M_{0,\infty}'\Omega_{\infty}^{-1}M_{0,\infty}a_{0}+M_{0,\infty}'\Omega_{\infty}^{-1}\widetilde{H}_{\infty}b_{0}) \\
					&\quad +b_{0}\widetilde{H}_{\infty}'\Omega_{\infty}^{-1/2}(I-P_{\Omega_{\infty}^{-1/2}M_{0,\infty}})\Omega_{\infty}^{-1/2}\widetilde{H}_{\infty}b_{0}\\
					&\quad -2e'\Omega_{\infty}^{-1}M_{0,\infty}(M_{0,\infty}'\Omega_{\infty}^{-1}M_{0,\infty})^{-1}(M_{0,\infty}'\Omega_{\infty}^{-1}M_{0,\infty}a_{0}+M_{0,\infty}'\Omega_{\infty}^{-1}\widetilde{H}_{\infty}b_{0}) \\
					&\quad -2e'\Omega_{\infty}^{-1/2}(I-P_{\Omega_{\infty}^{-1/2}M_{0,\infty}})\Omega_{\infty}^{-1/2}\widetilde{H}_{\infty}b_{0} + e'\Omega_{\infty}^{-1}e.
				\end{align*}
				By plugging in the formula for $(a_{0},b_{0})$ (note that $M_{0,\infty}'\Omega_{\infty}^{-1}M_{0,\infty}a_{0}+M_{0,\infty}'\Omega_{\infty}^{-1}\widetilde{H}_{\infty}b_{0}=M_{0,\infty}'\Omega_{\infty}^{-1}e$), we can get
				\begin{align*}
					&\min_{a,b}(M_{0,\infty}a+\widetilde{H}_{\infty}b-e)'\Omega_{\infty}^{-1}(M_{0,\infty}a+\widetilde{H}_{\infty}b-e)\\
					&\quad =
					e'\Omega_{\infty}^{-1/2}(I-P_{\Omega_{\infty}^{-1/2}M_{0,\infty}})\Omega_{\infty}^{-1/2}e  -e'\Xi_{\infty}\widetilde{H}_{\infty}(\widetilde{H}_{\infty}'\Xi_{\infty}\widetilde{H}_{\infty})^{-1}\widetilde{H}_{\infty}\Xi_{\infty}e 
				\end{align*}
				Therefore, the limit distribution of the test statistic is identical to
				\[
				e'\Xi_{\infty}\widetilde{H}_{\infty}(\widetilde{H}_{\infty}'\Xi_{\infty}\widetilde{H}_{\infty})^{-1}\widetilde{H}_{\infty}\Xi_{\infty}e,
				\]
				which has the $\chi^{2}_{1}$ distribution.

				\paragraph{Limit distribution of bootstrap estimator and test statistic}
				
				The derivation of the limit distributions of the bootstrap estimator and test statistic is almost identical to that of the asymptotic distributions of the sample estimator and test statistic.
				We need to replace $\delta_{0n}$ by $\delta_{0n}^{*}=\hat{\delta}_{0n}(\gamma_{0n})$, $\{\Delta\epsilon_{itn}\}$ by $\{\widehat{\Delta\epsilon}_{itn}\}$, and sample moments by bootstrap moments in the previous part of the proof regarding asymptotic analysis.
				Be mindful that we do not need to replace $\gamma_{0n}$ in the previous part of the proof as we focus on the grid bootstrap when $\gamma_{0n}^{*}=\gamma_{0n}$ to show that the grid bootstrap CI provides correct coverage rate.
				Lemmas \ref{lem:unif.boot ULLN}, \ref{lem:unif.boot.equicont}, \ref{lem:unif.boot.plim1}, and \ref{lem:unif.boot.plim2} are applied instead of Lemmas \ref{lem:unif.ULLN}, \ref{lem:unif.equicont}, \ref{lem:unif.plim1}, and \ref{lem:unif.plim2} in the places where the latter are used in the previous part of the proof.
				Moreover, Lemmas \ref{lem:unif.boot res LLN} and \ref{lem:unif.boot res CLT} are applied instead of the WLLN and CLT for triangular array applied to $\{z_{itn}\Delta\epsilon_{itn}:1\leq i\leq n, n\in\mathbb{N}\}$ in the places where the latter are used in the previous part of the proof.
			\end{proof}

			\subsection{Auxiliary Lemmas}
			
			\begin{lemma}
				\label{lem:unif.loc}
				Let $\{\phi_{0n}\in\Phi_{0}:n\geq 1 \}$ and $\pi_{n}(\phi_{0n})\rightarrow(\zeta_{1},\zeta_{2},\phi_{0,\infty})\in\Pi$.
				For any $\eta>0$, there is $h>0$ such that
				\begin{equation*}
					\lim_{n\rightarrow\infty}\sup_{\|\theta-\theta_{0n}\|<h}
					\frac{\sqrt{n}\left\|g_{0n}(\theta)-M_{0n}(\alpha-\alpha_{0n})-\widetilde{H}_{n}[(\delta_{10n}+\delta_{30n}\gamma_{0n})(\gamma-\gamma_{0n})+\frac{\delta_{30n}}{2}(\gamma-\gamma_{0n})^{2}]\right\|}
					{1+\sqrt{n}(\|\alpha-\alpha_{0n}\|+|(\delta_{10n}+\delta_{30n}\gamma_{0n})(\gamma-\gamma_{0n})|+(\gamma-\gamma_{0n})^{2})}
					<\eta.
				\end{equation*}
			\end{lemma}
			\begin{proof}
				Note that $g_{0n}(\theta)-M_{0n}(\alpha-\alpha_{0n}) = M_{0n}(\gamma)\alpha-M_{0n}\alpha_{0n}-M_{0n}(\alpha-\alpha_{0n}) = (M_{0n}(\gamma)-M_{0n})\alpha = (M_{20n}(\gamma)-M_{20n})\delta = (M_{20n}(\gamma)-M_{20n})[\delta_{0n}+(\delta-\delta_{0n})]$. 
				
				First, we derive a bound for $(M_{20n}(\gamma)-M_{20n})\delta_{0n}$ which is
				\begin{equation*}
					\begin{pmatrix}
						E[z_{it_{0}n}(\delta_{10n}+\delta_{30n}q_{it_{0}n})1\{\gamma \geq q_{it_{0}n}>\gamma_{0n}\}] -  E[z_{it_{0}-1,n}(\delta_{10n}+\delta_{30n}q_{it_{0}-1,n})1\{\gamma \geq q_{it_{0}-1,n}>\gamma_{0n}\}] \\
						\vdots \\
						E[z_{iTn}(\delta_{10n}+\delta_{30n}q_{iTn})1\{\gamma \geq q_{iTn}>\gamma_{0n}\}] -  E[z_{iT-1,n}(\delta_{10n}+\delta_{30n}q_{iT-1,n})1\{\gamma \geq q_{iT-1,n}>\gamma_{0n}\}]
					\end{pmatrix}.
				\end{equation*}
				Suppose $\gamma>\gamma_{0n}$, and the other case can be analyzed similarly.
				By Taylor expansion,
				\begin{multline*}
					E[z_{itn}(\delta_{10n}+\delta_{30n}q_{itn})1\{\gamma \geq q_{itn}>\gamma_{0n}\}] \\
					= E_{tn}[z_{itn}|\gamma_{0n}]f_{tn}(\gamma_{0n})\left\{(\delta_{10n}+\delta_{30n}\gamma_{0n})\cdot (\gamma-\gamma_{0n}) + \frac{\delta_{30n}}{2}(\gamma-\gamma_{0n})^{2}\right\} + R_{n},
				\end{multline*}
				where
				\begin{multline*}
					R_{n}=\frac{1}{2}\frac{d}{d\gamma}\left(E_{tn}[z_{itn}|\gamma]f_{tn}(\gamma)\right)|_{\gamma=\bar{\gamma}_{0n}}\times(\delta_{10n}+\delta_{30n}\bar{\gamma}_{0n})(\gamma-\gamma_{0n})^{2} \\
					+\frac{1}{2}\left\{
					E_{tn}[z_{itn}|\bar{\gamma}_{0n}]f_{tn}(\bar{\gamma}_{0n})-E_{tn}[z_{itn}|\gamma_{0n}]f_{tn}(\gamma_{0n})
					\right\}(\gamma-\gamma_{0n})^{2},
				\end{multline*}
				and $\bar{\gamma}_{0n} \in [\gamma_{0n},\gamma]$.
				Suppose $|\gamma-\gamma_{0n}|\leq h_{1}$.
				For sufficiently small $h_{1}>0$, there is $N$ such that if $n>N$, then
				$\|\frac{d}{d\gamma}\left(E_{tn}[z_{itn}|\gamma]f_{tn}(\gamma)\right)|_{\gamma=\bar{\gamma}_{0n}}\| \leq C_{1} < \infty$ for some $C_{1}<\infty$. 
                    There also exists $C_{2}<\infty$ such that
				$\delta_{10n}+\delta_{30n}\bar{\gamma}_{0n}\leq (\delta_{10n}+\delta_{30n}\gamma_{0n})+\sup_{n}|\delta_{30n}| h_{1} \leq (\delta_{10n}+\delta_{30n}\gamma_{0n})+C_{2}h_{1}$, and hence $\|\frac{d}{d\gamma}\left(E_{tn}[z_{itn}|\gamma]f_{tn}(\gamma)\right)|_{\gamma=\bar{\gamma}_{0n}}\times(\delta_{10n}+\delta_{30n}\bar{\gamma}_{0n})(\gamma-\gamma_{0n})^{2}\| \leq C_{1}((\delta_{10n}+\delta_{30n}\gamma_{0n})+C_{2}h_{1})h_{1}^{2}$ for sufficiently large $n$. Moreover, there exists $C_{3}<\infty$ such that $\|E_{tn}[z_{itn}|\bar{\gamma}_{0n}]f_{tn}(\bar{\gamma}_{0n})-E_{tn}[z_{itn}|\gamma_{0n}]f_{tn}(\gamma_{0n})\| \leq \sup_{\bar{\gamma}:|\gamma_{0n}-\bar{\gamma}|\leq h_{1}}\|\frac{d}{d\gamma}\left(E_{tn}[z_{itn}|\gamma]f_{tn}(\gamma)\right)\mid_{\gamma=\bar{\gamma}}\|h_{1} \leq C_{3} h_{1}$ for sufficiently small $h_{1}>0$ and sufficiently large $n$.
				Hence, $\|R_{n}\|< C ((\delta_{10n}+\delta_{30n}\gamma_{0n})h_{1}^{2}+h_{1}^{3})$ for some $C<\infty$ and for sufficiently small $h_{1}>0$ and sufficiently large $n$.
				Therfore, there exists $h_{1}>0$ such that if $|\gamma-\gamma_{0n}|\leq h_{1}$, then
				\begin{multline*}
					\Bigl\| E[z_{itn}(\delta_{10n}+\delta_{30n}q_{itn})1\{\gamma \geq q_{itn}>\gamma_{0n}\}] \Bigr.- E_{tn}[z_{itn}|\gamma_{0n}]f_{tn}(\gamma_{0n}) \\
					\left. \times\left\{(\delta_{10n}+\delta_{30n}\gamma_{0n})\cdot (\gamma-\gamma_{0n}) + \frac{\delta_{30n}}{2}(\gamma-\gamma_{0n})^{2}\right\} \right\| < C ((\delta_{10n}+\delta_{30n}\gamma_{0n})h_{1}^{2}+h_{1}^{3})
				\end{multline*}
				for some $C<\infty$ and for sufficiently large $n$.
				By similar computations for $E[z_{itn}(\delta_{10n}+\delta_{30n}q_{it-1,n})1\{\gamma \geq q_{it-1,n}>\gamma_{0n}\}]$, we can derive that there exists $h_{1}>0$ such that if $|\gamma-\gamma_{0n}|\leq h_{1}$, then $\left\|(M_{20n}(\gamma)-M_{20n})\delta_{0n} - \widetilde{H}_{n}[(\delta_{10n}+\delta_{30n}\gamma_{0n})(\gamma-\gamma_{0n})+\frac{\delta_{30n}}{2}(\gamma-\gamma_{0n})^{2}]\right\|< C((\delta_{10n}+\delta_{30n}\gamma_{0n})h_{1}^{2}+h_{1}^{3})$ for some $C<\infty$ and for sufficiently large $n$.
				
				Meanwhile, there exist $h_{1},h_{2}>0$ such that if $|\gamma-\gamma_{0n}|\leq h_{1}$ and $\|\alpha-\alpha_{0n}\|\leq h_{2}$, then $\|(M_{20n}(\gamma)-M_{20n})(\delta-\delta_{0n})\|<C h_{2}h_{1}$ for some $C<\infty$ and for sufficiently large $n$.
				This is because for sufficiently small $h_{1}>0$, $\|M_{20n}(\gamma)-M_{20n}\|<\sup_{\bar{\gamma}:|\bar{\gamma}-\gamma_{0n}|\leq h_{1}} \|\mathfrak{H}_{n}(\bar{\gamma})\| h_{1}$, where
				\begin{equation*}
					\mathfrak{H}_{n}(\gamma)=\begin{pmatrix}
						E_{t_{0}n}[z_{it_{0}n}(1,\gamma)|\gamma]f_{t_{0}n}(\gamma) - E_{t_{0}-1,n}[z_{it_{0}n}(1,\gamma)|\gamma]f_{t_{0}-1,n}(\gamma)\\
						\vdots\\
						E_{Tn}[z_{iTn}(1,\gamma)|\gamma]f_{Tn}(\gamma) - E_{T-1,n}[z_{iTn}(1,\gamma)|\gamma]f_{T-1,n}(\gamma)
					\end{pmatrix},
				\end{equation*}
				Note that if $h_{1}$ is sufficently small, $\sup_{\bar{\gamma}:|\bar{\gamma}-\gamma_{0n}|\leq h_{1}} \|\mathfrak{H}_{n}(\bar{\gamma})\|$ is bounded above by some nonnegative constant $C<\infty$, and $\|M_{20n}(\gamma)-M_{20n}\|<C h_{1}$.
				
				Hence, for any $\eta>0$, there exist $h_{1},h_{2}>0$ such that if $|\gamma-\gamma_{0n}|\leq h_{1}$ and $\|\alpha-\alpha_{0n}\|\leq h_{2}$, then 
				\begin{multline*}
					\Bigl\|(M_{20n}(\gamma)-M_{20n})[\delta_{0n}+(\delta-\delta_{0n})] \Bigr. \\
					\Bigl. - \widetilde{H}_{n}[(\delta_{10n}+\delta_{30n}\gamma_{0n})(\gamma-\gamma_{0n})+\frac{\delta_{30n}}{2}(\gamma-\gamma_{0n})^{2}]\Bigr\|< C(h_{1}h_{2}+(\delta_{10n}+\delta_{30n}\gamma_{0n})h_{1}^{2}+h_{1}^{3}),
				\end{multline*} 
				for some nonnegative $C<\infty$ and sufficiently large $n$.
				Therefore, for any $\eta>0$, we can set $h_{1}$ and $h_{2}$ sufficiently small such that $\sup_{|\gamma-\gamma_{0n}|\leq h_{1},\|\alpha-\alpha_{0n}\|\leq h_{2}}\sqrt{n}\|g_{0n}(\theta)-M_{0n}(\alpha-\alpha_{0n})-\widetilde{H}_{n}[(\delta_{10n}+\delta_{30n}\gamma_{0n})(\gamma-\gamma_{0n})+\frac{\delta_{30n}}{2}(\gamma-\gamma_{0n})^{2}]\|\leq \sqrt{n}(h_{2}+(\delta_{10n}+\delta_{30n}\gamma_{0n})h_{1}+h_{1}^{2})\eta$ for sufficiently large $n$, which completes the proof.
				
			\end{proof}
			
			\begin{lemma}
				\label{lem:unif.ULLN}
				Let $\{\phi_{0n}\in\Phi_{0}:n\geq 1 \}$ and $\pi_{n}(\phi_{0n})\rightarrow(\zeta_{1},\zeta_{2},\phi_{0,\infty})\in\Pi$.
				Then,
				\begin{equation*}
					\sup_{\gamma\in\Gamma}\|\bar{M}_{n}(\gamma)-M_{0n}(\gamma)\|\xrightarrow{p}0.
				\end{equation*}
			\end{lemma}
			
			\begin{proof}
				We show that the classes $\{z_{it}(1,q_{it})1\{q_{it}>\gamma\}:\gamma\in\Gamma\}$ and $\{z_{it}(1,q_{it-1})1\{q_{it-1}>\gamma\}:\gamma\in\Gamma\}$ are Glivenko-Cantelli uniformly in $\{P_{n}:n=1,2,...\}$, where $P_{n}$ is the probability law of $\omega_{in}=\{(z_{itn},y_{itn},x_{itn},\epsilon_{itn})_{t=1}^{T}\}$.
				We focus on the former class since the verification for the latter class is exactly identical.
				As it is sufficient to show that each element of $\{z_{it}(1,q_{it})1\{q_{it}>\gamma\}:\gamma\in\Gamma\}$, we additionally restrict our focus on $\mathcal{G}_{m\cdot index}=\{z_{it}q_{it}1\{q_{it}>\gamma\}:\gamma\in\Gamma\}$ and assume that $z_{it}$ is scalar without losing of generality.
				By Theorem 2.8.1 in \cite{van_der_vaart_weak_1996}, $\mathcal{G}_{m\cdot index}$ is Glivenko-Cantelli uniformly in $\{P_{n}\}$ if
				\begin{gather*}
					\sup_{n\in\mathbb{N}} E|G_{m\cdot index}(\omega_{in})|^{1+r} < \infty\text{ for some $r>0$, and} \\
					\sup_{Q}\log N(\varepsilon\|G_{m\cdot index}\|_{Q,1},\mathcal{G}_{m\cdot index},L_{1}(Q))<\infty \text{ for all $\varepsilon>0$},
				\end{gather*}
				where supremum is taken over all probability measures $Q$ such that $QG_{m\cdot index}<\infty$, and $G_{m\cdot index}=|z_{it}q_{it}|$ is an envelope of $\mathcal{G}_{m\cdot index}$.
				The first condition holds because $\sup_{n\in\mathbb{N}}E|z_{itn}q_{itn}|^{1+r} \leq \sup_{n\in\mathbb{N}} (E|z_{itn}|^{2+2r})^{1/2} (E|q_{itn}|^{2+2r})^{1/2} < C$ for some $C<\infty$ and $r>0$.
				The second condition holds as we have shown in the proof of \cref{lem:ULLN} that $\mathcal{G}_{m\cdot index}$ is a VC class that satisfies the uniform entropy condition. 
				Therefore, the ULLN with triangular array holds for $\{z_{it}q_{it}1\{q_{it}>\gamma\}:\gamma\in\Gamma\}$.
			\end{proof}
			
			\begin{lemma}
				\label{lem:unif.Wn}
				Let $\{\phi_{0n}\in\Phi_{0}:n\geq 1 \}$ and $\pi_{n}(\phi_{0n})\rightarrow(\zeta_{1},\zeta_{2},\phi_{0,\infty})\in\Pi$.
				Suppose that $|\hat{\theta}_{(1)n}-\theta_{0n}|\xrightarrow{p}0$.
				Then,
				\begin{equation*}
					\left\|\left\{\frac{1}{n}\sum_{i=1}^{n}[g(\omega_{in},\hat{\theta}_{(1)n})g(\omega_{in},\hat{\theta}_{(1)n})']-\bar{g}_{n}(\hat{\theta}_{(1)n})\bar{g}_{n}(\hat{\theta}_{(1)n})'\right\}-\Omega_{n}\right\|\xrightarrow{p}0,
				\end{equation*}
				where $\Omega_{n}=E[g(\omega_{in},\theta_{0n})g(\omega_{in},\theta_{0n})']-g_{0n}(\theta_{0n})g_{0n}(\theta_{0n})'$.
			\end{lemma}
			
			\begin{proof}
				We need to show $\|\bar{g}_{n}(\hat{\theta}_{(1)n})-g_{0n}(\theta_{0n})\|\xrightarrow{p}0$ and $\|\frac{1}{n}\sum_{i=1}^{n}g(\omega_{in},\hat{\theta}_{(1)n})g(\omega_{in},\hat{\theta}_{(1)n})'-E[g(\omega_{in},\theta_{0n})g(\omega_{in},\theta_{0n})']\|\xrightarrow{p}0$.
				$\mathcal{G}=\{g(\omega_{i},\theta):\theta\in\Theta\}$ is Glivenko-Cantelli class uniformly with respect to $\{P_{n}:n=1,2,...\}$, where $P_{n}$ is the probability law of $\omega_{in}=\{(z_{itn},y_{itn},x_{itn},\epsilon_{itn})_{t=1}^{T}\}$,
				as the proof of \cref{lem:unif.equicont} shows that the class is uniformly Donsker and pre-Gaussian.
				Therefore, $\|\bar{g}_{n}(\hat{\theta}_{(1)n})-g_{0n}(\theta_{0n})\|\xrightarrow{p}0$ when $|\hat{\theta}_{(1)n}-\theta_{0n}|\xrightarrow{p}0$.
				
				Let $\mathcal{G}^{2}=\{g(\omega_{i},\theta)g(\omega_{i},\theta)':\theta\in\Theta\}$.
				If $\mathcal{G}^{2}$ is Glivenko-Cantelli class uniformly with respect to $\{P_{n}\}$, then $\sup_{\theta\in\Theta}\|\frac{1}{n}\sum_{i=1}^{n}g(\omega_{in},\theta)g(\omega_{in},\theta)'-E[g(\omega_{in},\theta)g(\omega_{in},\theta)']\|\xrightarrow{p}0$.
				Then, $\|\frac{1}{n}\sum_{i=1}^{n}g(\omega_{in},\hat{\theta}_{(1)n})g(\omega_{in},\hat{\theta}_{(1)n})'-E[g(\omega_{in},\theta_{0n})g(\omega_{in},\theta_{0n})']\|\xrightarrow{p}0$ as $|\hat{\theta}_{(1)n}-\theta_{0n}|\xrightarrow{p}0$.
				By Theorem 2.8.1 in \cite{van_der_vaart_weak_1996}, $\mathcal{G}^{2}$ is Glivenko-Cantelli uniformly in $\{P_{n}\}$ if
				\begin{gather*}
					\sup_{n\in\mathbb{N}} E|G^{2}(\omega_{in})|^{1+r} < \infty\text{ for some $r>0$, and} \\
					\sup_{Q}\log N(\varepsilon\|G^{2}\|_{Q,1},\mathcal{G}^{2},L_{1}(Q))<\infty \text{ for all $\varepsilon>0$},
				\end{gather*}
				where supremum is taken over all probability measures $Q$ such that $QG^{2}<\infty$, and $G^{2}=[\sum_{t=1}^{T}\{C(\|z_{it}\Delta x_{it}\|+\|z_{it}(1,q_{it})'\|+\|z_{it}(1,q_{it-1})'\|)+\|z_{it}\Delta\epsilon_{it}\|\}]^{2}$ for some $C<\infty$ is an envelope of $\mathcal{G}^{2}$ as $G$ is an envelope of $\mathcal{G}$ as shown in the proof of \cref{lem:equicont}.
				The first condition $\sup_{n\in\mathbb{N}} E[G(\omega_{in})^{2+r}]<\infty$ holds because $\sup_{n\in\mathbb{N}}\max\{(E\|z_{itn}\|^{4+2r})^{1/2},(E\|x_{it-1,n}\|^{4+2r})^{1/2},(E\|x_{itn}\|^{4+2r})^{1/2},(E\|\Delta\epsilon_{itn}\|^{4+2r})^{1/2}\}<\infty$ for some $r>0$.
				The second condition holds because $\mathcal{G}$ satisfies the uniform entropy condition (see the proof of \cref{lem:equicont}) while pairwise product preserves uniform entropy condition, e.g., Theorem 2.10.20 in \cite{van_der_vaart_weak_1996}.

			\end{proof}

			\begin{lemma}
				\label{lem:unif.equicont}
				Let $\{\phi_{0n}\in\Phi_{0}:n\geq 1 \}$ and $\pi_{n}(\phi_{0n})\rightarrow(\zeta_{1},\zeta_{2},\phi_{0,\infty})\in\Pi$.
				If $h_{n}\rightarrow 0$, then
				\begin{equation*}
					\sup_{\|\theta_{1}-\theta_{2}\|<h_{n}}\sqrt{n}\|\bar{g}_{n}(\theta_{1})-\bar{g}_{n}(\theta_{2})-g_{0n}(\theta_{1})+g_{0n}(\theta_{2})\|=o_{p}(1).
				\end{equation*}
			\end{lemma}
			
			\begin{proof}
				Let $P_{n}$ be a probability law of $\omega_{in}=\{(z_{itn},y_{itn},x_{itn},\epsilon_{itn})_{t=1}^{T}\}$.
				We show that the class $\mathcal{G}=\{g(\omega_{i},\theta):\theta\in\Theta\}$ is pre-Gaussian uniformly in $\{P_{n}:n=1,2,...\}$ (see Section 2.8.2 in \cite{van_der_vaart_weak_1996} for its definition), which implies asymptotic equicontinuity uniform in $\{P_{n}\}$. 
				That is, for any $\epsilon>0$, $\sup_{m\in\mathbb{N}}P_{m}(\|\mathbb{G}_{n}\|_{\mathcal{G}_{h}}>\epsilon)\rightarrow 0$ if $h\rightarrow 0$ and $n\rightarrow\infty$, while $\mathcal{G}_{h}=\{g(\omega_{i},\theta_{1})-g(\omega_{i},\theta_{2}):\|\theta_{1}-\theta_{2}\|<h\}$.
				Let $G$ be an envelope of $\mathcal{G}$.
				By Theorem 2.8.3 in \cite{van_der_vaart_weak_1996}, it is sufficient to show that
				\begin{gather*}
					\sup_{n\in\mathbb{N}} E|G(\omega_{in})|^{2+r} < \infty\text{ for some $r>0$, and} \\
					\int_{0}^{\infty}\sup_{Q}\log N(\varepsilon\|G\|_{Q,2},\mathcal{G},L_{2}(Q))d\varepsilon<\infty,
				\end{gather*}
				where $Q$ ranges over all finitely discrete probability measures, which implies that $\mathcal{G}$ is Donsker and uniformly pre-Gaussian in $\{P_{n}\}$.
				
				Let $\widetilde{\mathcal{G}}^{(t)}=\{z_{it}\Delta\epsilon_{it}-z_{it}\Delta x_{it}\bar{\beta}-z_{it}1_{it}(\gamma_{1})'X_{it}\delta_{1}+z_{it}1_{it}(\gamma_{2})'X_{it}\delta_{2}:\|\bar{\beta}\|\leq K, \|\delta_{1}\|\leq K, \|\delta_{2}\|\leq K, \gamma_{1},\gamma_{2}\in\Gamma\}$.
				Suppose that $z_{it}$ is a scalar without losing of generality as it is sufficient to show the conditions hold for each element of $\mathcal{G}$.
				Note that $g_{t}(\omega_{i},\theta)=z_{it}(\Delta y_{it}-\Delta x_{it}'\beta-1_{it}(\gamma)'X_{it}\delta)=z_{it}\Delta\epsilon_{it}-z_{it}\Delta x_{it}(\beta-\beta_{0n})-z_{it}1_{it}(\gamma)'X_{it}\delta+z_{it}1_{it}(\gamma_{0n})'X_{it}\delta_{0n}$ is an element of $\widetilde{\mathcal{G}}^{(t)}$ for any $\theta_{0n}\in\Theta$.
				So it is sufficient to show $\widetilde{\mathcal{G}}^{(t)}$ is pre-Gaussian uniformly in $\{P_{n}\}$ instead of each element of $\mathcal{G}$.
				
				$\widetilde{G}(\omega_{i})=C(\|z_{it}\Delta x_{it}\| + \|z_{it}(1,q_{it})'\| + \|z_{it}(1,q_{it-1})'\|)+ \|z_{it}\Delta\epsilon_{it}\|$ is an envelope of $\widetilde{\mathcal{G}}^{(t)}$ for some $C<\infty$.
				The first condition for the uniform pre-Gaussianity $\sup_{n\in\mathbb{N}}E|\widetilde{G}(\omega_{in})|^{2+r} <\infty$ holds as $\sup_{n\in\mathbb{N}}\max\{(E\|z_{itn}\|^{4+2r})^{1/2},(E\|x_{itn}\|^{4+2r})^{1/2},(E\|x_{it-1,n}\|^{4+2r})^{1/2},(E\|\Delta \epsilon_{itn}\|^{4+2r})^{1/2}\}<\infty$ for some $r>0$.
				The second condition holds as $\widetilde{\mathcal{G}}^{(t)}$ is shown to satisfy the uniform entropy condition in the proof of \cref{lem:equicont}.
				
			\end{proof}
			
			\begin{lemma}
				\label{lem:unif.plim1}
				Let $\{\phi_{0n}\in\Phi_{0}:n\geq 1 \}$ and $\pi_{n}(\phi_{0n})\rightarrow(\zeta_{1},\zeta_{2},\phi_{0,\infty})\in\Pi$, and 
				suppose that $\zeta_{1}\neq\{\pm\infty\}$, and $\zeta_{2}=0$, i.e., it is (i) continuous or (ii) semi-continuous.
				Then,
				\begin{multline*}
					\frac{1}{\sqrt{n}}\sum_{i=1}^{n} z_{itn}(1_{itn}(\gamma_{0n})'-1_{itn}(\gamma_{0n}+\tfrac{b}{n^{\frac{1}{4}}})')X_{itn}\delta_{0n}\\ 
					\xrightarrow{p} \left\{E_{t,\infty}[z_{it,\infty}|\gamma_{0,\infty}]f_{t,\infty}(\gamma_{0,\infty})-E_{t-1,\infty}[z_{it,\infty}|\gamma_{0,\infty}]f_{t-1,\infty}(\gamma_{0,\infty})\right\}[\zeta_{1}b+\frac{\delta_{30,\infty}}{2}b^{2}]
				\end{multline*}
				uniformly over $b\in[-K,K]$ for any $K<\infty$.
			\end{lemma}
			
			\begin{proof}
				Note that
				\begin{align}
					\nonumber
					& \frac{1}{\sqrt{n}}\sum_{i=1}^{n} z_{itn}(1_{itn}(\gamma_{0n})'-1_{itn}(\gamma_{0n}+\tfrac{b}{n^{\frac{1}{4}}})')X_{itn}\delta_{0n}\\
					\label{eq:CLTlimit1.unif}
					&=\frac{1}{\sqrt{n}}\sum_{i=1}^{n} 
					\left\{
					z_{itn}(1_{itn}(\gamma_{0n})'-1_{itn}(\gamma_{0n}+\tfrac{b}{n^{\frac{1}{4}}})')X_{itn}\delta_{0n}-
					E[z_{itn}(1_{itn}(\gamma_{0n})'-1_{itn}(\gamma_{0n}+\tfrac{b}{n^{\frac{1}{4}}})')X_{itn}\delta_{0n}]
					\right\} \\
					\label{eq:deter.limit1.unif}
					&\quad+
					\sqrt{n}
					E[z_{itn}(1_{itn}(\gamma_{0n})'-1_{itn}(\gamma_{0n}+\tfrac{b}{n^{\frac{1}{4}}})')X_{itn}\delta_{0n}].
				\end{align}
				The stochastic term \eqref{eq:CLTlimit1.unif} converges in probability to zero uniformly with respect to $b\in[-K,K]$.
				This is because \cref{lem:unif.equicont} shows that when $h_{n}\downarrow0$, then
				\begin{equation*}
					\sup_{|\gamma-\gamma_{0n}|<h_{n}}\sqrt{n}\left\{\frac{1}{n}\sum_{i=1}^{n}z_{itn}(1_{itn}(\gamma_{0n})-1_{itn}(\gamma))'X_{itn}\delta_{0n}-E[z_{itn}(1_{itn}(\gamma_{0n})-1_{itn}(\gamma))'X_{itn}\delta_{0n}]\right\}=o_{p}(1)
				\end{equation*}
				as it can be expressed as $\sup_{|\gamma-\gamma_{0n}|<h_{n}}\| \bar{g}_{n}(\alpha_{0n},\gamma)- \bar{g}_{n}(\alpha_{0n},\gamma_{0n})-g_{0n}(\alpha_{0n},\gamma)+g_{0n}(\alpha_{0n},\gamma_{0n})\|$.
				
				Suppose $b>0$.
				The case for $b<0$ follows similarly.
				We will show that \eqref{eq:deter.limit1.unif} converges as follows:
				\begin{equation*}
					\begin{array}{l}
						\sqrt{n}E
						z_{itn}(1_{itn}(\gamma_{0n})'-1_{itn}(\gamma_{0n}+\tfrac{b}{n^{\frac{1}{4}}})')X_{itn}\delta_{0n} \\
						=\sqrt{n}\left\{
						E[z_{itn}(\delta_{10n}+\delta_{30n}q_{itn})1\{\gamma_{0n}+\tfrac{b}{n^\frac{1}{4}}\geq q_{itn}>\gamma_{0n}\}]
						\right. \\
						\left.\hspace{2cm}
						-E[z_{itn}(\delta_{10n}+\delta_{30n}q_{it-1,n})1\{\gamma_{0n}+\tfrac{b}{n^\frac{1}{4}}\geq q_{it-1,n}>\gamma_{0n}\}]
						\right\}
						\\
						\rightarrow 
						\left\{E_{t,\infty}[z_{it,\infty}|\gamma_{0,\infty}]f_{t,\infty}(\gamma_{0,\infty})-E_{t-1,\infty}[z_{it,\infty}|\gamma_{0,\infty}]f_{t-1,\infty}(\gamma_{0,\infty})\right\}
						[\zeta_{1}b+\frac{\delta_{30,\infty}}{2}b^{2}],
					\end{array}
				\end{equation*}
				uniformly with respect to $b\in[-K,K]$.
    
				Let
				\begin{equation*}
					\begin{array}{l}
						R_{n,b} = \left(\sqrt{n}E[z_{itn}(\delta_{10n}+\delta_{30n}q_{itn})1\{\gamma_{0n}+\tfrac{b}{n^\frac{1}{4}}\geq q_{itn}>\gamma_{0n}\}]
						\right.\\
						\left.\hspace{2cm}
						-\{E_{tn}[z_{itn}|\gamma_{0n}]f_{tn}(\gamma_{0n})\}(n^{1/4}(\delta_{10n}+\delta_{30n}\gamma_{0n})b+\frac{\delta_{30n}}{2}b^{2})\right),
					\end{array}
				\end{equation*}
                    which will be shown to converge to zero uniformly with respect to $b\in[-K,K]$.
				By Taylor epxansion, its formula can be derived as follows:
				\begin{multline*}
					R_{n,b} = \Bigl(\delta_{30n}\{E_{tn}[z_{itn}|\gamma_{n,b}]f_{tn}(\gamma_{n,b})-E_{tn}[z_{itn}|\gamma_{0n}]f_{tn}(\gamma_{0n})\}
					\Bigr.\\
					\Bigl. 
					+(\delta_{10n}+\delta_{30n}\gamma_{n,b})\frac{d}{d\gamma}\{E_{tn}[z_{itn}|\gamma]f_{tn}(\gamma)\}|_{\gamma=\gamma_{n,b}}
					\Bigl)\frac{b^{2}}{2},
				\end{multline*}
				where $\gamma_{n,b}\in[\gamma_{0n},\gamma_{0n}+\frac{b}{n^{1/4}}]$.
				Note that $|\gamma_{n,b}-\gamma_{0n}|\rightarrow 0$ unifromly with respect to $b\in[-K,K]$.
				Hence, for sufficiently large $n$, $\|\frac{d}{d\gamma}\{E_{tn}[z_{itn}|\gamma]f_{tn}(\gamma)\}|_{\gamma=\gamma_{n,b}}\|\leq C$ for some $C<\infty$. Moreover, $\delta_{10n}+\delta_{30n}\gamma_{n,b}\rightarrow 0$ and $E_{tn}[z_{itn}|\gamma_{0n}]f_{tn}(\gamma_{0n})-E_{tn}[z_{itn}|\gamma_{n,b}]f_{tn}(\gamma_{n,b})\rightarrow 0$ uniformly with respect to $b\in[-K,K]$.
				Therefore,
				$\|R_{n,b}\| \rightarrow 0$ uniformly with respect to $b\in[-K,K]$, i.e.,
				\begin{equation*}
					\begin{array}{l}
						\left(\sqrt{n}E[z_{itn}(\delta_{10n}+\delta_{30n}q_{itn})1\{\gamma_{0n}+\tfrac{b}{n^\frac{1}{4}}\geq q_{itn}>\gamma_{0n}\}]
						\right.\\
						\left.\hspace{2cm}
						-\{E_{tn}[z_{itn}|\gamma_{0n}]f_{tn}(\gamma_{0n})\}(n^{1/4}(\delta_{10n}+\delta_{30n}\gamma_{0n})b+\frac{\delta_{30n}}{2}b^{2})\right)\rightarrow 0
					\end{array}
				\end{equation*}
				uniformly with respect to $b\in[-K,K]$.
				We can derive a similar result for $\sqrt{n}E[z_{itn}(\delta_{10n}+\delta_{30n}q_{it-1,n})1\{\gamma_{0n}+\tfrac{b}{n^\frac{1}{4}}\geq q_{it-1,n}>\gamma_{0n}\}]$ that leads to
				\begin{equation*}
					\begin{array}{l}
						\Biggl\| 
						\sqrt{n}Ez_{itn}(1_{itn}(\gamma_{0n})'-1_{itn}(\gamma_{0n}+\tfrac{b}{n^{\frac{1}{4}}})')X_{itn}\delta_{0n} 
						\Biggr. \\
						\Biggl. 
						-\left\{E_{tn}[z_{itn}|\gamma_{0n}]f_{tn}(\gamma_{0n})-E_{t-1,n}[z_{itn}|\gamma_{0n}]f_{t-1,n}(\gamma_{0n})\right\}
						[n^{1/4}(\delta_{10n}+\delta_{30n}\gamma_{0n})b+\frac{\delta_{30n}}{2}b^{2}]
						\Biggr\|\rightarrow 0,
					\end{array}
				\end{equation*}
				uniformly with respect to $b\in[-K,K]$.
				As $\pi_{n}(\phi_{0n})\rightarrow(\zeta_{1},\zeta_{2},\phi_{0,\infty})$, 
				\begin{multline*}
					\left\{E_{tn}[z_{itn}|\gamma_{0n}]f_{tn}(\gamma_{0n})-E_{t-1,n}[z_{itn}|\gamma_{0n}]f_{t-1,n}(\gamma_{0n})\right\}
					[n^{1/4}(\delta_{10n}+\delta_{30n}\gamma_{0n})b+\frac{\delta_{30n}}{2}b^{2}]\\
					\rightarrow \left\{E_{t,\infty}[z_{it,\infty}|\gamma_{0,\infty}]f_{t,\infty}(\gamma_{0,\infty})-E_{t-1,\infty}[z_{it,\infty}|\gamma_{0,\infty}]f_{t-1,\infty}(\gamma_{0,\infty})\right\}[\zeta_{1}b+\frac{\delta_{30,\infty}}{2}b^{2}],
				\end{multline*}
				which completes the proof.
			\end{proof}
			
			\begin{lemma}
				\label{lem:unif.plim2}
				Let $\{\phi_{0n}\in\Phi_{0}:n\geq 1 \}$ and $\pi_{n}(\phi_{0n})\rightarrow(\zeta_{1},\zeta_{2},\phi_{0,\infty})\in\Pi$, and 
				suppose that $\zeta_{1}=\{\pm\infty\}$ and $\zeta_{2}=0$, i.e., it is (iii) semi-discontinuous.
				Then,
				\begin{multline*}
					\frac{1}{\sqrt{n}}\sum_{i=1}^{n} z_{itn}(1_{itn}(\gamma_{0n})'-1_{itn}(\gamma_{0n}+\tfrac{b}{\sqrt{n}(\delta_{10n}+\delta_{30n}\gamma_{0n})})')X_{itn}\delta_{0n}\\ 
					\xrightarrow{p} \left\{E_{t,\infty}[z_{it,\infty}|\gamma_{0,\infty}]f_{t,\infty}(\gamma_{0,\infty})-E_{t-1,\infty}[z_{it,\infty}|\gamma_{0,\infty}]f_{t-1,\infty}(\gamma_{0,\infty})\right\}b
				\end{multline*}
				uniformly over $b\in[-K,K]$ for any $K<\infty$.
			\end{lemma}
			
			\begin{proof}
				Note that
				\begin{align}
					\nonumber
					& \frac{1}{\sqrt{n}}\sum_{i=1}^{n} z_{itn}(1_{itn}(\gamma_{0n})'-1_{itn}(\gamma_{0n}+\tfrac{b}{\sqrt{n}(\delta_{10n}+\delta_{30n}\gamma_{0n})})')X_{itn}\delta_{0n}\\
					\nonumber
					&=\frac{1}{\sqrt{n}}\sum_{i=1}^{n} 
					\left\{
					z_{itn}(1_{itn}(\gamma_{0n})'-1_{itn}(\gamma_{0n}+\tfrac{b}{\sqrt{n}(\delta_{10n}+\delta_{30n}\gamma_{0n})})')X_{itn}\delta_{0n}
					\right. \\
					\label{eq:CLTlimit2.unif}
					& \left. \hspace{4cm} -
					E[z_{itn}(1_{itn}(\gamma_{0n})'-1_{itn}(\gamma_{0n}+\tfrac{b}{\sqrt{n}(\delta_{10n}+\delta_{30n}\gamma_{0n})})')X_{itn}\delta_{0n}]
					\right\} \\
					\label{eq:deter.limit2.unif}
					&
					\hspace{2cm}
					+\sqrt{n}
					E[z_{itn}(1_{itn}(\gamma_{0n})'-1_{itn}(\gamma_{0n}+\tfrac{b}{\sqrt{n}(\delta_{10n}+\delta_{30n}\gamma_{0n})})')X_{itn}\delta_{0n}].
				\end{align}
				The stochastic term \eqref{eq:CLTlimit2.unif} converges in probability to zero uniformly with respect to $b\in[-K,K]$ by \cref{lem:unif.equicont}, by an argument similar to the proof of \cref{lem:unif.plim1} that shows \eqref{eq:CLTlimit1.unif} converges to zero.
				
				Suppose $b>0$.
				The case for $b<0$ follows similarly.
				We will show that \eqref{eq:deter.limit2.unif} converges as follows:
				\begin{equation*}
					\begin{array}{l}
						\sqrt{n}E
						z_{itn}(1_{itn}(\gamma_{0n})'-1_{itn}(\gamma_{0n}+\tfrac{b}{\sqrt{n}(\delta_{10n}+\delta_{30n}\gamma_{0n})})')X_{itn}\delta_{0n} \\
						=\sqrt{n}\left\{
						E[z_{itn}(\delta_{10n}+\delta_{30n}q_{itn})1\{\gamma_{0n}+\tfrac{b}{\sqrt{n}(\delta_{10n}+\delta_{30n}\gamma_{0n})}\geq q_{itn}>\gamma_{0n}\}]
						\right. \\
						\left.\hspace{2cm}
						-E[z_{itn}(\delta_{10n}+\delta_{30n}q_{it-1,n})1\{\gamma_{0n}+\tfrac{b}{\sqrt{n}(\delta_{10n}+\delta_{30n}\gamma_{0n})}\geq q_{it-1,n}>\gamma_{0n}\}]
						\right\}
						\\
						\rightarrow 
						\left\{E_{t,\infty}[z_{it,\infty}|\gamma_{0,\infty}]f_{t,\infty}(\gamma_{0,\infty})-E_{t-1,\infty}[z_{it,\infty}|\gamma_{0,\infty}]f_{t-1,\infty}(\gamma_{0,\infty})\right\}
						b,
					\end{array}
				\end{equation*}
				uniformly with respect to $b\in[-K,K]$.
    
				Let
				\begin{equation*}
					R_{n,b} = \left(\sqrt{n}E[z_{itn}(\delta_{10n}+\delta_{30n}q_{itn})1\{\gamma_{0n}+\tfrac{b}{\sqrt{n}(\delta_{10n}+\delta_{30n}\gamma_{0n})}\geq q_{itn}>\gamma_{0n}\}]
					-\{E_{tn}[z_{itn}|\gamma_{0n}]f_{tn}(\gamma_{0n})\}b\right),
				\end{equation*}
                    which will be shown to converge to zero uniformly with respect to $b\in[-K,K]$.
				By Taylor expansion, its formula can be derived as follows:
				\begin{multline*}
					R_{n,b} = \frac{1}{\sqrt{n}(\delta_{10n}+\delta_{30n}\gamma_{0n})^{2}}\Bigl(\delta_{30n}\{E_{tn}[z_{itn}|\gamma_{n,b}]f_{tn}(\gamma_{n,b})\}
					\Bigr.\\
					\Bigl. 
					+(\delta_{10n}+\delta_{30n}\gamma_{n,b})\frac{d}{d\gamma}\{E_{tn}[z_{itn}|\gamma]f_{tn}(\gamma)\}|_{\gamma=\gamma_{n,b}}
					\Bigl)\frac{b^{2}}{2},
				\end{multline*}
				where $\gamma_{n,b}\in[\gamma_{0n},\gamma_{0n}+\frac{b}{\sqrt{n}(\delta_{10n}+\delta_{30n}\gamma_{0n})}]$.
				Note that $|\gamma_{n,b}-\gamma_{0n}|\rightarrow 0$ unifromly with respect to $b\in[-K,K]$.
				Hence, for sufficiently large $n$, $\|E_{tn}[z_{itn}|\gamma_{n,b}]f_{tn}(\gamma_{n,b})\|\leq C$ and $\|\frac{d}{d\gamma}\{E_{tn}[z_{itn}|\gamma]f_{tn}(\gamma)\}|_{\gamma=\gamma_{n,b}}\|\leq C$ for some $C<\infty$.
				Moreover, $\delta_{10n}+\delta_{30n}\gamma_{n,b}\rightarrow 0$ uniformly with respect to $b\in[-K,K]$.
				As $\sqrt{n}(\delta_{10n}+\delta_{30n}\gamma_{0n})^{2}\rightarrow\infty$,
				$\|R_{n,b}\| \rightarrow 0$ uniformly with respect to $b\in[-K,K]$, i.e.,
				\begin{equation*}
					\left(\sqrt{n}E[z_{itn}(\delta_{10n}+\delta_{30n}q_{itn})1\{\gamma_{0n}+\tfrac{b}{\sqrt{n}(\delta_{10n}+\delta_{30n}\gamma_{0n})}\geq q_{itn}>\gamma_{0n}\}]
					-\{E_{tn}[z_{itn}|\gamma_{0n}]f_{tn}(\gamma_{0n})\}b\right)\rightarrow 0
				\end{equation*}
				uniformly with respect to $b\in[-K,K]$.
				We can derive a similar result for $\sqrt{n}E[z_{itn}(\delta_{10n}+\delta_{30n}q_{it-1,n})1\{\gamma_{0n}+\tfrac{b}{\sqrt{n}(\delta_{10n}+\delta_{30n}\gamma_{0n})}\geq q_{it-1,n}>\gamma_{0n}\}]$ that leads to
				\begin{equation*}
					\begin{array}{l}
						\Biggl\| 
						\sqrt{n}Ez_{itn}(1_{itn}(\gamma_{0n})'-1_{itn}(\gamma_{0n}+\tfrac{b}{\sqrt{n}(\delta_{10n}+\delta_{30n}\gamma_{0n})})')X_{itn}\delta_{0n} 
						\Biggr. \\
						\Biggl. \hspace{4cm}
						-\left\{E_{tn}[z_{itn}|\gamma_{0n}]f_{tn}(\gamma_{0n})-E_{t-1,n}[z_{itn}|\gamma_{0n}]f_{t-1,n}(\gamma_{0n})\right\}
						b
						\Biggr\|\rightarrow 0,
					\end{array}
				\end{equation*}
				uniformly with respect to $b\in[-K,K]$. As $\pi_{n}(\phi_{0n})\rightarrow(\zeta_{1},\zeta_{2},\phi_{0,\infty})$, 
				\begin{multline*}
					\left\{E_{tn}[z_{itn}|\gamma_{0n}]f_{tn}(\gamma_{0n})-E_{t-1,n}[z_{itn}|\gamma_{0n}]f_{t-1,n}(\gamma_{0n})\right\}b\\
					\rightarrow \left\{E_{t,\infty}[z_{it,\infty}|\gamma_{0,\infty}]f_{t,\infty}(\gamma_{0,\infty})-E_{t-1,\infty}[z_{it,\infty}|\gamma_{0,\infty}]f_{t-1,\infty}(\gamma_{0,\infty})\right\}b,
				\end{multline*}
				which completes the proof.
			\end{proof}
			
			\begin{lemma}
				\label{lem:unif.boot res LLN}
				Let $\{\phi_{0n}\in\Phi_{0}:n\geq 1 \}$ and $\pi_{n}(\phi_{0n})\rightarrow(\zeta_{1},\zeta_{2},\phi_{0,\infty})\in\Pi$. Then,
				\begin{equation*}
					\hat{u}_{n}^{*}=\begin{pmatrix}
						\frac{1}{n}\sum_{i=1}^{n}z_{it_{0}n}^{*}\widehat{\Delta\epsilon}_{it_{0}n}^{*}\\
						\vdots\\
						\frac{1}{n}\sum_{i=1}^{n}z_{iTn}^{*}\widehat{\Delta\epsilon}_{iTn}^{*}
					\end{pmatrix}
					-
					\begin{pmatrix}
						\frac{1}{n}\sum_{i=1}^{n}z_{it_{0}n}\widehat{\Delta\epsilon}_{it_{0}n}\\
						\vdots\\
						\frac{1}{n}\sum_{i=1}^{n}z_{iTn}\widehat{\Delta\epsilon}_{iTn}
					\end{pmatrix}
					\xrightarrow{p^{*}}0 \text{ in $P$.}
				\end{equation*}
			\end{lemma}
			
			\begin{proof}
				Note that $\hat{u}_{n}^{*}=\frac{1}{n}\sum_{i=1}^{n}[g(\omega_{in}^{*},\hat{\theta}_{n})-E[g(\omega_{in},\hat{\theta}_{n})]]-\frac{1}{n}\sum_{i=1}^{n}[g(\omega_{in},\hat{\theta}_{n})-E[g(\omega_{in},\hat{\theta}_{n})]]$.
                    Let $P_{n}$ be the probability law of $\omega_{in}=\{(z_{itn},y_{itn},x_{itn},\epsilon_{itn})_{t=1}^{T}\}$.
				As $\mathcal{G}=\{g(\omega_{i},\theta):\theta\in\Theta\}$ is Glivenko-Cantelli uniformly in $\{P_{n}\}$, which is shown in the proof of \cref{lem:unif.equicont}, $\frac{1}{n}\sum_{i=1}^{n}[g(\omega_{in},\hat{\theta}_{n})-E[g(\omega_{in},\hat{\theta}_{n})]]$ is $o_{p}(1)$, and hence $o_{p}^{*}(1)$ in $P$ by \cref{lem:bootstrap order}.
				By \cref{prop:unif.boot GC}, $\frac{1}{n}\sum_{i=1}^{n}[g(\omega_{in}^{*},\hat{\theta}_{n})-E[g(\omega_{in},\hat{\theta}_{n})]]$ is also $o_{p}^{*}(1)$ in $P$, which completes the proof.
			\end{proof}
			
			\begin{lemma}
				\label{lem:unif.boot res CLT}
				Let $\{\phi_{0n}\in\Phi_{0}:n\geq 1 \}$ and $\pi_{n}(\phi_{0n})\rightarrow(\zeta_{1},\zeta_{2},\phi_{0,\infty})\in\Pi$. Then,
				\begin{equation*}
					\sqrt{n}\hat{u}_{n}^{*}
					=\sqrt{n}\left\{\begin{pmatrix}
						\frac{1}{n}\sum_{i=1}^{n}z_{it_{0}n}^{*}\widehat{\Delta\epsilon}_{it_{0}n}^{*}\\
						\vdots\\
						\frac{1}{n}\sum_{i=1}^{n}z_{iTn}^{*}\widehat{\Delta\epsilon}_{iTn}^{*}
					\end{pmatrix}
					-
					\begin{pmatrix}
						\frac{1}{n}\sum_{i=1}^{n}z_{it_{0}n}\widehat{\Delta\epsilon}_{it_{0}n}\\
						\vdots\\
						\frac{1}{n}\sum_{i=1}^{n}z_{iTn}\widehat{\Delta\epsilon}_{iTn}
					\end{pmatrix}\right\}
					\xrightarrow{d^{*}}N(0,\Omega_{\infty})\text{ in $P$.}
				\end{equation*}
			\end{lemma}
			
			\begin{proof}
				Note that $\sqrt{n}\hat{u}_{n}^{*}=\sqrt{n}\{\bar{g}_{n}^{*}(\hat{\theta}_{n})-\bar{g}_{n}^{*}(\theta_{0n})-\bar{g}_{n}(\hat{\theta}_{n})+\bar{g}_{n}(\theta_{0n})\}+\sqrt{n}\{\bar{g}_{n}^{*}(\theta_{0n})-\bar{g}_{n}(\theta_{0n})\}$.
				As $\|\hat{\theta}_{n}-\theta_{0n}\|=o_{p}(1)$ and $o_{p}^{*}(1)$ in $P$ by \cref{lem:bootstrap order}, $\sqrt{n}\{\bar{g}_{n}^{*}(\hat{\theta}_{n})-\bar{g}_{n}^{*}(\theta_{0n})-\bar{g}_{n}(\hat{\theta}_{n})+\bar{g}_{n}(\theta_{0n})\}$ is $o_{p}^{*}(1)$ in $P$.
				By applying \cref{lem:VW3.6.15}, $\sqrt{n}\lambda'\{\bar{g}_{n}^{*}(\theta_{0n})-\bar{g}_{n}(\theta_{0n})\}\xrightarrow{d^{*}}N(0,\lambda'\Omega_{\infty}\lambda)$ in $P$ for any real vector $\lambda$. 
				By Cram\'er-Wold, $\sqrt{n}\{\bar{g}_{n}^{*}(\theta_{0n})-\bar{g}_{n}(\theta_{0n})\}\xrightarrow{d^{*}}N(0,\Omega_{\infty})$ in $P$, and applying Slutsky theorem completes the proof.
			\end{proof}

			The \cref{lem:unif.boot ULLN} states uniform bootstrap probability limit of the following matrix:
			\begin{equation*}
				\bar{M}_{n}^{*}(\gamma) = \frac{1}{n}\sum_{i=1}^{n}
				\begin{pmatrix}
					z_{it_{0}n}^{*}\Delta x_{it_{0}n}^{*\prime} & z_{it_{0}n}^{*}1_{it_{0}n}^{*}(\gamma)'X_{it_{0}n}^{*} \\
					\vdots & \vdots \\
					z_{iTn}^{*}\Delta x_{iTn}^{*\prime} & z_{iTn}^{*}1_{iTn}^{*}(\gamma)'X_{iTn}^{*} 
				\end{pmatrix}.
			\end{equation*}
			
			\begin{lemma}
				\label{lem:unif.boot ULLN}
				Let $\{\phi_{0n}\in\Phi_{0}:n\geq 1 \}$ and $\pi_{n}(\phi_{0n})\rightarrow(\zeta_{1},\zeta_{2},\phi_{0,\infty})\in\Pi$.
				Then,
				\begin{equation*}
					\sup_{\gamma\in\Gamma}\|\bar{M}_{n}^{*}(\gamma)-M_{0n}(\gamma)\|\xrightarrow{p^{*}}0\text{ in $P$}.
				\end{equation*}
			\end{lemma}
			
			\begin{proof}	
				We apply \cref{prop:unif.boot GC} to prove the result. 
				First, we need to show that $\{z_{it}(1,q_{it})1\{q_{it}>\gamma\}:\gamma\in\Gamma\}$ and $\{z_{it}(1,q_{it-1})1\{q_{it-1}>\gamma\}:\gamma\in\Gamma\}$ are Glivenko-Cantelli uniformly in $\{P_{n}:n=1,2,...\}$, where $P_{n}$ is the probability law of $\omega_{in}=\{(z_{itn},y_{itn},x_{itn},\epsilon_{itn})_{t=1}^{T}\}$.
				It is shown in \cref{lem:unif.ULLN} that the functional classes are Glivenko-Cantelli uniformly in $\{P_{n}\}$.
				Second, the condition for envelope holds as $\sup_{n\in\mathbb{N}}E[\|z_{itn}(1,q_{itn})\|+\|z_{itn}(1,q_{it-1,n})\|]<\infty$, which is implied by $\sup_{n\in\mathbb{N}}\max\{(E\|z_{itn}\|^{2+r})^{1/2},(E\|q_{itn}\|^{2+r})^{1/2},(E\|q_{it-1,n}\|^{2+r})^{1/2}\}<\infty$ for some $r>0$.
			\end{proof}
			
			\begin{lemma}
				\label{lem:unif.boot.equicont}
				Let $\{\phi_{0n}\in\Phi_{0}:n\geq 1 \}$ and $\pi_{n}(\phi_{0n})\rightarrow(\zeta_{1},\zeta_{2},\phi_{0,\infty})\in\Pi$.
				If $h_{n}\rightarrow 0$, then
				\begin{equation*}
					\sup_{\|\theta_{1}-\theta_{2}\|<h_{n}}\sqrt{n}\|\bar{g}_{n}^{*}(\theta_{1})-\bar{g}_{n}^{*}(\theta_{2})-\bar{g}_{n}(\theta_{1})+\bar{g}_{n}(\theta_{2})\|=o_{p}^{*}(1)\text{ in $P$.}
				\end{equation*}
			\end{lemma}
			
			\begin{proof}
				Note that $\bar{g}_{n}^{*}(\theta_{1})-\bar{g}_{n}^{*}(\theta_{2})=\frac{1}{n}\sum_{i=1}^{n}(g(\omega_{in}^{*},\theta_{1})-g(\omega_{in}^{*},\theta_{2}))$ because $g_{in}^{*}(\theta)=g(\omega_{in}^{*},\theta)-g(\omega_{in}^{*},\theta_{0n}^{*})+g(\omega_{in}^{*},\hat{\theta}_{n})$, see \eqref{eq:unif.boot.momt}.
				Therefore, $\sqrt{n}\{\bar{g}_{n}^{*}(\theta_{1})-\bar{g}_{n}^{*}(\theta_{2})-\bar{g}_{n}(\theta_{1})+\bar{g}_{n}(\theta_{2})\}=\frac{1}{\sqrt{n}}\sum_{i=1}^{n}\{g(\omega_{in}^{*},\theta_{1})-g(\omega_{in}^{*},\theta_{2})-g(\omega_{in},\theta_{1})+g(\omega_{in},\theta_{2})\}$.
				Let $\widehat{\mathbb{G}}_{n}=\frac{1}{\sqrt{n}}\sum_{i=1}^{n}(\delta_{\omega_{in}^{*}}-\mathbb{P}_{n})$ and $\mathbb{P}_{n}=n^{-1}\sum_{i=1}^{n}\delta_{\omega_{in}}$, where $\delta_{\omega_{in}^{*}}$ and $\delta_{\omega_{in}}$ are dirac measures at $\omega_{in}^{*}$ and $\omega_{in}$.
				Then, it is sufficient to prove $\|\widehat{\mathbb{G}}_{n}\|_{\mathcal{G}_{h}}= o_{p}^{*}(1)$ in $P$ if $h\rightarrow0$ and $n\rightarrow\infty$
				
				For $h>0$, let $\mathcal{G}_{h}=\{g(\omega_{i},\theta_{1})-g(\omega_{i},\theta_{2}):\|\theta_{1}-\theta_{2}\|\leq h\}$ and $G_{h}$ be its envelope.
				Let $\widetilde{N}_{1},\widetilde{N}_{2},...$ be symmetrized Poisson random variables with parameter $1/2$.
				By \cref{lem:VW3.6.6},
				\begin{equation*}
					E^{*}\|\widehat{\mathbb{G}}_{n}\|_{\mathcal{G}_{h}} \leq 4E_{\widetilde{N}}\|\frac{1}{\sqrt{n}}\sum_{i=1}^{n}\widetilde{N}_{i}\delta_{\omega_{in}}\|_{\mathcal{G}_{h}}
				\end{equation*}
				conditionally on $\{\omega_{in}:1\leq i\leq n\}$.
				For all $1\leq n_{0}\leq n$, the last display is stochastically bounded upto constant by
				\begin{equation}
					\label{eq:unif.equicont.bound}
					(n_{0}-1)E_{\widetilde{N}}\max_{1\leq i\leq n}\frac{\widetilde{N}_{i}}{\sqrt{n}}PG(\omega_{in}) + \|\widetilde{N}_{1}\|_{2,1}\max_{n_{0}\leq j\leq n} E\|\frac{1}{\sqrt{j}}\sum_{i=n_{0}}^{j}\varepsilon_{i}\delta_{\omega_{in}}\|_{\mathcal{G}_{h}},
				\end{equation}
				by \cref{lem:VW2.9.1},
				where $G(\cdot)$ is an envelope function of $\mathcal{G}$.
				The first term is bounded above by $(n_{0}-1)2\sqrt{2}n^{-1/4}$, which converges to zero for any $n_{0}$ as $n\rightarrow\infty$, and $\|\widetilde{N}_{1}\|_{2,1}\leq 2\sqrt{2}$ (see proof of Theorem 3.6.3 in \cite{van_der_vaart_weak_1996}).
				By triangle inequality,
				\begin{align*}
					\max_{n_{0}\leq j\leq n}E\|\frac{1}{\sqrt{j}}\sum_{i=n_{0}}^{j}\varepsilon_{i}\delta_{\omega_{in}}\|_{\mathcal{G}_{h}}
					& \leq \max_{n_{0}\leq j\leq n} E\left(\|\frac{1}{\sqrt{j}}\sum_{i=1}^{j}\varepsilon_{i}\delta_{\omega_{in}}\|_{\mathcal{G}_{h}}+\|\frac{1}{\sqrt{j}}\sum_{i=1}^{n_{0}-1}\varepsilon_{i}\delta_{\omega_{in}}\|_{\mathcal{G}_{h}}\right)
					\\
					& \leq 2\max_{n_{0}-1\leq j\leq n}E\|\frac{1}{\sqrt{j}}\sum_{i=1}^{j}\varepsilon_{i}\delta_{\omega_{in}}\|_{\mathcal{G}_{h}},
				\end{align*}
				and the last display is bounded upto constant by
				\begin{multline*}
					\max_{n_{0}-1\leq j\leq n}\left(
					E\sup_{\|\theta_{1}-\theta_{2}\|\leq h}\|\frac{1}{\sqrt{j}}\sum_{i=1}^{j}\varepsilon_{i}(g(\omega_{in},\theta_{1})-g(\omega_{in},\theta_{2})-E[g(\omega_{in},\theta_{1})]+E[g(\omega_{in},\theta_{2})])\|
					\right.
					\\
					\left.+E\sup_{\|\theta_{1}-\theta_{2}\|\leq h}\|\frac{1}{\sqrt{j}}\sum_{i=1}^{j}\varepsilon_{i}(E[g(\omega_{in},\theta_{1})]-E[g(\omega_{in},\theta_{2})])\|
					\right).
				\end{multline*}
				For each $j$, by \cref{lem:VW2.3.6},
				\begin{multline*}
					E\sup_{\|\theta_{1}-\theta_{2}\|\leq h}\|\frac{1}{\sqrt{j}}\sum_{i=1}^{j}\varepsilon_{i}(g(\omega_{in},\theta_{1})-g(\omega_{in},\theta_{2})-E[g(\omega_{in},\theta_{1})]+E[g(\omega_{in},\theta_{2})])\|
					\\
					\leq 2 E\sup_{\|\theta_{1}-\theta_{2}\|\leq h}\|\frac{1}{\sqrt{j}}\sum_{i=1}^{j}(g(\omega_{in},\theta_{1})-g(\omega_{in},\theta_{2})-E[g(\omega_{in},\theta_{1})]+E[g(\omega_{in},\theta_{2})])\|.
				\end{multline*}
				The right hand side of the last display converges to zero uniformly with respect to $n$ as $j\rightarrow\infty$ and $h\rightarrow 0$ since the functional class $\mathcal{G}$ is shown to be pre-Gaussian uniformly in $\{P_{n}\}$ in the proof of \cref{lem:unif.equicont}.
				
				For each $j$,
				\begin{equation*}
					E\sup_{\|\theta_{1}-\theta_{2}\|\leq h}\|\frac{1}{\sqrt{j}}\sum_{i=1}^{j}\varepsilon_{i}(E[g(\omega_{in},\theta_{1})]-E[g(\omega_{in},\theta_{2})])\|
					\leq
					E|\frac{1}{\sqrt{j}}\sum_{i=1}^{j}\varepsilon_{i}|\cdot (E\|G_{h}(\omega_{in})\|),
				\end{equation*}
				and $E|\frac{1}{\sqrt{j}}\sum_{i=1}^{j}\varepsilon_{i}| <\infty$ by Hoeffding's inequality, e.g., Lemma 2.2.7 in \cite{van_der_vaart_weak_1996}.
				The following paragaph shows that $E\|G_{h}(\omega_{in})\|\rightarrow 0$ as $h\rightarrow0$ and $n\rightarrow\infty$.
				
				As it is sufficient to consider each element of $\mathcal{G}$, we focus on $g_{t}(\omega_{i},\theta)$, the $t$th term of $g(\omega_{i},\theta)$, and assume that $g_{t}(\omega_{i},\theta)$ is a scalar without losing of generality.  
				Note that
				\begin{align*}
					g_{t}(\omega_{i},\theta_{1})-g_{t}(\omega_{i},\theta_{2}) 
					& =
					-z_{it}\Delta x_{it}'(\beta_{1}-\beta_{2}) -z_{it}1_{it}(\gamma_{1}) 'X_{it}(\delta_{1}-\delta_{2}) \\
					&\quad 
					+z_{it}(1_{it}(\gamma_{2})'-1_{it}(\gamma_{1})')X_{it}\delta_{2}.
				\end{align*}
				Without losing of generality, let $\gamma_{1}\geq \gamma_{2}$, and $K$ be a constant such that $\|\theta\|\leq K/2$ for $\theta\in\Theta$. 
				Set
				\begin{multline*}
					G_{h,t}(\omega_{i}) = \|z_{it}\Delta x_{it}'\|\cdot h + (\|z_{it}(1,q_{it})\|+\|z_{it}(1,q_{it-1})\|)\cdot h \\
					+ K(\|z_{it}(1,q_{it})1\{\gamma_{1}\geq q_{it} > \gamma_{2}\}\|+\|z_{it}(1,q_{it-1})1\{\gamma_{1}\geq q_{it-1} > \gamma_{2}\}\|),
				\end{multline*}
				which is an envelope of $\{g_{t}(\omega_{i},\theta_{1})-g_{t}(\omega_{i},\theta_{2}):\|\theta_{1}-\theta_{2}\|<h\}$.
				$\sup_{n\in\mathbb{N}}E[\|z_{itn}\Delta x_{itn}'\|+ \|z_{itn}(1,q_{itn})\|+\|z_{itn}(1,q_{it-1,n})\|]<\infty$.
				Furthermore,
				\begin{equation*}
					E\|z_{itn}(1,q_{itn})1\{\gamma_{1}\geq q_{itn}> \gamma_{2}\}\| \leq (E\|z_{itn}(1,q_{itn})\|^{2})^{1/2}(E1\{\gamma_{1}\geq q_{itn}> \gamma_{2} \})^{1/2},
				\end{equation*}
				while $\sup_{n\in\mathbb{N}}(E\|z_{itn}(1,q_{itn})\|^{2})^{1/2}<\infty$, and
				\begin{equation*}
					E1\{\gamma_{1}\geq q_{itn}> \gamma_{2} \} = \int_{\gamma_{2}}^{\gamma_{1}}f_{tn}(q)dq = (\gamma_{1}-\gamma_{2})f_{tn}(\bar{\gamma})
				\end{equation*}
				for some $\bar{\gamma}\in[\gamma_{2},\gamma_{1}]$. 
				Hence, $E1\{\gamma_{1}\geq q_{itn}> \gamma_{2} \}<Ch$ for some $C<\infty$ uniformly over all $n$.
				Therefore, $E|G_{h,t}(\omega_{in})| < C\sqrt{h}$ for some $C<\infty$ and converges to zero as $h\rightarrow 0$.
				
				Recall that the first term in \eqref{eq:unif.equicont.bound} goes to zero for any fixed $n_{0}$ when $n\rightarrow\infty$.
				The second term in \eqref{eq:unif.equicont.bound} is bounded by $2\sqrt{2}\max_{n_{0}\leq j\leq n}Z_{jn}$, where $Z_{jn}=E\|\frac{1}{\sqrt{j}}\sum_{i=n_{0}}^{j}\varepsilon_{i}\delta_{\omega_{in}}\|_{\mathcal{G}_{h}}$. 
				It is shown in the previous paragraph that $Z_{jn}\rightarrow 0$ uniformly with respect to $n$ as $j\rightarrow\infty$ and $h\rightarrow0$.
				Therefore, for any $\epsilon>0$, there exists $n_{0}<\infty$ such that $\max_{n_{0}\leq j\leq n}Z_{jn}<\epsilon/2$ for all $n>n_{0}$.
				Then, there exists $N(n_{0})$ large enough such that the first term in \eqref{eq:unif.equicont.bound} is bounded by $\epsilon/2$ for $n>N(n_{0})$.
				In conclusion, $E^{*}\|\widehat{\mathbb{G}}_{n}\|_{\mathcal{G}_{h}}\rightarrow 0$ if $h\rightarrow0$ and $n\rightarrow\infty$.
				By applying the Markov inequality, we can complete the proof.
			\end{proof}

			\begin{lemma}
				\label{lem:unif.boot.plim1}
				Let $\{\phi_{0n}\in\Phi_{0}:n\geq 1 \}$ and $\pi_{n}(\phi_{0n})\rightarrow(\zeta_{1},\zeta_{2},\phi_{0,\infty})\in\Pi$, and 
				suppose that $\zeta_{1}\neq\{\pm\infty\}$, and $\zeta_{2}=0$, i.e., it is (i) continuous or (ii) semi-continuous.
				Then, for any $K<\infty$,
				\begin{multline*}
					\sup_{b\in[-K,K]}\Biggl\|\frac{1}{\sqrt{n}}\sum_{i=1}^{n} z_{itn}^{*}(1_{itn}^{*}(\gamma_{0n})'-1_{itn}^{*}(\gamma_{0n}+\tfrac{b}{n^{\frac{1}{4}}})')X_{itn}^{*}\delta_{0n}^{*}\\ 
					- \left\{E_{t,\infty}[z_{it,\infty}|\gamma_{0,\infty}]f_{t,\infty}(\gamma_{0,\infty})-E_{t-1,\infty}[z_{it,\infty}|\gamma_{0,\infty}]f_{t-1,\infty}(\gamma_{0,\infty})\right\}[\zeta_{1}b+\frac{\delta_{30,\infty}}{2}b^{2}]\Biggr\|
				\end{multline*}
				is $o_{p}^{*}(1)$ in $P$.
			\end{lemma}
			
			\begin{proof}
				As the proof is quite similar to the proofs of \cref{lem:boot plim1} and \cref{lem:unif.plim1}, we just explain direction of the proof heuristically.
				As $\delta_{0n}^{*}=\hat{\delta}_{n}(\gamma_{0n})$ is consistent to $\delta_{0n}$,
				\begin{equation*}
					\sup_{b\in[-K,K]}\left\|\frac{1}{\sqrt{n}}\sum_{i=1}^{n} z_{itn}^{*}(1_{itn}^{*}(\gamma_{0n})'-1_{itn}^{*}(\gamma_{0n}+\tfrac{b}{n^{\frac{1}{4}}})')X_{itn}^{*}(\delta_{0n}^{*}-\delta_{0n})\right\|=o_{p}^{*}(1)\text{ in $P$.}
				\end{equation*}
				By \cref{lem:unif.boot.equicont},
				\begin{multline*}
					\sup_{b\in[-K,K]}\Biggl\|\frac{1}{\sqrt{n}}\sum_{i=1}^{n} z_{itn}^{*}(1_{itn}^{*}(\gamma_{0n})'-1_{itn}^{*}(\gamma_{0n}+\tfrac{b}{n^{\frac{1}{4}}})')X_{itn}^{*}\delta_{0n}\\ 
					- \frac{1}{\sqrt{n}}\sum_{i=1}^{n} z_{itn}(1_{itn}(\gamma_{0n})'-1_{itn}(\gamma_{0n}+\tfrac{b}{n^{\frac{1}{4}}})')X_{itn}\delta_{0n}\Biggr\|=o_{p}^{*}(1)\text{ in $P$,}
				\end{multline*}
				as the last display can be expressed by $\sqrt{n}\|\bar{g}_{n}^{*}(\alpha_{0n},\gamma_{0n}+\frac{b}{n^{1/4}})-\bar{g}_{n}^{*}(\alpha_{0n},\gamma_{0n})-\bar{g}_{n}(\alpha_{0n},\gamma_{0n}+\frac{b}{n^{1/4}})+\bar{g}_{n}(\alpha_{0n},\gamma_{0n})\|$.
				Hence, 
				\begin{multline*}
					\sup_{b\in[-K,K]}\Biggl\|\frac{1}{\sqrt{n}}\sum_{i=1}^{n} z_{itn}^{*}(1_{itn}^{*}(\gamma_{0n})'-1_{itn}^{*}(\gamma_{0n}+\tfrac{b}{n^{\frac{1}{4}}})')X_{itn}^{*}\delta_{0n}^{*}\\ 
					- \frac{1}{\sqrt{n}}\sum_{i=1}^{n} z_{itn}(1_{itn}(\gamma_{0n})'-1_{itn}(\gamma_{0n}+\tfrac{b}{n^{\frac{1}{4}}})')X_{itn}\delta_{0n}\Biggr\|=o_{p}^{*}(1)\text{ in $P$,}
				\end{multline*}
				and applying \cref{lem:unif.plim1} completes the proof.
				
			\end{proof}
			
			\begin{lemma}
				\label{lem:unif.boot.plim2}
				Let $\{\phi_{0n}\in\Phi_{0}:n\geq 1 \}$ and $\pi_{n}(\phi_{0n})\rightarrow(\zeta_{1},\zeta_{2},\phi_{0,\infty})\in\Pi$, and 
				suppose that $\zeta_{1}=\{\pm\infty\}$ and $\zeta_{2}=0$, i.e., it is (iii) semi-discontinuous.
				Then, for any $K<\infty$,
				\begin{multline*}
					\sup_{b\in[-K,K]}\Biggl\|\frac{1}{\sqrt{n}}\sum_{i=1}^{n} z_{itn}^{*}(1_{itn}^{*}(\gamma_{0n})'-1_{itn}^{*}(\gamma_{0n}+\tfrac{b}{\sqrt{n}(\delta_{10n}+\delta_{30n}\gamma_{0n})})')X_{itn}^{*}\delta_{0n}^{*}\\ 
					- \left\{E_{t,\infty}[z_{it,\infty}|\gamma_{0,\infty}]f_{t,\infty}(\gamma_{0,\infty})-E_{t-1,\infty}[z_{it,\infty}|\gamma_{0,\infty}]f_{t-1,\infty}(\gamma_{0,\infty})\right\}b\Biggr\|
				\end{multline*}
				is $o_{p}^{*}(1)$ in $P$.
			\end{lemma}
			
			\begin{proof}
				We omit the proof as it is almost identical to the proof of \cref{lem:unif.boot.plim1}.
			\end{proof}
			
			The following proposition is bootstrap Glivenko-Cantelli theorem uniform in underlying probability measures $P\in\{P_{1},P_{2},...\}$.
			
			\begin{proposition}
				\label{prop:unif.boot GC}
				Let $\{X_{in}:1\leq i\leq n, n=1,2,...\}$ be a triangular array of random elements in a measurable space $(\mathcal{X},\mathcal{A})$ while $X_{in}$'s are independent to each other with probability law $P_{n}$, and $\mathcal{F}$ be a class of functions on $(\mathcal{X},\mathcal{A})$ with an envelope $F$.
				Suppose that $\mathcal{F}$ is a Glivenko-Cantelli class uniformly in $P\in\{P_{m}\}$, and $\sup_{n\in\mathbb{N}}P_{n}F<\infty$.
				For each $n$, let $W=(W_{1n},...,W_{nn})$ be an exchangeable nonnegative random vector independent of $X_{1n},X_{2n},...,X_{nn}$ such that $\sum_{i=1}^{n}W_{in}=1$ and $\max_{1\leq i\leq n}|W_{in}|$ converges to zero in probability.
				Then, for every $\epsilon >0$ and $\eta>0$, as $n\rightarrow\infty$,
				\begin{equation*}
					P_{n}\left(
					P_{W}\left(
					\|\sum_{i=1}^{n}W_{in}(\delta_{X_{in}}-P_{n})\|_{\mathcal{F}}>\epsilon
					\right)>\eta
					\right)\rightarrow 0,
				\end{equation*}
				where $\delta_{X_{in}}$ is a dirac measure at $X_{in}$.
			\end{proposition}
			
			Let $W=(W_{1n},...,W_{nn})$ be a multinomial vector divided by $n$ with parameters $n$ and probabilities $(1/n,...,1/n)$, which satisfies $\sum_{i=1}^{n}W_{in}=1$ and $\max_{1\leq i\leq n}|W_{in}|$ converges to zero in probability.
			Suppose that $\widehat{X}_{1n},...,\widehat{X}_{nn}$ are i.i.d. resampling draws from $\{X_{1n},...,X_{nn}\}$.
			Then, $\frac{1}{n}\sum_{i=1}^{n}(\delta_{\widehat{X}_{in}}-P_{n})=\sum_{i=1}^{n}W_{in}(\delta_{X_{in}}-P_{n})$, and the probability law of $W$ can be identified with the probability law of the empirical bootstrap conditional on the data.
			
			\begin{proof}
				Let $Z_{in}=(\delta_{X_{in}}-P_{n})$.
				By \cref{lem:VW3.6.7},
				\begin{multline}
					\label{eq:unif.boot.ulln}
					E_{W}\|\sum_{i=1}^{n}W_{in}Z_{in}\|_{\mathcal{F}} \leq 2(n_{0}-1)\frac{1}{n}\sum_{i=1}^{n}\|Z_{in}\|_{\mathcal{F}} E_{W}\max_{1\leq i\leq n}|W_{in}|
					\\ + 2n\|W_{1n}\|_{2,1}\max_{n_{0}\leq k\leq n} E_{R}\|\frac{1}{k}\sum_{i=n_{0}}^{k}Z_{R_{i}n}\|_{\mathcal{F}}.
				\end{multline}
				Note that $\frac{1}{n}\sum_{i=1}^{n}\|Z_{in}\|_{\mathcal{F}} \leq \frac{1}{n}\sum_{i=1}^{n}Z_{in}(F) \leq (\mathbb{P}_{n}-P_{n})F+2 P_{n}F$, while $(\mathbb{P}_{n}-P_{n})F\xrightarrow{p} 0$ and $\limsup_{n}\|P_{n}\|_{\mathcal{F}} \leq \limsup_{n} P_{n}F <\infty$.
				Moreover, $E_{W}\max_{1\leq i\leq n}|W_{in}|\rightarrow 0$ by dominated convergence theorem because $|W_{in}|\leq 1$.
				Hence, the first term in the right hand side of \eqref{eq:unif.boot.ulln} converges to zero in probability for fixed $n_{0}$ as $n\rightarrow\infty$.
				That is, for any $\epsilon>0$ and $n_{0}<\infty$,
				\begin{equation*}
					P_{n}\left(
					\left|(n_{0}-1)\frac{1}{n}\sum_{i=1}^{n}\|Z_{in}\|_{\mathcal{F}} E_{W}\max_{1\leq i\leq n}|W_{in}|\right| > \epsilon
					\right) \rightarrow 0\text{ as $n\rightarrow\infty$.}
				\end{equation*}
				
				Note that $n\|W_{1n}\|_{2,1} \leq n(EW_{1n}) = 1$ (see the proof of Theorem 3.6.16 in \cite{van_der_vaart_weak_1996}).
				Finally, we need to show $\max_{n_{0}\leq k\leq n} E_{R}\|\frac{1}{k}\sum_{i=n_{0}}^{k}Z_{R_{i}n}\|_{\mathcal{F}}\xrightarrow{p} 0$.
				By triangle inequality,
				\begin{align*}
					\max_{n_{0}\leq k\leq n}E_{R}\|\frac{1}{k}\sum_{i=n_{0}}^{k}Z_{R_{i}n}\|_{\mathcal{F}} 
					&\leq 
					\max_{n_{0}\leq k\leq n}\left\{E_{R}\|\frac{1}{k}\sum_{i=1}^{k}Z_{R_{i}n}\|_{\mathcal{F}} +E_{R}\|\frac{1}{k}\sum_{i=1}^{n_{0}-1}Z_{R_{i}n}\|_{\mathcal{F}}\right\} \\
					&\leq
					\max_{n_{0}-1\leq k\leq n} 2E_{R}\|\frac{1}{k}\sum_{i=1}^{k}Z_{R_{i}n}\|_{\mathcal{F}} \\
					&=
					\max_{n_{0}-1\leq k\leq n} 2 \|\frac{1}{k}\sum_{i=1}^{k}Z_{{i}n}\|_{\mathcal{F}}.
				\end{align*}
				The equality comes from $R$ being independent of $Z_{in}$.
				Note that $\sup_{n\in\mathbb{N}}P_{n}(\|\frac{1}{k}\sum_{i=1}^{k}Z_{{i}n}\|_{\mathcal{F}}>\epsilon)\rightarrow 0$ as $k\rightarrow \infty$ since $\mathcal{F}$ is Glivenko-Cantelli uniformly in $\{P_{m}\}$.
				Hence, the second term in the right hand side of \eqref{eq:unif.boot.ulln} converges to zero in probability as $n_{0}\rightarrow\infty$.
				That is, for any $\epsilon>0$,
				\begin{equation*}
					\sup_{n\geq n_{0}}P_{n}\left(
					\left|n\|W_{1n}\|_{2,1}\max_{n_{0}\leq k\leq n} E_{R}\|\frac{1}{k}\sum_{i=n_{0}}^{k}Z_{R_{i}n}\|_{\mathcal{F}}\right| > \epsilon
					\right) \rightarrow 0\text{ as $n_{0}\rightarrow\infty$.}
				\end{equation*}
				
				Therefore, for any $\epsilon>0$,
				\begin{equation*}
					P_{n}\left(
					E_{W}\|\sum_{i=1}^{n}W_{in}Z_{in}\|_{\mathcal{F}} > \epsilon
					\right)\rightarrow 0\text{ as $n\rightarrow\infty$.}
				\end{equation*}
				By applying the Markov inequality as follows, we can complete the proof:
                \begin{equation*}
                    P_{n}\left(P_{W}\left(\|\sum_{i=1}^{n}W_{in}Z_{in}\|_{\mathcal{F}}>\epsilon\right)>\eta\right)
                    \leq P_{n}\left(E_{W}\|\sum_{i=1}^{n}W_{in}Z_{in}\|_{\mathcal{F}}>\eta\epsilon\right).
                \end{equation*}

			\end{proof}
			
			\begin{lemma}[Lemma 3.6.6 \cite{van_der_vaart_weak_1996}]
				\label{lem:VW3.6.6}
				For fixed elements $x_{1},...,x_{n}$ of a set $\mathcal{X}$, let $\widehat{X}_{1},...,\widehat{X}_{k}$ be an i.i.d. sample from $\mathbb{P}_{n}=n^{-1}\sum_{i=1}^{n}\delta_{x_{i}}$, where $\delta_{x_{i}}$ is a dirac measure at $x_{i}$.
				Then,
				\begin{equation*}
					E_{\widehat{X}}\|\sum_{j=1}^{k}(\delta_{\widehat{X}_{j}}-\mathbb{P}_{n})\|_{\mathcal{F}} \leq 4E_{N,N'} \|\sum_{i=1}^{n}(N_{i}-N_{i}')\delta_{x_{i}}\|_{\mathcal{F}}
				\end{equation*}
				for every class $\mathcal{F}$ of functions $f:\mathcal{X}\rightarrow\mathbb{R}$ and i.i.d. Poisson variables $N_{1},N_{1}',...,N_{n},N_{n}'$ with mean $\frac{1}{2}k/n$.
			\end{lemma}
			
			\begin{lemma}[Lemma 2.3.6 \cite{van_der_vaart_weak_1996}]
				\label{lem:VW2.3.6}
				Let $Z_{1},...,Z_{n}$ be independent stochastic processes with mean zero.
				Then,
				\begin{equation*}
					E\|\sum_{i=1}^{n}\varepsilon_{i}Z_{i}\|_{\mathcal{F}} \leq 2 E\|\sum_{i=1}^{n}Z_{i}\|_{\mathcal{F}}
				\end{equation*}
				for i.i.d. Rademacher random variables $\varepsilon_{1},...,\varepsilon_{n}$ and any functional class $\mathcal{F}$.
			\end{lemma}
			
			\begin{lemma}[Lemma 2.9.1 \cite{van_der_vaart_weak_1996}]
				\label{lem:VW2.9.1}
				Let $Z_{1},...,Z_{n}$ be i.i.d. stochastic processes with $E\|Z_{i}\|_{\mathcal{F}}<\infty$ independent of the Rademacher variables $\varepsilon_{1},...,\varepsilon_{n}$.
				Then, for every i.i.d. sample $\xi_{1},...,\xi_{n}$ of mean-zero and symmetrically distributed random variables independent of $Z_{1},...,Z_{n}$ and $1\leq n_{0}\leq n$,
				\begin{equation*}
					E\|\frac{1}{\sqrt{n}}\sum_{i=1}^{n}\xi_{i}Z_{i}\|_{\mathcal{F}}
					\leq (n_{0}-1) E\|Z_{1}\|_{\mathcal{F}} E_{\xi}\max_{1\leq i\leq n}\frac{|\xi_{i}|}{\sqrt{n}} \\
					+\|\xi_{1}\|_{2,1}\max_{n_{0}\leq k\leq n}E\|\frac{1}{\sqrt{k}}\sum_{i=n_{0}}^{k}\varepsilon_{i}Z_{i}\|_{\mathcal{F}},
				\end{equation*}
				where $\|\cdot\|_{2,1}$ is $L_{2,1}$ norm such that $\|\xi\|_{2,1}=\int_{0}^{\infty}\sqrt{P(|\xi|>x)}dx$ for a random variable $\xi$.
			\end{lemma}
			
			\begin{lemma}[Lemma 3.6.7 \cite{van_der_vaart_weak_1996}]
				\label{lem:VW3.6.7}
				For arbitrary stochastic processes $Z_{1},...,Z_{n}$, every exchangeable random vector $(\xi_{1},...,\xi_{n})$ that is independent of $Z_{1},...,Z_{n}$, and any $1\leq n_{0}\leq n$,
				\begin{equation*}
					E_{\xi}\|\frac{1}{\sqrt{n}}\sum_{i=1}^{n}\xi_{i}Z_{i}\|_{\mathcal{F}} \leq 2(n_{0}-1)\frac{1}{n}\sum_{i=1}^{n}\|Z_{i}\|_{\mathcal{F}}E_{\xi}\max_{1\leq i\leq n}\frac{|\xi_{i}|}{\sqrt{n}}+2\|\xi_{1}\|_{2,1}\max_{n_{0}\leq k\leq n}E_{R}\|\frac{1}{\sqrt{k}}\sum_{i=n_{0}}^{k}Z_{R_{i}}\|_{\mathcal{F}},
				\end{equation*}
				where $(R_{1},...,R_{n})$ is a random vector uniformly distributed on the set of all permutations of $\{1,...,n\}$ and independent of $Z_{1},...,Z_{n}$.
				$\|\cdot\|_{2,1}$ is $L_{2,1}$ norm such that $\|\xi\|_{2,1}=\int_{0}^{\infty}\sqrt{P(|\xi|>x)}dx$ for a random variable $\xi$.
			\end{lemma}
			
			\begin{lemma}[Lemma 3.6.15 \cite{van_der_vaart_weak_1996}]
				\label{lem:VW3.6.15}
				For each $n$, let $(a_{1n},...,a_{nn})$ and $(B_{1n},...,B_{nn})$ be a vector of numbers and exchangeable random vector such that
				\begin{gather*}
					\frac{1}{n}\sum_{i=1}^{n}(a_{in}-\bar{a}_{n})^{2}\rightarrow{\sigma^{2}},\quad \lim_{M\rightarrow\infty}\limsup_{n\rightarrow\infty}\frac{1}{n}\sum_{i=1}^{n}a_{in}^{2}\{|a_{in}|>M\} = 0, \\
					\frac{1}{n}\sum_{i=1}^{n}(B_{in}-\bar{B}_{n})^{2}\xrightarrow{p}\alpha^{2},\quad
					\frac{1}{n}\max_{1\leq i \leq n}(B_{in}-\bar{B}_{n})^{2}\xrightarrow{p}0,
				\end{gather*}
				where $\bar{a}_{n}=\frac{1}{n}\sum_{i=1}^{n}a_{in}$ and $\bar{B}_{n}=\frac{1}{n}\sum_{i=1}^{n}B_{in}$.
				Then, $n^{-1/2}\sum_{i=1}^{n}(a_{in}B_{in}-\bar{a}_{n}\bar{B}_{n})\xrightarrow{d}N(0,\alpha^{2}\sigma^{2})$.
			\end{lemma}
			
			Let $B=(B_{1n},...,B_{nn})$ be a multinomial vector with parameters $n$ and probabilities $(1/n,...,1/n)$.
			Then, $\bar{B}_{n}=1$, and conditions for $B$ in \cref{lem:VW3.6.15} hold.
			
		\end{appendices}

	\end{document}